\documentclass[final,3p,times]{elsarticle}
\journal{Physics Reports}

\usepackage{graphicx,epsfig}
\usepackage{a4,amsmath,bm,amssymb}
\usepackage{amsthm,epsf,psfrag,verbatim}
\usepackage{fancyhdr,sectsty,titlesec}
\usepackage{array}
\usepackage{cancel}
\usepackage{soul}
\usepackage[normalem]{ulem}
\usepackage{fancybox}
\usepackage[dvips,pdftex]{color}
\usepackage{color}
\usepackage[english]{babel}
\usepackage{underscore}
\usepackage[utf8]{inputenc}
\usepackage[english]{babel}

\newtheorem{mydef}{Definition}
\newcommand{\ee}{\end{equation}}
\newcommand{\be}{\begin{equation}}
\biboptions{sort&compress}

\def\la{\langle}
\def\ra{\rangle}
\def\ur{\hat {\bf r}}

\def\hkpm{{\bf h}^{s_k}_\bk}
\def\hppm{{\bf h}^{s_p}_\bp}
\def\hqpm{{\bf h}^{s_q}_\bq}

\def\bk{{\bf k}}
\def\Dk{\Delta k}
\newcommand{\bp}{{\bf p}}

\def\bbf{{\bf f}}
\def\br{{\bf r}}
\def\bq{{\bf q}}
\def\bu{{\bf u}}
\def\bbt{\tilde{\bf b}} 
\def\but{\tilde{\bf u}} 
\def\ut{\tilde{ u}} 
\def\utpk{\tilde{ u}^+_{\bf k}(t)} 
\def\utmk{\tilde{ u}^-_{\bf k}(t)} 
\def\utpmk{\tilde{ u}^\pm_{\bf k}(t)} 
\def\utpmp{\tilde{ u}^\pm_{\bf p}(t)} 
\def\utpmq{\tilde{ u}^\pm_{\bf q}(t)} 
\def\bgtk{\tilde{\bf g}_{\bf k}(t)} 

\def\bx{{\bf x}}
\def\bX{{\bf X}}
\def\bB{{\bf B}}
\def\bb{{\bf b}}
\def\ba{{\bf a}}
\def\btau{{\bm \tau}}
\def\bnabla{{\bm \nabla}}
\def\bz{{\bf z}}
\def\bw{{\bm w}}
\def\hm{{\bf h}^-_{\bf k}}
\def\hp{{\bf h}^+_{\bf k}}
\def\hpm{{\bf h}^\pm_{\bf k}}
\def\hmp{{\bf h}^\mp_{\bf k}}
\def\bg{{\bf g}}
\def\Red{{Re}}
\def\Rhd{{R_\alpha}}
\def\Reu{{Re}}
\def\Sm{{Sm}}
\def\Rhu{{R_\alpha}}
\def\lin{{\ell_{in}}}

\def\tnl{{\tau_{nonlin}}}

\def\twv{{\tau_{wave}}}

\def\iepsilon{\epsilon}       
\def\sepsilon{\varepsilon}    
\def\ein{{ \iepsilon_{in}}}
\def\enu{{ \iepsilon_{\nu}}}
\def\emu{{ \iepsilon_{\alpha}}}
\def\vein{{\iepsilon}_{in}}
\def\eF{{{ \iepsilon}_{_F}^{<k}}}
\def\eUV{{{\iepsilon}_\nu^{<k}}}
\def\eIR{{{\iepsilon}_\alpha^{<k}}}
\def\bein{{\iepsilon}_{in}}
\def\benu{{\iepsilon}_{\nu}}

\def\oein{{\iepsilon}_{in}}
\def\oenu{{\iepsilon}_{\nu}}
\def\oemu{{\iepsilon}_{\alpha}}

\def\thetain{{\chi_{in}}}
\def\thetachi{{\chi_{\kappa}}}
\def\thetaalpha{{\chi_{\alpha}}}

\def\bfo{{\bf f}}

\def\hin{{\mathit{h}_{in}}}
\def\hnu{{\mathit{h}_{\nu}}}
\def\halpha{{\mathit{h}_{\alpha}}}

\def\wein{{\zeta_{in}}}
\def\wenu{{\zeta_{\nu}}}
\def\wemu{{\zeta_{\alpha}}}
\def\bwein{{\zeta}_{in}}
\def\bwenu{{\zeta}_{\nu}}
\def\bwemu{{\zeta}_{\alpha}}
\def\kin{{k_{in}}}
\def\knu{{k_{\nu}}}
\def\kmu{{k_{\alpha}}}
\def\lnu{{\ell_{\nu}}}

\def\curl{{\nabla \times}}
\def\kmax{k_{max}}
\def\cE{{\mathcal E}}
\def\cH{{\mathcal H}}
\def\cZ{{\mathcal Z}}
\def\cA{{\mathcal A}}
\def\cB{{\mathcal B}}
\def\cM{{\mathcal M}}
\newcommand{\hr}{{\hat {\bf r}}}
\newcommand{\obv}{\hat{{\bf u}}^\ell}
\newcommand{\obu}{\hat{{\bf u}}^\ell}

\newcommand{\ou}{\hat{ u}^\ell}
\newcommand{\ov}{\hat{ u}^\ell}

\newcommand{\oP}{ P^\ell}
\newcommand{\oS}{\hat{{\bf S}}^\ell}
\newcommand{\by}{{\bf y}}
\def\p{\partial}
\def\bnabla{{\bm \nabla}}

\begin{document}

\begin{frontmatter}



  \title{Cascades and transitions in turbulent flows\footnote{Post print version of the article published on Physics Reports 2018. \\
      https://doi.org/10.1016/j.physrep.2018.08.001}}


\author{A. Alexakis$^a$, L. Biferale$^b$}

\address{$^a$  Laboratoire de  physique  statistique,  D{\'e}partement  de  physique  de  l'
{\'E}cole  normale  sup{\'e}rieure,  PSL Research University, Universit{\'e} Paris Diderot, Sorbonne Paris Cit{\'e}, 
Sorbonne Universit{\'e}s, UPMC Univ. Paris 06, CNRS, 75005 Paris, France}
\address{$^b$  Department of Physics and INFN University of Rome 'Tor Vergata', Via della Ricerca Scientifica 1, 00133 Rome, Italy}
\date{\today}

\begin{abstract}
    Turbulent flows are characterized by the non-linear cascades of energy and other inviscid invariants across a huge range of scales, from where they are injected to where they are dissipated. Recently, new experimental, numerical and theoretical works have revealed that many turbulent configurations deviate from the ideal three and two dimensional homogeneous and isotropic cases  characterized by the presence of  a strictly direct and  inverse energy cascade, respectively. New phenomena appear that alter the global and local transfer  properties. In this review,  we provide a critical summary of historical and  recent works from a unified point of view and we present a classification of all known  transfer mechanisms. Beside the classical cases of direct and inverse energy cascades,  the different scenarios include:  split cascades for which an invariant flows both to small and large scales simultaneously, multiple/dual cascades of different quantities, bi-directional cascades where direct and inverse transfers of the same invariant coexist in  the same scale-range and finally equilibrium states where no cascades are present, including the case when a large scale condensate is formed. We classify all possible transitions from one scenario to another as the control parameters are changed and we analyse when and why different configurations are observed. Our discussion is based on a set of paradigmatic applications: helical turbulence, rotating and/or stratified flows, magnetohydrodynamics (MHD) turbulence, and passive/active scalars where the transfer properties are altered as one changes the embedding dimensions, the thickness of the domain or other relevant control parameters, as, e.g.,  the Reynolds, Rossby, Froude, P\'eclet, or Alfv\'en numbers. We briefly discuss the presence of anomalous scaling laws in 3D hydrodynamics and in other configurations, in connection with the intermittent nature of the energy dissipation in  configuration space. A quick overview is also provided concerning the importance of cascades in other applications such as  bounded flows, quantum fluids, relativistic and compressible turbulence, and active matter, together with a discussion of the implications for turbulent modelling. Finally, we present a series of  open problems and challenges that future work needs to address. \\

\end{abstract}

\begin{keyword}
Homogeneous and isotropic turbulence, two dimensional turbulence, wave turbulence, rotating flows, thick layers, MHD, turbulent diffusion, passive and active scalars, stratified flows, helicity, convection, Lagrangian turbulence, Richardson cascade, inverse energy cascade, direct energy cascade, intermittency, anomalous scaling laws, energy condensate, absolute equilibrium.
\end{keyword}

\end{frontmatter}

\newpage
\tableofcontents
\newpage








\section{Introduction}


{\it ``Big whorls have little whorls that feed on their velocity, and little whorls have smaller whorls and so on to viscosity - in the molecular sense''}. This is the celebrated poem composed by Lewis Fry Richardson in 1922 \cite{Richardson}, where it was proposed  that turbulent cascades are the fundamental driving mechanism of the atmosphere, moving energy from the large injections scales down to the dissipative small scales.  It is a visionary way to summarize many fundamental aspects of the turbulent energy transfer which is empirically observed in the three dimensional Navier-Stokes equations (NSE).  The Richardson cascade  description, first quantified  by A.N. Kolmogorov   \cite{Kolmogorov},  constitutes the most fundamental concept of turbulence theory. In more words, in a turbulent flow, the energy externally injected is redistributed among length scales due to non-linear eddy (whorls) interactions. If this energy is removed from the flow (or accumulated) at a scale $\ell_{out}$ significantly different than the injection scale,  $\lin$, a cascade can build up with a continuous transfer of energy  from $\ell_{in}$ to $\ell_{out}$. Depending on the system, the energy transfer can be towards the small scales and/or towards  the large scales, leading to what it is referred to as a forward or inverse cascade, respectively. 
In three dimensional (3D) homogeneous and isotropic  turbulence  (HIT) the  energy is transferred to the small scales while in two dimensions (2D)  it  cascades to the large scales.
Despite the simplicity of the cascade description, after almost 100  years, we are still fighting  to define the exact terms of the game and fail to have a complete statistical description even for the simplest case of HIT.
When and why Richardson's cascade is correct? 
What happens when it fails? 
These and many other questions have challenged mathematicians, physicists and engineers for more than a century without reaching, to this date, clear answers, rightfully titling turbulence as {\it the last open problem of classical physics}.
There are various text books written over the years reviewing both findings and open questions of different aspects of turbulence \cite{Frisch,davidson2015turbulence,pope2001turbulent}.
A recent overview can be found in \cite{davidson2011voyage}, where the historical developments in engineering,  mathematical, and physical sciences have been analysed from the "shoulders of twelve historical fathers of turbulence research".  \\

%
The situation, however, can be very complex. 
First, any inviscid invariant of the system is also subject to a non-linear transfer.  The interaction of the transfer of different quantities plays an important role in determining the direction of their cascade. For example, it is the conservation of enstrophy that forces an inverse cascade of energy in 2D. In 3D, the second invariant is helicity and it is not sign-definite and all experimental investigations, numerical simulations and phenomenological theories indicate that both energy and helicity have a simultaneous mean transfer to the small scales. However, this is an empirical observation and it is  not proven from basic principles yet.
The dual cascade of helicity and energy is important for 3D hydrodynamical turbulence \cite{brissaud1973, moffatt1992helicity, chen2003joint, chen2003intermittency}. In the latter case,  recent studies have  shown that even in the idealized case of 3D HIT, there exists a bi-directional transfer of energy with some helical-Fourier channels that cascade energy forward  and others that transfer energy backward  \cite{biferale2012inverse, biferale2013split, biferale2013global, sahoo2015role, kessar2015non, stepanov2015hindered, sahoo2015disentangling, sahoo2017helicity, alexakis2017Helically, rathmann2017pseudo, sahoo2017discontinuous}, with potential applications to rotating turbulence also. Multiple transfers of competing invariants also occurs  in other flow configurations. A paradigmatic example is given by  MHD flows that conserve three invariants in 3D \cite{woltjer1958stability, chandrasekhar1958force, montgomery1982two,brandenburg2001inverse, Alexakis2006mhelicity, Malapaka2013mhelicity, linkmann2016helical, linkmann2017effects, linkmann2016large, linkmann2017triad}.  {\it In fact, it is fair to say that except for some very idealized situations, we cannot predict the direction of the energy transfer in homogeneous turbulence}. \\

\begin{figure*}[htbp]                                                                                               
\centering                                                                                                          
\includegraphics*[width=0.45\textwidth]{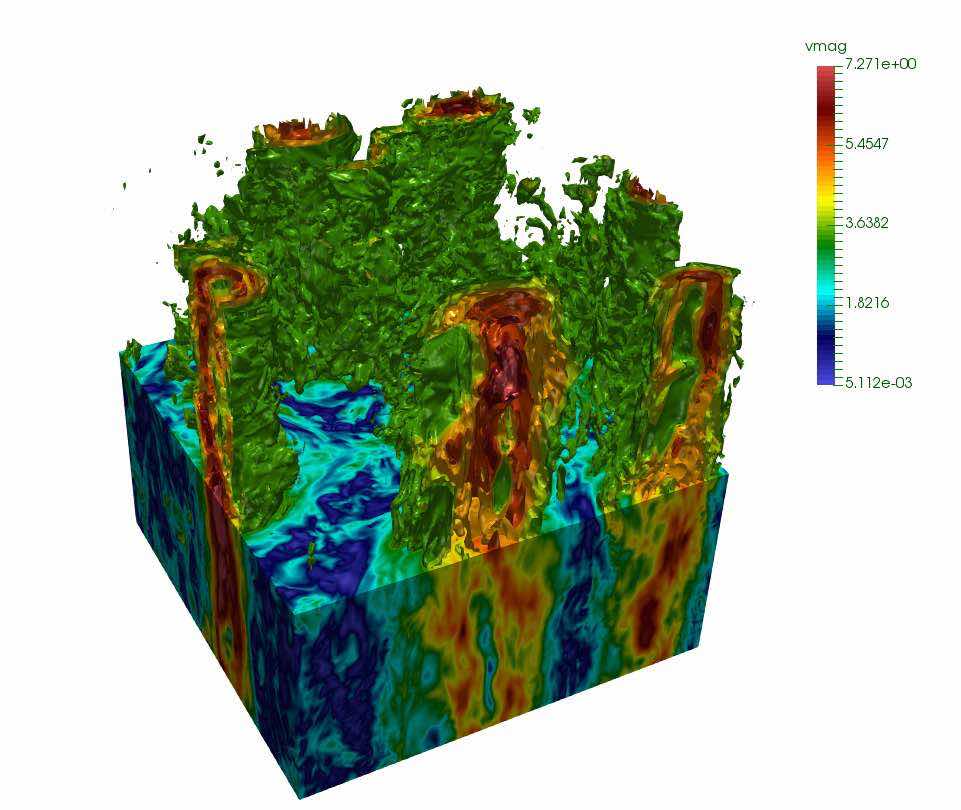}                                                        
\includegraphics*[width=0.40\textwidth,height=0.40\textwidth]{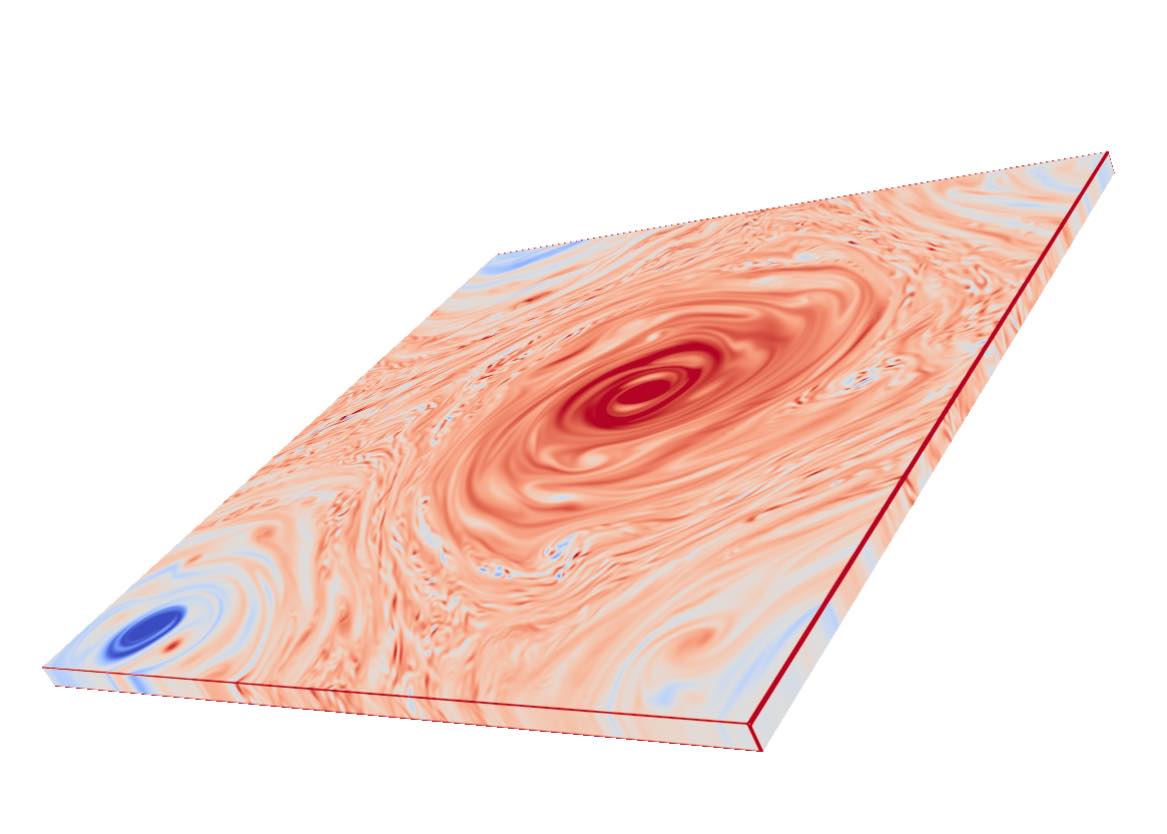}                                 
\caption{Left panel: a 3D representation of velocity amplitude in a fast rotating flow                              
(adapted from \cite{biferale2016coherent}). Right panel: Vorticity  of a flow in a thin layer.}                     
\label{fig:rotation}                                                                                                
\end{figure*}

Second, there exist many important turbulent configurations that  deviate  from the idealized situation of HIT. e.g., in the presence of external mechanisms such as rotation, stratification, confinement, shear, or magnetic fields where the direction of the energy cascade might  -and indeed it does-  change. In many of these systems energy is transferred with a split-cascade, i.e.  simultaneously forward and inversely in fractions that depend on the value of a control parameter (rotation rate, magnetic field strength, aspect ratio etc). This is demonstrated in Fig. (\ref{fig:rotation}) for two paradigmatic examples of a fast rotating flow (left panel) and flow constrained in a thin layer (right panel), where structures at both large and small scales coexist. 
Split cascades have been shown to exist in different physical situations, in numerical simulations and experiments 
of thin/thick layers \cite{Celani2010turbulence,benavides2017critical,Shats2010turbulence,Xia2011upscale,francois2013inverse}, 
in rotating and stratified turbulence \cite{Deusebio2014dimentional, Marino2013invers, aluie2011joint, pouquet2013geophysical, rorai2013helicity, marino2014large, rorai2014turbulence, Sozza2015dimensional, Rosenberg2015evidence, rorai2015stably, Marino2015resolving, herbert2016waves, Staplehurst2008, Bokhoven2009experiments, Yoshimatsu2011, duran2013turbulence, Machicoane2016Two, yeung1998numerical, Smith1999transfer, godeferd1999direct, chen2005resonant, Thiele2009, mininni2009, Mininni2010Rotating, favier2010space, Sen2012anisotropy, alexakis2015rotatingTG, biferale2016coherent,valente2017spectral}
and  in MHD  turbulence \cite{Alexakis2011two,sujovolsky2016tridimensional, Seshasayanan2014edge, Seshasayanan2016critical, sundar2017dynamic, favier2010two, reddy2014anisotropic, reddy2014strong, baker2018inverse, baker2017controlling, potherat2014why}.
They have been observed in geophysical flows, e.g. where  the atmosphere acts like a 2D flow at large scale and  as a 3D  flow at small scales \cite{nastrom1984kinetic, nastrom1985climatology, gage1986theoretical, charney1971geostrophic, Byrne2011robust, Byrne2013height, brunner2014upscale, tang2015horizontal, callies2014transition} and in the ocean \citep{arbic2013eddy,king2015upscale}. Similar behaviour has been attributed to astrophysical flows (like the atmosphere of Venus and Jupiter  \cite{Izakov2013Venus,young2017forward}, and accretion discs \cite{Lesur2011accretion}), in plasma flows \cite{miloshevich2018direction} and in  industrial applications (like  in tokamak \cite{Diamond2005plasma}) either due to the thinness of the layer, to fast rotation or to the presence of strong magnetic fields. Split cascades have also been observed in wave systems \cite{lvov2015formation}, multi-scale optical turbulence \cite{malkin2018transition}, acoustic turbulence \cite{ganshin2008observation} and capillary turbulence on the surface of liquid hydrogen and helium \cite{abdurakhimov2015bidirectional,abdurahimov2015formation}. In many of  these systems, a  change from a split to a strict forward cascade has also been detected with a critical transition  behaviour. \\


Furthermore in many situations there is a vanishing, or very weak, net flux at scales larger than the forcing. In some cases like in 3D turbulence this leads to an equipartition of energy among the large scale modes  \cite{kraichnan1973helical, cichowlas2005effective, krstulovic2009cascades, ray2015thermalized, michel2017observation, herbert2014restrictedp, herbert2014restrictede, herbert2014nonlinear, zhu2014note, zhu2014purely, ditlevsen1996cascades,Dallas2015statistical,cameron2017effect,cameron2016large}. In other cases, for finite box size,  the presence of inverse transfers  might lead to  the formation of a large-scale condensate, with a strong feedback on the whole flow \cite{smith1993bose, xia2009spectrally, gallet2013two, laurie2014universal, farrell2007structure,bouchet2008simpler, venaille2009statistical, venaille2011solvable, bouchet2012statistical,  falkovich2016interaction, woillez2017theoretical, frishman2017culmination, frishman2017jets}.  Similarly, ultraviolet effects might produce  an accumulation of energy (or of another inviscid invariant) at small scales. The coexistence in the same scale range of different  channels can also drive the system  to an {\it exotic} flux-loop state,  with a perfect balance among direct and inverse transfer, zero flux and non-equilibrium properties,  as for the case of stratified 2D turbulence \cite{boffetta2011flux}. In some of these zero-flux cases the flow is close to a quasi-equilibrium state while others remain strongly out of equilibrium.   \\

Finally, deviations from a perfect self-similar cascade are known to exist even in HIT. For example,  we know that kinetic energy in 3D  tends to be dissipated in spiky and  intermittent events, and we do not know if this is due to the presence of coherent structures or just because of an enhancement of statistical fluctuations  \cite{Frisch}.  On the contrary, the 2D inverse energy cascade is close to be Gaussian, without any intermittent properties. We only have a loose phenomenological understanding of why fluctuations grow for the 3D case, based on the Richardson's idea and of multi-fractal processes \cite{Frisch}, but we do not control the connection with the equations of motion. Intermittency, anomalous scaling, multi-fractal energy dissipation are still the subject of many investigations (see \cite{sinhuber2017dissipative,iyer2017reynolds} for recent experimental and numerical state-of-the-art results).

The present work  attempts to address all the above issues by reviewing some recent numerical, experimental and theoretical advancements achieved in the field. 
Our goal is to clarify and categorize the long list of subtle scenarios that might occur in turbulence and view different turbulence systems from a unified point of view.
We would like thus to distinguish  between turbulent configurations with the same mean properties but different spectra (e.g. with and without large scale condensate), the same spectral behaviour but with different transfer directions (e.g. with direct or inverse cascades) and  even the same transfer properties but with a different statistical ensemble (e.g. a zero flux system in a equilibrium or out-of-equilibrium state). In particular, we want to emphasize the key distinguishable properties that characterize a non-linear out-of-equilibrium system, by looking at the behaviour of single-point global quantities, e.g.   energy or enstrophy, or two-points spectral properties and transfer terms. 

To do that, we start in Sec. (\ref{sec:theoretical}) with the basic theoretical set-up. In sec. (\ref{sec:directinverse}) we give a short review of the classical idealized turbulence cascades in three and two dimensions and proceed in the same section by providing a series of precise definitions for different flow states based on the cascade properties. In many cases, the same system can display a (phase) transition among two or more of the above flow states. We thus also classify the different paths that such transitions can occur by changing the dimensionless control parameters, e.g. the Reynolds number for hydrodynamic turbulence or the Rossby, Froude and Alfv\'en numbers for rotating, stratified and conducting flows. \\

In Sec. (\ref{sec:applications}) we examine  different systems in detail. We follow a simple-to-complex path, discussing how turbulence  changes when additional ingredients are added into the system, by breaking certain symmetries, e.g. mirror, rotational and scale invariance,  by adding confinement, by changing the non-linearities, or by coupling the flow to active components. We review empirical findings for a set of paradigmatic applications where the above cascade realizations are encountered. The examples vary from  helical turbulence, turbulence in confined domains,  turbulence under rotation, stable and unstable stratified turbulence, turbulence of conducting fluids (MHD) to  passive and active scalar advection, just to cite the most important cases.  We review how these different realizations  can drastically modify the properties of turbulence by altering the conserved quantities,  their cascade directions and what  the consequences for large and small scales flow properties are. \\

Finally, in Sec. (\ref{sec:further}) we present a short overview of cascades in turbulent models and in other  flow configurations, as for the case of quantum flows, wave-turbulence, bounded flows, relativistic and compressible flows, active fluids, Shell Models, Large Eddy Simulations, EDQNM approximation and  elastic materials. In Sec. (\ref{sec:conclusions})  we conclude with a series of open questions and challenges in the field with the hope that these issues will attract the interest of future research.\\ 

  All applications are cherry-picked with the only aim to highlight the aspects related to the cascades and their transition, without hoping to be self-contained and exhaustive for each particular subject. We tried to be as precise as possible concerning the classification of different cascade scenarios and we kept an empirical mood whenever we need to report about numerical and experimental findings, due to the absence of rigorous results for most of the cases treated here.

\section{ {Theoretical Setup}    \label{sec:theoretical} }       
In this section we provide a short theoretical background and the notation that will be used  throughout the review. We discuss the  balance of globally inviscid conserved quantities in the configuration and Fourier space. We introduce the concept of direct and inverse inertial-ranges, defined as those interval of  scales where the main
driving mechanisms are given by the non-linear transfer of the cascading quantities. We address also the scale-by-scale energy budget for the two-point correlation function in configuration space, for the energy spectrum in Fourier space and for the mixed scale-filtered representation often used for small-scale modelling.
\subsection{{Dynamical equations and control parameters}} 
%
To start the discussion we consider the incompressible Navier-Stokes equation in the presence of a large-scale drag term:
\be
\label{eq:GNS}
\partial_t {\bf u} + {\bu \cdot \bnabla \bu} = -\bnabla P + \nu  \Delta {\bf u} -  \alpha  \bu + \bbf.
\ee
Here ${\bf u}$ is the divergence free  velocity field, $(\bnabla \cdot {\bf u}=0)$, with constant unit density, $P$ is the pressure per unit density that enforces incompressibility, $\nu$ is the kinematic  viscosity, ${\bf f}$ and  $\alpha$ are the forcing term and  the coefficient of a large scale drag mechanisms, respectively. 
In some systems, the large scale drag is required to reach a steady state in the presence of an inverse cascade. 
It simplifies the theoretical discussion that follows, avoiding   the need to discuss quasi-stationary states.
For most of this review  we are going to assume the flow to be confined in a periodic box of size $L$, and that the external forcing is acting  on a band limited range of scales centered around  $\ell_{in}$. The energy injection rate by the forcing will be denoted as $\ein$.  In some cases, it is useful to study the flow evolution when the Newtonian  viscosity is replaced with a hyper-viscous term, $ -\nu_n (-\Delta)^n {\bf u} $ and/or when the drag force is replaced with a  hypo-viscous sink, $\alpha_m  (-\Delta)^{-m} {\bu} $, such as to  confine the energy dissipation due to viscous effects to very small scales (by increasing $n$ and decreasing $\nu_n$) and/or the one due to the drag mechanisms to very large scales  (by increasing $m$ and decreasing $\alpha_m$). There are three non-dimensional control parameters in the system:   the ratio of the box size to the forcing length scale, $L/\lin$, and two numbers that measure the relative amplitude of the non-linearities compared to the two dissipation terms defined above. The first one is the Reynolds number, $Re$, which compares advection with the viscous dissipation. The second one is the equivalent of $Re$  but for the large scale drag, $\Rhu$,  obtained by replacing the viscous term  with the drag term
\be
\Reu= \frac{ u_f \lin  }{\nu},
\quad
\Rhu = \frac{ u_f }{\alpha \lin}
\label{eq:Re}
\ee
where $u_f$ is the root-mean-square velocity measured at the injection scale. 
In many experimental and numerical set-ups, it is not $u_f$ that is controlled but rather the energy injection rate. In  these cases it is useful to define the dimensionless numbers in terms of $\ein$:  
\be
\Red = \frac{ \ein^{1/3} \lin^{4/3}  }{\nu},
\quad
\Rhd = \frac{ \ein^{1/3} }{\alpha \lin^{2/3}}.
\label{eq:Red}
\ee
The two definitions (\ref{eq:Re}-\ref{eq:Red})  become equivalent up to a multiplicative constant  for fully developed turbulence
where the relation $\ein \propto u_f^3/\ell$ holds \cite{Frisch}.  There are other cases, e.g. for wave turbulence,  where the two definitions do not agree (see Sec. \ref{sec:WWT}).  

\subsection{{Inviscid Invariants and balance equations} } 
It is straightforward to show that in 3D for $\nu=\alpha={\bf f}=0$ 
(and if the flow remains smooth) equation (\ref{eq:GNS}) has two global invariants, the total Energy $\cE$ and total Helicity $\cH$:
\be
\label{eq:inviscid3D}
\cE(t) = \frac{1}{2} \langle \bu \cdot \bu \rangle   \,; \; \qquad   \cH(t) = \frac{1}{2} \langle \bu \cdot \bw \rangle
\ee
where $\bw=\curl \bu$ is the vorticity. The angular brackets stand for spatial average that in $d$ dimensions is defined as
\be
\label{eq:avg}
\langle g(t) \rangle \equiv \frac{1}{ L^d}\int_{L^d} g(\bx,t)  dx^d.
\ee
In $2D$, helicity is identically zero and  the second  quadratic  invariant is the enstrophy:
\be
\label{eq:inviscid2D}
   \cZ(t) = \frac{1}{2} \langle \bw \cdot \bw \rangle 
\ee
which like the energy, and unlike helicity, it is positive definite. \\
In the presence of forcing and for finite values of $\alpha$ and $\nu$  a balance is 
reached where the injected energy is absorbed by the two dissipation terms.
By writing the evolution equation for the energy
(\ref{eq:inviscid3D}) we get:
\be
\label{eq:globalbalance}
 \partial_t \cE(t) =  - \enu(t) -  \emu(t) + \ein(t) 
\ee 
where $\enu(t) = \nu \langle |\nabla \bu|^2 \rangle$ is the energy dissipation due to viscosity, 
$\emu(t) = \alpha \langle |\bu|^2 \rangle$ is the energy dissipation due to the drag and $\ein(t)  =  \langle \bu \cdot {\bf f}   \rangle$ is the energy injection rate. 
It is worth noting that the terms $\enu(t),\emu(t)$ are strictly positive for any non zero $\bu(t)$ while in principle 
$\ein(t)$ can fluctuate taking both positive and negative values. Assuming stationarity and  performing a long time average we obtain
\be
\label{eq:globalbalance2}
 \ein =  \enu +  \emu.  
\ee
Where above and from hereafter a long-time average is implied  whenever  time does not explicitly appear. 
The interpretation of eq. (\ref{eq:globalbalance2}) is clear: the energy injected by the forcing equals, in average,  the energy
dissipated by viscosity at the small scales plus the energy dissipated by the drag  at the large scales. A similar balance holds for all the other
invariants of the system.

\subsection{{Fourier space representation}}                                                     
In order to better disentangle the scale-by-scale dynamics it is useful to define the NSE in  Fourier space.
We thus decompose the velocity field in Fourier modes $\but(\bk,t)$ as:
\be
\label{eq:fourier}
\but(\bk,t) = \frac{(2\pi)^d}{L^d} \int \bu(\bx,t) e^{-i \bk \cdot \bx} dx^d \quad \mathrm{with} \quad
\bu(\bx,t)  = \sum_{\bk} \but(\bk,t) e^{i \bk \cdot \bx }
\ee
where $\bk = k_0 {\bf n}$, ${\bf n} \in \mathbb{Z}^3$ and  $k_0=2\pi/L$ being the smallest wavenumber in the system (in what follows we will always assume that there is no mean flow, 
$\but(\bk,t)=0$ for $\bk=0$). 
The Navier-Stokes equation in Fourier space can then be written as 
\be
\label{NSF}
\partial_t \tilde{u}_i (\bk, t)  = -
i \sum_{\bf p+q=k}  P_{ij}(\bk)   q_l \tilde{u}_l ({\bf p}, t) \tilde{u}_j ({\bf q}, t)
- \nu k^2  \tilde{u}_i (\bk, t) - \alpha    \tilde{u}_i (\bk,t)
+ \tilde{f}_i(\bk,t),
\ee
where we have defined the projector on incompressible fields, $P_{ij}(\bk)= \left(\delta_{ij} - \frac{k_i k_j}{k^2}  \right)$,
and $k = |\bk|$. Hereafter,  we will always make use
of the Einstein notation for summation over repeated indices, unless otherwise stated. 
The components of each Fourier mode  are linked by the incompressibility condition,
$\bk \cdot \but=0$, that reduces the degrees of freedom by one. In 2D we can  write these amplitudes in terms of a single stream function mode.  
In 3D  we have the freedom to choose a basis of two eigenvectors for each  $\bk$.
One possible option will be discussed in terms of helical components in section (\ref{sec:Helicity})

\begin{figure*}[htbp]                                                                     %
\centering                                                                                %
\includegraphics*[width=0.45\textwidth, angle=0]{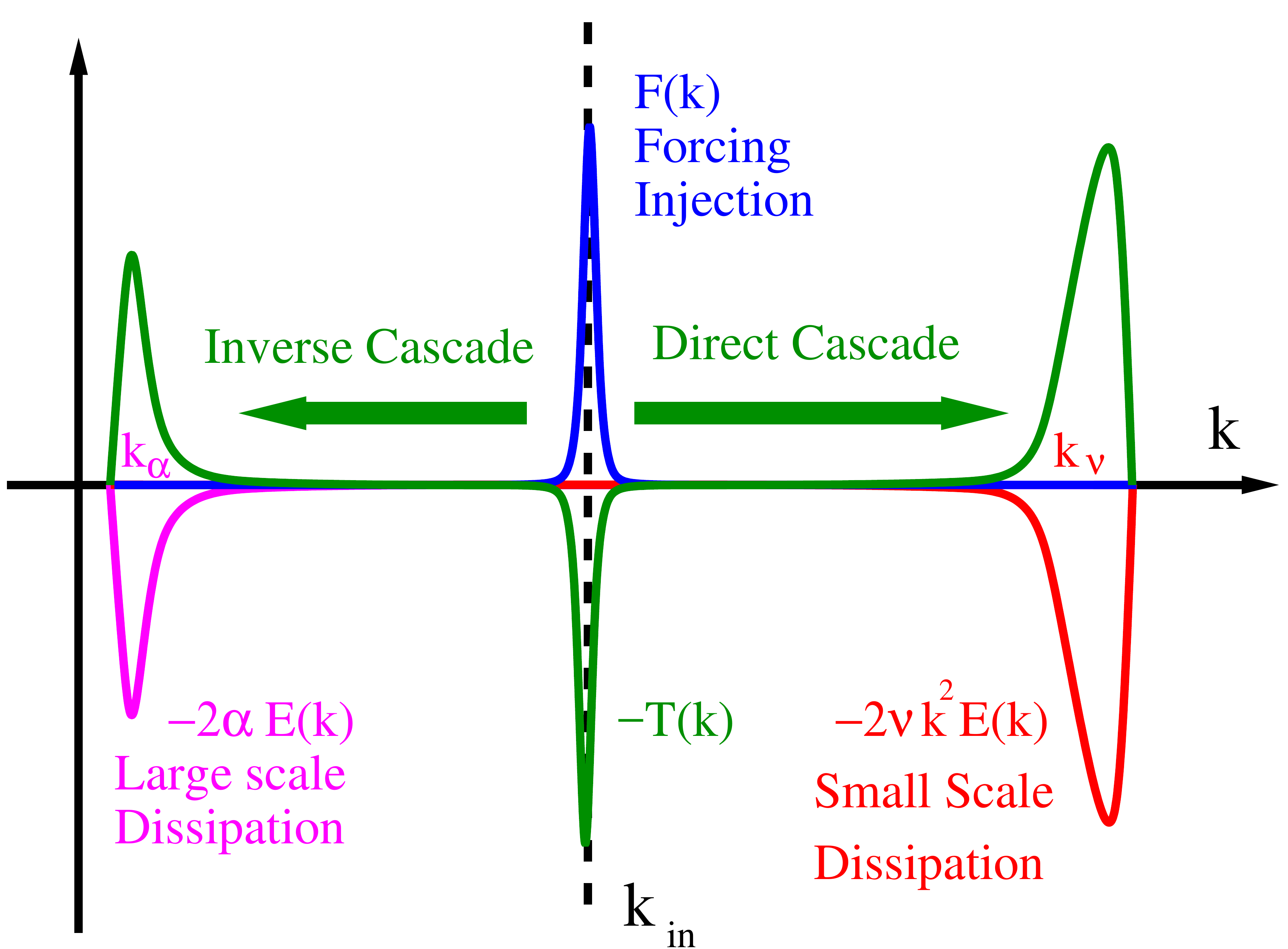}                        %
\includegraphics*[width=0.45\textwidth, angle=0]{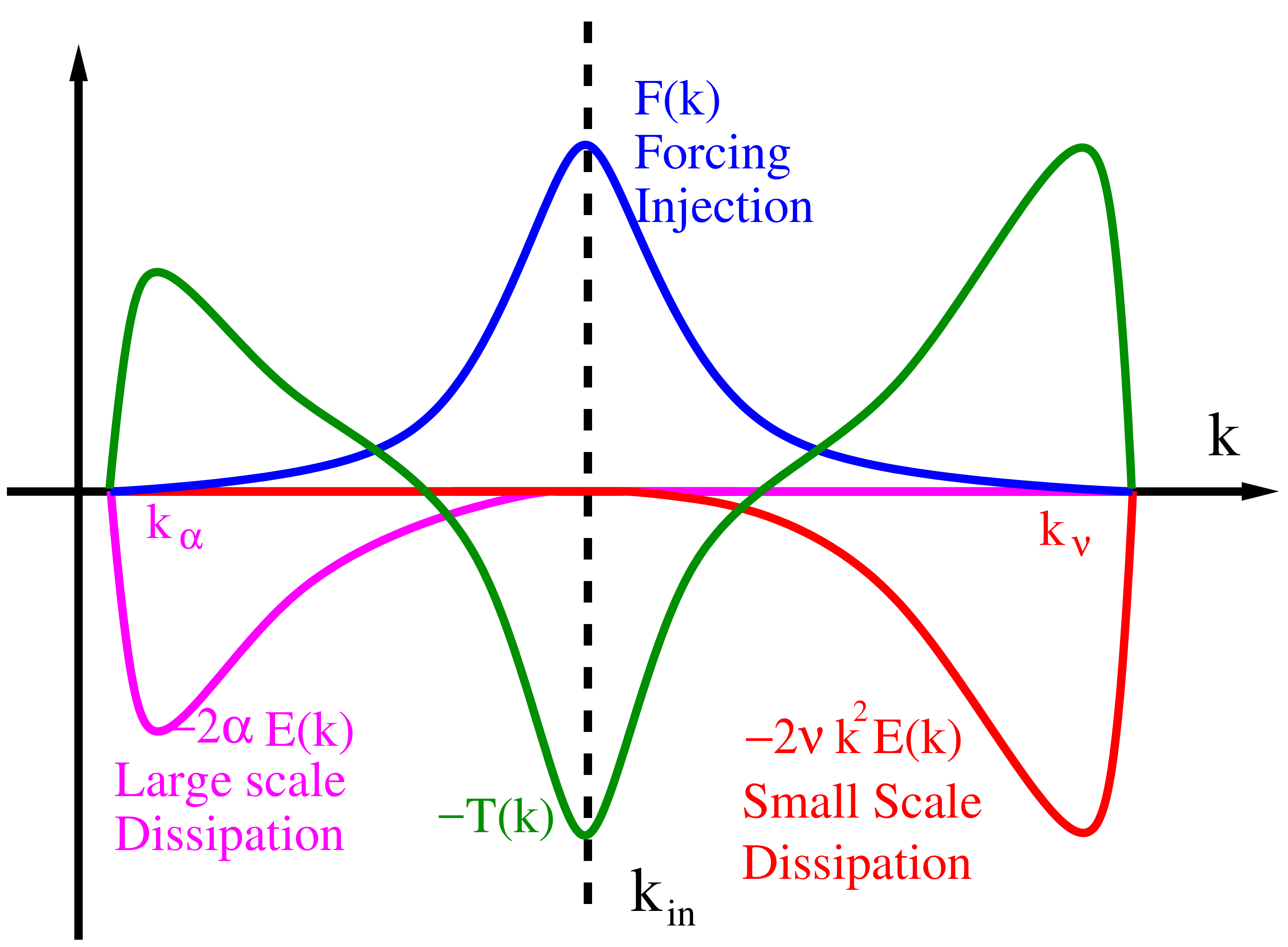}                        %
\caption{                                                                                 %
  Left: qualitative sketch of the stationary scale-by-scale energy balance in             %
the presence of an injection term confined to a band-limited range  around $\kin$, and    %
of two well separated scales, $k_\alpha \ll \kin \ll k_\nu$ fixing the onset of infra-red  %
drag terms and of ultraviolet viscous dissipation, respectively.  Right:  the same of     %
left panel but in the absence of scale separation, as in the presence of a condensate or of     %
a small-scale bottleneck as discussed in Sec. (\ref{sec:statequil}). In both cases the   %
balance  (\ref{NSFspectrumstationary})  is still exactly satisfied. The presence of a     %
cascade requires that there exists a range of scales where all terms on the RHS of        %
(\ref{NSFspectrumstationary}) are vanishingly small. }\label{fig:transfer}                %
\end{figure*}

\begin{figure*}[htbp]                                                                
\centering                                                                           
\includegraphics*[width=0.65\textwidth, angle=0]{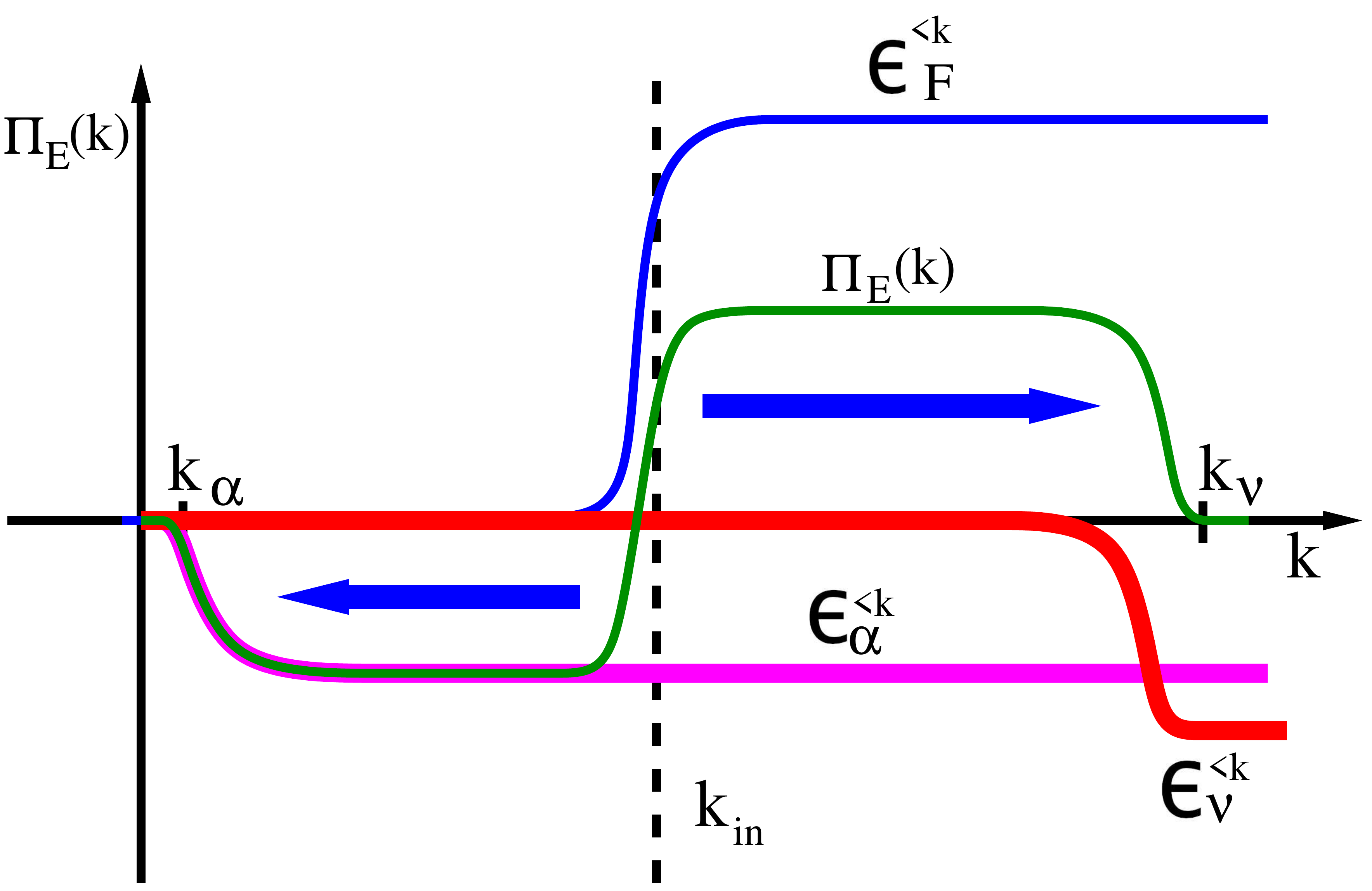}                   
\caption{
Qualitative sketch of the stationary scale-by-scale energy balance for the energy    
flux, $\Pi_E(k)$. Notice that the assumption of scale separation,                    
$ k_\alpha \ll \kin \ll k_\nu$ predicts the existence of two {\it inertial ranges}   
where the energy flux is constant  and due only to the non-linear triadic            
interactions. }\label{fig:flux}                                                      
\end{figure*}

The distribution of energy among scales is given by the  spectrum $E(k,t)$ averaged over a spherical shell of width $\Dk = 2\pi/L$:
\be
\label{eq:spectrum}
E(k,t) = \frac{1}{2\Dk} \sum_{k \le |\bk| < k +\Dk}  |\but(\bk,t)|^2 \, .
\ee
Summing over all shells  we obtain the total energy $\cE(t)= \Dk \sum_k E(k,t)$. 
From (\ref{NSF}) we   can also derive the evolution of the  energy spectrum as: 
\be
\label{NSFspectrum}
\partial_t  E(k,t) = -T(k,t) 
 - 2 \nu  k^2 E(k,t)  - 2 \alpha  E(k,t)  + F(k,t).
\ee
Where we have introduced the notation for the instantaneous non-linear energy transfer, $T(k,t)$,   across $k$:
\be
T(k,t) =  \Im \sum_{k < |\bk| <k+1} \sum_{\bf p+q=k}  \ut_i^*(\bk,t) P_{ij}(\bk)   q_l \tilde{u}_l ({\bf p},t) \tilde{u}_j ({\bf q},t),
\label{eq:T}
\ee
and of the scale-by-scale energy injection:
\be
\label{eq:F}
F(k,t) =  \Re \sum_{k < |\bk| <k+1}  \but(\bk,t)\cdot \tilde{{\bf f}}^*(\bk,t),
\ee
where in (\ref{NSFspectrum}-\ref{eq:F})  and hereafter we  assume $\Dk=1$  for the sake of simplicity.
We will consider the explicit dependency on $\Dk$ only  when the infinite volume limit is  considered. From the above expression, we expect that  in the limit of high $Re$ and $\Rhu$, i.e. by sending $\nu, \alpha \to 0$ at fixed $\ein$ and $\lin \sim 1/\kin$, the viscous dissipation term will  play a role only if $k^2 E(k)$ becomes larger and larger for  high $k$. Similarly, for the large-scale drag to be active we need $E(k)$ to grow for $k \to 0$. In other words, for any fixed viscosity we expect the existence of a wavenumber $k_\nu$  defining the onset of viscous effects  in the  ultraviolet (UV) range  of the spectrum while the drag term will be dominant  in the infra-red (IR) limit, i.e.  for wavenumbers smaller than another reference scale, $k_\alpha$. In such a scenario, at steady state, three well distinguished scales, $k_\alpha \ll \kin \ll k_\nu$, must exist and  the  balance (\ref{NSFspectrum}) tells us that the non-linear transfer term must be vanishingly small except on those ranges where it balances the viscous dissipation terms ($k > k_\nu$), the drag term ($k < k_\alpha$)
and the  injection term ($k \sim \kin$):
\be
\label{NSFspectrumstationary}
{T}(k)=   - 2 \nu   k^2  E(k)  - 2 \alpha     E(k)  +  F(k) 
\ee
In the left panel of Fig. (\ref{fig:transfer}) we summarize the balance (\ref{NSFspectrumstationary}). It is important to stress
that the condition to have two well-defined scale separations is not guaranteed {\it a priori}, even in the limit of $Re, \Rhu \to \infty$ and that all consequences and predictions that might follow 
from this assumption must be checked self-consistently {\it a posteriori} by studying the resulting dependency of $\kmu$ and $\knu$ on the control parameters. In  the right panel of the same figure, we show  how the balance (\ref{NSFspectrumstationary}) would look in the absence of scale separation. \\
For what follows, it is very important to define also the net energy transfer
 across a wavenumber $k$. It can be derived by looking at the evolution
of the total energy contained inside a sphere of radius $k$. Following the notation of \cite{Frisch}
we denote with  $\bu^{<k}(\bx,t)$ the  velocity field $\bu(\bx,t)$ low-pass filtered so that all wave numbers outside the sphere of radius $k$ are set to zero:
\be
\bu^{<k} (\bx,t)  = \sum_{|\bk|\le k} \but(\bk,t) e^{i \bk \cdot \bx }
\label{eq:lowpass}
\ee
The change of the total  energy across the sphere with radius $ <k$ is then obtained by taking the inner product of $\bu^{<k}$ with (\ref{eq:GNS})
and averaging over the whole space:  
\be
 \label{eq:fluxbalance}
 \partial_t \sum_{k'<k} E(k',t) = - \Pi_E(k,t) - 2 \nu  \sum_{k'<k} (k')^2 E(k',t) -2 \alpha  \sum_{k'<k} E(k',t)  
 +  \sum_{ k' <k } F(k',t)  
\ee
where the quantity $\Pi_E(k)$ is the non-linear energy flux across a  sphere of radius $k$ in Fourier space:
\be
\label{eq:fluxdef}
\Pi_E(k,t) = \sum_{k'<k} T(k',t) = \langle \bu^{<k} \cdot  [(\bu \cdot \bnabla) \bu] \rangle.
\ee
Because of the inviscid conservation of the total energy we have $\Pi_E(\infty,t)=0$, i.e. the total exchange of energy due to all triadic Fourier interactions is zero.  The second and the third terms on the RHS of (\ref{eq:fluxbalance}) give the total energy dissipation inside the Fourier sphere due to the viscous terms and  the large-scale drag, respectively.  The fourth term gives the  energy injection rate.
In   the $k \to \infty$ limit, the relation (\ref{eq:fluxbalance}) coincides with  (\ref{eq:globalbalance}):
\be
\label{eq:fluxbalancek}
\oenu(t) =  2 \nu \sum_{k=0}^{\infty} k^2 {E}(k,t);
\qquad  \oemu(t) =  2 \alpha \sum_{k=0}^{\infty}  {E}(k,t);
\qquad  \oein(t) =  \sum_{ k=0 }^{\infty}  F(k,t).
\ee
For any fixed $k$ we can rewrite the stationary balance as follows:
\be
 \label{eq:fluxbalancek2}
 {\Pi}_E(k) = -\eUV -\eIR  + \eF
 \ee
 where we have denoted the  time averaged total energy injection, viscous dissipation and viscous drag  inside the sphere of radius $k$, as:
 $$\eF =   \sum_{ k' <k }  F (k'); \qquad 
\label{eq:UV}
\eUV =  - 2 \nu  \sum_{k'<k} (k')^2 {E}(k'); \qquad \eIR =  -2 \alpha  \sum_{k'<k} {E}(k').$$
 Up until now, all manipulations  leading to the global and to the scale-by-scale energy
 balances  (\ref{eq:globalbalance2}) and (\ref{NSFspectrumstationary}-\ref{eq:fluxbalance}) are exact. In order
 to proceed further we need to make some assumptions. \\
 We first assume  that the forcing is concentrated around a thin window $k \sim \kin$. In this case,  we must have $ \eF = 0$ if $k<\kin$ and $ \eF \to \oein$ for $k>\kin$.\\
 Then, assuming the existence of two scales, $k_\alpha$ and $k_\nu$ where the infra-red drag and the viscous dissipation are predominant as in the left panel of Fig. (\ref{fig:transfer}), we can  estimate the asymptotic matching of the different terms entering in  (\ref{eq:fluxbalancek2}). 
 We  examine separately  the long-time stationary relation (\ref{eq:fluxbalancek2})  in the two range of scales, $ k_\alpha  \ll k \ll \kin$ or $ \kin \ll
 k \ll \knu$ that is summarized in Fig. (\ref{fig:flux}).\\
 \noindent 
\paragraph{Inverse cascade: wavenumbers  smaller  than the forcing scale.}
 By referring to the balance (\ref{eq:fluxbalancek2}) and to the left panel of  Fig. (\ref{fig:transfer})  we conclude that for $ k \sim k_\alpha$ 
 the non-linear transfer, $\Pi_E(k)$,  must be negative and matching the contribution {due to $\eIR$, because $\eF$ and $\eUV$  are negligible at those scales.} Moreover,  {the integrated drag contribution up to the wavenumber} $\eIR$  must saturate to a constant,  equal to the total drag dissipation if $k \gg k_\alpha$. As a result,  there  exists an intermediate range of scales where:
   \be
\label{eq:inverseflux}
    \Pi_E(k) = -\eIR \sim -\oemu; \qquad k_\alpha \ll  k \ll \kin.
   \ee
Because only inertial terms play an active  role in the transfer,   we will call this set of scales the {\it inverse-cascade inertial-range}. \\
 \noindent  
\paragraph{ Direct cascade: wavenumbers larger than the forcing scale. }
 For $ k \gg \kin$, both the  integrated contribution of the drag terms and of the total injection have reached its asymptotic values, $\eIR \sim \oemu$ and  $\eF \sim  \vein $ while the viscous dissipation is still not active, $\eUV \sim 0$,  if $k \ll k_\nu$.  As a result, there exist an intermediate range of scales where the balance gives:
 \be
 \label{eq:directflux}
    \Pi_E(k) = - \oemu + \vein = \oenu \qquad \kin \ll  k \ll k_\nu .
   \ee
We will call this set of scales the {\it direct-cascade inertial-range}.   
In Section (\ref{sec:direct}) we will clarify the above picture and  to which extent it can be pushed to rigorous and quantitative statements.
Before entering in these aspects, we move back to  configuration space to re-derive the previous results from a different perspective.

\subsection{{Configuration  space representation}}                              
\begin{figure*}[htbp]                                                                        
\centering                                                                                   
\includegraphics*[width=0.6\textwidth, angle=0]{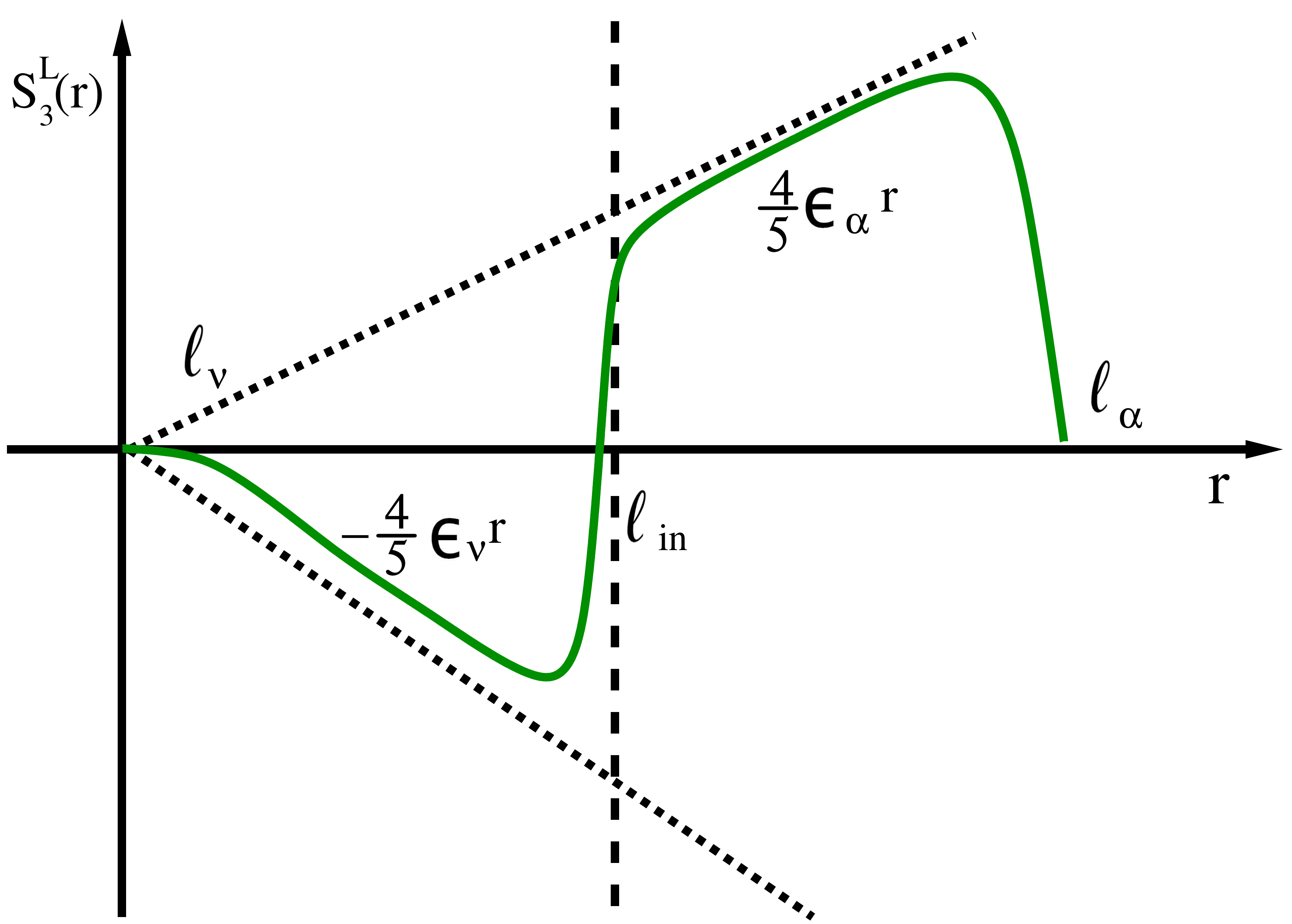}                             
\caption{ Sketch of the scaling properties for the third order longitudinal structure        
function, following the relations (\ref{eq:45direct}) and (\ref{eq:45inverse}) in the       
hypothetical presence of simultaneous split energy cascade, both forward and inverse.  }     
\label{fig:fluxrealspace}                                                                    
\end{figure*}

The energy distribution among scales and the energy flux can also be defined directly in configuration  space by considering velocity correlation
between two  points $\bx$ and $\bx'$ with  $\br=\bx'-\bx$ and then averaging over all possible points $\bx$ for a fixed $\br$. Here, we
follow the discussion  presented in \cite{Frisch}, referring to the textbook for details. We first introduce the second-order correlation function:
\be
C_2(\br,t) = \la \bu({\bf x},t)\cdot \bu({\bf x}+\br,t) \ra.
\ee
Assuming homogeneity (but not isotropy)
it is possible to derive from the NSE the K{\'a}rm{\'a}n-Howarth-Monin relation \cite{monin2013statistical}:
\be
\frac{1}{2} \partial_t C_{2}(\br,t) = \frac{1}{4} {\bnabla}_{\br} \cdot \la |\delta_{\br} \bu|^2 \delta_{\br} \bu \ra 
+ \nu \Delta_{\br} C_{2}(\br,t) 
- \alpha  C_2(\br,t) + F(\br,t)
\label{eq:KHM}
\ee
where we have introduced the notation for the velocity increment over a distance $\br$: 
$\delta_{\br} \bu = \bu(\bx+\br) -\bu(\bx)$ and the forcing-velocity correlation $F(\br,t) = \la \bu(\bx+\br,t) \cdot {\bf f}(\bx,t) \ra$. In the presence of a 
stationary statistics we can derive the law
for the total energy balance by setting  $\br =0$, averaging over time
and putting to zero all time derivatives:
\be
\nu \Delta_{\br} {C}_2(\br)|_{\br=0}  - \alpha   {C}_2(\br)|_{\br=0} =   F(\br)|_{\br=0}  = \oein
\ee
where we exploited that  for $\br=0$ the non-linear  contribution ${\bnabla}_{r} \cdot \la |\delta_{\br} \bu|^2 \delta_{\br} \bu \ra$ is identically zero because of the inviscid  energy conservation. It is easy to see that the above relation is equivalent to (\ref{eq:globalbalance2}) by noticing that $    \Delta_{\br} {C}_2(\br)|_{\br=0} = - \langle |\nabla {\bu}|^2 \rangle$ and
$  {C}_2(\br)|_{\br=0} = \langle |{\bu} |^2 \rangle$. 
Assuming stationarity and keeping the distance fixed with  $\br \neq 0$ we get from (\ref{eq:KHM}):
\be
\label{eq:KHMstationary}
-\frac{1}{4} {\bnabla}_{\br} \cdot \la {|\delta_{\br} \bu|^2 \delta_{\br} \bu} \ra =  \nu \Delta_{\br} {C}_2(\br) +   F(\br) 
-\alpha  {C}_2(\br) ,
\ee
which is the  configuration-space equivalent of the scale-by-scale energy balance (\ref{NSFspectrumstationary}).
Let us now suppose to have an isotropic forcing-velocity correlation that is peaked at one given scale, $\lin \sim 1/\kin$:
$$  F (r) \sim \ein \exp^{- (r/\lin)^2},$$
where $r = |\br|$. As we did for the Fourier space we need to distinguish two different scaling regimes.\\
\noindent 
{\bf  Direct cascade: scales smaller than the forcing scale. }
In such  a range, for any fixed $r$ one can fix the viscosity small enough to make the dissipative term vanishingly small, i.e there  exists
a dissipative scale, $\ell_\nu \sim 1/k_\nu$  such that for $  \ell_\nu \ll  r $ the viscous dissipation is negligible. Moreover, both the energy injection and the two-point correlation function are smooth for  $ r \ll \lin$: 
$$  F (r)  \sim \ein; \qquad  \alpha  C _2(\br) \sim \oemu. $$
As a result, in this range of scales we have 
\be
-\frac{1}{4} {\bnabla}_{\br} \cdot \la {|\delta_{\br} \bu|^2 \delta_{\br} \bu} \ra = \ein -\oemu = \oenu ,
\ee
by further assuming isotropy, it is possible to express the left hand side in terms of the longitudinal third order structure functions,
$S^L_3(r) =   \la (\delta_{\br} \bu \cdot \ur)^3\ra$ to finally obtain the celebrated $4/5$ law of  Kolmogorov \cite{Frisch}  for the inertial range of the direct cascade that reads (in 3D):
\be
\label{eq:45direct}
{S}^L_3(r) = -\frac{4}{5} \oenu \, r; \qquad \ell_\nu \ll  r  \ll \lin. 
\ee
The above relation can be dimensionally summarised as
\be
\frac{(\delta_r u)^3}{r} \sim \oenu
\label{eq:flux} 
\ee
\noindent 
{\bf   Inverse cascade: scales larger than the forcing scale. } 
 In the other limit, $ r \gg \lin$, we have no direct injection of energy from the forcing,
 $  F (\br)  \sim 0$, we can consider  the viscous term to be vanishingly small, and  the third-order correlation on the LHS of (\ref{eq:KHMstationary}) is balanced by the drag-term contribution only:  
\be
\label{eq:45inverse}
{S}^L_3(r) = \frac{4}{5} \oemu \, r, \qquad \lin \ll  r  \ll \ell_\alpha,
\ee
where we have assumed that  for $ r \ll \ell_\alpha$, the correlation $C_{2}(\br)$  has already saturated to its constant value.
Eq. (\ref{eq:KHMstationary}) is exact, while in order to get the direct and inverse scaling range in eqs. (\ref{eq:45direct}-\ref{eq:45inverse})
we need to  assume the existence of the two scales, $\ell_\alpha, \ell_\nu$,  fixing the onset  of the viscous effects (for $r \ll \ell_\nu$) and
the onset of the drag term (for $r \gg \ell_\alpha$), as already done for the same quantities in the Fourier space. Depending on the existence of such scales and
on their dependency on the control parameters $\Reu,\Rhu$, one might
end up in a situation where the direct/inverse energy transfer develops  a true asymptotically inertial direct and/or inverse {\it scaling} range, i.e. a
set of scales which becomes more and more extended when $Re \to \infty$   and/or $\Rhu \to \infty$, as depicted in Fig. (\ref{fig:fluxrealspace})

 Expressions (\ref{eq:45direct}-\ref{eq:45inverse}) are also important because  they give a clear signature of the breaking of 
time-reversibility and of Gaussianity (in both cases any third order velocity correlation function   vanishes identically). 
The latter observation is connected to the existence of a net non-vanishing energy flux.

\subsection{{Scale-filtered representations} \label{sec:scalefiltered}}
In order to control the multi-scale properties of the flow  in  configuration space,  it is often useful to study
the  velocity field evolution coarse-grained
at  scale $\ell \sim 1/k$.
In order to do that, we need to define a filtering
operation on a generic velocity field, $\hat{\bullet}^\ell$:
\be
\label{eq:filtering}
\obv(\bx,t) = \int  dy^d G^\ell(\bx-\by) \bu(\by,t) = \sum_{\bk} \tilde  G^\ell(\bk) \but^*(\bk,t) e^{i \bk \cdot \bx }
\ee
where we have  $\tilde G^\ell(\bk)=\tilde G^{\ell,*}(-\bk)$ to ensure that the velocity field is a real function. If the filter is a projector, then 
$\tilde G^\ell(\bk) \otimes \tilde G^\ell(\bk)    = \tilde G^\ell(\bk)$. For example, when
$\tilde G^\ell(\bk) = 1$ if $|\bk| < k = 2\pi/\ell$ and $\tilde G^\ell(\bk)=0$ otherwise, the filtered field, $\obv$ would coincide with the {low-pass field (\ref{eq:lowpass})}.  Another popular choice  is a Gaussian filter $\tilde{G}^\ell(\bk) = exp(-\ell^2 k^2/2)$. Other  possible options for $\tilde{G}^\ell(\bk)$ are discussed in \cite{pope2001turbulent}.\\ 
 The NSE equations for the filtered field are:
\be
\p_t \obv  + (\obv \cdot \bnabla) \obv = -\bnabla \oP -\bnabla {\btau}^\ell+ \nu \Delta \obv   -\alpha \obv + \hat{\bf f }^\ell
\label{eq:les}
\ee
where $\oP$ is a pressure term that enforces  incompressibility of $\obv$ and  the sub grid scale (SGS) stress tensor is given by:  
\be
 \tau^\ell_{ij} =  \widehat{u_iu_j}^\ell - \ou_i\ou_j
\ee
Notice that (\ref{eq:les}) is  exact but not closed, i.e. the evolution of the filtered field does depend on the filtered scales via the SGS  term. Large eddy simulations are based on the idea of modelling  $\tau^\ell_{ij}$  in terms of the resolved field and it  will be briefly discussed in Sec. (\ref{sec:LES}).\\
It is useful to define the SGS energy-transfer, entering in the dynamical
evolution of the resolved kinetic energy:
\be
\frac{1}{2}\p_t |\obv|^2   +  \bnabla  \cdot {\bf A}^\ell   =  - \Pi^\ell + \nu \obv \cdot (\Delta \obv) -\alpha |\obv|^2 + \obv \cdot \hat {\bf f} ^\ell 
\label{eq:sg-ene}
\ee
where ${A}^\ell_i = \ov_i(\frac{1}{2}|\obu|^2 + \oP) + \ov_j \tau^\ell_{ij}$ is a globally conservative flux term that redistributes the resolved energy among different spatial positions and 
\be
\label{eq:SGSE}
\Pi^\ell(\bx,t) = - \tau^\ell_{ij}(\bx,t)\oS_{ij}(\bx,t) 
\ee
is the SGS energy flux and $\oS_{ij} = \frac{1}{2}(\p_i \ou_j + \p_j \ou_i) $ is the resolved strain-rate tensor.
The energy transfer between resolved and sub-filter scale is controlled by $\Pi^\ell$ and it is straightforward
to show that for the case of a sharp Fourier projector  ($\tilde G^\ell(\bk) = 0$  if $ |\bk| > k = 1/\ell$; $\tilde G^\ell(\bk)=1 $ otherwise) we have that:
\be
\label{eq:piaverage}
\langle \Pi^\ell(\bx,t) \rangle  = \Pi_E(k,t); \qquad \text{with} \qquad k=2\pi/\ell, 
\ee
i.e. the space average of the SGS energy transfer coincides with the nonlinear energy  flux in Fourier space (\ref{eq:fluxdef})
at the corresponding wavenumber.  An important issue is connected to the characterization of the space-time fluctuations of $\Pi^\ell(\bx,t)$ and to the possibility of defining  a local-in-space  {\it Richardson cascade}, to  identify regions where a breakdown of eddies into smaller and smaller eddies is coherently observed within some space and time domains, as discussed in Sec. (\ref{sec:LES}). The SGS tensor also becomes very relevant in quasi-2D flows where in the presence of a split cascade and in the presence of a large-scale condensate, some regions in space act locally like 3D cascading energy forward while other regions act like 2D cascading energy inversely (see e.g. section \ref{sec:CritThin},\ref{sec:Rotation})  . \\
 
\section{Definitions of Turbulent Cascades  \label{sec:directinverse}}          
In this section we give a precise definition of what do we mean by  {\it cascade}. Although the meaning of forward and  inverse cascades has been stated many times in the literature, and it might be simple  for  3D and 2D turbulence, in the case where a split transfer to both large and small scales is developed (as will be the case for many thin layers configurations examined in Sec. \ref{sec:applications}) one needs to be careful because the amplitude of the  cascade  can be very small and needs to be distinguished from any transient transfer. Furthermore, even in presence of a simple unidirectional transfer, ultraviolet and infra-red cut-off might play an important role in the limit of large $\Red, \Rhd$ numbers. E.g., in the presence of a finite volume and in the limit $\Rhd \to \infty$, the infrared cutoff will stop the inverse cascade, leading to an accumulation of energy at the smallest wavenumber and to the formation of a large-scale condensate (see Sec. \ref{sec:statequil} and the discussion about  Fig.  \ref{fig:thermalflux2}). The large-scale structures might  finally induce a strong feedback on the inverse inertial range, break the scaling properties and bring the system close to a quasi-equilibrium state. Our definitions will be  tailored to take into account all basic ingredients.   Here we only consider the cases of a steady state, thus in all arguments the long time limit is considered before any other limiting procedure is taken.

The best way to proceed is by first examining a few  idealized examples, that demonstrate some basic concepts before reaching the exact definitions.
We start by examining the Kolmogorov theory for the forward 3D energy cascade (Sec. \ref{sec:direct}) and the Batchelor-Kraichnan theory for the 2D inverse energy and direct enstrophy transfers (Sec. \ref{sec:2DinvCsd}). These two examples will allow us to introduce also the concept of
scale-invariance and a first important set of dimensional estimates for the dependency of $\ell_\alpha$ and $\ell_\nu$ on $\Reu,\Rhu$.  We proceed  by discussing the case of: a split cascade (Sec. \ref{sec:split}),  a statistical equilibrium (Sec. \ref{sec:statequil}), multiple simultaneous transfers of different inviscid invariants (Sec. \ref{sec:multiple}). Finally we conclude this chapter in Sec. (\ref{sec:definitions}) with a series of precise definitions for all the above cases,  including bidirectional and flux-loop cascades, and with a classification of cascade transitions in Sec. (\ref{sec:Classification}). 

\subsection{3D Direct Energy Cascade and the Kolmogorov 1941 theory \label{sec:direct}}                   
As a paradigmatic  example of a direct cascade we discuss the case of 3D HIT.
It is an empirical fact that in 3D there is no inverse energy cascade and therefore we start by putting  the  large scale drag $\alpha=0$ and $\oemu=0$. Here we will only consider the case of zero helicity injection, postponing to  Sec. (\ref{sec:Helicity}) the discussion of the opposite case.\\  
From the stationary relation (\ref{eq:globalbalance2}),   the  energy injection rate needs to be equal to the 
 the averaged energy dissipation rate ${\ein} = {\enu}$.
 We note that these equalities hold independently of the value of $\Red$, and their validity does not imply the presence of a cascade.  
 To define the cascade we need to compare the flux to the amplitude of the fluctuations and we need to make sure that there exists a well-defined scale separation  between the wavenumbers where we have the maximum of the injection rate and the maximum of  viscous dissipation. 
Thus, in the limit $\Red \rightarrow \infty$  we consider the dimensionless ratio 
\be
\label{eq:dragcoefficient}
D_\nu = \frac{  {\enu} }{ {\cE}^{3/2} \kin  }, 
\ee
sometimes called the {\it drag coefficient}. It is a fact based on numerical simulations and experimental data that this 
quantity remains finite even  in the limit of infinite Reynolds number:
\be
\label{eq:zerothlaw}
\lim_{\Red \rightarrow \infty} D_\nu > 0. 
\ee
where the constant value  might depend on the forcing mechanism \cite{ishihara2009study}. 
Let us notice that both definitions of the  Reynolds number given by
expressions (\ref{eq:Re}) or (\ref{eq:Red}) would be here  equivalent.
In the first case one considers $\enu=\ein$ fixed and (\ref{eq:zerothlaw}) implies that as $\nu \to 0$,  ${\cE}$ remains bounded from above
while in the second case one considers $ \cE$ fixed and it implies that as $\nu \to 0$ the dissipation rate $\enu$ remains bounded from below. The existence of a non-zero energy dissipation even in the limit $\nu \rightarrow 0$ is one of the major fingerprints of 3D turbulence and goes under the name of  {\it dissipative anomaly} \cite{onsager1949statistical,eyink2006onsager,duchon2000inertial}. The fact that $D_\nu$ remains an order one quantity, tell  us that the direct energy cascade in 3D HIT is always strongly out-of-equilibrium, and far from a perturbative  quasi-equilibrium solution which  would require $D_\nu =0$ or $ D_\nu \ll 1$. \\
\subsubsection{Fourier space}
\begin{figure*}[htbp]                                                        %
\centering                                                                   %
\includegraphics*[width=0.45\textwidth]{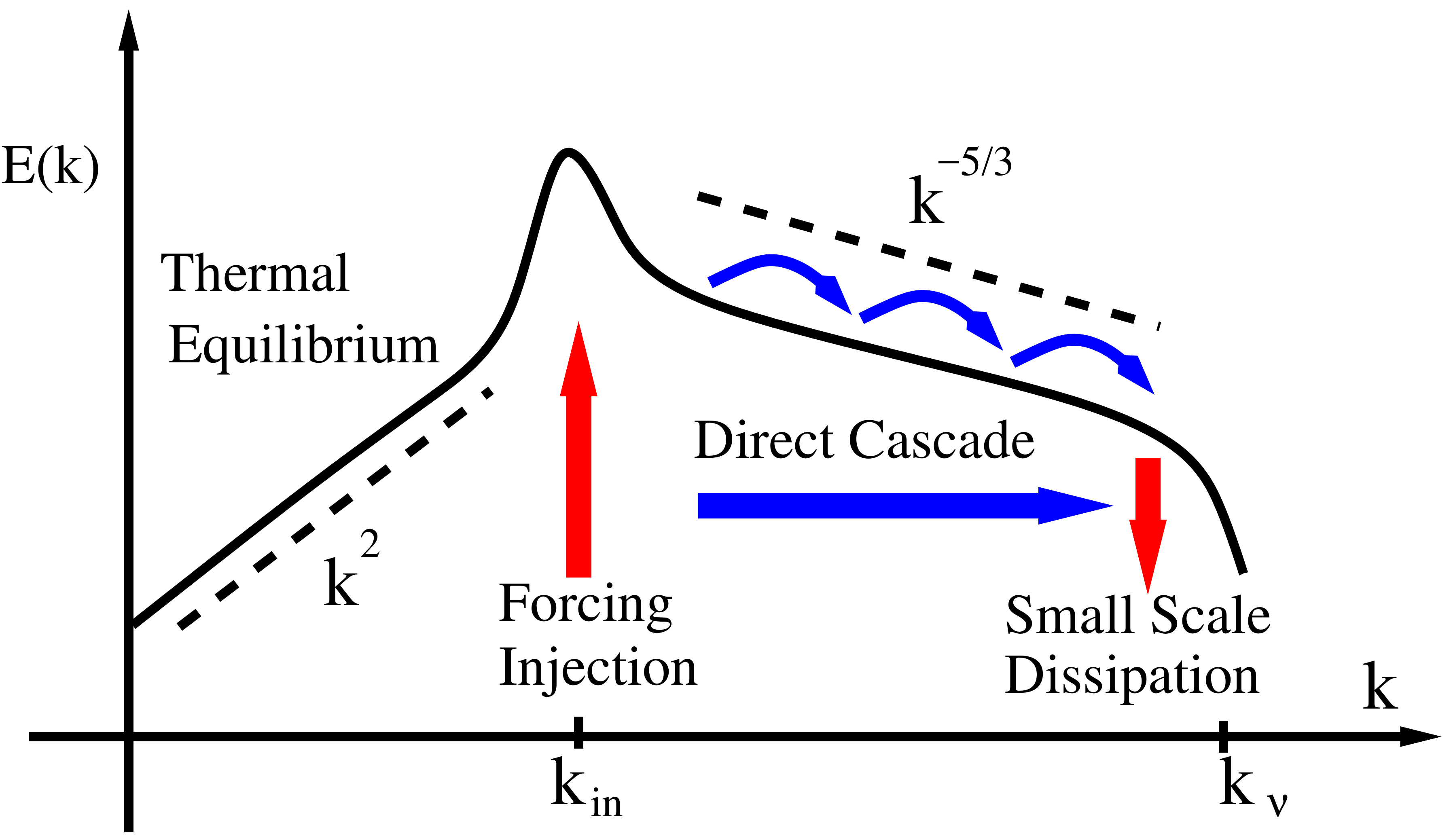}                    %
\includegraphics*[width=0.45\textwidth]{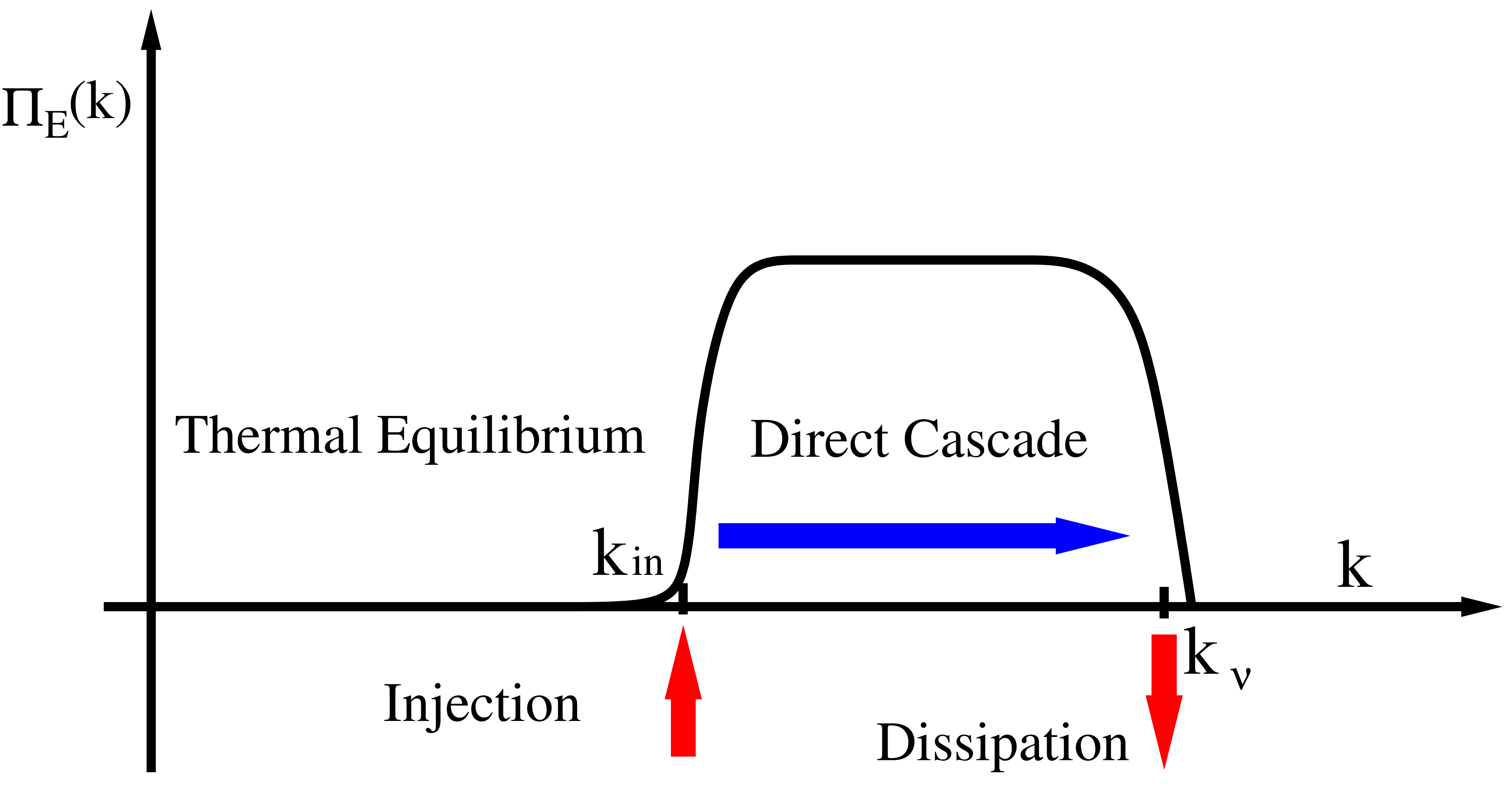}                       %
\caption{ 
Left: sketch of the spectrum for the direct cascade regime, with             %
the $k^2$ thermal equilibrium in the inverse range for $k < k_{in}$ %
and the $-5/3$ Kolmogorov slope in the direct cascade range (\ref{eq:K41}).  %
Right: the same as in the left panel but for the energy flux. Notice that in the          %
equilibrium range we have $\Pi_E(k)=0$ as discussed in Sec.                  %
(\ref{sec:statequil}).}  \label{fig:direct}                                  %
\end{figure*}

If (\ref{eq:zerothlaw}) holds and considering the expression (\ref{eq:fluxbalancek}) for the energy dissipation,  we must have that the peak of the enstrophy spectrum, $k^2 E(k)$, is centered around a wavenumber $\knu \to \infty$  when $\nu\to 0$ leading to the formation of the inertial range,  $\kin \ll k \ll \knu$, where viscous effects can be neglected and the energy flux is constant, ${\Pi}_E(k)={\ein}$. This is equivalent to the statement made in  configuration space, leading to (\ref{eq:45direct}). \\
A phenomenological prediction for the spectral property  was developed by  Kolmogorov in 1941 (K41) based on the idea that  only the mean flux, $\oein$,  plays a statistical role in the inertial range \cite{Kolmogorov}. In such a case, Kolmogorov derived the  celebrated $-5/3$ power law inertial-range behaviour for the spectrum:
\be
\label{eq:K41}
{E}(k) = C_K {\ein}^{2/3} k^{-5/3}, \qquad \kin \ll k \ll k_\nu.
\ee
where $C_K$ is the so-called Kolmogorov constant. By plugging the K41 spectrum into the expression for the total viscous dissipation
given after eq. (\ref{eq:fluxbalancek2}), we can finally define the dependency of the UV cut-off, $k_\nu$, on $Re$. In order to do that, we  define  $\knu$  as the scale where the inertial range flux  becomes comparable with the dissipative term:
\be
\Pi_E(k_\nu) \sim  \enu; \qquad \to \qquad \oein \sim  \nu \oein^{2/3} \knu^{4/3}, 
\ee
where to get to this estimate we have plugged (\ref{eq:K41}) in  (\ref{eq:fluxbalancek2}) and summed up to $k_\nu$. By using (\ref{eq:Red}) we finally obtain:
\be
\label{eq:knuRe}
k_\nu \sim \kin Re^{3/4}.
\ee
The above result is very important, it shows the self-consistency of the K41 theory: by assuming that for large Reynolds viscosity does not play a role
for $k \ll k_\nu$,  one derives a prediction for the spectrum that consistently defines a viscous cut-off which becomes larger and larger for $Re \to \infty$. In Fig. (\ref{fig:direct}) we summarize the results fro spectrum and flux within the K41 theory.  This result, together with the existence of the dissipative anomaly are the two fingerprints of the {\it forward energy cascade} in HIT.  On the other hand, when viscosity sets-in
we must have that in the far dissipative region, $ k \gg k_\nu$, an asymptotic matching between transfer and viscous terms  develop, leading to an exponential or super exponential fall off with an exponent $\beta >0$:
$$
 E(k) \sim \exp{-(k/k_\nu)^\beta}; \qquad k \gg k_\nu.
\label{eq:K41UV}
$$
Before concluding this section, we briefly discuss what happens in the IR range, $ k < \kin$, where no net inverse transfer of energy is observed in 3D HIT. 
As said, if  $\alpha=0$, at equilibrium  we must have no transfer at all,  ${\Pi}_E(k) = 0$, and the spectral distribution of energy is strongly depleted. It has been suggested that these scales reach a statistical (thermal) equilibrium due to local and non-local energy diffusion across the wavenumbers, with all Fourier modes feeling a sort of thermal bath described by a Gibbs-ensemble equipartition distribution  $E(k) \propto k^{2}$ that is discussed in Sec. (\ref{sec:statequil}). The validity of such assumption is  still unclear  and it is the subject of current investigations \cite{cichowlas2005effective, krstulovic2009cascades,Dallas2015statistical,ray2015thermalized,bos2006dynamics, cameron2017effect}. If we consider the case of small but non-zero $\alpha$ we expect similar conclusions. Since in 3D there is no evidence of an inverse cascade, the flux of energy to the large scales although non-zero ($\emu$ is a strictly positive quantity if $\alpha >0$) it will be still sub-dominant. The spectra could however be strongly modified even for small values of $\alpha$ and become steeper than $k^2$. The absence of a range of scales where there exists a constant inverse flux implies that we
cannot expect the existence of a large-scale dissipative anomaly similar to (\ref{eq:zerothlaw}). If we define the drag for the inverse transfer as: 
\be
\label{eq:inversezerothlaw} 
D_\alpha = \frac{\oemu}{{\cE}^{3/2} \kin}
\ee
we must have:
 \be
\label{eq:inversezerothlaw2} 
\lim_{\Rhd \rightarrow \infty} D_\alpha = 0.
\ee

\subsubsection{Configuration space \label{sec:configurational}} 
Generalising the results obtained  in  Fourier space, the K41 theory assumes that the whole probability distribution function (PDF) of the velocity increments in the inertial range is fully determined by the mean energy injection only, if isotropy holds.
Introducing the longitudinal and transverse  velocity increments as:
\be
\delta_\br^L u = \delta_{\br} \bu \cdot \hat \br; \qquad
\delta_\br^T \bu = \delta_\br \bu - (\delta_\br^L u)  \hat \br
\label{eq:sfdef}
\ee
where with $\hr $ we indicate the unit vector in the direction of the displacement,  Kolmogorov predicted for
the $n$-th order longitudinal and transverse structure function: 
\be 
S_n^L(r) = \langle  {(\delta^L_{\br} u)^n} \rangle  \qquad
\label{eq:SFk41SF}
 S_n^T(r) = \langle  {|\delta_{\br}^T \bu|^n} \rangle 
\ee
the celebrated $n/3$ law for the scaling exponents in the inertial range:
\be
  S_n^L(r) \sim C^L_n  \oein^{n/3} r^{n/3}; \qquad
\label{eq:k41SF}
 S_n^T(r) \sim C^T_n  \oein^{n/3} r^{n/3}; \qquad \ell_\nu \ll r \ll \lin
\ee
where  $C^L_n,C^T_n$ are  dimensionless constants.   From  (\ref{eq:knuRe}) we have $\lin/\ell_\nu \sim \knu/\kin \sim Re^{3/4}$. Similarly, the exponential decay in the Fourier viscous range corresponds to an  analytical smooth
behaviour for the velocity field in the range $ r \ll \ell_\nu$, i.e.  the longitudinal and transverse increments  must be $ \propto  r^n$ if $ r \ll \ell_\nu $.  It is important to notice that in (\ref{eq:k41SF}) the scaling of each single component  of the transverse increments would be strictly zero for odd moments if isotropy holds \cite{biferale2005anisotropy}. This is not the case for longitudinal increments as seen from the $4/5$-law (\ref{eq:45direct}). From the scaling (\ref{eq:k41SF}) one can build the longitudinal skewness, a dimensionless measure of the intensity of the flux normalized with the root-mean-squared velocity fluctuations at scale $r$:
\be
K_3^L(r)  = \frac{S_3^L(r)}{(S_2^L(r))^{3/2}} =  C^L_3/(C^L_2)^{3/2} = const. 
\label{eq:skewness}
\ee
which can be considered a scale-by-scale generalisation of (\ref{eq:dragcoefficient}), i.e. a proxy
of the out-of-equilibrium nature of the forward energy cascade. \\
On one side, K41 theory is considered a  milestone for building  a {\it theory of turbulence} and it is known to describe many important quantitative and qualitative realistic properties. On the other side, there is now  a long enough series of experimental and numerical evidence to know that it is not exact, even limiting the discussion to the case of HIT.
The most important point where it fails is linked to the existence of anomalous corrections to the inertial-range scaling exponents. Overwhelming evidence shows that velocity increments are characterized by a whole spectrum of anomalous inertial range exponents, see, e.g., \cite{benzi2008intermittency,ishihara2009study,iyer2017reynolds,sinhuber2017dissipative}:
\be
S_n^L(r) \propto  (\frac{r}{\lin})^{\zeta_n^L};  \qquad S_n^T(r) \propto
(\frac{r}{\lin})^{\zeta_n^T};
\label{eq:sf}
\ee
with $\zeta_n^L,\zeta_n^T \neq n/3$ except for the case $\zeta_3^L=1$ where  (\ref{eq:45direct}) holds exactly. The departure from a linear dimensional behaviour of the scaling exponent goes under the name of {\it intermittency} and it will be further discussed in Sec. (\ref{sec:intermittency}). Theoretical considerations based on isotropy \cite{biferale2005anisotropy} would also require $\zeta_n^L =\zeta_n^T$, a fact that is not fully realized at the Reynolds numbers available nowadays \cite{iyer2017reynolds}.  For what has been  said until now,  the most important consequence of intermittency is connected  to the fact that by using (\ref{eq:sf})  the skewness is not  constant any more:
\be
K_3^L(r)  \sim (\frac{r}{\lin})^{\zeta_3^L-\frac{3}{2} \zeta_2^L}  =  r^{-0.06}; \qquad  \ell_\nu \ll r \ll \lin
\label{eq:skewnessanomalous}
\ee
where the value $0.06$ of the exponent is taken from the most updated numerical and experimental results \cite{iyer2017reynolds}. The fact that the exponent is small and negative is very important. It means that at moderate Reynolds numbers, the skewness can be considered almost constant, because $r$ cannot vary too much due to the limited extension of the inertial range. On the other hand, for large enough Reynolds, because of (\ref{eq:knuRe}) we have $\ell_\nu \to 0$ and we can send
the ratio $r/\lin \to 0$, still remaining in the  scaling region. In this limit, the  skewness will become larger and larger, indicating stronger and stronger out-of-equilibrium properties and  a stronger and stronger departure from quasi-Gaussian statistics (where $K_3(r) =0$). This is often visualized by plotting the energy dissipation field, which turns out to be highly spotty and spiky in configuration space, with vast regions with very small values and a few isolated islands of high intensity, i.e. an intermittent spatial distribution (see Fig. \ref{fig:intermittency}). \\

\subsection{2D Inverse Energy Cascade and the Batchelor-Kraichnan theory  \label{sec:2DinvCsd} }            
2D turbulence is a paradigmatic example where an inverse energy transfer is observed. { It was predicted in a series of historical works by Kraichnan, Leith and Batchelor \cite{kraichnan1967inertial,leith1968diffusion,batchelor1969computation} and has been reproduced in numerical simulations \cite{boffetta2007energy,boffetta2010evidence,vallgren2011infrared,xiao2009physical} and experiments \cite{ouellette2012turbulence,Xia2011upscale}. A review can be found in \cite{Boffetta2012two,kraichnan1980two}. } In two dimensions,  the NSE (\ref{eq:GNS}) can be written in terms of the out-of-plane vorticity $w=\partial_x u_y - \partial_y u_x$ as an advection-diffusion equation with the forcing term $f_w=\partial_x f_y - \partial_y f_x$:
\be 
\partial_t w + \bu \cdot \bnabla w = -\alpha w + \nu \Delta  w + f_w \label{eq:2DNS}.
\ee 
 As a result,
the nonlinearity conserves  the enstrophy $\cZ(t)=\frac{1}{2}\la w^2 \ra $ and all moments of the vorticity $\la w^n \ra $. 
Unlike for the  3D flow, where we considered for simplicity  the case of zero helicity injection, enstrophy injection  can not be set to zero  without setting the energy injection also to zero. 
In 2D, for  enstrophy we have the exact balance: 
 \be
\label{eq:globalbalanceW}
 \partial_t \cZ(t) =  - \wenu(t) -  \wemu(t) + \wein(t) 
\ee 
here $\wenu(t) = \nu \langle |\nabla w|^2 \rangle$ is the enstrophy dissipation due to viscosity, 
      $\wemu(t) = \alpha \langle |w|^2\rangle$ is the enstrophy dissipation due to the large scale drag  and 
      $\wein(t) =     \langle w {f_w} \rangle \propto \ein/\lin^2$ is the enstrophy injection rate. The long-time average leads to 
\be
\label{eq:globalbalance2H}
 \bwein =  \bwenu +  \bwemu  .
 \ee
\begin{figure*}[htbp]                                                      %
\centering                                                                 %
\includegraphics*[width=0.45\textwidth]{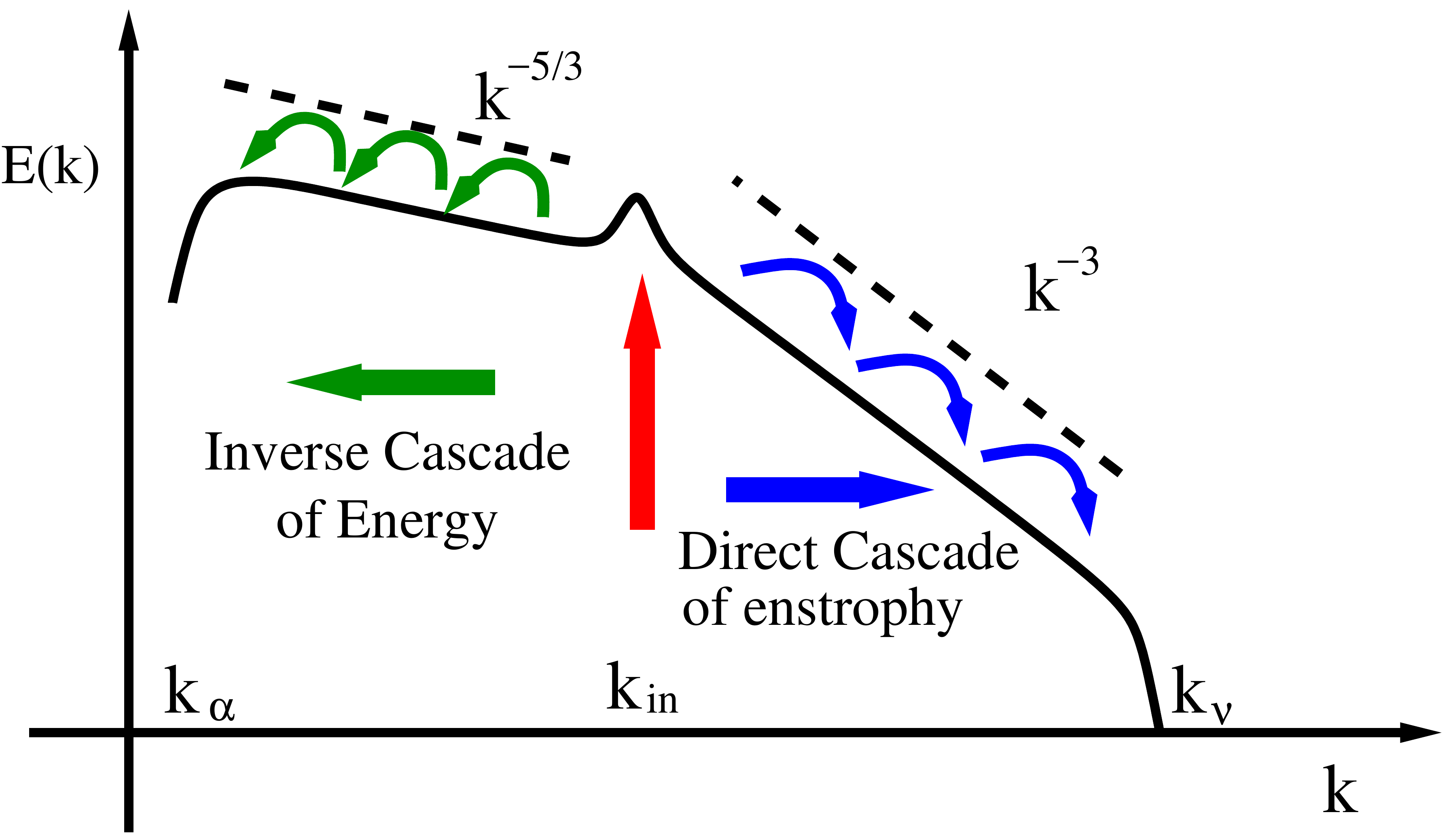}                  %
\includegraphics*[width=0.45\textwidth]{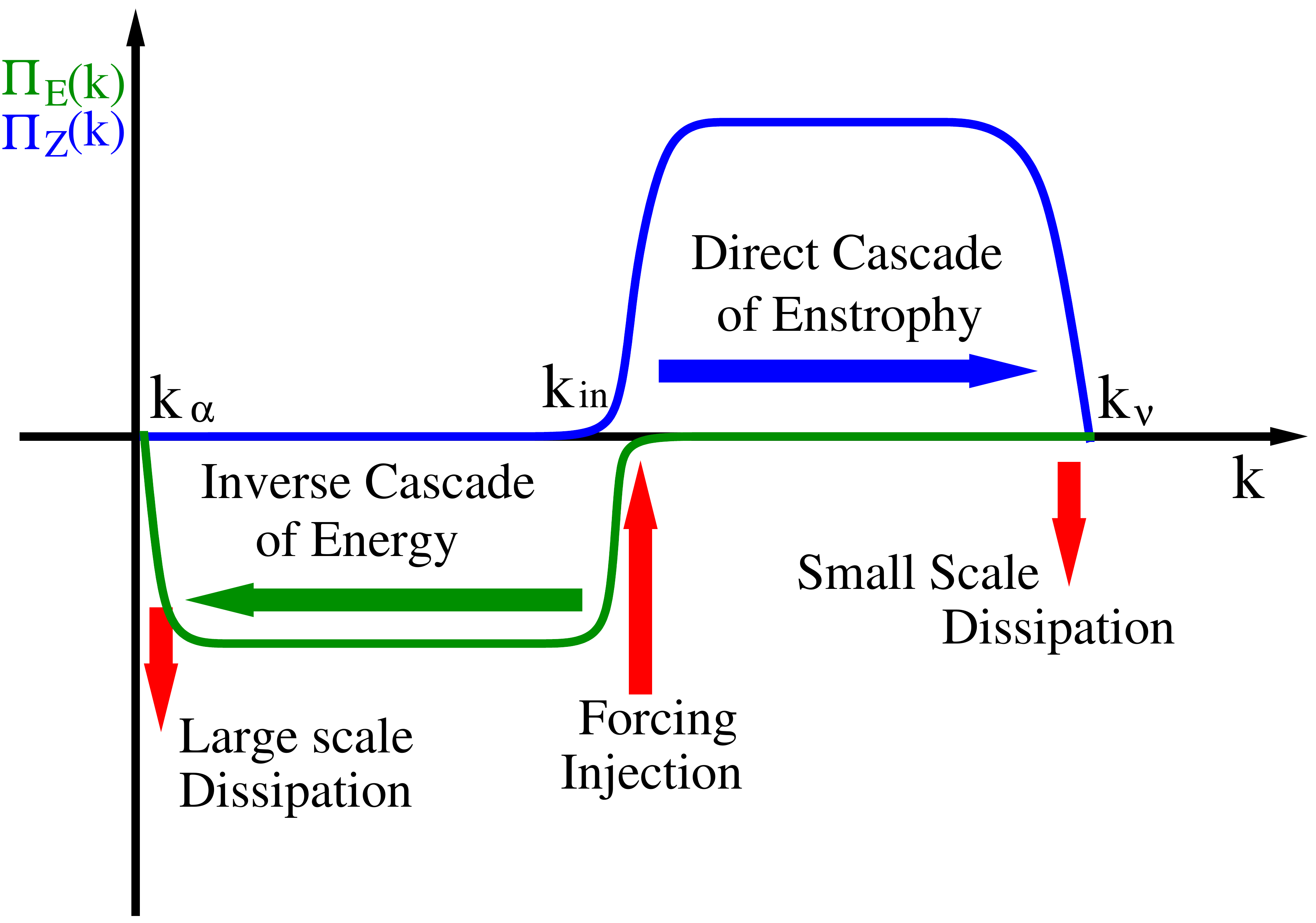}                 %
\caption{ Log-log sketch of the energy spectrum (left) and of energy and enstrophy fluxes (right) for   %
 the 2D Batchelor-Kraichnan  %
theory  (\ref{eq:2denstrophy}-\ref{eq:2dkraichnan})                        %
}\label{fig:inverse}                                                       %
\end{figure*}                                                              %

 \subsubsection{Fourier space} \label{sec:2Df}
 Considering the forcing acting only at a limited range of wavenumbers around a specific scale, $\kin$, the energy and enstrophy injections are
 related by $\wein = \kin^2 \ein$. Similarly,  the energy spectrum $E(k)$ and the enstrophy spectrum $E_\cZ(k)$  are connected by the  relation  $E_\cZ(k)\simeq k^2E(k) $. 
Thus, the cascade of the two ideal invariants can not be discussed separately, but leads to a situation where two fluxes 
coexist, referred often as a {\it dual cascade}.  
Defining  a low-pass vorticity field similarly to (\ref{eq:lowpass}), we introduce the  2D enstrophy flux:
\be
 \label{eq:wflux}
 {\Pi}_\cZ(k) =  \langle w^{<k}  (\bu \cdot \bnabla)  w  \rangle 
\ee
and using the same arguments as for the 3D  energy cascade we distinguish a direct  inertial range  for the enstrophy transfer if
there exists a window at high  wavenumbers where the  non-linear flux is constant  
\be
 \label{eq:hwfluxbalanceforward}
 {\Pi}_\cZ(k)  \simeq    \wein   -  \wemu =   \wenu, \qquad \kin \ll k \ll \knu 
\ee
while for the inverse enstrophy cascade we should  have in the low $k$ region:
\be
 \label{eq:hwfluxbalanceinverse}
 {\Pi}_\cZ(k)  \simeq    \wein   -  \wenu =   \wemu, \qquad  \kmu \ll k \ll \kin.
\ee 
To study the inverse cascade range,  there are two limiting procedures that need to be considered and that lead  to two different situations.
The first limit corresponds to the large box limit $\kin L\to \infty$ (or $k_0/\kin \to 0$) and the second is the 
limit $\Rhd\to \infty$. If the large box limit is taken first then the system saturates in a state with  a finite  inverse energy flux, while if the large $\Rhd$ limit is taken first  the system saturates to a condensate state that resembles to a statistical equilibrium (see also Sec. \ref{sec:statequil}).\\
We begin with the case when the large-box limit is taken first.
Assuming the working hypothesis that there exists a separation of scales,  $\kmu \ll \kin \ll \knu$ (an assumption that as we saw for the
K41 theory can depend on the existence of a finite dissipation limit and whose self-consistently can be only checked aposteriori) we have that for any wavenumber in the inverse 
range  the energy and enstrophy flux will satisfy 
 (\ref{eq:inverseflux}) and  (\ref{eq:hwfluxbalanceinverse}) respectively, while in the forward  range  (\ref{eq:directflux}) and (\ref{eq:hwfluxbalanceforward}) hold.\\
In the forward range,  $\kin \ll k  \ll \knu$,  we can write: 
\be
\label{eq:fluxConstraint}
\Pi_E(k) \simeq 2 \nu \sum_{k'=k}^\infty  (k')^{2} E(k')  \le 2 \nu k^{-2} \sum_{k'=k}^\infty  (k')^{4} E(k') 
= k^{-2} {\Pi}_{_\cZ}(k)=\wein k^{-2}
\ee
where the first equality is obtained by using (\ref{eq:directflux}) and estimating  $\oenu \sim  2 \nu \sum_{k'=k}^\infty  (k')^{2} E(k')$ if  $k < \knu$ because the contribution to dissipation must be concentrated at $ k \sim \knu$. \\
Similarly, in the inverse range,  $\kmu \ll k  \ll \kin$, we must have: 
\be
\label{eq:fluxConstraint2}
-{\Pi}_{_\cZ}(k) \simeq 2 \alpha \sum_{k'=0}^k  (k')^{2} E(k')  \le  2 \alpha k^{2} \sum_{k'=0}^k E(k') 
= k^{2} \Pi_E(k)=\ein k^{2}.
\ee
Thus, neither in the large  (\ref{eq:fluxConstraint}) nor in the small (\ref{eq:fluxConstraint2})  wavenumber range we can have a constant enstrophy  flux together  with a constant energy flux: the two quantities must have a {\it dual-counter-directional cascade} (see Definition 9 in Sec. \ref{sec:definitions}).   The original version of this argument was presented in  \citep{fjortoft1953changes}. It  has been reformulated and applied to different contexts in the literature \cite{fjortoft1953changes,leith1968diffusion, eyink1993lagrangian, constantin1994effects, scott2001evolution, tran2002constraints, gkioulekas2005adouble, gkioulekas2005bdouble, alexakis2006energy, gibbon2007estimates, gkioulekas2007new}  and we will generalize it in Sec. (\ref{sec:multiple}). It indicates  that energy is transferred towards large scales  with a constant flux ${\Pi_E}(k)=-\bein $ and a  vanishing  enstrophy flux, $\lim_{ k \to 0} {\Pi_\cZ}(k)=0 $,   while at small scales enstrophy cascades forward with a constant flux ${\Pi_\cZ}(k)=-\wein $ and a vanishing  energy flux $\lim_{k \to \infty} {\Pi_E}(k)=0 $. Note that, unlike in the 3D case for which the strict forward cascade of energy is an empirical result, the simultaneous conservation of energy and enstrophy  allows us to predict the inverse cascade of energy (and a forward cascade of enstrophy) in the 2D case. It is also important to note that these arguments work because enstrophy is a sign-definite invariant.
Once the inverse energy cascade is established, one can reproduce the same phenomenological arguments put forward in K41 for the 3D case
leading to the same spectrum slope, $-5/3$, and a  different prefactor $C_K'$:  
\be
{E}(k) = C_K' \bein^{2/3} k^{-5/3}; \qquad  \kmu \ll k \ll \kin. 
\label{eq:2dkraichnan}
\ee Repeating the same reasoning done for the range dominated by the energy transfer, one can assume  that in the direct cascade
only the mean enstrophy flux plays a role, resulting in  the Batchelor-Kraichnan prediction  for the energy spectrum in the 
enstrophy cascade range:
\be
\label{eq:2denstrophy}
{E}(k) = C_B \bwein^{2/3} k^{-3} [\log(k/k_*)]^{-1/3}; \qquad \qquad  \kin \ll k \ll \knu
\ee
where the logarithmic correction $[\log(k/k_*)]^{-1/3}$ comes from
a more detailed self-consistent analysis \cite{kraichnan1971}. This correction can be neglected for most of the purpose of this discussion.
With the same arguments and by balancing the energy flux with the 
drag term,  $\bein \sim \alpha \sum_k E(k)\sim \alpha \bein^{2/3} k_\alpha^{-2/3}$  we obtain that
\be
\label{kalpha2d}
\kmu \propto \kin \Rhd^{-3/2}
\ee
where we have used the fact that in the inverse cascade regime (\ref{eq:2dkraichnan})
the large scale drag is dominated by the contribution at $k_\alpha$: $\sum_k E(k) \sim k_\alpha E(k_\alpha)$. 
Similarly, for the direct enstrophy cascade range by  balancing the enstrophy flux with the enstrophy dissipation rate, $ \bwein \sim \nu \sum_k k^4 E(k) \sim \nu k_\nu^5 E(k_\nu)$  we obtain 
\be
\label{eq:knu2D}
\knu \propto \kin \Red^{1/2}.
\ee
 Both estimates (\ref{kalpha2d}) and (\ref{eq:knu2D}) are important because they self-consistently close the assumptions made, indicating that the hypothesis to have an inverse inertial range dominated by the energy flux and a direct inertial range dominated by the enstrophy flux lead to the definition of two cut-off scales such that $ \kmu \ll \kin \ll \knu$. 
In Fig (\ref{fig:inverse}) we summarize the 2D phenomenology, by plotting the spectra for the energy (inverse) and enstrophy (direct) cascade together with the relative fluxes. 
\\
\begin{figure*}[htbp]                                                               %
\centering                                                                          %
\includegraphics*[width=1\textwidth]{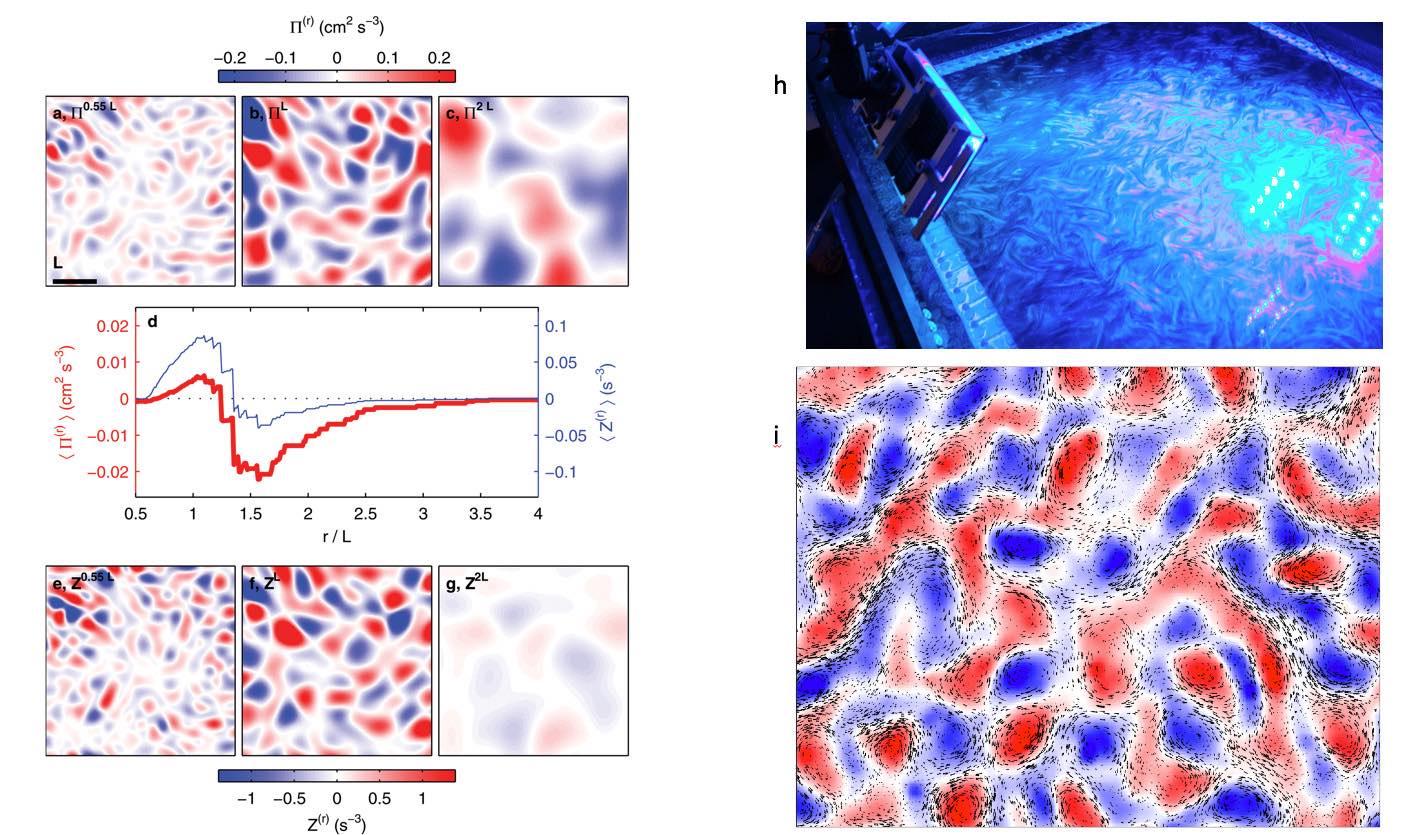}                          %
\caption{Realization of inverse energy cascade and direct enstrophy cascade in      %
experimental 2D turbulence. Energy and enstrophy fluxes from                        %
\cite{kelley2011spatiotemporal}. (a-c), Spatial variation of energy flux, here      %
indicated with the symbol $\Pi(r)$, through scales $r= 0.55 L, L, 2L$ respectively, %
where $L$ here is the injection scale. Color indicates flux intensity. All panels   %
show the same region and time. (d) Mean energy flux $\langle \Pi (r)\rangle$        %
(thick line) and enstrophy flux $\langle Z(r)\rangle$  (thin line) as a function    %
of length scale $r$ and  using a scale filtered representation                      %
(\ref{eq:SGSE}). Energy moves away from the injection scale $L$ and                 %
primarily toward large scales, whereas enstrophy moves away from $L$ and primarily  %
toward small scales. (e-g): spatial variation of enstrophy flux through the same    %
scales as in (a-c). (h): kalliroscope visualization of a 2D turbulent flow          %
\cite{ouellette2012turbulence}. (i): rendering of a 2D turbulent flow with velocity %
vectors where the tracer particles were found and vorticity shown by the colors     %
(red is positive, blue is negative). The Reynolds number (based on rms velocity     %
and  the magnet length scale of the electromagnetic forcing) is 185                 %
\cite{kelley2011spatiotemporal}. Courtesy of                                        %
N. Ouellette.} 
\label{fig:2doullette}                                                              %
\end{figure*}                                                                       %
There are two more things that we need to discuss at this point. 
First, we can not have some energy dissipation without having some enstrophy dissipation too and vice versa. Thus, along with the inverse energy  cascade some enstrophy  has to arrive at the large scales also and similarly some energy needs to  arrive at the small scales due to the enstophy direct cascade. 
In particular,     given the scaling (\ref{eq:2dkraichnan}) and (\ref{eq:2denstrophy}) the enstrophy dissipation by the drag force $\bwemu \sim \alpha \sum_k k^2 E(k) $ is dominated by the forcing scale $\kin$ and   
\be
\bwemu \propto  \, \alpha \kin^3 {E}(\kin) \propto \bwein \Rhd^{-1}
\ee
and therefore $\bwemu/\bwein \to 0$ as $\Rhd\to \infty.$
One arrives at the  same conclusion  using a high-order hypo-viscosity  $-\alpha_n (-\Delta)^{-n} \bu$,
so that most of its dissipation is limited to the small wavenumbers $k \sim k_\alpha$ then $\bwemu \sim  (k_\alpha/\kin)^2 \bwein$ where $k_\alpha\to 0$ as $\alpha_n\to 0$. 
Thus, the amplitude of the  inverse  enstrophy flux  compared to the total enstrophy injection rate goes to zero in the limit $\Rhd\to \infty$.
In the same way, by  estimating the viscous energy dissipation at the viscous scale as 
$\benu = \nu \sum_k k^2 E(k) \sim \nu k_\nu^3 E(k_\nu)$  we reach:
\be
\benu \propto  \bein \Red^{-1} \log[\Red]^{2/3} \label{eq:fwflxEn2D}
\ee
where the logarithmic correction (\ref{eq:2denstrophy}) was taken in to account. If a high-order hyper-viscosity  $-\nu_n (-\Delta)^{n} \bu$ is used to limit dissipation at  $k \sim k_\nu$ we obtain $\enu \sim  (\kin/\knu)^2 \ein$ with
$k_\nu \to \infty$  as $\nu_n\to 0$. Therefore the forward  energy flux decreases to zero in the limit $\Red \to \infty$.
A beautiful experimental verification of the  2D energy and enstrophy  cascades is provided in Fig. (\ref{fig:2doullette}).\\
Second, we need to note that in the presence of the inverse cascade the total energy in the system is dominated by the large scales and grows  with $\Rhd$ as  $\cE \sim k_\alpha E(k_\alpha) \propto \Rhd$ where we have used (\ref{kalpha2d}).  Thus, the drag coefficient due to the large scale friction $D_\alpha$ if defined as in   (\ref{eq:inversezerothlaw})  decreases with $\Rhd$. The system however remains far from a statistical  equilibrium state. The reason is that  the peak of the energy spectrum is  around $\kmu$, and the latter should be used instead of  $\kin$ in the definition (\ref{eq:inversezerothlaw}). 
Therefore, we can redefine the two drag coefficients as:  
\be
\label{eq:dragcoefficientD3}
D_\nu = \frac{  {\enu} }{\cE^{3/2} \kmu  } 
\quad \mathrm{and} \quad
D_\alpha = \frac{  {\emu} }{\cE^{3/2} \kmu  } .
\ee 
Alternatively, we can replace the energy used in the definition of the dimensionless drags as the energy $\cE_{in}$ contained in a neighborhood of $\kin$  (of fixed width independently of $\alpha$ and $\nu$) e.g.
\be
\cE_{in}  =  \sum_{k'=\kin/2}^{2\kin} {E} (k')
\ee   
and define the small and large-scale drag coefficients  as
\be
\label{eq:dragcoefficientD2}
D_\nu = \frac{  {\enu} }{ {\cE_{in}}^{3/2} \kin  } 
\quad \mathrm{and} \quad
D_\alpha = \frac{  {\emu} }{ {\cE_{in}}^{3/2} \kin  } .
\ee
The two definitions (\ref{eq:dragcoefficientD2}) and (\ref{eq:dragcoefficientD3}) are equivalent up to an order one coefficient and lead to $D_\nu \to 0$ and $D_\alpha \to \mathcal{O}(1)$ as $\Rhd,\Red\to \infty$.
Similarly we can define the enstrophy drag coefficients as:
\be
\label{eq:dragcoefficientZ2}
G_\nu = \frac{  {\wenu} }{ {\cE_{in}}^{3/2} \kin^3  } 
\quad \mathrm{and} \quad
G_\alpha = \frac{  {\wemu} }{ {\cE_{in}}^{3/2} \kin^3  } .
\ee
With these definitions and within the 2D Batchelor-Kraichnan phenomenology, we obtain that also in 2D we have two {\it dissipative anomalies} both at large and small scales, i.e. a  finite dissipation limit: 
\be
\lim_{\Rhd\to\infty} D_\alpha >0 \quad \mathrm{and} \quad \lim_{\Red\to\infty} G_\nu >0 \, .
\ee 

We now discuss the other case when the limit $\alpha \to 0$ is taken before taking the large box limit $L\to \infty$. In this case the inverse cascade that develops at early times will eventually reach the scale of the box, and energy will start to accumulate at the largest scales of system forming a condensate. The transition to the condensate regime will occur when the expression  (\ref{kalpha2d})
for  $\kmu$ becomes smaller than the minimum wavenumber $k_0 \sim 2\pi/L$  i.e. there will be a condensate if 
\be k_0/\kin >  \Rhd^{-3/2 }. \ee
The saturation of the condensate  occurs when the large scales become efficient at dissipating the energy either by the drag term  or  by  the viscous term $ \bein \sim (\alpha k_0 +\nu k_0^3) E(k_0)$. This implies that at steady state the energy of the condensate is given by
\be \cE \simeq k_0 E(k_0) \propto \frac{\ein }{\alpha + \nu k_0^2 }, \label{eq:ampcond1} \ee 
which therefore tends to $\infty$ when $\Red, \Rhd \to \infty$.
To estimate the ratio of the inverse flux to the velocity fluctuations we need to evaluate the drag coefficients at the large scales using (\ref{eq:dragcoefficientD3}) with $\kmu=k_0$ and $\cE\sim k_0E(k_0)$: 
\be D_\alpha \sim \frac{\emu}{\cE^{3/2} k_0} 
\label{eq:ampcond2}
\ee
which tends to 0 as as $\Red,\Rhd \to  \infty $ and where we must also consider that $\ein \sim \emu$. In such a case, the amplitude of the velocity fluctuations at the large scales are much larger than the injection energy rate, the system does not have any more a clear scale separation, as summarized by the right panel of Fig. (\ref{fig:transfer}), and it is close to an equilibrium state as discussed in Sec. (\ref{sec:statequil}) and by  Definition  (\ref{def13}) in Sec. (\ref{sec:definitions}). 
We note that these arguments assume a constant energy injection rate $\ein$. It was pointed out in \cite{tsang2009forced,gallet2013two} that if one uses a constant forcing amplitude, although  (\ref{eq:ampcond1}) and (\ref{eq:ampcond2}) still hold, $\ein$ depends on the condensate amplitude $k_0E(k_0)$, it might tend to 0 for $\Rhd \to \infty$ and can thus alter the  scaling properties.
In conclusion,  when one investigates the properties of the steady state inverse cascade, the limit $L\to \infty$ has to be taken before the $\Rhd\to \infty$ while when the properties of a condensate are studied the limits need to be taken in the inverse order. 

\subsubsection{Configuration space}
\begin{figure*}[htbp]                                                 
\centering                                                            
\includegraphics*[width=0.6\textwidth]{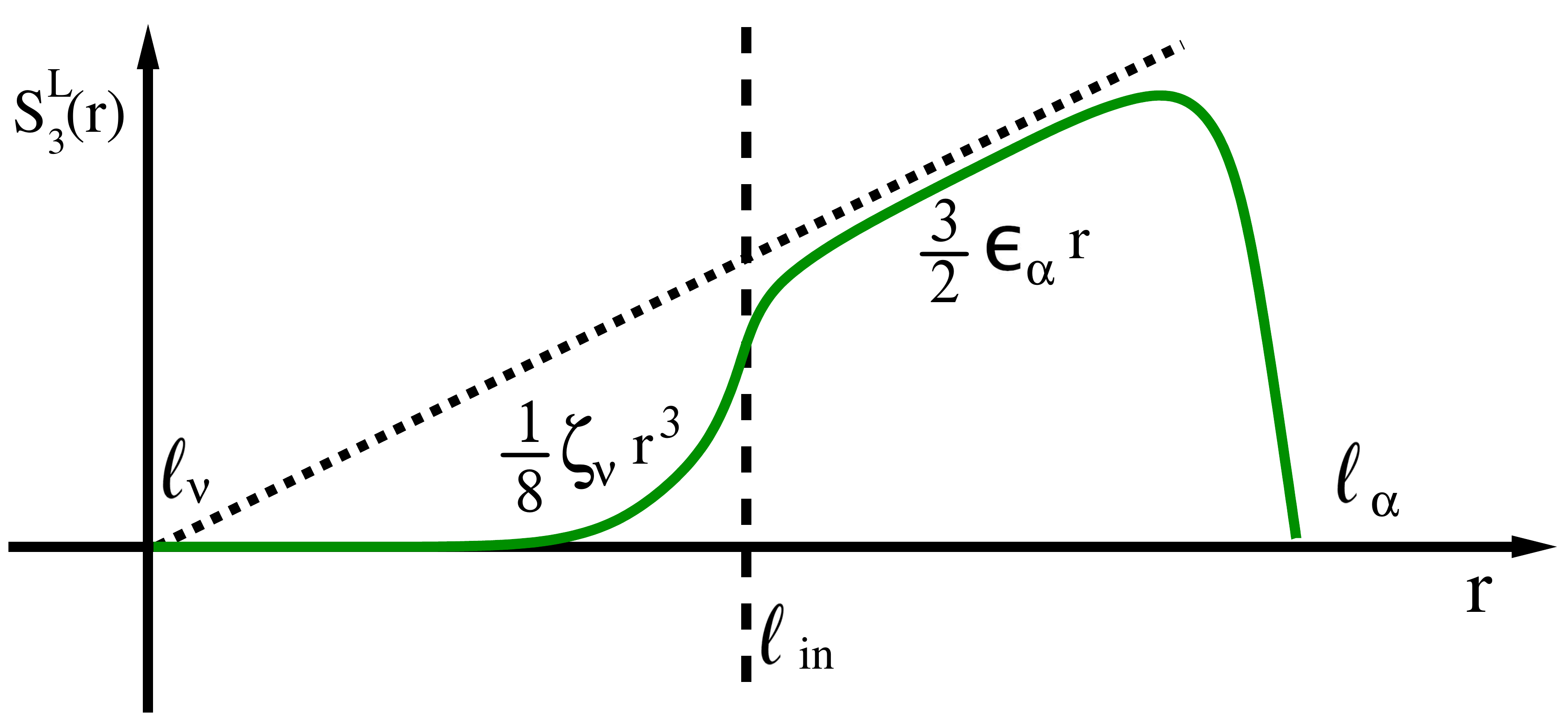}             
\caption{Lin-lin plot of the scaling for the third-order longitudinal 
structure function with the two scaling ranges: $\propto r^3$ in the  
smooth enstrophy cascade regime and $\propto r$ for the inverse       
energy cascade (\ref{eq:452d}).  } \label{fig:2dreal}                 
\end{figure*}

As done for the 3D case, one can move to  configuration space and extend the Batchelor-Kraichnan theory by assuming self similarity of 
the velocity increments PDF in the direct and inverse inertial ranges.
Concerning the inverse energy cascade range,  one obtains the same $n/3$ scaling prediction for the $n$th order  longitudinal and 
transverse velocity structure functions as in  (\ref{eq:k41SF}). In the 2D inverse cascade regime however the pre-factor of the exact 
law (\ref{eq:45inverse}) is $3/2$ instead of $4/5$. For the direct enstrophy cascade, only the enstrophy flux might enter in the description 
and  one obtains the prediction of a smooth velocity behavior \cite{bernard1999three,lindborg1999can,yakhot1999two}:
\be 
\begin{cases}
S_3^L(r) = \frac{3}{2} \ein r; \qquad \ell_{in} \ll r \ll \ell_\alpha\\
S_3^L(r) = \frac{1}{8} \bwein  r^3; \qquad \ell_\nu  \ll r \ll \ell_{in}
\label{eq:452d}
\end{cases}
\ee
which can be dimensionally generalized to all orders and to transverse increments as:
\be
\begin{cases}
S_n^L(r) \propto   \ein^{n/3} r^{n/3}; \qquad S_n^T(r) \propto  \ein^{n/3} r^{n/3} \qquad \ell_{in} \ll r \ll \ell_\alpha \\
  S_n^L(r) \propto   \bwein^{n/3} r^{n}; \qquad S_n^T(r) \propto  \bwein^{n/3} r^{n} \qquad \ell_\nu  \ll r \ll \ell_{in}.   
\end{cases}
\ee
At difference from what is observed in 3D, there is no signature of anomalous corrections to the  scaling in either the inverse energy cascade 
or the direct enstrophy cascade for what  concerns the velocity statistics, while the scaling of the vorticity field is anomalous and intermittent in the direct inertial range \cite{nam2000lagrangian,bernard2000influence,boffetta2002intermittency}. Both intermittency and the energy 
spectrum are eventually affected by the intensity of the large-scale drag  as briefly discussed at the end of section (\ref{sec:activescalarcascades}). It is important to stress that
even if the inverse energy cascade is not intermittent, it remains far from equilibrium, with a constant $O(1)$ skewness.
In Fig. (\ref{fig:2dreal}) we summarize the scaling for the longitudinal third-order structure function for the Batchelor-Kraichnan theory. \\
%
\subsection{{Split Energy Cascade}  \label{sec:split}  }                        
Next, we move to the case where we are in the presence of a  split energy transfer,
such that part of the injected energy goes to small and part to large scales. 
This situation appears in many systems, e.g. thin layers, rotating or conducting fluids as will be  examined
below. In these examples, some external mechanisms (e.g. rotation, confinement, stratification, etc...)  introduce a breaking of global symmetries and result in the system behaving  as if it is 2D at some scales and 3D at others. The phenomenology becomes much more complex, because the external mechanisms often break statistical isotropy and scale invariance.  The transition from a forward to an inverse cascade however does not only occur due to a reduction of dimensionality. As we will discuss  in Sec. (\ref{sec:Helicity}) a transition  from direct to inverse energy transfer is possible in 3D flows, under some constraints on the helical content of all Fourier modes and  in pure 2D flows in the presence of magnetic fields (see Sec. \ref{sec:MHD}). Therefore a change of dimensionality is only one of the possible mechanisms that can be responsible for the reversal of  the turbulent energy flux.   We will go back to the most important applications later in the review, here we give a general reasoning for the presence of split cascades and list their basic properties.
{We first connect the presence  of a split cascade with the  existence of
  different  physical features at small and large scales.
  As an example, let us  consider a  thin layer of vertical height $H$, where the  non-dimensional control parameter
  is the relative height $\kin H$. In this system the large scales, $\gg H$, are constrained to have a 2D behaviour while the small scales, $ \ll H$, are free to evolve in all three dimensions.} Thus, provided that some energy arrives at large scales, it  will continue to cascade inversely becoming { more and more  2D-like} as the inverse cascade proceeds. Similarly,  if some energy arrives at the small scales,  it will continue to cascade forward since the effect of confinement becomes less and less important. If both cases are feasible, and the non-linearities are efficient enough to transport energy to small and large scales,  then a fraction of the energy cascades forward and a fraction backward.
 
\begin{figure*}[htbp]                                                              
\centering                                                                         
\includegraphics*[width=0.45\textwidth]{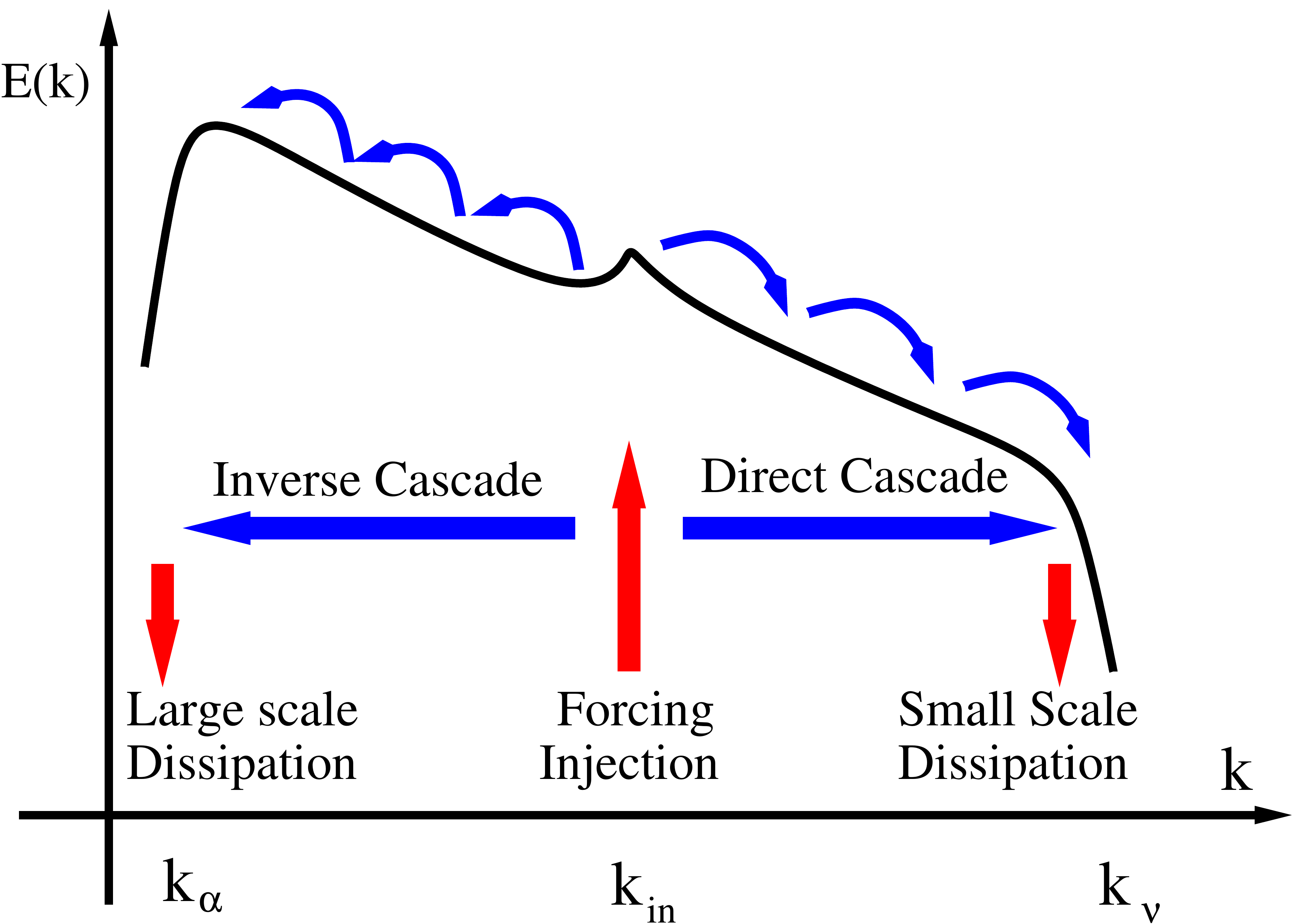}                          
\includegraphics*[width=0.45\textwidth]{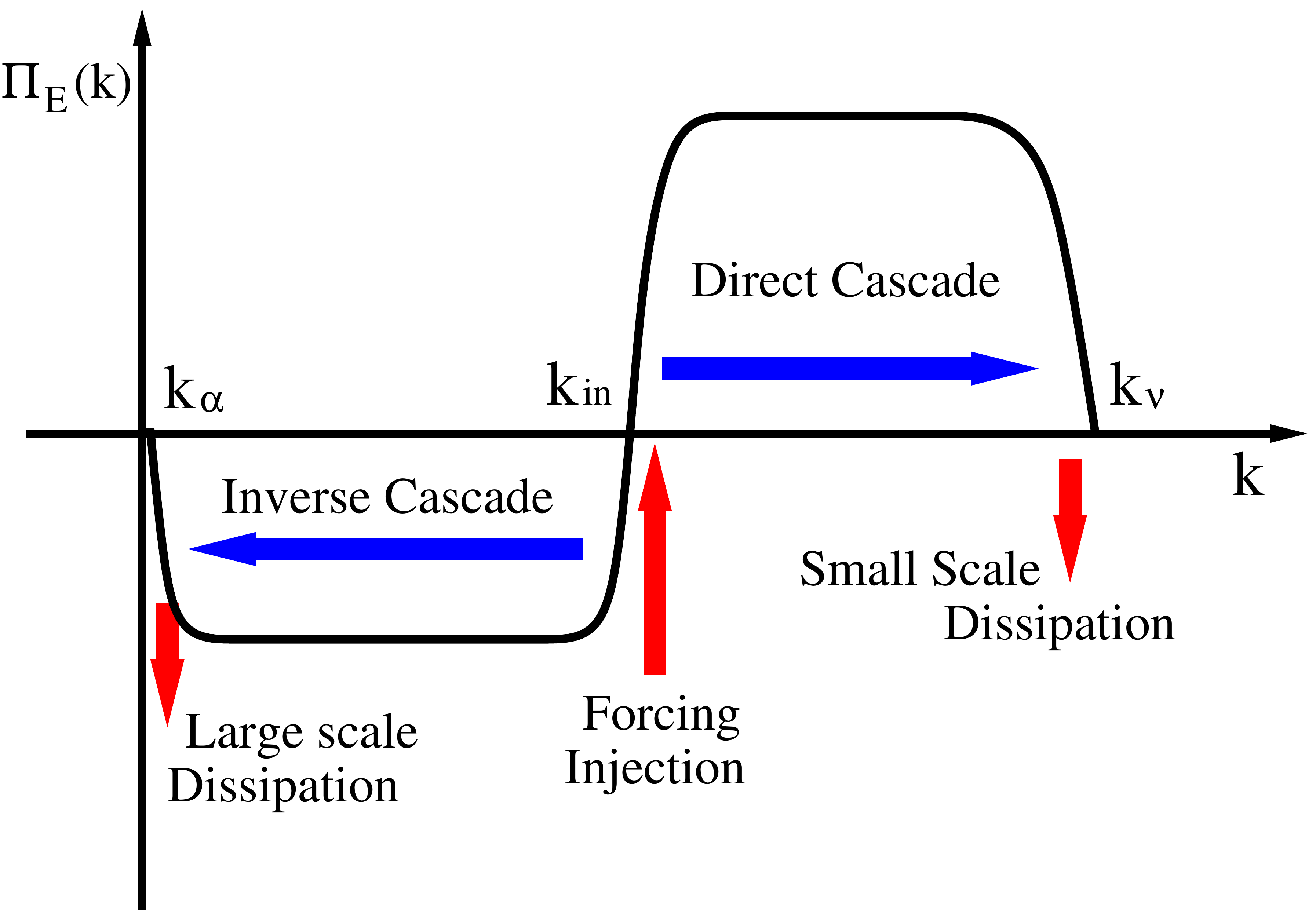}                          
\caption{ Log-log sketch of the spectra (left) and fluxes (right) in the presence  
of a split energy cascade.  }\label{fig:split}                                     
\end{figure*}

  One can therefore generalize and conjecture that if the  dynamical properties of the flow at scales $r \gg \lin $ are such to support an inverse transfer, and at the same time at  scales $r \ll \lin$ they support a  forward  cascade then a split cascade is observed, at least for some range of parameters. 
If a split cascade exists,  the flux relations are given by  the ones derived in 
(\ref{eq:inverseflux}-\ref{eq:directflux}). The difference from the 3D and 2D HIT is that neither $\oenu$ nor $\oemu$ are negligible in the limits $\Red,\Rhd\to \infty$. In this limit we expect that the energy flux
will be constant and equal to $-\oemu\ne0 $ in the range $\kmu \ll k < \kin$ while constant and equal to $\oenu\ne0 $ in the range $\kin < k \ll \knu$
\be
\begin{cases}
{\Pi}_E(k) =  -\oemu, \qquad \kmu \ll k < \kin  \\
{\Pi}_E(k) =  +\oenu, \qquad \kin < k \ll \knu .\\
\end{cases}
\ee
{The asymptotic form of the spectra at $\kmu \ll k \ll \kin$ and $\kin \\ k \ll \knu$ will be connected to the
  properties of the corresponding cascade,  but at intermediate wavenumbers different superpositions of   power laws can appear.} 
The idealized situation is qualitatively summarized in Fig. (\ref{fig:split}).
Similarly in configuration space the third order structure function $S_3(r)$ will be as the one depicted in  Fig. (\ref{fig:fluxrealspace}) where the two slopes are proportional to the two opposite fluxes.
In the presence of a split cascade, it is important to determine what is the fraction of the injected energy that cascades to the large scales  and the one that goes to small scales:
\be
\label{Q}
Q_\alpha=\frac{\oemu}{\oein},\qquad Q_\nu=1-Q_\alpha=\oenu/\oein.
\ee
The fractions $Q_\nu,Q_\alpha$ might depend on the control parameters and in some cases it is  possible that either  $Q_\alpha$ or $1-Q_\alpha$ can take very small values. In the latter situation,  one must be careful to distinguish the system
from a flow in equilibrium, as it will be  discussed in the next section.

\subsection{{Statistical  Equilibrium} \label{sec:statequil}}    
 Here, we examine the
cases where  the  flux is exactly zero or close to zero, so that the system reaches a stationary  equilibrium. 
These zero flux solutions were examined in 
\cite{lee1952td,hopf1952statistical,kraichnan1973helical} and have been the subject of many theoretical work in turbulence \cite{kraichnan1989there, bos2006dynamics, zhu2014note, zhu2014purely,ditlevsen1996cascades, ray2015thermalized}. They have provided important insight for the behavior of large-scale flows \cite{Dallas2015statistical, cameron2017effect, cameron2016large, shukla2016statistical}, they have been used to explain the bottleneck effect at small scales \cite{cichowlas2005effective, krstulovic2009cascades, frisch2008hyperviscosity} and applied to a variety of different systems \cite{herbert2014restrictedp, herbert2014nonlinear, mininni2011large, michel2017observation}. They have also been conjectured to predict the direction of a cascade based on whether the resulting distribution of energy of these equilibrium solutions peaks at small or large scales.  

Strictly speaking the equilibrium is thought to describe the truncated Euler equations defined in a finite set of wavenumbers so that there is a minimum  $k_0$ and a maximum $k_{max}$. 
The truncated Euler equations are given by
\be
\label{eq:TE}
\partial_t \bu^{<k_{max}} + {\mathbb{P}}_{k_{max}}[\bu^{<k_{max}} \cdot \bnabla \bu^{<k_{max}}  + \bnabla P] = 0
\ee 
where the projector operator  ${\mathbb{P}}_{k_{max}}$ removes all wave numbers larger than $k_{max}$ as in (\ref{eq:lowpass}) and there exists a finite IR cut-off, $k_0>0$. This idealized  system conserves exactly energy and helicity in 3D and energy and enstrophy in 2D. In the truncated Euler system all stationary fluxes are equal to zero, since ${\ein} = {\enu}={\emu}=0$. Thus, there  is no notion of a cascade in  steady state.  \\

At large times this system is said to {\it thermalize} in the sense that its degrees of freedom can be described  to a good approximation by a Gibbs-ensemble, such that the probability $\mathcal{P}$ that the system is found in a state $\bu^{<k_{max}}$ 
is given by 
\be
\label{eq:Gibbs}
\mathcal{P}[\bu^{<k_{max}}]= \frac{1}{Z} e^{-\gamma \cE - \beta \cH} \quad \mathrm{in\,\, 3D \, \, and \quad}
\mathcal{P}[\bu^{<k_{max}}]= \frac{1}{Z} e^{-\gamma \cE - \beta \cZ} \quad \mathrm{in\,\, 2D }
\ee 
where $Z$ is a normalization factor  and $\gamma$ and $\beta$ are two constants that determine the energy and the helicity of the system in 3D
or the energy and the enstrophy in 2D. 
 Using this assumption, the time averaged energy spectra can be evaluated \cite{kraichnan1973helical},  resulting  in 3D to
\be
\label{eq:thermal3D}
{E}(k) = \frac{4\pi k^2}{ \gamma^2 - \beta^2 k^2 }, \quad k_0 \le k \le k_{max}
\ee
while in 2D one obtains:
\be
\label{eq:thermal2D}
{E}(k) = \frac{2 \pi k }{ \gamma + \beta k^2 }, \quad k_0 \le  k \le k_{max}.
\ee
\begin{figure*}[htbp]                                                       
\centering                                                                  
\includegraphics*[width=0.4\textwidth]{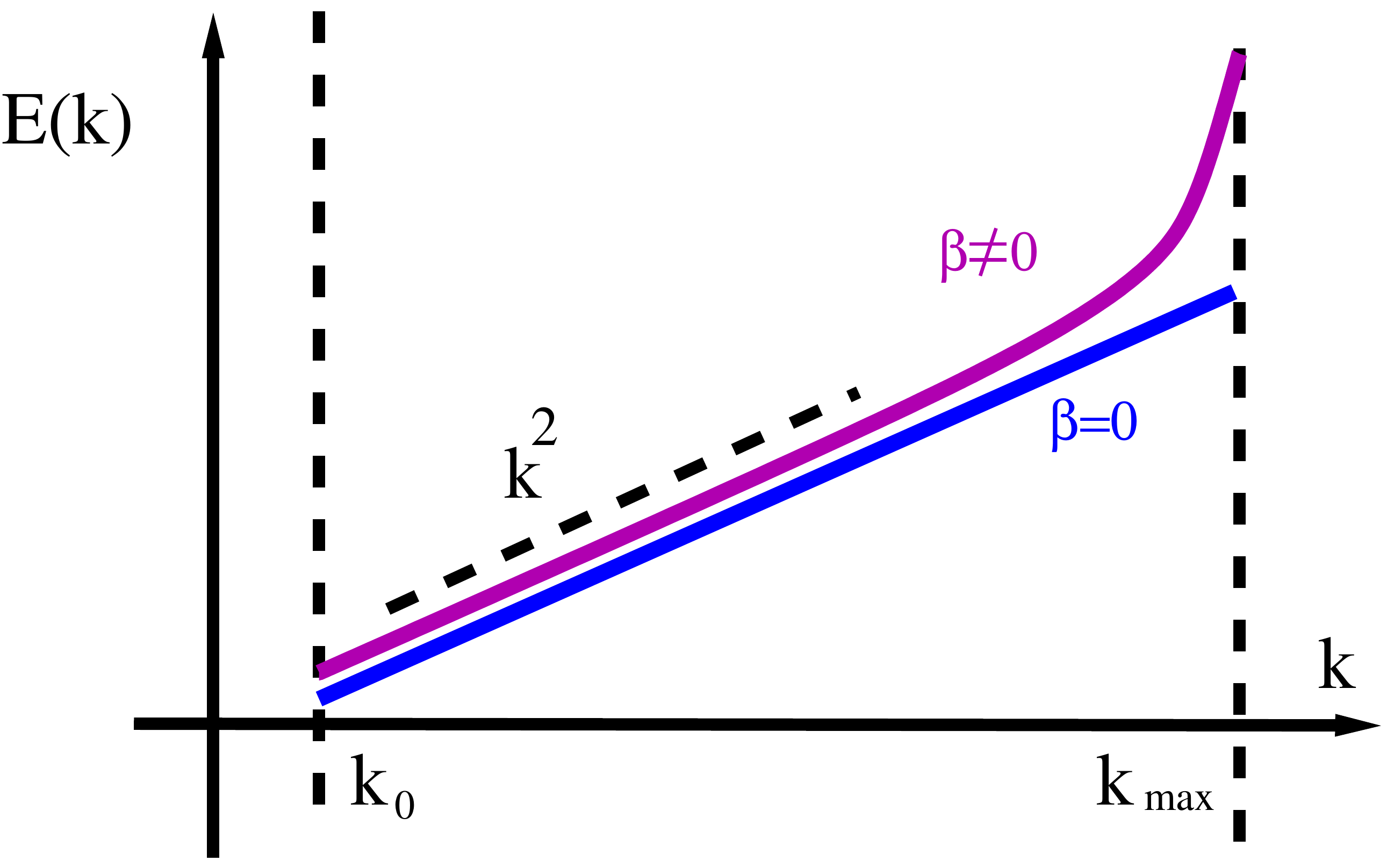} \qquad        
\includegraphics*[width=0.4\textwidth]{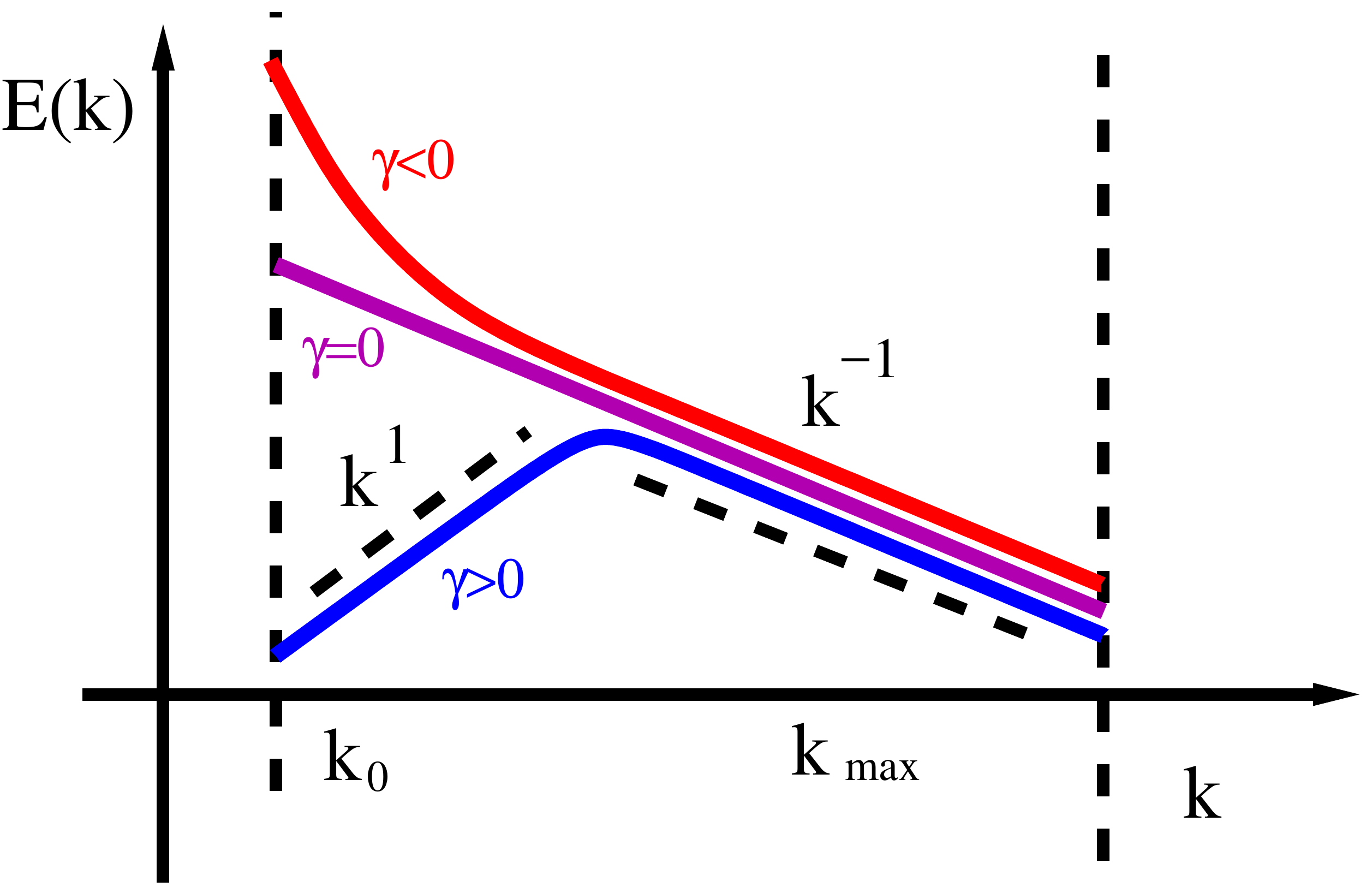}               
\caption{ Log-log sketch of the absolute equilibrium spectra in 3D (left)   
and in 2D (right).}\label{fig:inverse3}                                     
\end{figure*}                                                               
{Note that in the 3D case, the  energy spectrum always peaks at the largest wave number $k_{max}$ and for zero helicity, $\beta=0$,  the expression (\ref{eq:thermal3D}) corresponds to the equipartition of energy in all Fourier modes, which is also often called the {\it thermal} spectrum.  For non-zero helicity  the system has a singularity at $k=\gamma/\beta$ which however is larger than $k_{max} < \gamma/\beta$. 
The stronger the helicity the closer $k_{max} $ is to $ \gamma/\beta$ and energy is concentrated close to $k_{max}$. In 2D,  depending on the relative value and sign of $\gamma$ and $\beta$,  the spectrum can peak either at  $k_{max}$ or at  $k_{0}$. A singularity exists also in this case if $\gamma \beta < 0$ at $k=\sqrt{-\gamma /\beta}$  that lies outside the range of available wavenumbers  $ \sqrt{-\gamma /\beta} \notin [k_0,k_{max}]$. When $\gamma<0$ and $ \sqrt{-\gamma /\beta}$ is close (but smaller) to $k_0$ then most of the energy is concentrated at the smallest wavenumber $k_0$. This is typically referred as a condensate state and corresponds to a ``{\it negative temperature}" state because $\gamma < 0$. On the contrary when $\gamma>0$ and $ \sqrt{-\gamma /\beta}$ is close to (but larger than) $k_{max}$ corresponds to a maximal enstrophy state where most of the energy and enstrophy is concentrated close to $k_{max}$.}

\begin{figure*}[htbp]                                                                     %
\centering                                                                                %
\includegraphics*[width=0.45\textwidth,height=0.3\textwidth]{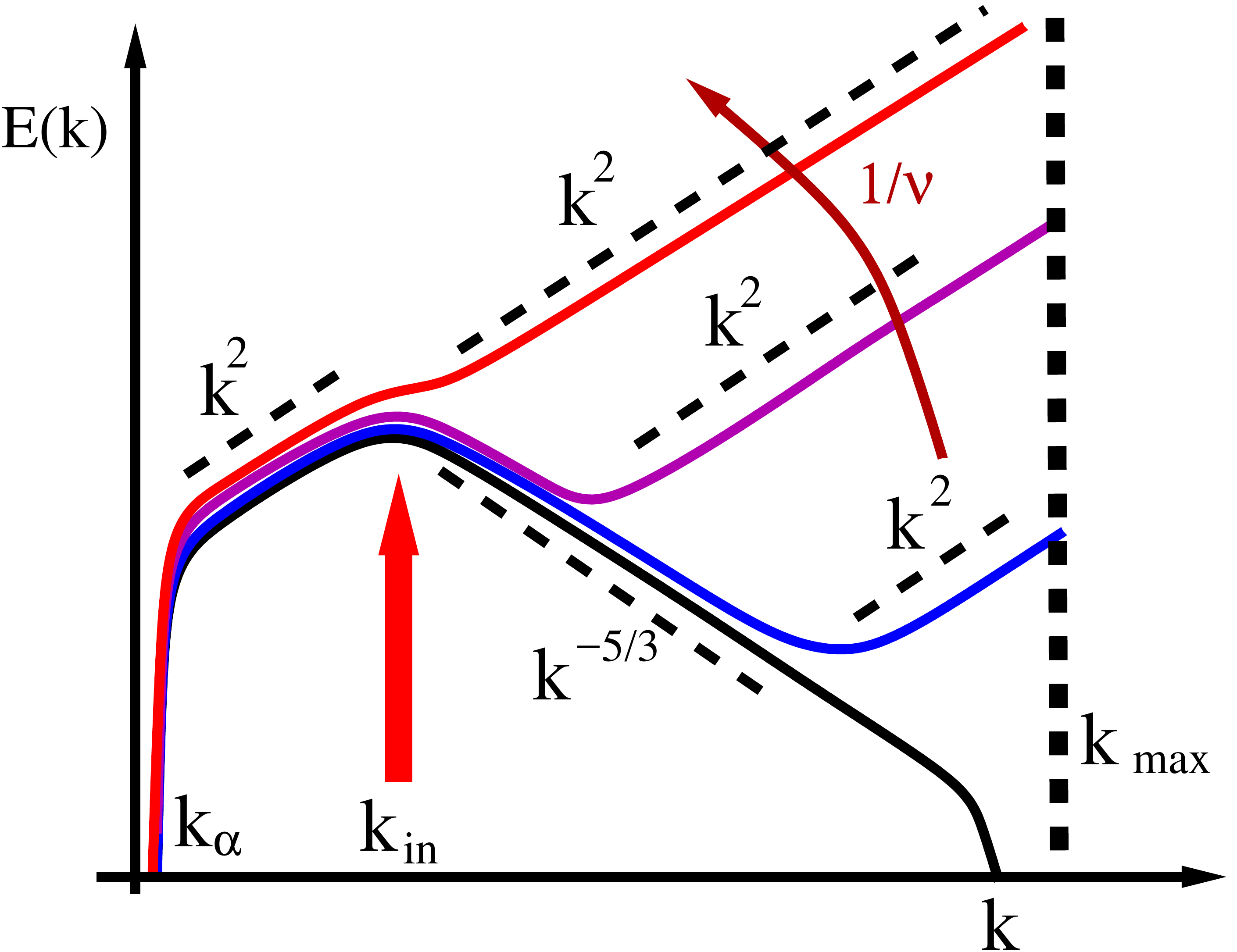}           %
\includegraphics*[width=0.45\textwidth,height=0.3\textwidth]{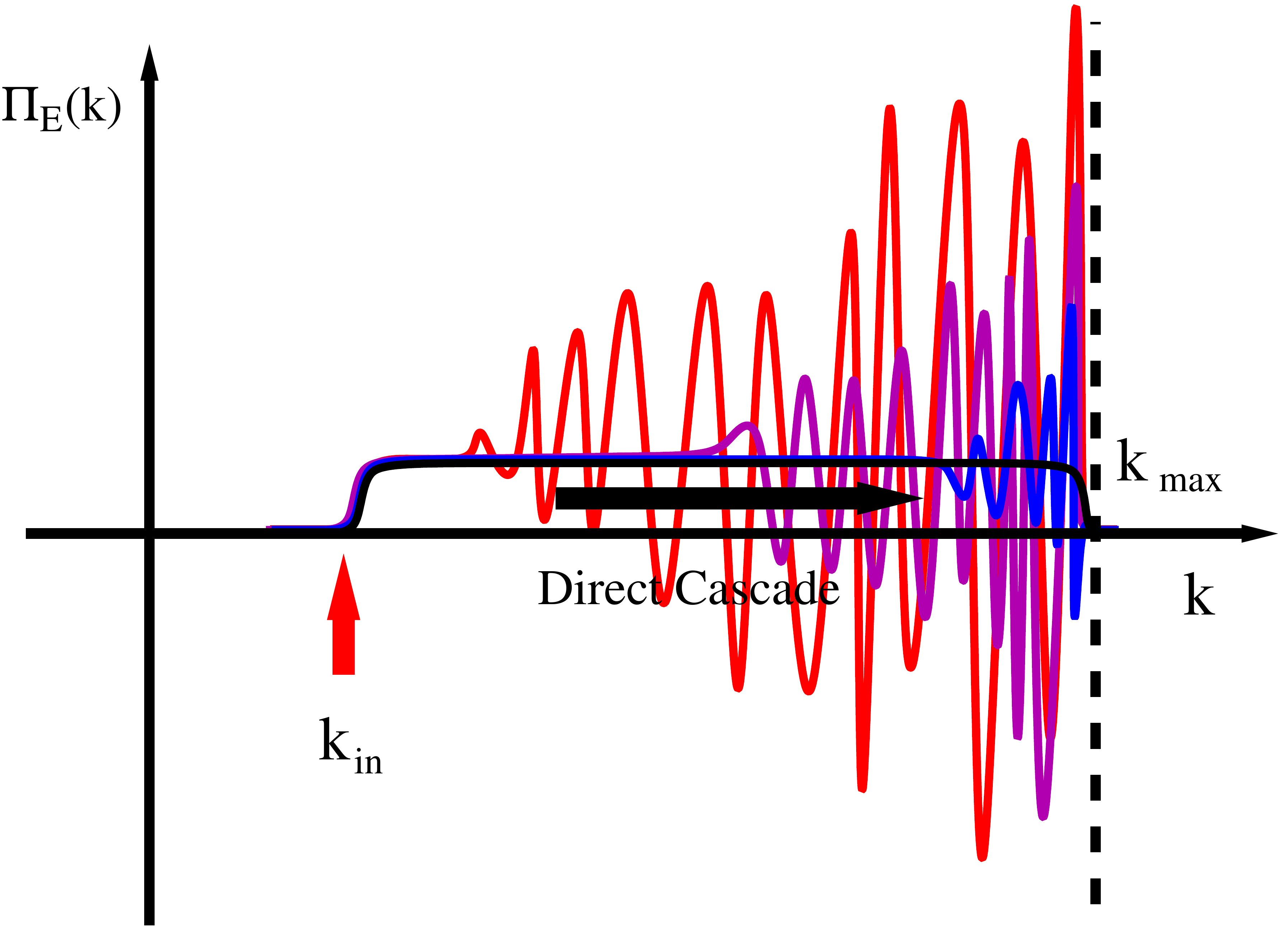}
\caption{ Left: Change of the spectrum  from a Kolmogorov to a thermal spectrum in the presence of an ultraviolet cutoff $k_{max}$ at decreasing the  viscosity           %
as indicated by the arrow. Right: the instantaneous energy flux for the same system.      %
As the thermal equilibrium is approached the  time-averaged flux (black line)         %
remains unchanged independently of $\nu$  but its scale-by-scale and temporal fluctuations become larger. }           %
\label{fig:thermalflux}                                                                   %
\end{figure*}

{The functional form of the energy spectrum at statistical equilibrium has been  used to describe the energy spectrum of 3D turbulent flows at scales larger than $\lin$ where zero flux is observed \cite{Dallas2015statistical, ray2015thermalized, bos2006dynamics, cameron2017effect}. Another case where it is applicable is the 3D NSE in the presence of a maximum wavenumber $k_{max}$ beyond which the cascade cannot proceed, either because new physics emerges (coupling with other active fields, e.g.  polymer solutions, temperature in presence of buoyancy, quantum vortices etc...) or because of the presence of a high order hyper-viscosity \cite{frisch2008hyperviscosity} or due to an artificial cut-off in numerical applications.  The last case is realized in under-resolved numerical simulations when $k_{max}$ is much smaller than the required viscous wave number $k_\nu\propto \kin \Red^{3/4}$. For values of  $\Red$ such that $\knu \ll \kmax$ the system has enough degrees of freedom to provide a good approximation to the original NSE without any UV cut-off. However as $\Red$ is increased and $\knu$ becomes of the same order or larger than $\kmax$ the system is not as efficient in dissipating energy. As a result energy will  pile up at the small scales, leading to the formation of a local quasi-equilibrium with a $ E(k) \sim C k^2$ spectrum. At very large values of $\Red$, the $k^2$  spectrum will dominate all available wavenumbers. The system then reaches a global state  close to thermal  equilibrium  where there is a sub-dominant flux of energy. This is displayed in the left panel of
  Fig. (\ref{fig:thermalflux}). Note that since the spectrum has a positive slope, the energy dissipation is still restricted to wavenumbers close to $\kmax$ thus the energy flux in the range $\kin < k \ll \kmax$ will still be  approximately constant in average ${\Pi}_E(k) ={\ein}$ and  developing strong temporal fluctuations. The thermalised wavenumbers have been shown to act as an eddy viscosity and remove energy from the part of the spectrum displaying a $k^{-5/3}$ power-law \cite{cichowlas2005effective,krstulovic2009cascades}. }
%

\begin{figure*}[htbp]
\centering
\includegraphics*[width=0.45\textwidth,height=0.3\textwidth]{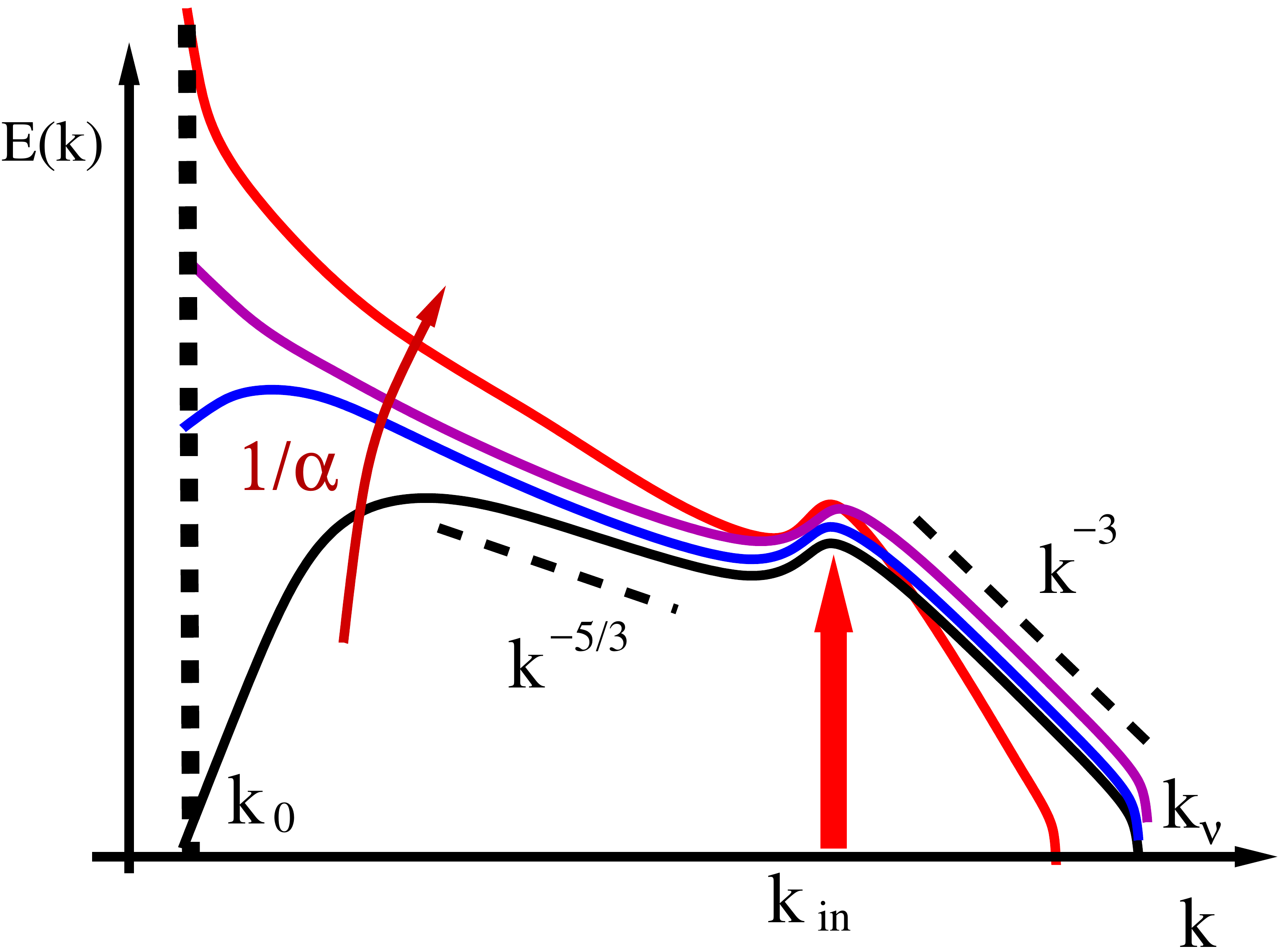}
\includegraphics*[width=0.45\textwidth,height=0.3\textwidth]{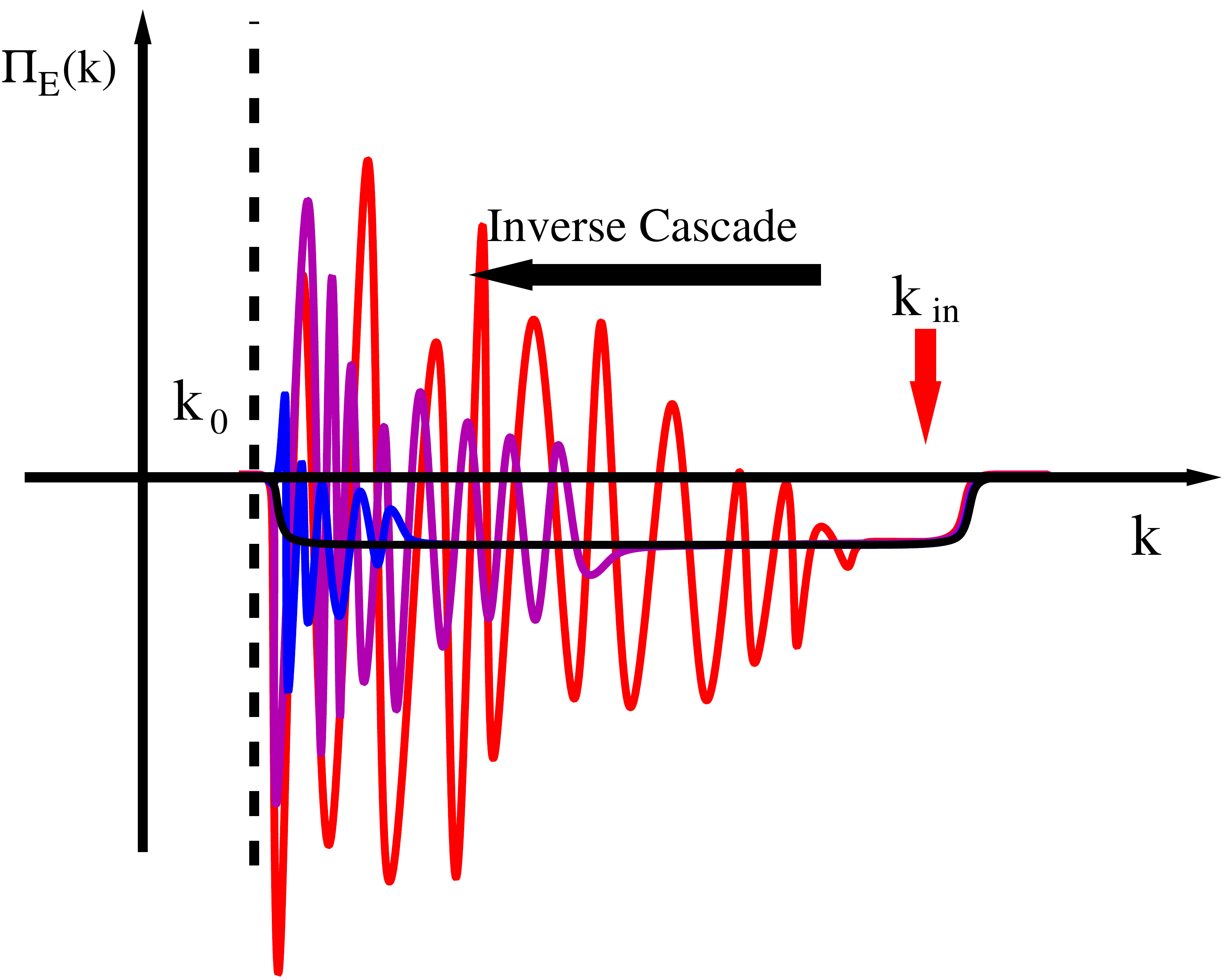}
\caption{   Left: Change of the spectrum  in a 2D inverse cascade from Kolmogorov $-5/3$ to  a condensate as the value of the drag coefficient $\alpha$ is decreased in the presence of an infrared cutoff $k_0$. Right: The instantaneous energy flux for the same system. As the thermal equilibrium is approached the time averaged flux (black line) remains unchanged independently of $\alpha$ but its scale-by-scale and temporal  fluctuations become larger.  }
\label{fig:thermalflux2}
\end{figure*}

{Another more physically motivated situation occurs in 2D when $\alpha$ in (\ref{eq:GNS}) is not large enough {(see end of section \ref{sec:2Df})}. For any finite domain if $\alpha$ is reduced beyond a certain value a similar process as in truncated 3D NSE occurs but now with the energy pilling up at the smallest wavenumbers as shown in Fig. (\ref{fig:thermalflux2}) forming a spectrum close to the one given by the thermal equilibrium (\ref{eq:thermal2D}). In this case, at scales $\ell > \lin$ the flux of energy plays a sub-dominant role and system behaves as in a quasi-equilibrium. There have been many attempts to describe and predict these condensates by means of statistical physics \cite{robert1991statistical,laurie2014universal, bouchet2012statistical, shukla2016statistical, herbert2014restrictede, herbert2014restrictedp,  laurie2014universal, farrell2007structure,bouchet2008simpler, venaille2009statistical, venaille2011solvable, bouchet2012statistical,  falkovich2016interaction, woillez2017theoretical, frishman2017culmination, frishman2017jets,cha1,cha2}.}

\noindent
Despite that the energy flux can be constant in the range $\kin \ll k \ll \kmax$ in 3D  (or in the range $k_0 \ll k \ll \kin$ in 2D) the energy transfer  does not qualify as a cascade because the limit (\ref{eq:zerothlaw}) is not satisfied. Constancy of the flux implies the existence of a cascade only if a finite dissipation limit at infinite $\Red$ and $\Rhd$ exists. For the 3D case the total dissipation can be estimated as $\propto \nu \sum_{k <k_{max}} k^2E(k)  \sim \nu   k_{max}^3 E(k_{max})  \sim \nu   k_{max}^5$ and it must balance  $\oein$.  This implies that the energy scales like ${\cE} \sim k_{max}E(k_{max}) \propto {\ein} \nu^{-1}\kmax^{-2}$, and the drag coefficient (\ref{eq:dragcoefficient}) goes to zero at large Reynolds number: $$D_\nu  \sim \Red^{-3/2}.$$ 

\noindent
The spectrum shape at absolute equilibrium has also been used to predict the cascade direction  for the full NSE.
According to this argument,  in 3D energy cascades always forward because the inertial dynamics wants to approach a state where  most of the energy is concentrated in the large wavenumbers while in 2D the energy cascade can be also inverse.  However, we note that this argument is not enough to make an exact statement. As it will be shown for the case of rotating turbulence in Sec. (\ref{sec:Rotation})
a transition from direct to a  split cascade can exist despite of the fact that the statistical equilibrium does not depend on the rotation rate. 

Finally, we  note that having ${\Pi}_E(k)=0$ or equivalently ${S}^{(3)}_L(r)=0$ does not imply that the system is in an equilibrium state. As shown in Sec. (\ref{sec:2Dstratification}) there are cases for which the net flux can be zero because of the balancing among interactions that move consistently energy up and others that move it down.  As a result, the energy distribution is not the one predicted by (\ref{eq:Gibbs}) and the spectrum is different from the equilibrium one.


\subsection{Multiple invariants \label{sec:multiple}}

Finally we would like to examine turbulent flows for which multiple invariants  that cascade backward or forward coexist. This is in fact the most common scenario  in many real applications.  We will see in section (\ref{sec:Helicity}) that in 3D both energy and helicity (which is non sign-definite) cascade forward, while in 2D  (see section \ref{sec:2DinvCsd}) the presence of a second  positive-definite invariant, the enstrophy, leads to an inverse cascade of energy and to a direct transfer of enstrophy. We remind that the arguments that constrain energy to cascade inversely in 2D rely on the relation among  the energy and enstrophy  spectrum leading to the  inequality (\ref{eq:fluxConstraint}).  In 3D,  Helicity is sign indefinite and a similar inequality cannot   be constructed. Nevertheless, we show  in Sec. (\ref{sec:inversehomochiral}) that when helicity becomes sign-definite,  a scenario similar to the 2D case can develop in 3D also.

In the following, we  generalize the arguments made in (\ref{eq:fluxConstraint},\ref{eq:fluxConstraint2}) for two arbitrary invariants,  to demonstrate the necessary conditions required to restrict the cascades directions. The same conclusions can be obtained by following different arguments as shown in  \cite{fjortoft1953changes,leith1968diffusion, eyink1993lagrangian, constantin1994effects, scott2001evolution, tran2002constraints, gkioulekas2005adouble, gkioulekas2005bdouble, alexakis2006energy, gibbon2007estimates, gkioulekas2007new}. Consider two invariants $\cA$ ,$\cB$ with energy spectra $E_\cA(k),E_\cB(k)$, viscous dissipation spectra $\nu k^2E_\cA(k), \nu k^2E_\cB(k)$ and drag-force spectra $\alpha E_\cA(k), \alpha E_\cB(k)$ respectively. In general we can not determine the direction of cascade of the two invariants unless there are special relations between the two. More precisely for the forward cascade we can state that 
\be
\mathrm{if \, there \, exist \,} c> 0, \, n>0 \mathrm{\,such\, that\, } \, |E_\cA(k) | \le c \, k^{-n}  E_\cB(k) \mathrm{\, then\,} \cA \mathrm{\, cannot\, cascade\, forward}.
\label{eq:fluxConstraintAB1}
\ee 
The sketch of the proof follows the same arguments discussed in section (\ref{sec:2DinvCsd}). Suppose  that both invariants cascade forward and the viscous dissipation is dominated by the large wavenumbers.  We can then show that for any $k$ such that $\kin< k\ll \knu$  the flux $\Pi_\cA(k)$ of the invariant $\cA$ must satisfy:
\be
\label{eq:fluxConstraintAB1b}
\left| {\Pi}_{_\cA}(k) \right| \simeq 
\nu          \left| \sum_{k'=k}^\infty  (k')^{2}   E_\cA(k') \right|  \le 
\nu k^{-n}   \left| \sum_{k'=k}^\infty  (k')^{2+n} E_\cA(k') \right|  \le
\nu k^{-n} c        \sum_{k'=k}^\infty  (k')^{2}   E_\cB(k')          \simeq 
c \iepsilon_{\nu \cB} k^{-n}. 
\ee
where $ \iepsilon_{\nu \cB} $ is the viscous dissipation rate of the invariant $\cB$.  
Therefore, unless $ \iepsilon_{\nu \cB} $ is diverging the flux of $\cA$  can not be constant (either positive or negative)
and has to decrease at least as $k^{-n}$. 
Similarly for the inverse cascade we can state that for any $k<\kin$
\be
\mathrm{if \, there \, exist \,} c> 0, \, n>0 \mathrm{\, such\,  that\, }  |E_\cB(k)| \le c k^n E_\cA(k) \mathrm{\, then\,} \cB \mathrm{\, cannot\, cascade\, inversely}.
    \label{eq:fluxConstraintAB2}
\ee 
This again can be shown by considering  a wavenumber $k$ in the inertial range of the inverse cascade $k_\alpha  \ll k \ll \kin $.
 Then  if the dissipation by the drag force of the two invariants is dominated by the small wavenumbers, the 
flux $\Pi_\cB(k)$  satisfies 
\be
\label{eq:fluxConstraintAB2b}
\left| {\Pi}_{_\cB}(k) \right| \simeq 
\alpha                    \left| \sum_{k'=0}^k             E_\cB(k') \right| \le 
\alpha k^{ n}             \left| \sum_{k'=0}^k  (k')^{-n}  E_\cB(k') \right| \le
\alpha k^{ n}          c         \sum_{k'=0}^k             E_\cA(k')         \simeq 
                       c  \iepsilon_{\alpha \cA} k^{n}. 
\ee 
  Note that for (\ref{eq:fluxConstraintAB1}) to be true, $ E_\cB(k)$ has to be strictly positive, while for (\ref{eq:fluxConstraintAB2})  to hold, we need $ E_\cA(k)$ to be strictly positive. For the energy and enstrophy 2D cascades we have $\cB=\cZ $ and $\cA=\cE$, which means $ k^2 E_\cE(k)  =  E_\cZ(k)$, and both conditions (\ref{eq:fluxConstraintAB1}-\ref{eq:fluxConstraintAB2}) are true as an equality. Therefore, we can conclude that enstrophy cascades forward while energy  cascades backward. For 3D turbulence, we have helicity $\cB=\cH$ and energy $\cA=\cE$ but $\cH$ is not sign definite and only (\ref{eq:fluxConstraintAB2}) is true, $| \, E_\cH(k) | \le  k E_\cE(k) $, while (\ref{eq:fluxConstraintAB1}) does not hold.
  Therefore, while we can conclude that $\cH$ can not have an inverse cascade we can not restrict the direction of the energy cascade. If however $\cH$ is a sign-definite quantity (as for the cases  discussed in Sec. \ref{sec:inversehomochiral}) both  (\ref{eq:fluxConstraintAB1}) and (\ref{eq:fluxConstraintAB2}) are satisfied and energy cascades backward even in a 3D isotropic turbulence.  

Given the transfer direction  of each invariant one  might ask the question which invariant determines the spectrum.
Clearly when they cascade in opposite directions it is the cascading invariant that determines the spectra in the relevant scales. When however there is a coexistence of cascades at the same scales there is no trivial answer and
the spectrum can be the outcome of either cascading process. It is also possible that the spectrum is a superposition of power-laws and each cascade determines the energy spectrum at different range of scales.  We note furthermore that coexistence of cascades at the same scales does not imply that both invariants cascade in the same direction. If there is energy injection at multiple scales a forward and an inverse cascade can overlap. This has been recently demonstrated in \cite{cencini2011nonlinear} for 2D turbulence forced at both small and large scales.

It is also important to discuss the notion of the direction of cascade for  a sign-indefinite quantity like the helicity. For a positive quantity, the direction of a cascade is simply given by the sign of the flux indicating a forward cascade if it is positive and an inverse cascade if it is negative. While the flux of helicity can be uniquely defined (see section \ref{sec:Helicity}) its interpretation can be confusing. Positive values of $\Pi_{H}$ imply that the non-linearities decrease helicity in the large scales and increase helicity in the small scales. Therefore if the helicity is positive at all scales this can be interpreted as transfer of helicity from large scales to small, and thus a forward cascade. If however the helicity is negative at all scales the large scale helicity will increase in magnitude at the large scales and thus it signifies an inverse cascade (of negative helicity). It is even harder to give an interpretation in terms of a cascade when the helicity is not of the same sign at all scales. 

\subsection{Definitions  \label{sec:definitions}  }                  

In this section we proceed by giving general definitions of the terminology used in the rest of this paper. Our aim is to provide a glossary and the basic tools that distinguish in a precise and clean manner different situations such as split cascades, strictly forward/inverse cascades and systems in statistical equilibrium. The definitions presented here are tailored for the needs of this review and are not meant to be complete. We restrict the discussion to the  energy cascade case to have a specific example, however they can be easily replicated with appropriate modifications for the cascade of any ideal quadratic invariant. Furthermore we have neglected some other aspects that are typically (and justifiably) attributed to cascades as for example locality of interactions. Putting further constraints on our definitions to capture also these aspects goes beyond the main objectives of this review, and would make the definitions more complex without contributing to the oncoming discussions. However, we remind the reader that this is certainly not the end of the story.

\subsubsection{Forward, Inverse \& Split cascades\label{sec:CascadeDef} }   
We start by giving a precise definition of what we mean by an out-of-equilibrium energy cascade. This is important in order to distinguish it from other cases where the energy transfer from the injection scale to the dissipative scale is not the dominant mechanism in the system, as for the case of quasi-equilibrium flows. 

As usual, we consider a flow forced at a wavenumber $\kin$, in a box domain of size $L$  that results in a minimum wavenumber $k_0=2\pi/L$. For quasi-2D systems (thin layers, fast rotating flows, etc...) for which one particular direction (vertical) plays a special role, we will use  $H$ to  denote the box size in this direction while $L$ will stand for the box size in the remaining two (horizontal) directions. Furthermore we also consider the possibility that  a maximum wave number $k_{max}$ exists in the system, which will be considered infinite in most cases,  except for some special instances where it is interesting to consider the effects introduced by an ultraviolet cut-off, i.e. for under-resolved numerical simulations. With this in mind, we  give the following definition for an energy cascading system. 

\begin{mydef}
\label{def1}{\bf Out-of-equilibrium energy cascade.}\\
\noindent
If ${\cE}_{in}$ is the energy in a neighborhood of $\kin$ 
\be
{\cE}_{in}  =  \sum_{k'=k_1}^{k_2} {E} (k')
\label{enin}
\ee
with $ k_{1} < \kin < k_{2} $ we will say that a system is 
{\bf \it energy cascading} 
(or {\bf \it   out of equilibrium } )
if for any fixed wavenumbers $k_{1},k_{2}$ the limit
\be
D_{in} \equiv \lim_{\scriptsize \begin{array}{c} \Red\to\infty,  \\ \Rhd \to \infty \end{array} }  
\lim_{\scriptsize \begin{array}{c}   k_0 \to 0, \\ k_{max} \to \infty \end{array}} 
\frac{ \oein } 
{{\cE}_{in}^{3/2} \kin } > 0
\label{def:out}
\ee
exists and is finite. Otherwise if the limit in (\ref{def:out}) is zero
we will say that the system is in an {\it asymptotically equilibrium state}.  
\end{mydef}
\noindent
The condition guaranties that (for fixed energy injection rate) the energy in the neighbourhood of $\kin$ remains bounded.  Since the energy dissipation rate inside this neighbourhood $(k_1,k_2)$ is smaller than $\cE_{in} (\alpha + \nu k_2^2) $, it becomes vanishingly small in the limit of $\Red,\Rhd \to \infty$. This implies  that the injected energy has to be transferred either to small or large scales to be dissipated. 

The order of the limits considered in this definition are crucial. The quantities used are time averaged implying that first the long time limit is taken then the limits $L\to\infty$ and $\kmax\to \infty$ are taken and finally the large $\Red,\Rhd$ are taken. Reverting the ordering of these limits can lead to different results. We discuss the case that the time limit is taken last at the end of this subsection.

Given that a system is energy cascading we can define the relative forward/inverse energy cascade rate as follows
\begin{mydef}{\bf Relative inverse/forward energy-cascade rates.} \\
\label{def2}
\noindent In an energy cascading system the relative rate of forward cascade $Q_\nu$  is defined as
\be
\lim_{\scriptsize \begin{array}{c} \Red\to\infty, \\ \Rhd \to \infty \end{array} }  
\lim_{\scriptsize \begin{array}{c}   k_0 \to 0, \\ k_{max} \to \infty \end{array}} 
\frac{ \oenu }{\oein} =Q_\nu >0,
\label{def:realtive1}
\ee
while the relative rate of inverse cascade $Q_\alpha$ is given by
\be
\lim_{\scriptsize  \begin{array}{c} \Red\to\infty, \\ \Rhd \to \infty \end{array} } 
\lim_{\scriptsize \begin{array}{c}   L \to \infty, \\ k_{max} \to \infty \end{array}}
\frac{ \oemu }{\oein} =Q_\alpha = 1-Q_\nu.
\label{def:relative2}
\ee
\end{mydef}
%
\noindent
Definitions (\ref{def2}) is based on global injection and dissipation rates of energy (or other invariants). Given however the relations (\ref{eq:inverseflux}-\ref{eq:directflux}) that equate the dissipation rates with the energy flux we can equivalently  say: 
\begin{mydef}{\bf Forward and inverse cascades. }\\
\label{def3}
\noindent   We will say that in a cascading system there is a  forward energy cascade
  for $\kin\ll k \ll \knu$ if in this range the energy flux  satisfies:
\be
\lim_{\scriptsize \begin{array}{c} \Red\to\infty,  \\ \Rhd \to \infty \end{array} }  
\lim_{\scriptsize \begin{array}{c}   k_0 \to 0, \\ k_{max} \to \infty \end{array}} 
\frac{\Pi_E(k)}{\ein}=Q_\nu> 0
\label{def:forwardinverse}
\ee
and similarly we will say that in a cascading system there is an inverse energy cascade for $\kmu\ll k \ll \kin$ if the energy flux in this range satisfies: 
\be
\lim_{\scriptsize \begin{array}{c} \Red\to\infty,  \\ \Rhd \to \infty \end{array} }  
\lim_{\scriptsize \begin{array}{c}   k_0 \to 0, \\ k_{max} \to \infty \end{array}} 
-\frac{\Pi_E(k)}{\ein}=Q_\alpha> 0
\label{def:forwardinverse2}
\ee
\end{mydef}
\noindent
Note that the definitions (\ref{def2}-\ref{def3}) 
are equivalent for  defining $Q_\nu,Q_\alpha$, and that both require the system to be energy cascading based on Def. (\ref{def1}) as a prerequisite.  
However, the first ones make a  statement for the whole system while the second define a cascade in a particular (inertial) range of scales.

Constancy of flux only implies that the non-linearities conserve energy and that forcing and dissipative effects can be neglected in the inertial range. 
 Given the above definitions  we can now proceed to further distinguish what we call a strictly forward, strictly inverse or split cascade regime.
\begin{mydef}{\bf Strictly forward (direct), strictly inverse (backward) and split cascades.}\\
\label{def4}
\noindent
An energy cascading system will be called  
    {\bf \it strictly  forward (direct) energy cascading              } if $Q_\nu  =1$,  
    {\bf \it strictly inverse (backwards)  energy cascading           } if $Q_\nu  =0$, 
or {\bf \it split energy cascading system                             } if $0< Q_\nu < 1$.
\noindent
\end{mydef}
%
We  stress one more time the importance of the limiting procedures 
for the split cascade. This is because close to a transition point the amplitudes of $Q_\nu$ or $Q_\alpha$ 
can be very small and it is important to distinguish them from any transient transfer caused, e.g., by large-scale instabilities \cite{frisch1987large,sulem1989generation}.

Finally it is also worth considering the case that the long time limit is taken last.
In this case the system (particular in the presence of an inverse cascade)
does not necessarily reach a steady state, and many of the results
derived so far (as the equality of energy injection with energy dissipation) are no longer valid.
It is worth, however, considering it because it corresponds to the case of many numerical simulations.
Due to computational costs numerical experiment are run up to a time such that
all small scale quantities have reached a quasi-steady state while the large-scale quantities are
still growing, without the spectrum reaching the smallest wavenumber. The amplitude of the 
inverse cascade is then not measured by the dissipation rate due to the drag term (which in most cases is not even used) 
but by the rate that the inversely cascading quantity increases in time. This method is computationally
less expensive and it has been shown in some cases to lead to the same conclusions with the steady 
state experiments \cite{benavides2017critical}.

\subsubsection{ Partial Fluxes and bidirectional cascades\label{sec:fluxes} }    
In many systems discussed in this review,  there are multiple processes that cascade energy or other invariants in different directions. In many cases these processes can be {\it rigorously isolated} so that each one leads to a partial flux energy  defined  as follows.

\begin{mydef}{\bf Partial Fluxes.}\label{def5}\\
\noindent We will say that the total flux $\Pi(k)$ of a given invariant 
can be decomposed in $N$ partial fluxes $\Pi_i(k)$ (for $i=1,\dots,N$) if $\Pi(k)$ can be written as: 
\be 
\Pi(k) = \sum_{i=1}^N \Pi_i(k)
\ee 
and each $\Pi_i(k)$ is (a) constant for $k$ in the inertial range, 
(b)  conserves globally the invariant
\be 
 \lim_{k\to\infty } \Pi_i(k) =\lim_{k\to 0} \Pi_i(k)  = 0.
\ee 
and (c)  remains finite in the limit
\be 
 \lim_{\scriptsize \begin{array}{c} \Red\to\infty, \\ \Rhd \to \infty \end{array} } \Pi_i(k)/\ein  \ne 0.
\ee 
\noindent
\end{mydef}
Given that there can be partial fluxes, 
we need to distinguish whether they are unidirectional or bidirectional. 

\begin{mydef}{\bf Bidirectional  cascades.  }\\
\label{def6}
  \noindent
  We will say that a system displays a bidirectional cascade in a given
range of scales if the total flux $\Pi(k)$ of a given invariant can be decomposed  in multiple partial fluxes $\Pi(k)$ such that at least two of them have opposite signs. 
\noindent
\end{mydef}
This implies that in a bidirectional cascade there exist different processes that transfer energy forward and backward in the same range of scales.  We note that sometimes in the literature the name ``bidirectional" is used in a different way, to denote a split cascade (as given in definition \ref{def4}).
\begin{mydef}{\bf Flux-less or flux-loop cascades.}\\
\label{def7}
  \noindent
  We will say that a system displays a  flux-less cascade  or a flux-loop cascade  at a given range of scales if it develops a bidirectional cascade and  the total flux vanishes, $\Pi(k)=0$. 
\noindent
\end{mydef}
We note that a flux-less cascade is very different from a statistical equilibrium state that also has zero global flux. Examples of flux-less cascade can be found in secs. (\ref{sec:2Dstratification} and \ref{sec:finitehorizontalsize}).

\subsubsection{ Dual and Multiple cascades \label{sec:Dual}       }   
We now proceed in examining the case for which there are multiple invariants.
\begin{mydef}{\bf Dual/multiple cascades.}\\
\label{def8}
\noindent We will say that a system displays a dual/multiple cascade  if 
it has two/multiple cascading invariants. 
\noindent
\end{mydef}
This is in fact the case in almost all systems that we are going to
meet in this 
review. Note that both invariants need to satisfy a condition equivalent to def.(\ref{def1}) 
for the system to display a cascade. Dual cascading systems can be further categorized in two
branches co-directional and counter-directional dual cascades.
\begin{mydef}{\bf Co-directional and counter-directional dual cascades.  }\\
  \noindent
  \label{def9}
  We will further distinguish  a dual   cascade to be  
 a   co-directional dual cascade  if the two transfers
   are in the same direction 
and a  counter-directional dual cascade if one invariant cascades strictly forward  and the other strictly inversely.
\noindent
\end{mydef}
For example 2D turbulence displays a counter-directional dual energy-enstrophy cascade while 3D turbulence  displays a co-directional energy-helicity  dual cascade.

\subsubsection{Finite domains and Asymptotically  Equilibrium states }  

Some of the systems described so far do not lead to a finite dissipation in the limit of infinite $\Red,\Rhd$. They correspond to systems that are dominated by thermal fluctuations and we refer  to them as reaching asymptotically a statistical equilibrium state. Here by statistical equilibrium we mean the case of a  system with many degrees of freedom, with ergodic phase-space trajectories and without  energy injection or dissipation as the ones described in Sec. (\ref{sec:statequil}). If the equilibrium distribution can  be approximated by a Gibbs ensemble, we will also refer to them as
a thermal equilibrium state. Strictly speaking,  in the turbulent configuration examined  here, there is always some injection and  dissipation  mechanisms and thus  do not fall in this category. Nonetheless, in certain limits, the flow  can asymptotically approach a state very close to a statistical  equilibrium. 

This happens in particular when finite domains are considered ($k_0>0$) or when a maximum wavenumber  $k_{max}$ exists. To investigate such systems we can not consider the limits $k_0 \to 0$ and $\kmax \to \infty$ in  definition \ref{def1}. Then we need to make the following distinctions.  If both $k_0$ and $k_{max}$ are finite then the system will always reach asymptotically an equilibrium state at the limit $\Red,\Rhd\to \infty$ because there cannot be an energy cascade either to small or large scales. 
For a system with $k_0$ finite and $k_{max}\to \infty$, or $k_{max}$ finite and $k_0 \to 0$ we will say it reaches an asymptotically equilibrium state based on the following definition.
\begin{mydef}{\bf Asymptotically  statistical equilibrium state in finite domains }\\
\noindent
\label{def10}
For a system of finite size ($k_0>0$ and $\kmax\to \infty$ ) we will say that it approaches {\it asymptotically a statistical energy equilibrium state} if for any $k_2>\kin$ and defining $\cE_{in}=\sum_{k_0}^{k_2}E(k)$
\be
D_{in} \equiv \lim_{\scriptsize \begin{array}{c} \Red\to\infty,  \\ \Rhd \to \infty \end{array} }  
\frac{ \oein } {{\cE_{in}}^{3/2} \kin } = 0.
\label{Definition10}
\ee
Similarly for a system with  $k_{max} < \infty$ (and $k_0=0$)  we will say that it approaches {\it asymptotically a statistical energy equilibrium state} if for any $k_1 < \kin$ and defining $\cE_{in}=\sum_{k_1}^{\kmax}E(k)$
\be
D_{in} \equiv \lim_{\scriptsize \begin{array}{c} \Red\to\infty,  \\ \Rhd \to \infty \end{array} }   
\frac{ \oein } {{\cE_{in}}^{3/2} \kin } = 0.
\label{Definition10b}
\ee
\end{mydef} 
We note again that the system being at equilibrium does not imply necessarily being at thermal equilibrium.
\begin{mydef}{\bf Asymptotically thermal equilibrium state. }\label{def11}\\
\noindent
An asymptotically statistical equilibrium will be called a { thermal equilibrium} if its dynamical variables can be approximated by Gibbs statistics (\ref{eq:Gibbs}). 
\end{mydef}
In many situations only a range of wave numbers is approaching an equilibrium state.
This happens when all invariants cascade in the same direction leaving the scales at opposite direction to have zero flux.
The latter  range  can possibly be described asymptotically by a statistical equilibrium. 
\begin{mydef}{\bf Statistical equilibrium range. }\\
\label{def12}
\noindent
A range of wavenumbers will be called to be in a { asymptotically statistical equilibrium} if 
the fluxes of all ideal invariants, and all their partial fluxes are zero in the limit $\Red\to\infty$ and $\Rhd\to \infty$. 
If the dynamical variables in this range follow Gibbs statistics we will call this range as being in an  {  asymptotically  thermal equilibrium}.
\end{mydef}
\noindent 
An example of a statistical equilibrium range are expected to be the scales larger than the forcing scale in 3D HIT discussed in Sec. (\ref{sec:direct}). 
Note that a system does not have to be in a statistical equilibrium state to have a statistical equilibrium range as the latter provides information for only a particular range of wavenumbers.

A special case in a statistical equilibrium state is when a condensate is formed for systems with $k_0$ finite. 
In such a case,   energy is also concentrated in the lowest wavenumber. These condensate states appear in many different situations but not all can be described by a statistical equilibrium state. As it will be discussed in Sec. (\ref{sec:Rotation}) the amplitude of the condensate can be
so large that it can alter the dynamics leading to a flux-less cascade rather than to a  statistical equilibrium. 
It is thus  hard to give a definition that covers all possible cases. Here, we decided to define a condensate as follows: 
\begin{mydef}{\bf Condensate.  }\\ 
\label{def13}
\noindent
We will say that  system with an inverse transfer approaches {\it asymptotically a condensate state} in the limit $\Red,\Rhd\to\infty$  if for fixed $L$ most of its energy (or an other invariant) is concentrated in the smallest wavenumbers $k_0 \propto 1/L$ of the  system.  
\end{mydef} 
In this situation, there is not any more clear scale separation, as shown by the right panel of Fig. (\ref{fig:transfer}). We will distinguish the condensates in two classes: equilibrium condensates and flux-loop condensates that are distinguished based on the scaling properties of their amplitude 
with $\Red,\Rhd$ as follows. 

\begin{mydef}{\bf Equilibrium condensate.  }\\ 
\label{def14}
\noindent 
For fixed $\ein$ and $k_0$ we will say that an inverse cascading system approaches {\it asymptotically an equilibrium condensate state } if its amplitude $k_0E(k_0)$ increases without bound as $\Red,\Rhd\to\infty$ {\it i.e.} 
\be \lim_{\Rhd\to\infty} \lim_{\Red\to\infty}  \frac{ \ein  }{ E(k_0)^{3/2} k_0^{5/2} } = 0. \label{Definition14} \ee 
\end{mydef}
\noindent 
while  
\begin{mydef}{\bf Flux-loop condensate.  }\label{def15}\\
\noindent
we will say that an inverse cascading system approaches {\it asymptotically a flux-loop condensate state}  if its amplitude $k_0E(k_0)$ has a finite value in the limit $\Red,\Rhd\to\infty$ {\it i.e.}
\be \lim_{\Rhd\to\infty} \lim_{\Red\to\infty}  \frac{  \ein  }{ E(k_0)^{3/2} k_0^{5/2} } > 0. \label{Definition15} \ee
\end{mydef}
\noindent 
In the first case most of the dissipation occurs at the condensate scale $k_0$ while in the second case since $k_0E(k_0)$ remains finite, one must imagine that long-range Fourier interactions  become efficient enough to redirect the energy back to the small scales where it can be dissipated. Thus, the first system is  closer and closer to a statistical equilibrium state, while the second develops also a forward non-local flux triggered by the presence of the large-scale condensate and it is not in a statistical equilibrium state. The amplitude of the condensate in the latter case depends on other control parameters like the Rossby number for rotating turbulence (see  sec.\ref{sec:finitethin} and \ref{sec:finitehorizontalsize} for two example cases). One must also notice that the presence of strong large-scale structures can influence the way energy is injected in the system, with a non-trivial feedback on the dependency of  $\ein$
on the control parameters. 
%

\subsection{Classification of Cascade transitions  }  
\label{sec:Classification}
The idea that a turbulent system may change the direction of the energy transfer (or of other quantities)  as a system parameter varies is not new. Theoretical ideas were introduced  already by  \cite{Frisch1976crossover,Fournier1978ddimensional, Giuliani2002critical, Yakhot2001mean,lvov2002quasi}  where a phase transition from forward to inverse cascade was conjectured to occur at a critical fractal dimension $d^*$ with $2<d^*<3$. These ideas were pursued with the use of EDQNM models \cite{Fournier1978ddimensional} and shell models \cite{Giuliani2002critical,bell1977nonlinear} while more recently fractal dimensional turbulence has been modelled by decimated Navier-Stokes models \cite{buzzicotti2016intermittency,Frisch2012fractal,lanotte2016vortex} that will be also further discussed in Sec.
(\ref{sec:fractal}). In all of these examples, the transition  was considered to be from a forward to an inverse cascade or vice versa. The physics of the transition however can be more  rich if
   the possibility of having a split cascade as an intermediate step is also considered.  In this section we  generalize and provide a list of different scenarios for all possible  cascade (phase) transitions. 

   More precisely we discuss the way a system can pass from a state that cascades energy forward to a state that cascades energy inversely and vice versa as a critical control parameter $\mu$ is varied.  We note that as discussed in the previous subsection,  to be able to clearly state the direction of a cascade the large $\Red,\Rhd,L,k_{max}$ limit needs to be made first, before any limiting procedure for $\mu$ takes place. This guaranties that as $\mu$ is varied the flow remains in a turbulent state.  Without any loss of generality  we consider a case for which at $\mu=0$ the system cascades energy strictly forward $Q_\alpha=0$,  while for $\mu \to \infty$ the system cascades energy strictly inversely $Q_\alpha=1$. For example, $\mu$ can be the rotation rate, or the thickness of the fluid volume or the intensity of the external magnetic field, to cite just a few key control parameters that will be discussed later.  We want thus to find out how the value of the ratio $Q_\alpha$
   (\ref{def:relative2}) is varying away from 0 as $\mu$ is changed.

\begin{figure*}[h!]                                                                                         %
\centering                                                                                                  %
\includegraphics[width=0.9\textwidth]{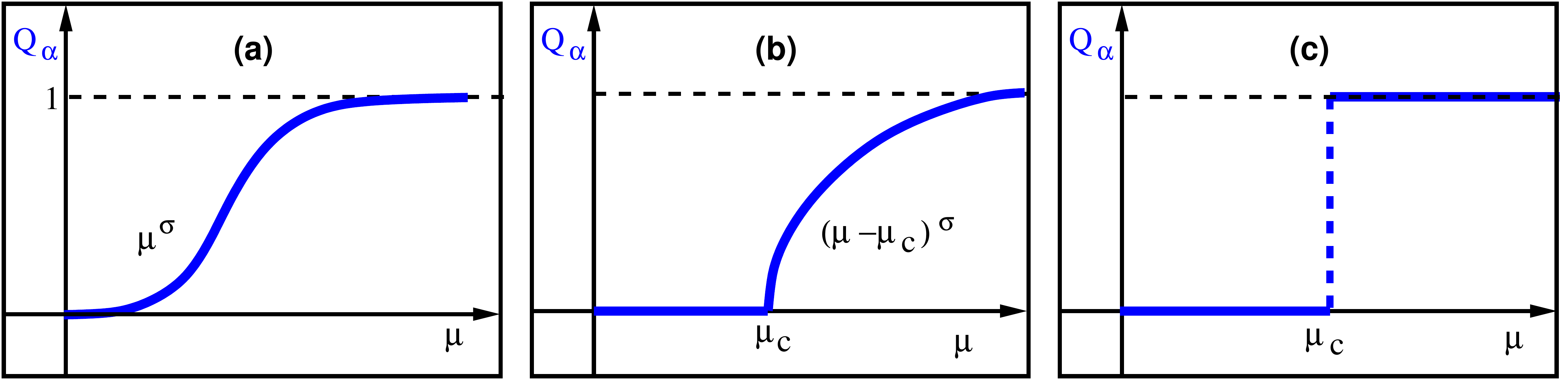}                                                  %
\caption{                                                                                                   %
 Three possible ways for the transition from  a strictly forward cascade, $Q_\alpha=0$, to a split cascade, %
 $Q_\alpha>0$, or to an inverse cascade, $Q_\alpha=1$, as the control parameter $\mu$ is varied:            %
(a)  smooth way,                                                                                            %
(b) continuous non-smooth (second order phase transition),                                                  %
(c) discontinuous (first order phase transition). }                                                         %
\label{fig:classification}                                                                                  %
\end{figure*}

\begin{figure*}[h!]                                                                         %
\centering                                                                                  %
\includegraphics[width=0.99\textwidth]{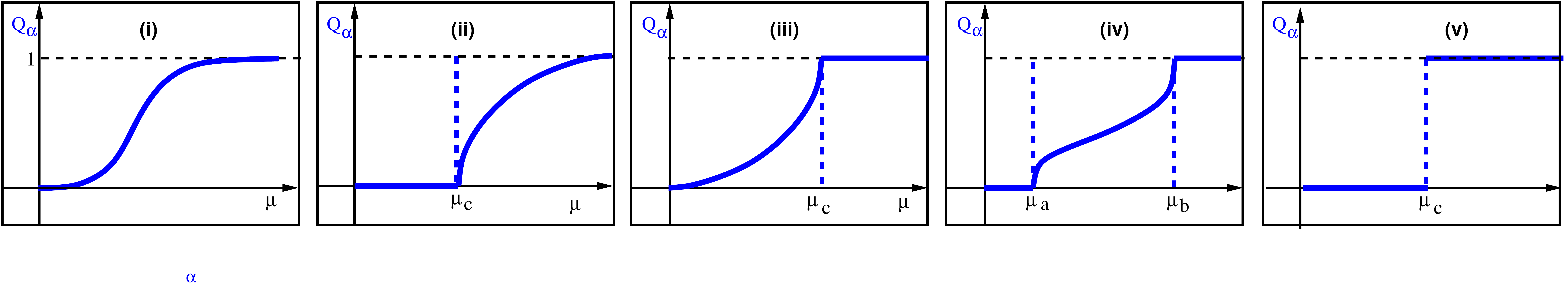}                                 %
\caption{    All possible transitions from a strictly forward cascade as a function of      %
         the control parameter $\mu$  for  monotonically increasing $Q_\alpha$.}            %
\label{fig:classII}                                                                         %
\end{figure*}

Three possible scenarios can be envisaged for the way this transition takes place that are depicted  in  figure  \ref{fig:classification}.
%
\begin{itemize}
\item (a) The transition happens in a smooth way.  
The relative amplitude of the inverse cascade $Q_\alpha$ increases smoothly  as the control parameter is increased from $Q_\alpha=0$ at $\mu=0$. 
Note that in this case there is always a forward and an inverse cascade for any finite value of $\mu>0$. 
The behavior of $Q_\alpha$ close to zero will then possibly scale like a power-law $Q_\alpha \propto \mu^{\sigma}$ for which the exponent 
$\sigma$ needs to be determined. We will refer to these transitions as smooth. %

Examples of systems that display smooth transitions are given by the advection of a passive scalar in confined geometries 
(see Sec. \ref{compadvection}) and,  possibly, rotating stratified turbulence (see Sec. \ref{sec:RotationStratification} and Fig. \ref{fig:RSphase2}).
\item (b) The system transitions from a strictly forward cascade to a split cascade 
at a critical value of $\mu=\mu_c$. The amplitude of the inverse cascade remains exactly zero for all $\mu < \mu_c$
and beyond the critical point $\mu_c$ the system starts to cascade inversely with discontinuous or diverging derivatives $dQ_\alpha/d\mu$
at the critical point $\mu_c$,  much like a super-critical instability or a second order phase transition. 
Close to the critical point we thus expect the behavior $Q_\alpha \propto (\mu-\mu_c)^\sigma$ where $\sigma$ 
is again an exponent that needs to be determined. Using a terminology borrowed from equilibrium statistical mechanics we will refer to these transitions as continuous 2nd order
(phase) transitions.

Examples of systems that display this kind of transitions are given by thin layer flows (Sec. \ref{sec:Thin}), rotating flows  
(Sec. \ref{sec:Rotation}), stratified flows (Sec. \ref{sec:stratification}), MHD (Secs. \ref{sec:MHD}, \ref{sec:MHD2D}) and  possibly rotating stratified turbulence (Sec.
\ref{sec:RotationStratification}).

\item (c) The transition happens from a strictly forward cascade $Q_\alpha=0$ to a strictly inverse cascade $Q_\alpha=1$ at a critical value of $\mu=\mu_c$ in a discontinuous way, much like a sub-critical instability or a first-order phase transition. In this case thus the system goes from a forward to an inverse cascade without passing through a split cascade. In such a scenario it is possible that a hysteresis curve is present such that  for a range of parameters $\mu_h < \mu <\mu_c$ the system can find itself in a forward cascading state or an inversely cascading state depending on initial conditions. We will refer to these transitions as discontinuous or 1st-order (phase) transition.
 
Examples of systems that display discontinues transitions are given by the helical decomposition models in Sec. (\ref{sec:discontinuous}), the fractal dimension transition discussed in \cite{Giuliani2002critical} and the Surface-Quasi-Geostrophic system at changing the {\it locality} parameter $z$ as discussed in Sec.(\ref{sec:activescalarcascades}).
 
\end{itemize}
For the second transition scenario, the  critical points do not need to be limited to the transition from  a strictly inverse cascade to a split cascade but we can have a second critical point for the  transition from a  strictly forward cascade to a split cascade, and in general there might be  a multiple of such a critical points.  For example, assuming  that $Q_\alpha$ monotonically increases with $\mu$ the transition from a forward to an inverse cascade can happen in one of the five more general ways summarized in  figure \ref{fig:classII}. 
%
In many systems there is more than  one critical parameter (e.g. box-height, rotation and stratification \dots) that we here denote as $\mu_1,\mu_2,\dots$. In such cases, the critical points become critical lines or surfaces.  Following the example of  phase transitions in statistical mechanics one can anticipate that it is also possible that depending on the path followed  in the parameter space ($\mu_1,\mu_2,\dots$)  one can pass through a  critical point or not. The simple description thus given in Fig. (\ref{fig:classII}) for the case where only one critical parameter exists might be more complex in general. 

Finally we note that in a finite domain and in  the presence of an inverse cascade a condensate is formed and a different classification must  be made. It can not be  based on $Q_\alpha$ that is not a good measure of the condensate state. {It is better to use  the amplitude of the condensate itself, quantified for example as the energy contained in the smallest wavenumbers normalized by the total energy as 
\be 
Q_0 =  
         \lim_{\Rhd\to \infty}  \lim_{\Red\to \infty}
         \frac{E(k_0)k_0}{\cE} .
\ee 
The ratio $Q_0$ is small in the absence of a condensate 
(it becomes zero in the limit $k_0\to 0$) while it takes a finite value if a condensate is present. 
As before the transition from a condensate to a no-condensate state can be either smooth, 1st order or 2nd order.
It is important to stress that a (phase) transition for the normalized energy flux $Q_\alpha$ in the infinite domain limit and the (phase) transition for the amplitude of the condensate $Q_0$
are two different situations. Only few examples have been investigated 
of transitions to the condensate state  in rotating turbulence \cite{alexakis2015rotatingTG,yokoyama2017hysteretic,seshasayanan2018condensates} and are discussed in Sec. (\ref{sec:finitehorizontalsize}).


\section{Applications \label{sec:applications}} 
In this section we review different systems that deviate from the statistically homogeneous, isotropic and mirror-symmetric flows and examine the effect of these deviations. The choice of the systems examined is based on their simplicity and direct applicability in natural systems.   We start by discussing the case of homogeneous and isotropic turbulence with helicity input, i.e. the case where only mirror symmetry is broken in the NSE. We consider this case important because it allows us to further dissect the turbulent dynamics in terms of the helical-Fourier decomposition. Then, we proceed to discuss a set of real turbulent configurations appearing in many applications in nature and engineering as:  thin layers, turbulence under rotation, turbulence in the presence of stable and unstable stratification, 2D and 3D MHD; and we end with the transfer properties of passive and active scalars, advected by a turbulent flow.
\subsection{ {Helicity}   \label{sec:Helicity}     } 
We start by reviewing the statistical properties of helical homogeneous and isotropic turbulence. Helicity is an invariant  that is connected to  the topological structure of vortex lines, which can be studied and characterized in terms of their twist, writhe and linking number  by numerical, experimental and theoretical tools \cite{scheeler2014helicity, kedia2016weaving, laing2015conservation, moffatt1992helicity, kerr2015simulated} (see Figure \ref{fig:helicity} for an experimental example of  an helical vortex evolution).
\begin{figure*}[htbp]                                                   %
\centering                                                              %
\includegraphics*[height=0.35\textwidth]{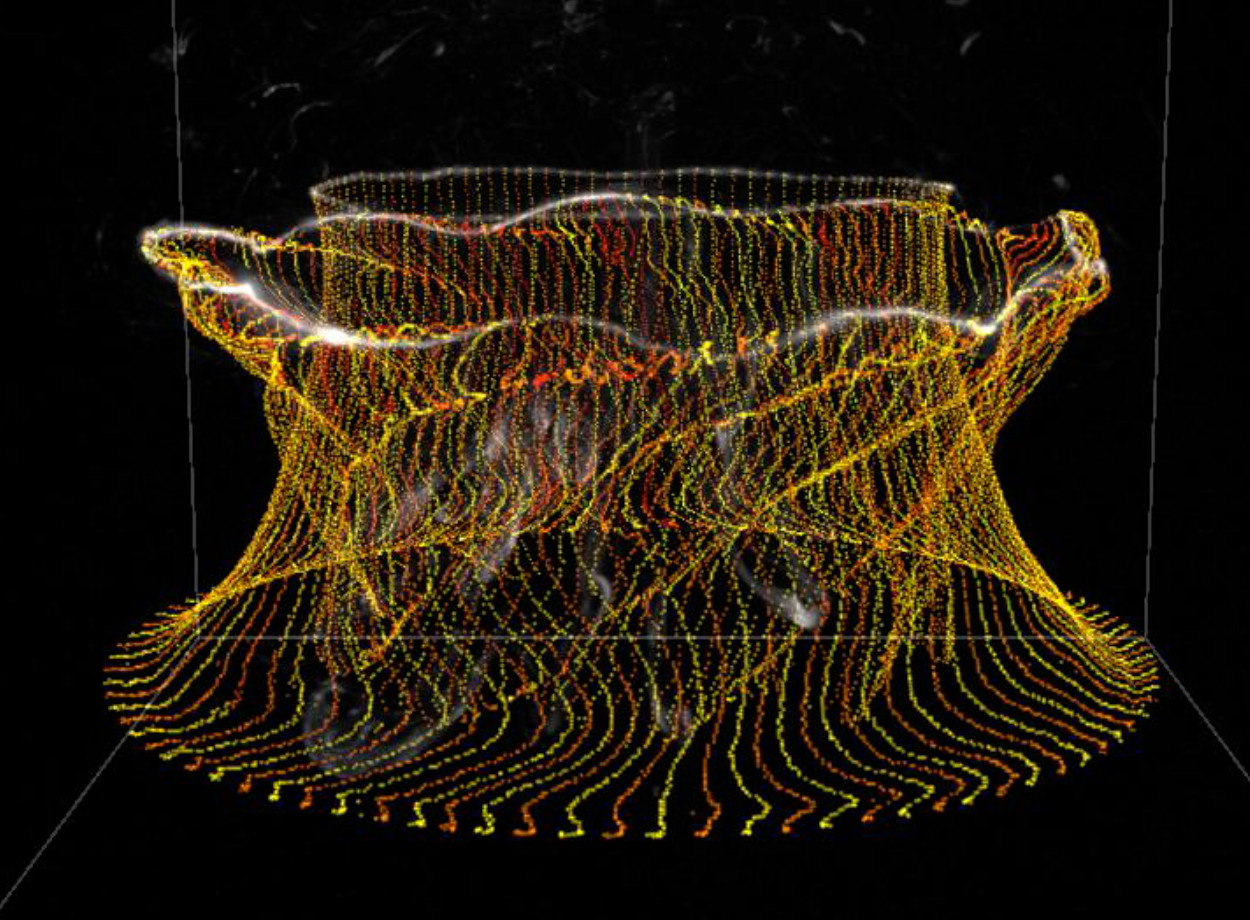}    %
\includegraphics*[height=0.35\textwidth]{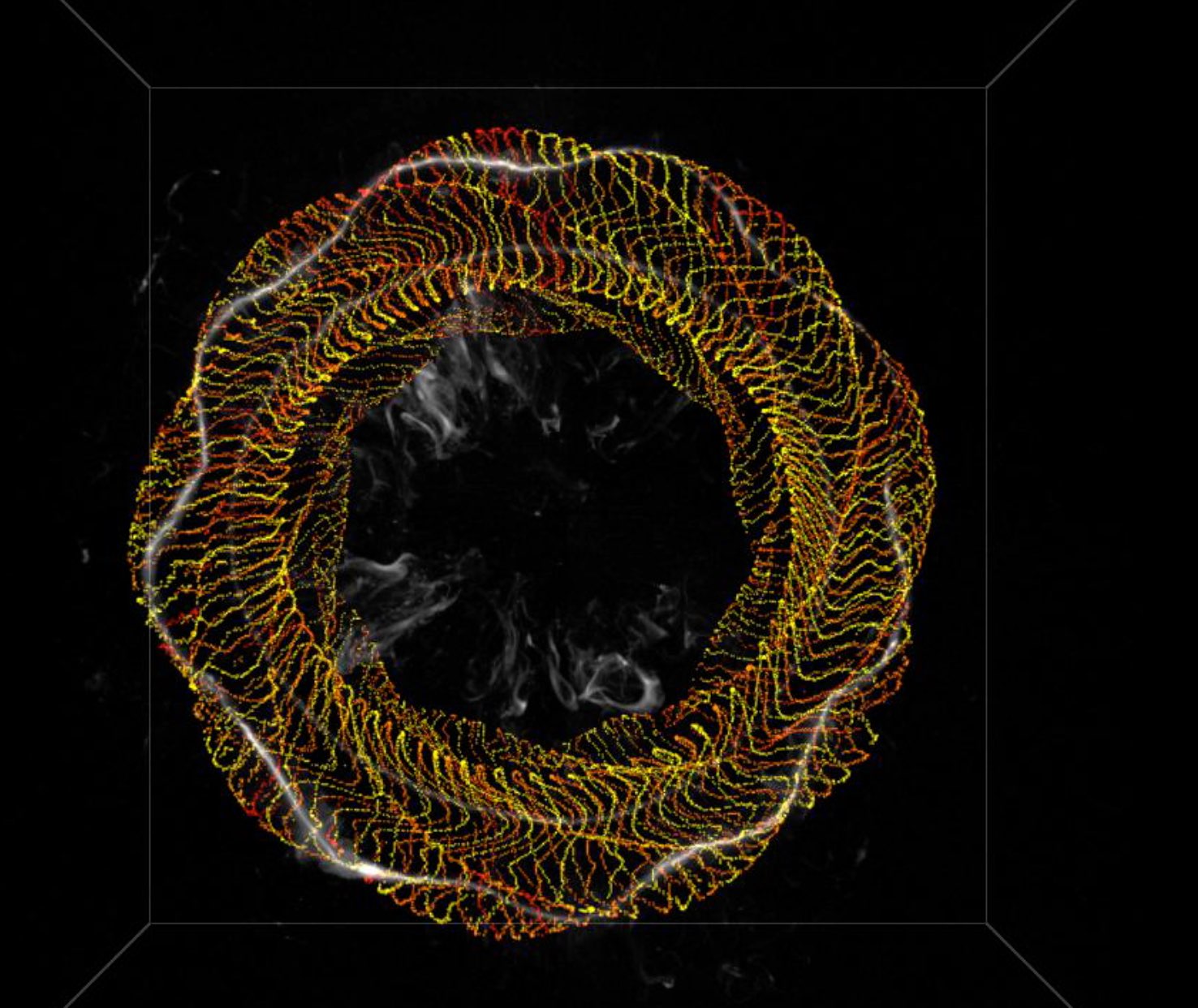}    %
\caption{A leap-frogging helical vortex ring, visualized by placing     %
evenly spaced blobs of dye in the vortex core (white). The trajectories %
of each blob are shown in autumn colors. The measurement of the vortex  %
shape was combined with the dye trajectories to yield a measurement of  %
the total helicity and its dynamics. Courtesy of W. Irvine              %
\cite{scheeler2017complete}. }                                          %
  \label{fig:helicity}                                                  %
\end{figure*}                                                           %

Helicity can be introduced in a flow by the presence of large-scale stirring mechanism that breaks mirror symmetry. It is considered important for many flow configurations as shown in the pioneering studies reported in \cite{borue1997spectra, pelz1985velocity, kerr1987histograms, kholmyansky1991some, kit1987experimental, moffatt1992helicity}. In nature, helicity is believed to play a key role in the atmospheric Ekman layer, where theoretical and numerical arguments support  a  turbulent helicity cascade of right-handed helicity in the logarithmic range of the boundary layer \cite{deusebio2014helicity,koprov2005experimental,kurgansky2017helicity}. Helical vortices of different prescribed topology have also been produced in the lab \cite{kleckner2013creation}. Furthermore local and non-local effects under strong shear have been studied experimentally in von-K{\'a}rm{\'a}n flows  \cite{herbert2012dual} and numerically in axisymmetric turbulence \cite{qu2018cascades}. Finally, a helical flow, constructed by homochiral helical waves with constant $k$ is a Beltrami flow that is an exact solution of the Euler equation. Suitable superposition (gluing) of  such local Beltrami flows  has been used to build up singular solutions of the  NSE and the Euler equations.    \cite{moffatt2014helicity,buckmaster2015anomalous,buckmaster2017nonuniqueness}. The 
NSE  projected on the homochiral sector can be proven to admit global solutions \cite{biferale2013global}. Models and closure of turbulence have also been widely applied to helical flows starting from the early works of \cite{andre1977influence,lesieur20003d}  (see also  \cite{briard2017dynamics} for a recent contribution). In this section we review recent advancements on the field achieved by decomposing the flow in helical Fourier modes.
This results in a classification of the non-linear terms in homochiral and heterochiral triads. In terms of this decomposition we discuss how the flow reacts to  mirror-symmetry breaking by the forcing, the rate of recovery of mirror-symmetry at smaller and smaller scales and what happens to the energy cascade in the presence of an explicit helicity-breaking mechanism at all wavenumbers. Furthermore this decomposition has revealed that depending on the relative weight of heterochiral and homochiral  triads classes, energy can be transferred either to small or to large scale. Thus, homochiral energy transfer is a pure 3D mechanism that provides an isotropic  reversal of the energy cascade  and in the absence of any two-dimensionalization of the flow. 

\subsubsection{Homogeneous and Isotropic Helical Turbulence \label{sec:HIHT}}    
All phenomenological approaches to 3D turbulence are based on the concept of direct energy cascade.  However, as already said after its introduction (\ref{eq:inviscid3D}), helicity  is a second quadratic inviscid invariant of the 3D NSE and its  mean value is  exactly zero if the flow is statistically invariant under mirror symmetry, $\bw$ being a pseudo-vector. In the presence of forcing, viscous and large scale dissipation a global  helicity balance exists, similar to  the one for energy (\ref{eq:globalbalance}):
\be
\partial_t \cH(t) = -\hnu(t) -  \halpha(t) + \hin(t)
\label{eq:globalbalanceH}
\ee
where $\hnu(t) = \nu \langle \partial_i u_j \partial_i \omega_j \rangle$ is the helicity dissipation due to viscosity, $\halpha(t) = \alpha  \la \bu  \cdot \bw \ra$ is the dissipation due to the large scale drag and $\hin(t) = \langle   \bw \cdot \bfo  \rangle$ is the helicity injection by the external forcing mechanism. Performing a long time average, we end up with the global balance for the average input and output similarly to  (\ref{eq:globalbalance2}):
\be
\hin = \hnu + \halpha.
\label{eq:globalbalanceH2}
\ee  Since its discovery \cite{moffatt69,moffatt1992helicity,brissaud1973}, helicity has been the object of many speculations. In particular, it has been argued that the  presence of a non-zero mean helicity, globally or locally, can affect the statistical properties of the forward energy cascade. On one side, the non-linear term of the NSE is locally proportional to the solenoidal component of $\bu \times \bw$, and flows with a non-zero helicity might have a strongly depleted non-linearity and consequently a small energy transfer \cite{kraichnan1988depression, moffatt2014helicity}. On the other hand, helicity is not sign-definite and cancellations among close regions with opposite helicity might strongly reduce this energy {\it blocking} mechanism. As a result, its influence on the turbulent transfer mechanisms is still unclear \cite{borue1997spectra, pelz1985velocity, kerr1987histograms, chen2003joint,chen2003intermittency}.\\

We have already seen that the conservation of energy leads to the exact $4/5$ law (\ref{eq:45direct})  for the third-order longitudinal velocity increment in the direct energy cascade regime. Similarly, if the injection of helicity is non-zero, we can derive  another exact law for velocity-vorticity mixed  third-order correlation functions that are not invariant under mirror symmetry \cite{Frisch, GomezPolitano, kurien2004isotropic, Chkhetiani1996}, the two exact constraints are summarised below: 
\begin{align}
&  \langle (\delta_\br u)^3\rangle = -\frac{4}{5} \ein\, r \, , \label{eq:45}\\
&  \langle \delta_\br u (\delta_\br \bu \cdot \delta_\br \bw)\rangle - 
\frac{1}{2}\langle \delta_\br w (\delta_\br \bu \cdot \delta_\br \bu)\rangle  = -\frac{4}{3} \hin\, r \, ,
\label{eq:45bis} \end{align}
where we have also assumed homogeneity and isotropy (but not mirror symmetry) and where we have introduced the short hand notation, $\delta_\br u$ and $\delta_\br w$ for  the longitudinal velocity and vorticity increments, defined in terms of the projection on the unit  vector $\hat \br$: $\delta_\br X = \delta_\br \bm X \cdot \hat \br$
and the generic vector increment between two points is $\delta_\br \bm X = \bm
X(\br +\bx ) -\bm X(\bx)$. The  equation (\ref{eq:45bis}) is different from zero only in presence of a mirror-symmetry breaking forcing mechanism.  The two exact scaling relations (\ref{eq:45}-\ref{eq:45bis}) are valid in the  inertial range, i.e., when the increment $r$ is such that both forcing and dissipative effects can be neglected.  
Helicity  is not sign-definite. Thus, as discussed in Sec. (\ref{sec:multiple}), it is not possible to  predict the energy transfer direction in 3D turbulence.  According to the definitions given in Sec. (\ref{sec:definitions}), either a co-directional cascade of energy and helicity toward small-scales or a counter-directional cascade with energy flowing upward and helicity downward would be possible \cite{brissaud1973, kraichnan1971, chen2003joint, kurien2004cascade, biferale2013split, herbert2012dual}. It is useful to adopt the helical-Fourier decomposition,  in order to disentangle in a systematic way the statistical properties under mirror symmetry. 
{This decomposition was proposed early on \cite{craya1958contributiona,herring1974approach,Lesieur72} and
has been used by numerous authors \cite{constantin1988beltrami,waleffe1992nature,cambon1989spectral}. It allows us to write a Fourier mode of the velocity as:}
\be
{\bf \tilde{u}}(\bk,t) = \tilde{u}^{+}_{\bk}(t) \hp 
                  + \tilde{u}^{-}_{\bk}(t) \hm.
\label{upm}                  
\ee
where $\hpm$ are the eigenvectors of the curl, i.e., $i {\bk} \times \hpm = \pm k \hpm$. 
We choose $\hpm = {{\bf e}}(\bk) \times (\bk/k) \pm i {{\bf e}}(\bk)$, where $ {\bf e}(\bk)$ is a
unit vector orthogonal to ${\bk}$,  e.g.
${{\bf e}}(\bk) = {\bz} \times {\bk}/|| {\bz} \times {\bk} ||$, with
any arbitrary vector ${\bz}$ not parallel to $\bk$ \cite{waleffe1992nature}.
They satisfy, $ \hpm \cdot  {\hpm}^* = 2$  and $\hpm \cdot \hmp = 0$, where the asterisk
denotes complex conjugation. They thus form a complete base for incompressible vector fields.
The velocity field ${\bf \tilde{u}_k}$ for each Fourier mode $\bk$
is then determined by the two scalar complex functions
$\tilde{u}^\pm_{\bk}(t)=({\hpm}^* \cdot \but(\bk,t))/2$. 

We can write  the Navier Stokes equation
in terms of the scalar amplitudes $\tilde{u}^\pm_{\bk}(t)$ as \cite{waleffe1992nature}
\be
\label{NSH}
\partial_t \tilde{u}^{s_k}_\bk  =  \sum_{\bf p+q+k=0}  \sum_{s_p,s_q} \mathcal{C}^{s_k,s_q,s_p}_{\bp,\bq,\bk} (\tilde{u}^{s_p}_\bp)^* (\tilde{u}^{s_q}_\bq)^* 
 - \nu k^2  \tilde{u}^{s_k}_\bk  - \alpha   \tilde{u}^{s_k}_\bk + \tilde{f}^{s_k}.
\ee
where  we further have $(s_k,s_p,s_q) = (\pm,\pm,\pm)$ and 
$\mathcal{C}^{s_k,s_q,s_p}_{\bk,\bq,\bp}= -1/4 (s_q p -s_p q) [\hppm \times \hqpm \cdot \hkpm]^* $. \\
In terms of  such  decomposition the total energy and
helicity can be rewritten as
\begin{align}
&    \cE(t) = \sum_{\bk} |\utpk|^2 + |\utmk|^2 \, , \label{eq:Etot}\\
&  \cH(t) = \sum_{\bk} k\,(|\utpk|^2 - |\utmk|^2) \,. \label{eq:Htot}
\end{align} 
It is useful to further distinguish the
energy content of the positive and negative helical modes, 
$E^\pm(k,t) =  \sum_{k \le |\bk| < k+1} |\utpmk|^2$, such that we have for  the energy and helicity spectra
\cite{chen2003joint}: 
\be
  E(k,t)  = E^+(k,t) + E^-(k,t) \, ,\qquad  \label{eq:epm}
    H(k,t)  = k\,[E^+(k,t) - E^-(k,t)]\,.
\ee 
It is straightforward to realize that the equivalent in  configuration space of the decomposition (\ref{eq:epm})  is given by the  second-order structure  functions  decomposed  in terms of the fields 
\be
\label{eq:helicadecomposition}
\bu^\pm (\bx,t) =  \sum_{\bk}  
\utpmk \hpm \exp^{i \bk \cdot \bx}
\ee 
as follows:
\begin{align}
\label{eq:epm2}
&   \langle \delta_\br \bu \cdot  \delta_\br \bu \rangle   =  \langle \delta_\br \bu^+ \cdot \delta_\br \bu^+\rangle   + \langle \delta_\br \bu^- \cdot \delta_\br \bu^-\rangle \, , \\
&   \langle \delta_\br \bu \cdot \delta_\br \bw \rangle   =  \langle \delta_\br \bu^+ \cdot \delta_\br \bw^+\rangle   
+  \langle \delta_\br \bu^- \cdot \delta_\br \bw^-\rangle \, ,
\end{align} 
where we have exploited the fact that  both  mixed terms
$\langle \delta_\br \bu^\pm \cdot \delta_\br \bw^\mp \rangle$ and $\langle \delta_\br
\bu^\pm \cdot \delta_\br \bu^\mp \rangle$  vanish, due to the orthogonality of $\hpm$. 

It is not possible to derive
a  prediction for the mean energy and helicity spectra from (\ref{eq:45}-\ref{eq:45bis}) alone, because there exists a continuum of
possible combinations of $\ein, \hin$ and $k$ with the correct dimensional
properties:
\begin{align}
E(k) = \ein^{\frac{2}{3}-a} 
\hin^a k^{-\frac{5}{3}-a} \, .
\end{align} 
Different possibilities have been proposed, based on different closures of 
the spectral equations, depending on the dynamical time-scale 
that drives the energy and helicity transfers. One possibility is 
based on the idea that the  only relevant time-scale is the 
one given by the energy fluctuations,
$\tau^E_r \sim r/\langle |\delta_\br \bu|^2\rangle^{1/2} \sim 
\ein^{-1/3} r^{2/3}$ \cite{brissaud1973}. In this case we have the dimensional  
estimate for the (mirror invariant) energy flux:
\begin{align}
\ein \sim \langle |\delta_\br \bu|^2 \rangle /\tau_r^E \rightarrow \langle |\delta_\br \bu|^2\rangle  \sim \ein^{2/3}r^{2/3} \, ,
\label{eq:fluxe}
\end{align}
while for  the chiral term we get:
\begin{align}
  \hin  \sim \langle \delta_\br \bu \cdot \delta_\br \bw \rangle /\tau_r^E \rightarrow \langle \delta_\br \bu \cdot
  \delta_\br \bw \rangle  \sim \hin \ein^{-1/3}r^{2/3} \, .
\label{eq:fluxh}
\end{align}
Translating back to Fourier space we get for the semi-sum  (mirror-symmetric) and the semi-difference (mirror-antisymmetric)   of the spectral components \cite{chen2003joint}:
\begin{align}
&   E^+(k) +   E^-(k)  \sim C_E \ein^{2/3} k^{-5/3} \, ,\label{eq:chen1}\\
&  E^+(k) - E^-(k) \sim C_H \hin \ein^{-1/3} k^{-8/3}  \, ,\label{eq:chen2}
\end{align} 
where $C_E$ and $C_H$ are two dimensionless constants. Hence, the two energy components can be written as: 
\begin{align}
E^\pm(k)  \sim C_E\ein^{2/3} k^{-5/3} \pm C_H \hin \ein^{-1/3} k^{-8/3}  \, . \label{eq:epm-chen}
\end{align} 
Another possible dimensional closure  employs the   helicity 
time-scale, $\tau_r^H \sim \hin^{-1/3} r^{1/3}$, to  evaluate both  fluxes 
(\ref{eq:fluxe})-(\ref{eq:fluxh}). In this case we have 
: 
\begin{align}
E^\pm(k)  \sim C_E \ein \hin^{-1/3} k^{-4/3} \pm C_H \hin^{2/3}  k^{-7/3}  \, . \label{eq:epm-kurien}
\end{align}
Relation (\ref{eq:epm-kurien}) breaks the $-5/3$ law for the energy spectrum and has been proposed to be valid only in the high $k$-region of strongly helical turbulence, to explain the bottleneck observed close to the viscous scale \cite{kurien2004cascade}.
Indeed, relation (\ref{eq:epm-kurien}) diverges for $\hin \rightarrow 0$ and therefore cannot be considered a good option if helicity is
sub-leading. A third possible scenario is the one described by  a counter-directional dual cascade (see Def. (\ref{def9}) of Sec. \ref{sec:definitions}), where energy flows upward and helicity downward \cite{brissaud1973}. In the latter  case, for the forward-helicity cascade range, only $\hin$ appears, while the inverse energy cascade is driven by  $\ein$, leading to the two predictions:
\begin{equation}
  \begin{cases}
    E(k) \sim \hin^{2/3} k^{-7/3} \, , \qquad H(k) \sim \hin^{2/3} k^{-4/3},  \qquad k > k_{in}\\
   E(k) \sim \ein^{2/3} k^{-5/3} \, , \qquad H(k) \sim \ein^{2/3} k^{-2/3}, \qquad k < k_{in}.\\
    \end{cases}
\label{eq:split}
\end{equation}
This last scenario has never been observed in isotropic turbulence, 
unless a dynamical mode reduction on helical modes with the same sign 
is imposed (see  subsection \ref{sec:inversehomochiral}
and \cite{waleffe1992nature,biferale2012,biferale2013split}). \\
In Fig. (\ref{fig:helicitycascades}) we summarise the three scenarios in a graphical way. 
\begin{figure*}[htbp]                                                 %
\centering                                                            %
\includegraphics*[width=0.4\textwidth]{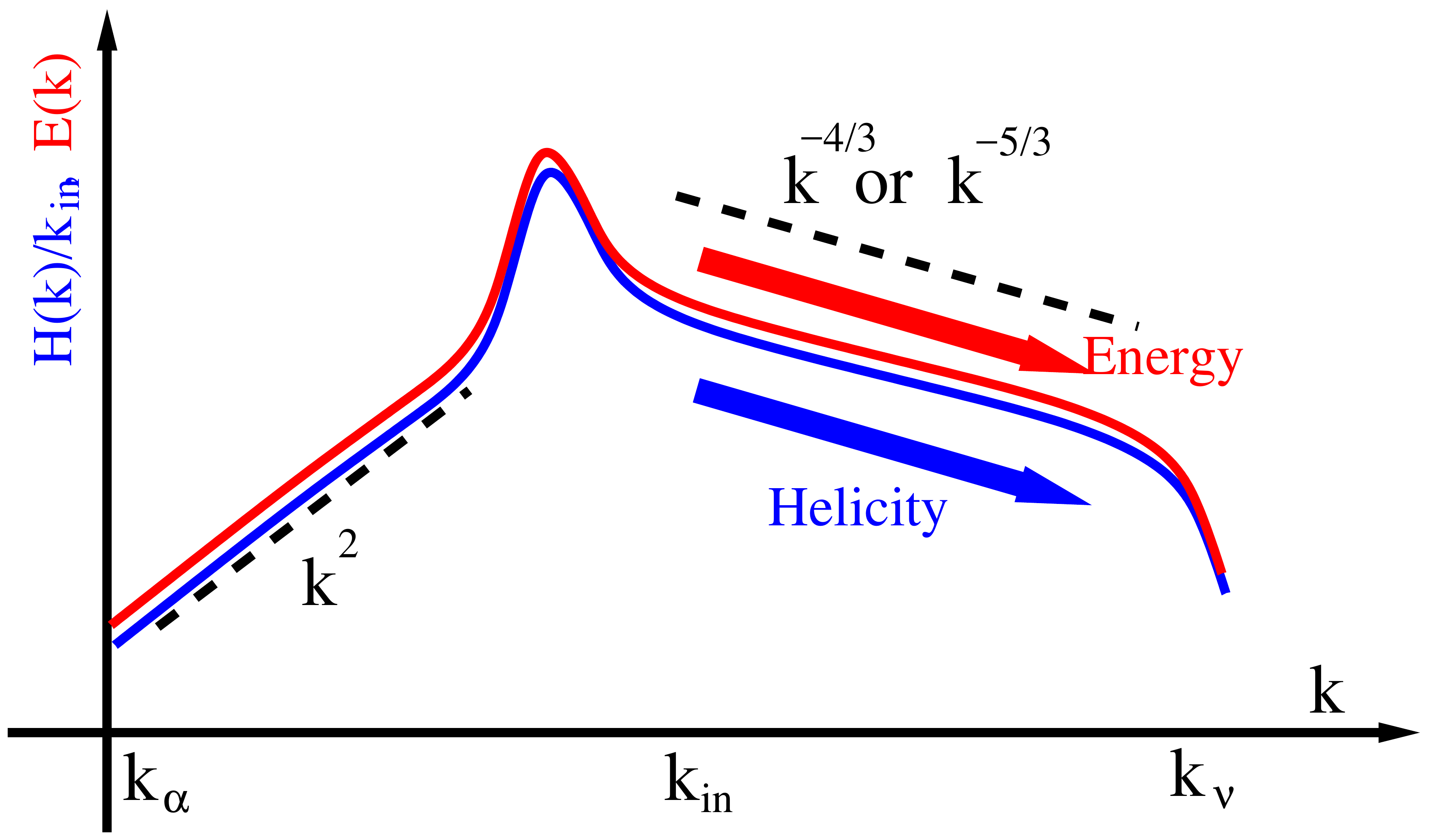} \quad         %
\includegraphics*[width=0.4\textwidth]{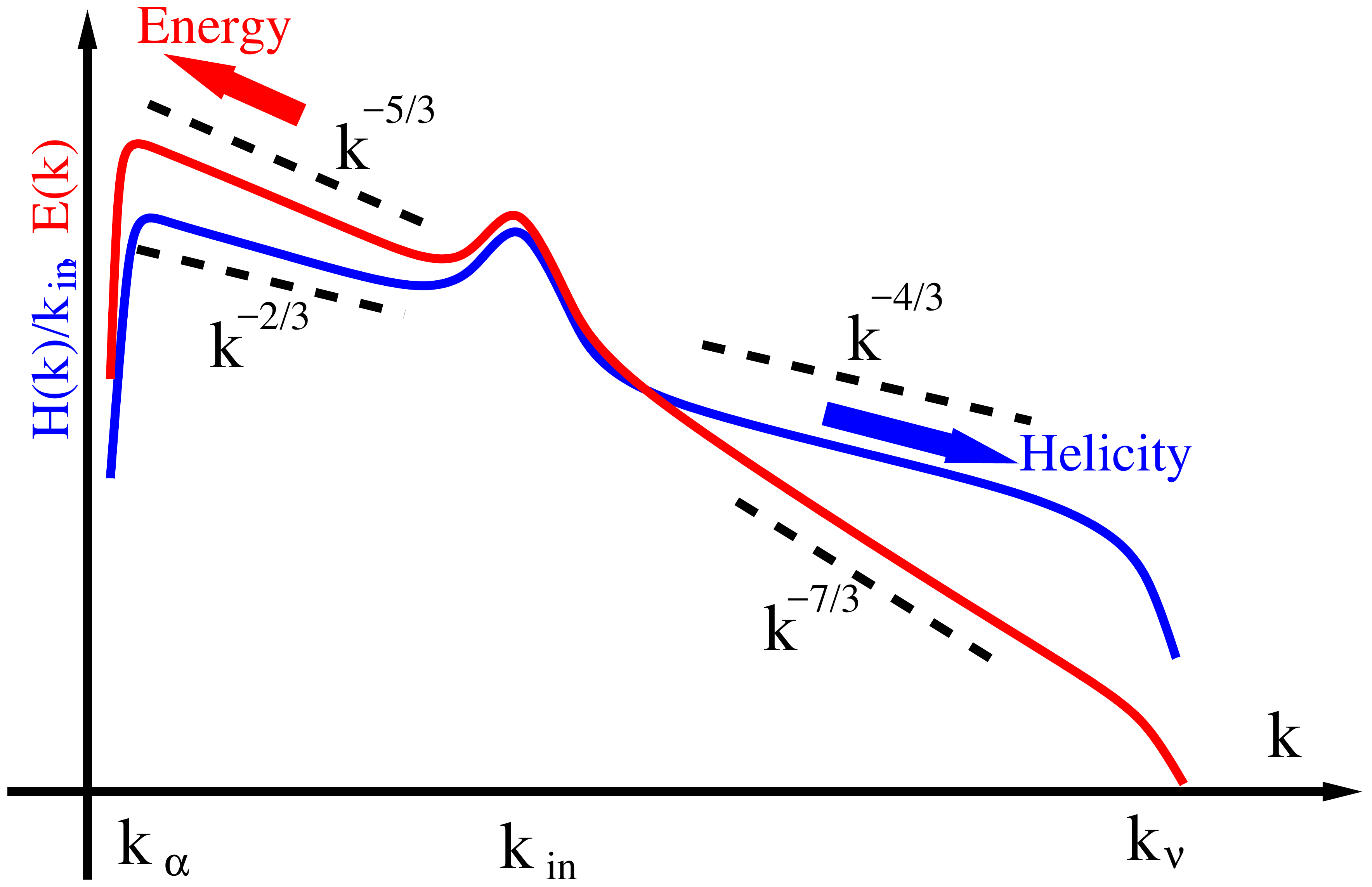}               %
\caption{Left: log-log sketch of the spectra for the energy   %
(\ref{eq:epm-chen}) or helicity  (\ref{eq:epm-kurien}) driven forward  %
cascades, leading to a $-5/3$ or $-4/3$ slopes, respectively. Right:  %
the same but for the split-dual case (\ref{eq:split}).}               %
  \label{fig:helicitycascades}                                        %
\end{figure*}                                                         %
Nowadays, there is overwhelming evidence that in the presence of a large-scale helical forcing mechanism, there is a simultaneous transfer of both energy and helicity toward small scales as suggested in (\ref{eq:epm-chen}).
Two forward cascades coexist
with helicity introducing only sub-leading corrections to the $-5/3$ Kolmogorov K41 scaling (see Fig. \ref{fig:spectrahelicity}).  Consequently,  there is a 
strong recovery of mirror symmetry by going to smaller and smaller scales \cite{chen2003intermittency, sahoo2017helicity}. The latter  can be quantified  by introducing the  dimensionless ratio:
\[
R_{\cH} = \frac{|E^+(k) -E^-(k)|}{E^+(k) +E^-(k)} \sim k^{-1}
\]
\begin{figure*}[htbp]                                                 %
\centering                                                            %
\includegraphics*[width=0.8\textwidth]{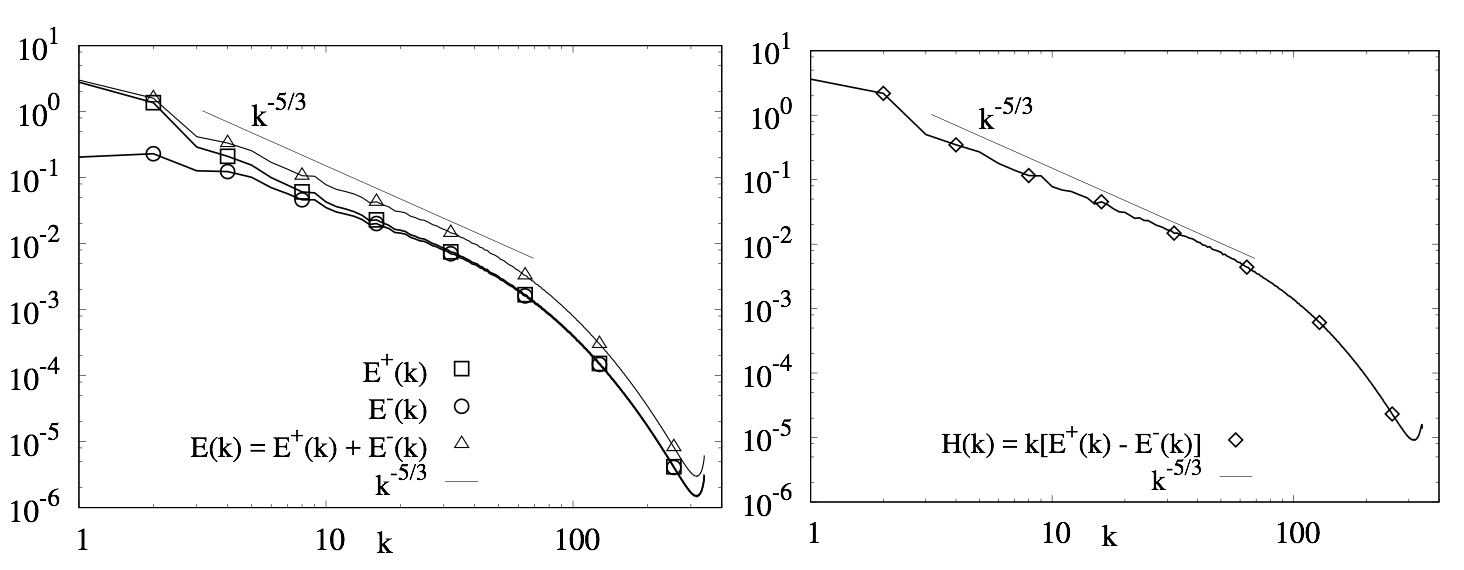}   %
\caption{Left: Log-log plot (adapted from \cite{sahoo2017helicity})   %
of the energy spectra for the positive and negative helical waves     %
together with the total spectrum, $E(k)$. Data are taken from a       %
direct numerical simulations at $1024^3$ collocation points.  Notice  %
that the mismatch at small $k$ between $E^+(k)$ and $E^-(k))$ is due  %
to the fact that  turbulence is forced  at large scales and with     %
positive helicity. Right: the same for the helicity spectrum. Both    %
plots show also the $-5/3$ prediction corresponding to the scenario   %
(\ref{eq:epm-chen}). }\label{fig:spectrahelicity}                     %
\end{figure*}                                                         %
As a result, even in the presence of coherent large-scale injection of helical fluctuations, we must expect a strong homogenization of small-scales positive and negative helical components. It is important to notice that such recovery of mirror symmetry can be violated if the forcing mechanisms is multi-scale and strongly helical as shown by using both Navier-Stokes equations and helical shell models in \cite{biferale2017scatamacchia,kessar2015non}.\\
Much less is known about the existence of the helical dissipative anomaly, i.e. if the limit 
$ \lim_{Re \to \infty} \hnu \to const.
$
Very few numerical simulations exist and there are no experimental data for this specific issue, due to the difficulty to measure velocity derivatives. Moreover, due to the fact that helicity is not positive definite, there are subtle issues about how to define the viscous dissipative range. The latter point has been discussed in detail using shell models for turbulence \cite{biferale2013split,ditlevsen2000,ditlevsen2001dissipation,ditlevsen2010turbulence,ditlevsen1996cascades} which will be briefly discussed  in Sec. (\ref{sec:shell}). Before concluding, let us notice that because of the absence of 
an inverse helicity cascade one can always assume in the balance (\ref{eq:globalbalanceH2}),  $\halpha =0$ without any loss of generality. 

\subsubsection{Inverse energy cascades in homochiral turbulence \label{sec:inversehomochiral}}
In this section we further refine the energy-helicity dynamics by analyzing the turbulent  transfer  as function of the helical contents of the Fourier triads (\ref{NSH}). It is easy to realize that, in terms of the helical decomposition, 
{all triads of interacting Fourier wavenumbers  $[\but_{\bf k},\but_{\bf q},\but_{\bf p}]$ are}
split in 4 (8 by considering the symmetry that changes the sign of all helical modes) possible classes of helical interactions, 
$[\utpmk,\utpmq,\utpmp]$, as depicted in Fig. (\ref{fig:NSheterochiral}). Out of the four classes  indicated by [I-IV], class-I is homochiral, i.e. includes interactions by three Fourier waves with the same sign of helicity, while classes II-III-IV are heterochiral.  One can also easily show that energy and helicity are conserved triad-by-triad \cite{waleffe1993inertial,waleffe1992nature}. Assuming without any loss of generality that $ q \le k \le p$  we can interpret the NSE decomposition  depicted in (\ref{fig:NSheterochiral}) as the superposition of 4 classes of interactions distinguished on the basis of their helical contents. In two seminal papers  \cite{waleffe1993inertial,waleffe1992nature}, Waleffe analyzed the stability properties of each single class as a proxy of the energy transfer direction and discovered that the homochiral class I and the heterochiral class-II are such that the intermediate wavenumber tends to loose energy toward small and large scale simultaneously, while class-III and class IV are always such that the smallest wave number is the most unstable. On the basis of this observation, Waleffe argued that even for the NSE case, when all wavenumbers are coupled together,  class-I and class-II are responsible of some three-dimensional inverse energy transfer while classes III and IV should be the channels responsible of the energy transfer to small scales  (instability hypothesis). On a physical ground, one must notice that the homochiral class I (if taken alone)  corresponds to a fluid where also helicity is sign definite. As a result, it is not totally surprising that the  dynamics develops some inverse energy transfer, similarly to what happens  for 2D turbulence where the second sign definite invariant is the enstrophy. The fact that in the analysis of Waleffe, also the heterochiral class-II shows  some tendency to transfer energy upscale can also be interpreted in a similar way considering that the two largest wavenumbers are homochiral, i.e. any mechanism that transports fluctuations from small to large wavenumbers would produce a systematic increase of helicity also for  class-II. 

Summarizing, the single triad analysis performed in \cite{waleffe1992nature} demonstrates that the basic building blocks of the helical-Fourier
decomposition have distinct stability properties that indicate that the forward energy transfer is mainly mediated by triads where  the two largest legs are heterochiral.
{ However care needs to be taken when extrapolating the  stability properties of isolated triads to the full network present in the  NSE.
As noted in \cite{moffatt2014note}, the NSE restricted to any single triad (homo or heterochiral) can be mapped to a 2D3C fluid evolution, because triads forms closed triangles in Fourier space, $\bp+\bq+\bk=0$. As a  result, the NSE can be seen as the 3D intricate coupling of 2D3C sub-systems, some that transfer energy forward and some that transfer energy backward,  and the connections among the full 3D dynamics and the one of its subsystems must  be clarified. Nonetheless, the analysis of isolated triads of \cite{waleffe1992nature} has motivated new directions for turbulence research as we describe below.} 

\begin{figure*}[htbp]                                                                                   %
\centering                                                                                              %
\includegraphics*[width=0.9\textwidth, angle=0]{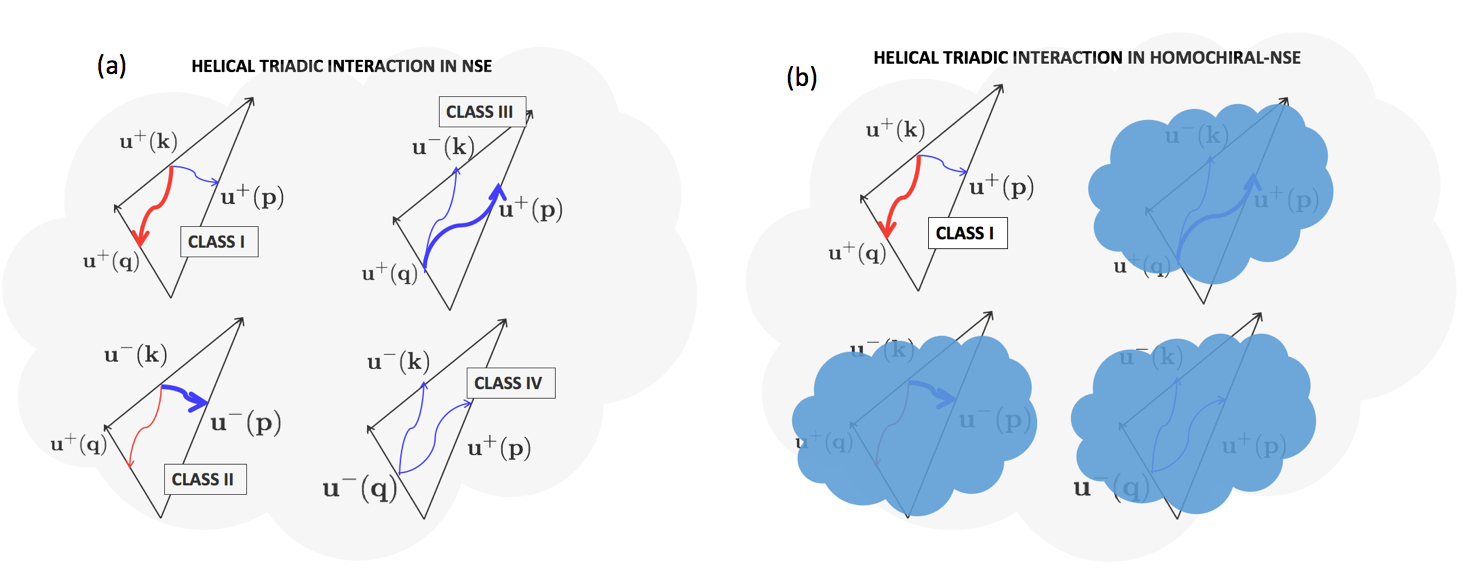}                       %
\caption{Sketch of the NSE triadic interactions cloud in terms of the Helical-Fourier decomposition     %
(\ref{NSH}). (a) The full NSE set of triads is divide in 4 classes, according to the helicity           %
contents in the three interacting wavenumbers and where we do not show the equivalent 4 classes         %
obtained by changing the sign of helicity in all modes. The color and thickness of arrows inside        %
each triad correspond to the direction (red: inverse, blue: direct) and intensity of the most           %
unstable energy transfer mode according to the single triad analysis  \cite{waleffe1992nature}.         %
As one can see, Classes I and and II present a tendency to have a backward energy transfer              %
(red arrows) from the middle wavenumber to the smallest one. Class I is also the only one               %
homochiral. (b) Sketch of the decimation to only homochiral dynamics performed using the projector      %
(\ref{eq:decomp}) with $s=+$ and leading to the H-NSE (\ref{eq:GNS+}).}\label{fig:NSheterochiral}  %
\end{figure*}                                                                                           %

The idea to decompose the NSE into helical interactions was further explored  in a series of papers \cite{biferale2012,biferale2013split} where direct numerical simulations of homochiral turbulence were performed for the first time. The idea
is that one can introduce two Galerkin projectors to positive and
negative helical modes as 
\begin{align}
  \label{eq:proj}
 \tilde{{\mathcal P}}^{\pm}_{\bk} \equiv \frac {\hpm \otimes {\hpm}^*} {{\hpm}^* \cdot \hpm}.
\end{align}
The Fourier modes  are  given by 
$\but_\bk^\pm = \tilde{{\mathcal P}}^{\pm}_{\bk} \but_\bk = \frac{1}{2}(\but_\bk \pm i \bk \times \but_\bk/k)$. Using the above definition, one defines the projection on the positive/negative helical sector of any real-valued divergence free vector field in three dimensions:
\be
\bg^s(\bx,t) \equiv {\mathcal P}^s  \bg(\bx,t) \equiv \sum_{\bk}  \tilde{{\mathcal P}}^s_{\bk} \, \bgtk e^{i \bk \cdot \bx}.
\label{eq:decomp}
\ee
where $s = \pm$.
Finally, we define the  
{ {\it  homochiral Navier-Stokes equation (H-NSE)} 
  that describe the evolution of the velocity field of  one chirality ($\bu^+$   or $\bu^-$) interacting with itself only. For the positive modes it reads:} 
\be 
\label{eq:GNS+}
\partial_t {\bf u}^+ + {\mathcal P}^+[{\bf u}^+ \cdot {\bm \nabla} {\bf u}^+] =  \nu  \Delta {\bf u}^+ -  \alpha  \bu^+ + {\bf f}^+.
\ee
The  linear operator ${\mathcal P}^+$  in front of the non-linear terms combines the projection into a divergenceless manifold and  in to the homochiral sector. It is needed because  the interactions between two positive helical waves produce also negative helical modes.
The resulting dynamics involve only interactions of class-I, as depicted in Fig. (\ref{fig:NSheterochiral}) panel (b). \\ 
From (\ref{eq:Etot}-\ref{eq:Htot}) it is evident that
 both energy and helicity are sign definite for the  H-NSE dynamics. As a result, one might expect that turbulence will develop a counter-directional dual cascade, with energy flowing upscale and helicity downscale.
The corresponding spectra will be determined by the requirement of having a  constant energy flux in the inverse energy cascade  range and a constant helicity flux in the forward helicity  cascade range as predicted by (\ref{eq:split}).
Figure (\ref{fig:dire-inverse}) shows that the above behaviour is indeed  observed in the DNS of (\ref{eq:GNS+}) \cite{biferale2012,biferale2013split}. \\
\begin{figure*}[htbp]                                                           %
  \includegraphics*[width=0.9\textwidth, angle=0]{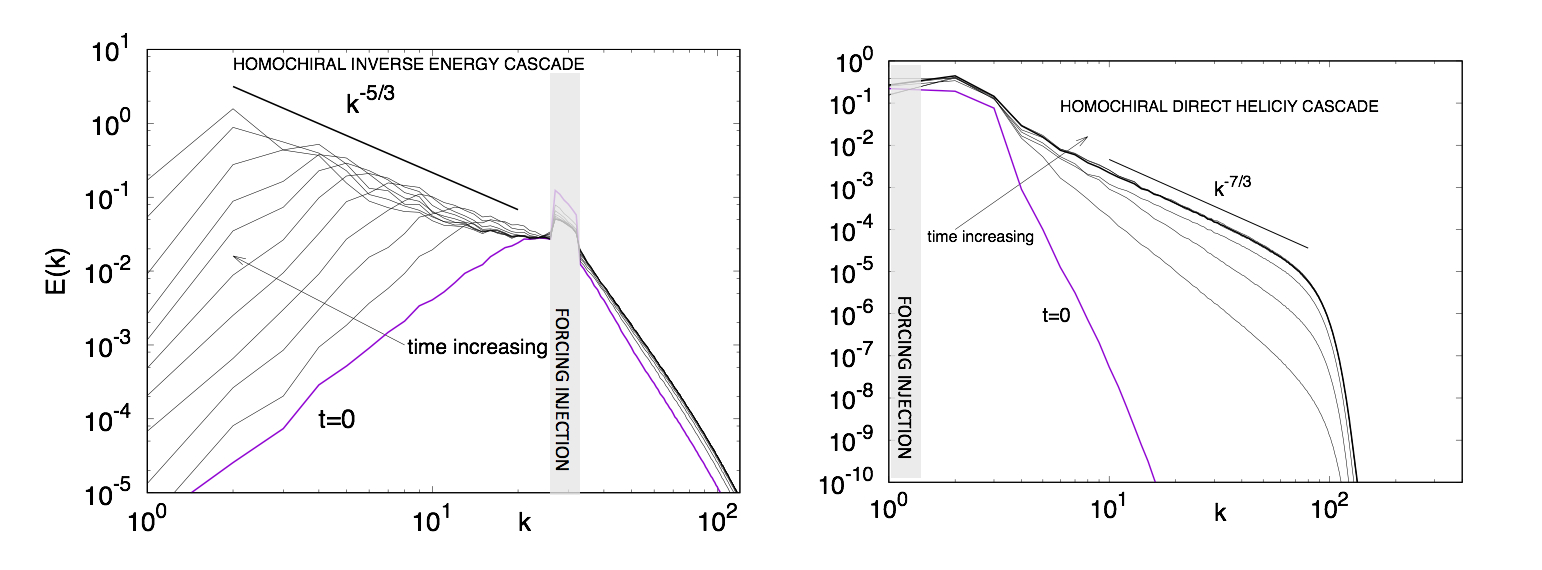}          %
\centering                                                                      %
\vspace{-0pt}                                                                   %
\caption{Log-log plot for the time evolution of the  energy spectra for two     %
homochiral simulations of  (\ref{eq:GNS+}) with small scale forcing (left)      %
to highlight the inverse energy cascade and with large scale forcing (rigth) for        %
the direct helicity cascade. Initial configuration at $t=0$  is marked          %
with a red curve. The two straight lines with slopes  $-5/3$ (left) and       %
$-7/3$ (right) represent the  two predictions  (\ref{eq:split}),           %
respectively. The grey line marks the forcing injection scales.
For details about the numerical simulations see                   %
\cite{biferale2012,biferale2013} }\label{fig:dire-inverse}                      %
\end{figure*}                                                                   %
The observation of a stable inverse energy cascade in the DNS of the  H-NSE  has provided the first evidence that there exists a new mechanism to transfer energy up-scale in a turbulent flow, beside the one of being two-dimensional, or {\it quasi}  two-dimensional.

More importantly, in \cite{kessar2015non,alexakis2017Helically} it was shown that homochiral triads persistently transfer energy upward even in the original case of the full NSE where all modes are active. By analysing a-posteriori the contribution of the total flux induced only by homochiral triads, it was found that it corresponds to an inverse cascade, with a sign opposite to that of  the heterochiral contributions. Indeed, even in the fully coupled NSE one can distinguish partial fluxes which conserve independently the energy contents in the positive or negative helical modes (and similarly for the helicity) as described in our Def. (\ref{def5}) in Sec. (\ref{sec:definitions}). This can be seen by rewriting the helical decomposition of the NSE (\ref{NSH}) in  configuration space following  \cite{alexakis2017Helically}:
\be
\label{eq:alex2016}
\partial_t {\bf u}^{s_1} =  {\mathcal P}^{s_1} \sum_{s_2,s_3}[{\bf u}^{s_2} \times \bw^{s_3}]  + \nu  \Delta {\bf u}^{s_1} -  \alpha  \bu^{s_1} + {\bf f}^{s_1},
\ee
where $(s_1,s_2,s_3)  = (\pm,\pm,\pm)$ indicates all possible permutations of projections on the two helical sectors. The evolution of each of the two helical projections $\partial_t {\bf u}^{\pm}$  is driven by four different non-linear terms that have different properties. 
{
This is revealed by looking at the evolution of the energy of the two helical fields $\cE^\pm(t)=1/2 \langle (\bu^\pm)^2 \rangle$ that, e.g.,  for the positive case reads 
\be
\label{eq:alex2}
\partial_t \cE^+(t) = \langle  \bu^+  {\mathcal P}^{+} [{\bf u}^{+} \times \bw^{+}] \rangle +   \langle  \bu^+  {\mathcal P}^{+} [{\bf u}^{+} \times \bw^{-}] \rangle +
  \langle  \bu^+  {\mathcal P}^{+} [{\bf u}^{-} \times \bw^{+}] \rangle +   \langle  \bu^+  {\mathcal P}^{+} [{\bf u}^{-} \times \bw^{-}] \rangle   .
 \ee
 Out of all terms, }
only the first two on the RHS of (\ref{eq:alex2}) are conservative, meaning they individually conserve $\cE^+$.
These terms are responsible for the exchange of energy among positive modes only or by the sweeping of positive modes by negative ones. The last two terms  describe the exchange of energy with the negative components (and similarly for the total energy in the negative modes if we reverse all helical signs). We can also split  the total flux in Fourier space by first introducing the quantities $$ \Pi^{s_1s_2s_3}_E(k) = \langle \bu^{ <k} {\mathcal P}^{s_1} [{\bf u}^{s_2} \times \bw^{s_3}]  \rangle,  $$
 where $ \bu^{ <k}(\bx,t)$ is the low-pass filtered field defined in (\ref{eq:lowpass}) and then writing:
 \be
 \label{eq:decPi}
 \Pi_E(k)=\Pi^{homo}_E(k) + \Pi^{hete,1}_E(k)+  \Pi^{hete,2}_E(k)
 \ee
 where
 \begin{eqnarray}
   \label{eq:decompojfm}
   \Pi^{homo}_E(k)= \Pi^{+++}_E(k) + \Pi^{---}_E(k),& \\
   \Pi^{hete,1}_E(k) =  \Pi^{++-}_E(k) + \Pi^{--+}_E(k),&  \nonumber \\
   \Pi^{hete,2}_E(k) =
   \Pi^{+-+}_E(k) + \Pi^{-+-}_E(k)& +\Pi^{+--}_E(k) + \Pi^{-++}_E(k). \nonumber
   \end{eqnarray}
 The $\Pi^{hete,1}_E$ flux corresponds to the two contributions given by the sweeping of positive (negative) modes by the negative (positive) ones, while  $\Pi^{hete,2}_E$ describes the transfer of energy among the two chiral sectors. In Fig. (\ref{fig:alexandrosjfm2017}) we show the measurements of the different contributions in (\ref{eq:decPi}) from a DNS of fully homogeneous turbulence. The different sign brought by the homochiral channel is  clearly detectable. This implies that within the forward cascade  of 3D turbulence there is a process that transfers energy inversely. Thus the cascade in   3D hydrodynamic  turbulence is bidirectional (based on the definitions \ref{def6} given in section \ref{sec:fluxes}) as some processes bring energy upscale and some downscale with the latter ones dominating for HIT.   
 \begin{figure*}[htbp]                                                                  %
\centering                                                                              %
\includegraphics*[width=0.7\textwidth]{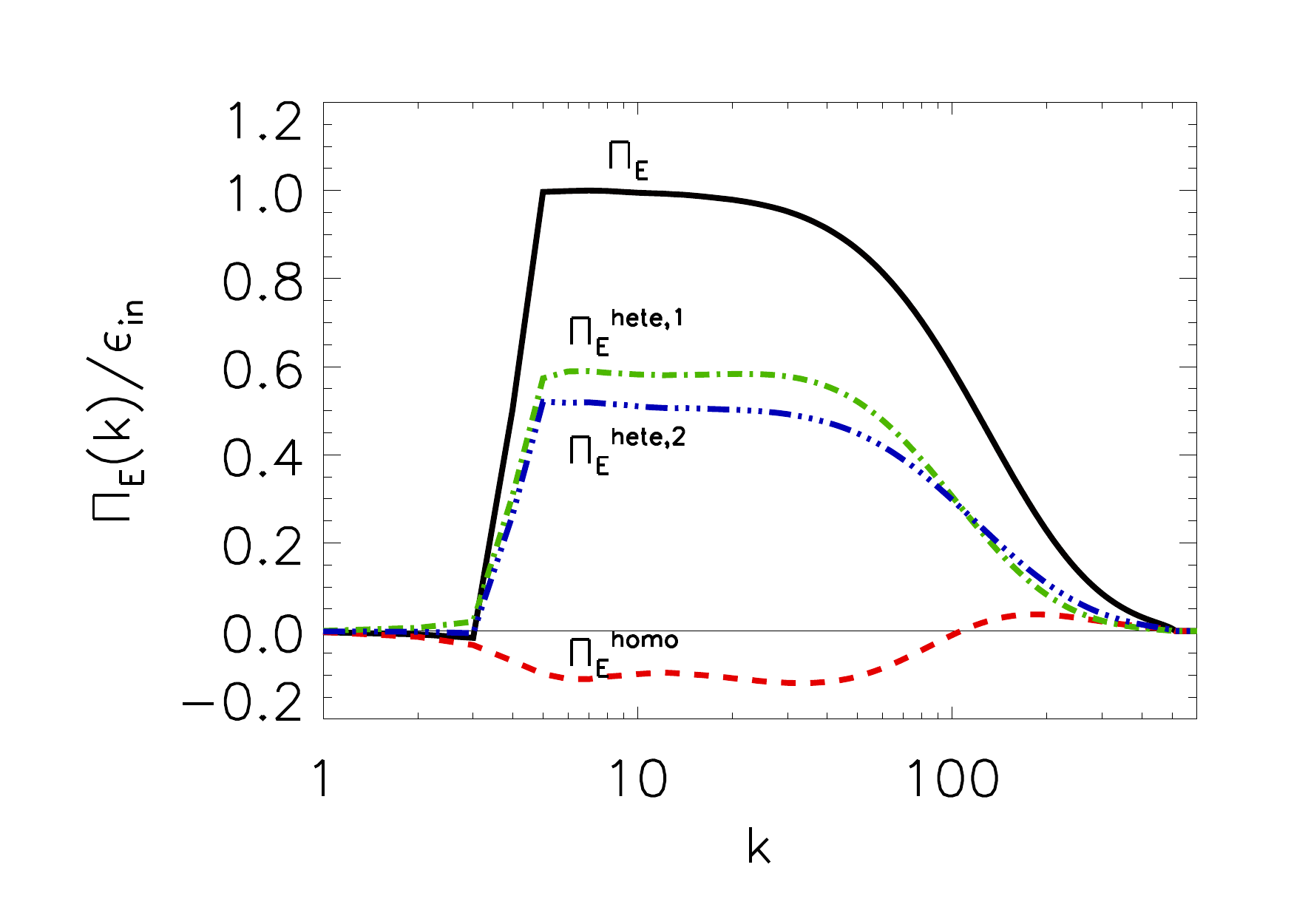}                              %
\caption{Lin-log plot of the fluxes entering in the decomposition (\ref{eq:decompojfm}) %
  from a direct numerical simulation of HIT. Notice the presence of the homochiral channel
  (red dashed line) %
with an inverse energy flux. This is a clear demonstration that the energy transfer   %
for 3D HIT must be classified as a bi-directional case according to Definition          %
(\ref{def6}) of Sec. (\ref{sec:definitions}). Details of the numerical simulation can  %
be found in \cite{alexakis2017Helically}.}\label{fig:alexandrosjfm2017}                %
\end{figure*}                                                                           %
 A similar analysis can be also repeated for the helical conservation law \cite{alexakis2017Helically}. \\

It is important to note that the decomposition of the NSE in  configuration space given by (\ref{eq:alex2016}) is different from that in 4 helical Fourier
classes (I-IV) described in Fig. (\ref{fig:NSheterochiral}). Only the homochiral case coincides. The difference is due to the fact that for the Fourier case the interactions classes (I-IV) are defined in terms of triad structure and as such they  cannot be re-expressed as a product of fields in configuration space (except for the homochiral case where triads are formed by Fourier components with one sign of helical modes only).

The existence of  a subset of fully three-dimensional and isotropic interactions in the NSE that are capable to coherently produce an inverse energy cascade promises to open new directions in the interpretation of many fundamental and applied configurations. Some of them are connected to the case of rotating turbulent at not too small Rossby numbers and  will be discussed in Sec.
(\ref{sec:Rotation}), some others concern  back-scatter events observed in the interactions among sub-grid and resolved-scale fluctuations in Large Eddy Simulations (see Sec. \ref{sec:LES}).  One needs also to remark that {\it being homochiral} is not a property preserved by the  NSE, i.e. without the extra projector on the non-linear evolution of (\ref{eq:GNS+}) any homochiral initial condition, $\bu(\bx,0) = \bu^+(\bx,0)$ would immediately produce also fluctuations with opposite (negative) helical components, due to the coupling among $\bu^+$ and $\bu^-$ introduced by classes II-IV. From this perspective, being {\it homochiral} is more fragile than being {\it two-dimensional three-component (2D3C)}, $\bu(\bx,t) = (u_x(x,y,t),u_y(x,y,t),u_z(x,y,t))$,  or of being {\it fully anti-symmetric}, $\bu(-\bx,t) = -\bu(\bx,t)$, which are two symmetries exactly preserved by the time evolution of NSE. 
\\
Before ending this section, we mention that similar conclusions concerning the existence of inverse energy cascades for homochiral turbulence can also be obtained by studying absolute equilibrium spectra for the H-NSE
with $\alpha=\nu=0$ and $\bfo=0$, if restricted to a finite number of Fourier modes \cite{herbert2014restrictedp} or by using second-order closures based on the so-called Eddy Damped Quasi Normal Markovian Approximation \cite{briard2017closure}. Recently it was also found that the dynamics of triads belonging to class-II have an extra quadratic invariant whose dimension depends on the triad's geometry \cite{rathmann2017pseudo,rathmann2016role}.  As a result, one might argue that this class of triads should lead to a direct or inverse energy cascade depending if they represent local or strongly non-local interactions (if the triad is close to be equilateral or $ q \ll  p, k$) as also confirmed by shell models based on the same helical structure \cite{biferalehelical1996,de2015inverse}.  The helical decomposition as studied in this section can also be applied to 3D MHD as will be  discussed in Sec. (\ref{sec:MHD}). 
\subsubsection{Energy cascade transitions in Helical variants of the Navier-Stokes equations \label{sec:discontinuous}}
In this section we summarize recent attempts made to investigate the balancing between forward and backward energy transfer of heterochiral and homochiral triads in the 3D NSE. 
In order to continuously move from a forward transfer, typically of the original NSE, to a backward transfer, observed for the H-NSE  (\ref{eq:GNS+}), one might follow  different protocols.

One direction was pursued in \cite{kessar2015non}, where the effect of different interactions in the Navier–Stokes equations was investigated by suppressing the negative helicity modes using a dynamical forcing function. The dynamical forcing controlled the amount of helicity at all scales. These authors showed that interactions from three positive helical modes transfer energy to the large scales. However, in the presence of negative helical modes, with even weak amplitude, the cascade of energy remained forward. Their results were also quantified by calculating the energy fluxes due to the different interactions among helical modes.
 
Another  possibility is to vary the number of Fourier modes that are projected to one homochiral sector (say the positive one to fix the notation) so if all modes are projected one obtains (\ref{eq:GNS+}) while if none 
is projected one returns to the NSE. This can be achieved by defining an
operator ${D}^\lambda$ that projects  only a percentage $\lambda$ of randomly chosen Fourier modes on positive helical components.
\begin{figure*}[htbp]                                                                     %
\centering                                                                                %
\includegraphics*[width=1\textwidth, angle=0]{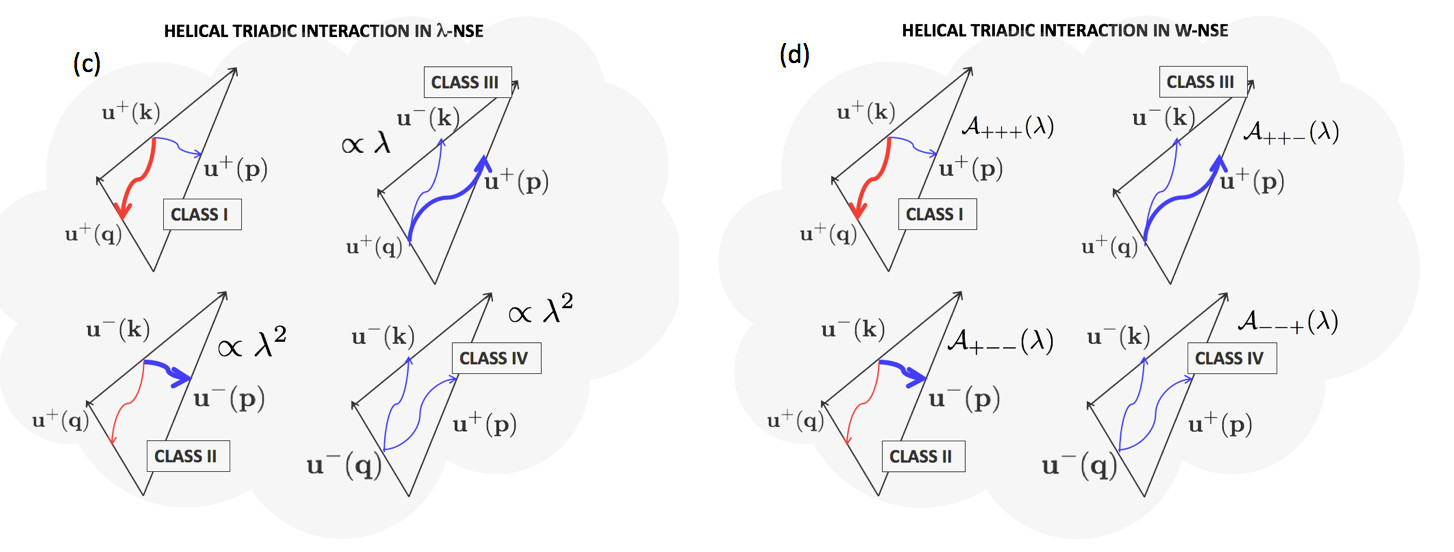}           %
\caption{Sketch of the NSE interacting triadic cloud in terms of the Helical-Fourier      %
decomposition (\ref{NSH}) with  helical decimation protocols different from the one on    %
homochiral triads depicted in panel (b) of figure \ref{fig:NSheterochiral}.              %
(c): Quenched stochastic surgery of NSE interactions performed using the projector        %
(\ref{eq:projv}).  (d) Sketch of the reweighing protocol described by                     %
(\ref{eq:NSEprl}). }\label{fig:NSheterochiral2}                                           %
\end{figure*}                                                                             %
More precisely, ${D}^\lambda$ action on a  generic function $\bg(\bx,t)$ is given by:
\begin{equation} 
  \label{eq:projv}
        {\bg}^\lambda(\bx,t) \equiv {\cal D}^{\lambda} {\bg}(\bx,t) \equiv \sum_{\bk} \tilde {\cal  D}^\lambda_{\bk} \tilde \bg(\bk,t) \, e^{i{\bm k}\bx}\ 
\end{equation}
{ where the operator $\tilde {\cal  D}^\lambda_{\bk}$ is either
 $\tilde {\cal  D}^\lambda_{\bk} = \tilde {\cal P}^+_{\bk}  $  
 with probability $\lambda$ or $\tilde {\cal  D}^\lambda_{\bk} =1$ with probability $1-\lambda$  for any  given wave number $\bk$.
 }  
The $\lambda$-decimated Navier-Stokes equation ($\lambda$-NSE) is then given by: 
\begin{equation} 
  \label{eq:lambdaNSE}
  \partial_t \bu^\lambda = {\cal D}^{\lambda}[- \bu^\lambda \cdot {\bm \nabla} \bu^\lambda -{\bm \nabla} P^\lambda] 
  +\nu \Delta \bu^\lambda -\alpha  \bu^\lambda + \bfo^\lambda.
\end{equation}
Clearly,  for $\lambda=0$ we have $\tilde {\cal  D}^\lambda_{\bk} =1$ for every wavenumber
and we are back to the original NSE, while for $\lambda=1$ 
we have  ${D}^\lambda ={\cal P}^+ $ and return to  the homochiral case (\ref{eq:GNS+}). The system is such that we always have the backbone of positive helical modes, while depending on the value of $\lambda$, a fraction of $1-\lambda$ negative helical modes are also allowed to participate in the dynamics (see panel (c) in Fig. \ref{fig:NSheterochiral2}). The transition from direct ($\lambda=0$) to inverse ($\lambda=1$) energy cascade has been studied in a series of recent papers
\cite{sahoo2015,sahooEPJE}, where it was shown that the system develops a {\it discontinuous (1st order)} transition} as that depicted in panel (c) of  Fig. (\ref{fig:classification}). Moreover, the transition appears to be also quasi singular, in the sense that the critical value of the control parameter is very close to the full homochiral value, $\lambda_c \sim 1$. In other words, it is enough to maintain a very small percentage of modes of opposite helicity to establish a direct energy cascade. A summary of the spectral properties for both $E^+(k)$ and $E^-(k)$ is shown in figure (\ref{fig:lambdaNSE}). The critical role of heterochiral interactions was already conjectured in the original papers by Waleffe \cite{waleffe1992nature} on the basis of the single instability of each triadic class. In terms of helical-Fourier interactions, it means that heterochiral triads are much more efficient than homochiral ones in the energy transfer and that in a sea of homochiral modes, the introduction of modes of the other chirality acts as a catalyzer for the forward energy transfer. Similar conclusions where also obtained by looking at the statistical equilibrium in \cite{herbert2014restrictedp}. \\ 

\begin{figure*}[h!]                                                                         %
\centering                                                                                  %
\includegraphics[width=0.92\textwidth]{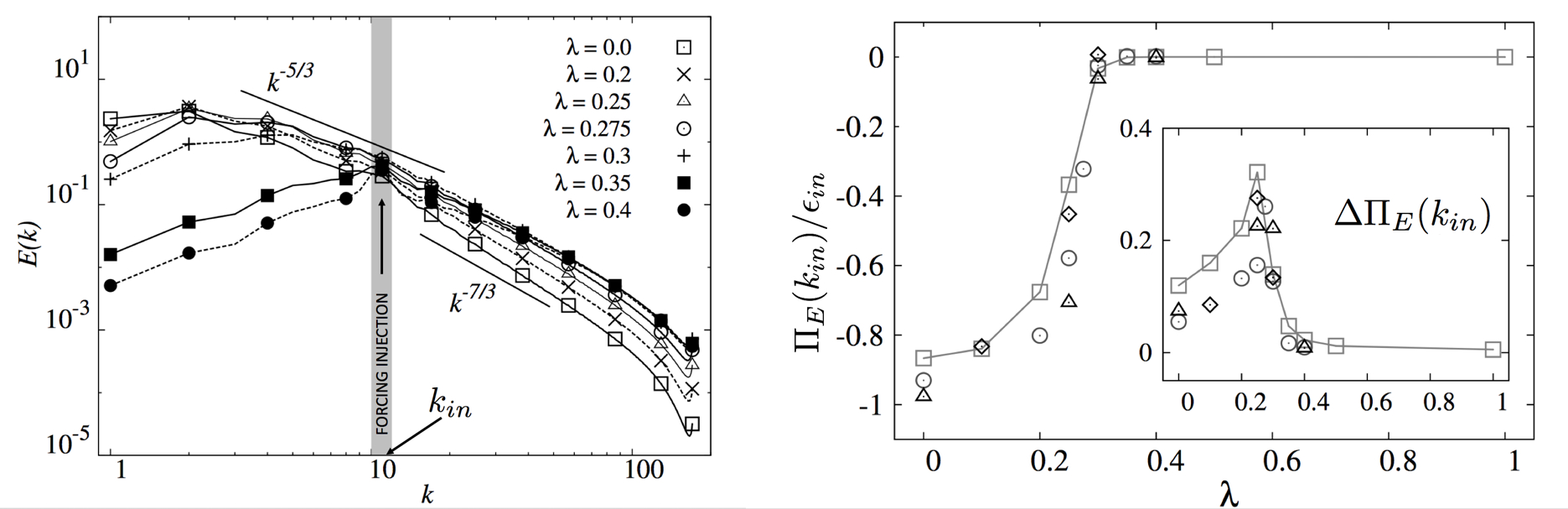}                       %
\caption{Log-log plot of spectra (left) and lin-lin plot of fluxes (right) at changing the  %
weighting parameters $\lambda$ in the simulations of  (\ref{eq:NSEprl}). Transition from   %
zero inverse energy flux (and equipartition spectrum) at $k < k_{in}$  to an inverse energy %
cascade $-5/3$ spectrum is observed around  $\lambda_c \sim 0.3$. Grey region indicates the %
scales where the energy is injected. Different symbols in the right panels correspond to    %
different Reynolds numbers. In the inset of the right                                       %
panel we show the fluctuations of the flux at the injection scale, calculated from the      %
standard deviations of measurements at different times, normalized with $\ein$.  Notice the %
increasing fluctuations by approaching $\lambda_c$. For details about the numerical         %
simulations see  \cite{sahoo2017helicity}. }                                                %
\label{fig:NSEprl}                                                                          %
\end{figure*}

Another alternative way to study the transition from the original NSE (\ref{eq:GNS}) to their homochiral limit (\ref{eq:GNS+}) was performed in \cite{sahoo2017helicity} following the idea of partitioning the  nonlinear term of the NSE in subset of interactions with different coupling properties among homochiral and heterochiral triads \cite{alexakis2017Helically}.
\begin{figure*}[h!]                                                                        
\centering                                                                                
\includegraphics[width=0.92\textwidth]{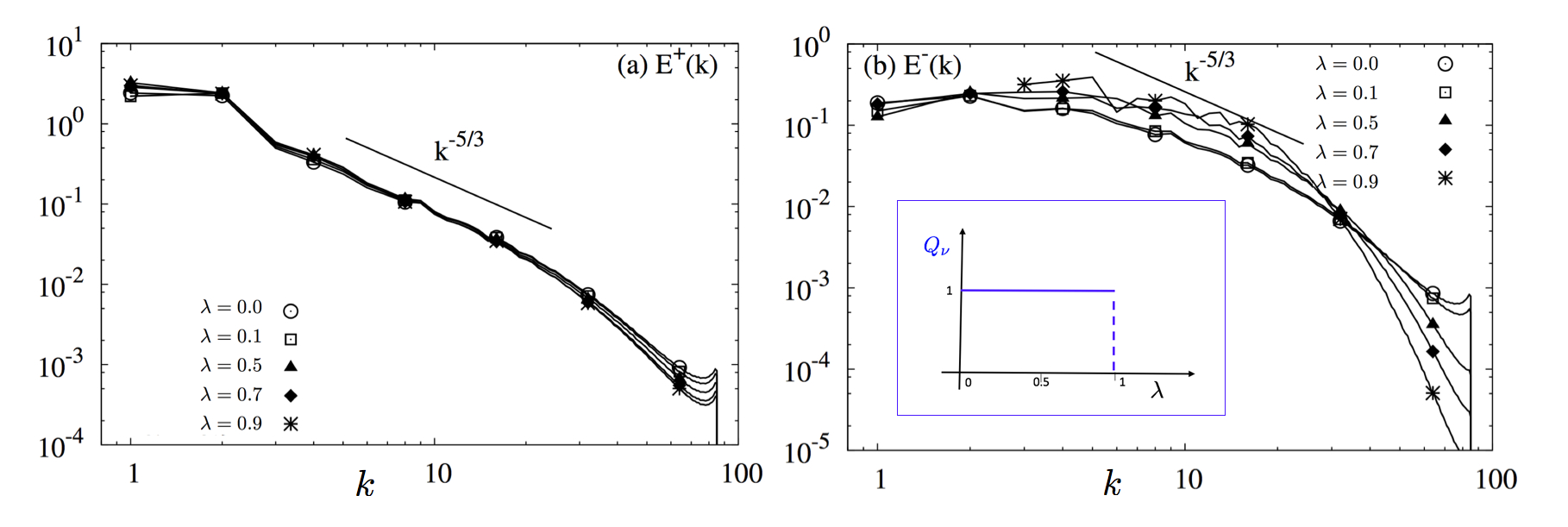}                                       
\caption{(a) Log-log plot of the positive helical spectrum $E^+(k)$ at changing the fraction $\lambda$ of negative helical modes for the decimation protocol depicted in the left panel of Fig. (\ref{fig:NSheterochiral2}). (b) The same but for the fraction of negative helical modes $E^-(k)$. In the inset we sketch the phase diagram of $Q_\nu$ for the transition from direct to inverse cascade at changing $\lambda$ corresponding to  case (c) of Fig. (\ref{fig:classification}). From panel (b) one notices  the catalytic role of modes with opposite chirality, which become larger and larger while decreasing their total number.   For details about the numerical set up see \cite{sahoo2015}. }
\label{fig:lambdaNSE}                                                                             
\end{figure*}                                                                               %
The idea in \cite{sahoo2017helicity} is to start from  the helical decomposition (\ref{eq:alex2016}) giving different weights to homochiral and heterochiral non-linear interactions:
\be
\label{eq:NSEprl}
\partial_t {\bf u}^{s_1} =  {\mathcal P}^{s_1} \sum_{s_2,s_3}{\cal A}_{s_1,s_2,s_3}(\lambda)[{\bf u}^{s_2} \times \bw^{s_3}]  + \nu  \Delta {\bf u}^{s_1} -  \alpha  \bu^{s_1} + {\bf f}^{s_1}.
\ee
where the prefactor is
\be
\begin{cases}
  {\cal A}_{s_1,s_2,s_3}(\lambda) = 1, \qquad \mbox{for homochiral triads} \\
  {\cal A}_{s_1,s_2,s_3}(\lambda) = \lambda, \qquad \mbox{for heterochiral triads} \\
\end{cases}
\ee
In such a way  for $\lambda=1$ we are back to NSE while for $\lambda=0$ we have  two independent homochiral-NSE (see panel (d) of Fig. \ref{fig:NSheterochiral2}). Unlike (\ref{eq:lambdaNSE}) the modified NSE given by (\ref{eq:NSEprl}) does not break helical symmetry  for any $\lambda$. Furthermore,  (\ref{eq:NSEprl}) remains invariant under dilatation as the original NSE.  In Fig. (\ref{fig:NSEprl}) we show that (\ref{eq:NSEprl}) displays a discontinuous energy transfer transition from inverse to direct cascade for $\lambda_c \sim 0.3$. Indeed, the  flux transition becomes more and more abrupt by increasing the Reynolds number. In the limit of infinite $Re$  it is expected to converge to a sharp jump (first-order transition) as the one depicted in panel (c) of  Fig. (\ref{fig:classification}).  This empirical observation was justified in \cite{sahoo2017helicity} by noting that for any value of $\lambda$ the nonlinearity in (\ref{eq:NSEprl}) remains invariant under a dilatation transformation both at scales larger and  smaller than the forcing scale.  This observation is hardly  compatible with the establishment of a Re independent split-cascade. Indeed, for the latter to exist there must be some mechanism that acts differently at small and large scales such as to push the system to move the energy in both directions. Thus, an important requirement for a flow to display a split-cascade is to break scale invariance. This is indeed what  happens in all the other applications where one observes a split-energy cascade, as  will be  discussed in the next sections for the case of flows in thin layers, under strong rotation or stratification.

\subsubsection{Summary}  In this section we have discussed the joint energy-helicity cascade dynamics in homogeneous and isotropic turbulence.
We have shown that in the  presence of an injection concentrated in a limited range of  characteristic scales, energy and helicity are cascading forward in HIT, leading to a co-directional dual cascade according to definitions (\ref{def8}-\ref{def9}) in Sec. (\ref{sec:definitions}).
If the NSE are restricted to homochiral sectors, we have a counter-directional dual cascade, with  an inverse  energy transfer and a  forward helicity cascade (see definition (\ref{def9}) in Sec. \ref{sec:definitions}).
Homochiral interactions  give an inverse contribution to the flux even in the full NSE, confirming that inside all 3D turbulent flows there are channels able to
transfer energy in both directions, a clean example of our definition of a bidirectional cascade (\ref{def6}) in Sec. (\ref{sec:definitions}). The inverse stable energy cascade prevails also when the heterochiral triads are re-weighted. A transition from direct to inverse cascade as that depicted in case (c) of Fig. (\ref{fig:classification}) might be triggered by suitable manipulation of non-linear terms. 
  
 \subsection{ {Turbulence in layers of finite thickness} \label{sec:Thin}}     

 We begin with flows in layers of finite height $H$ (which we fix for the sake of reference to be  in the $z$-direction) much smaller than the box length $L$ in the other two directions ($x$ and $y$). 
 It comprises possibly the simplest case that can display a split cascade.
\begin{figure*}[h!]                                                                         %
\centering                                                                                  %
\includegraphics[width=1.0\textwidth]{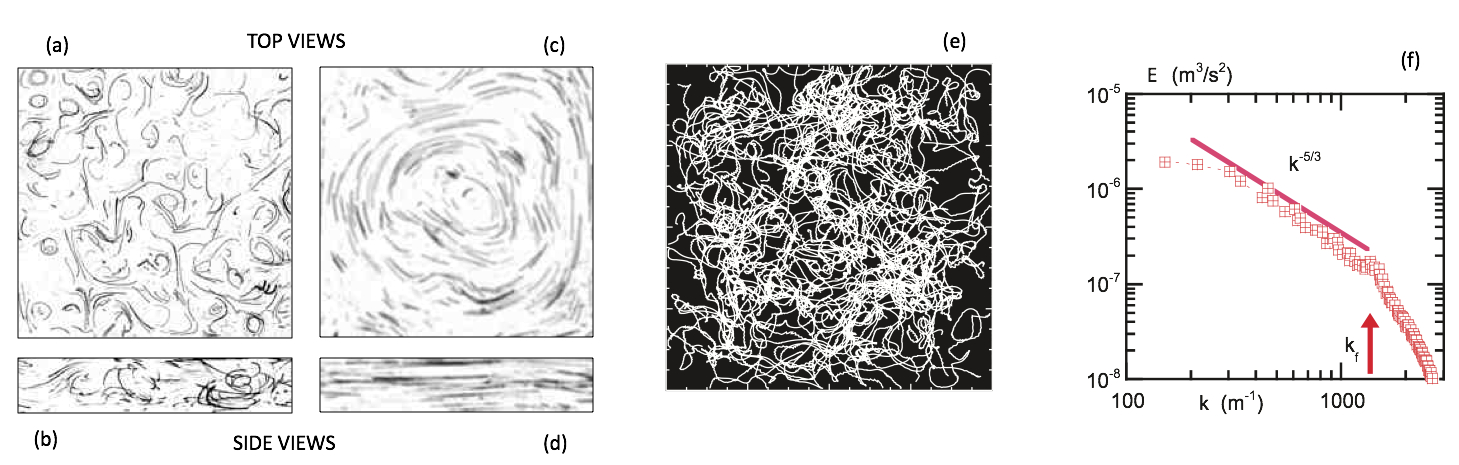}                                    %
\caption{ Turbulence forced electromagnetically by the array of magnetic dipoles,           %
in the top layer of electrolyte resting on a layer of non-conducting heavier liquid.        %
The forcing scale is comparable with the layer thickness                                    %
$\ell_{in} \sim H$ and $L \sim 10 H$.  Top and later view before (a-b) and after the        %
formation of the condensate (c-d)  where particle streaks are visualized.                   %
Plots reprinted by permission from \cite{Xia2011upscale}.                                   %
(e-f): Lagrangian trajectories and an inverse energy spectrum in 2D turbulence generated by %
a Faraday-wave experiment, reprinted  with permission  from \cite{xia2013lagrangian}.}      %
\label{fig:shats}                                                                           %
\end{figure*}                                                                               %
 Experimentalists refer to these layers as {\it thick layers} \cite{Xia2011upscale}
 being more accustomed to very thin layers that approximate a 2D flows
 while researchers using numerical simulations refer to them as {\it thin layers} \cite{benavides2017critical}
 since most simulations are performed in cubic boxes and therefore anything flatter than a cube is thin.
 We refer to them here as layers of finite thickness to reconcile  the two points of view.
 Flows in confined layers are important  because they are close to atmospheric flows for which the pressure scale height is hundreds/thousands times smaller than the  turbulent {\it horizontal} large scale motions 
 \cite{houghton2002physics}. For this reason, such layers have attracted a lot of attention from experimental \cite{sommeria1982and,sommeria1986experimental,tabeling1987instability,Shats2010turbulence,ecke2017From} and theoretical groups \cite{Celani2010turbulence,musacchio2017split,benavides2017critical}. Here, we are interested in discussing  why under certain conditions they display a transition in the energy cascade direction. Figure (\ref{fig:shats})  shows one experimental visualization of the flow on a thick (left panels) and thin (middle-left panels) layer. The  2D condensate is visible in the latter. In the same figure we also show the inverse energy cascade spectrum (right panel) and the visualization of tracers trajectories (middle-right panel) in a different experiments where the quasi-2D turbulent regime is generated by Faraday-waves \cite{xia2013lagrangian}. 
  
 \subsubsection{Formulation}
\begin{figure*}[h!]                                                                         %
\centering                                                                                  %
\includegraphics[width=0.70\textwidth]{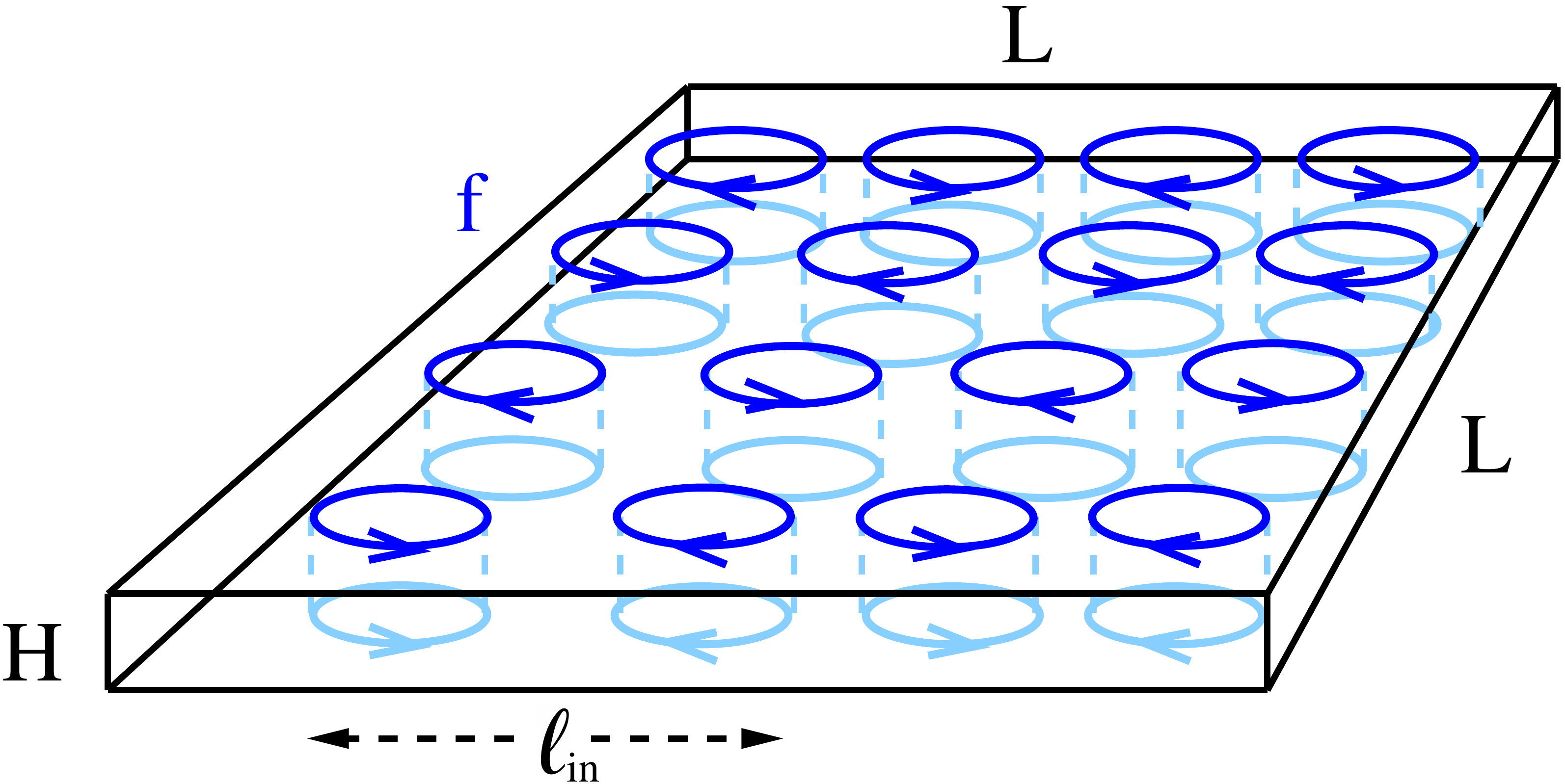}                                       %
\caption{A 3D sketch of a domain of finite thickness where the horizontal    %
extension and forcing scale are much larger than the vertical one $L\gg H$.                                    %
Here, the forcing is represented  to inject energy  at a characteristic length              %
scale $\ell_{in}$ only, and to be z-invariant acting only on the two horizontal components.}%
\label{fig:box}                                                                             %
\end{figure*}

\begin{figure*}[h!]                                                                         %
\centering                                                                                  %
\includegraphics[width=0.45\textwidth]{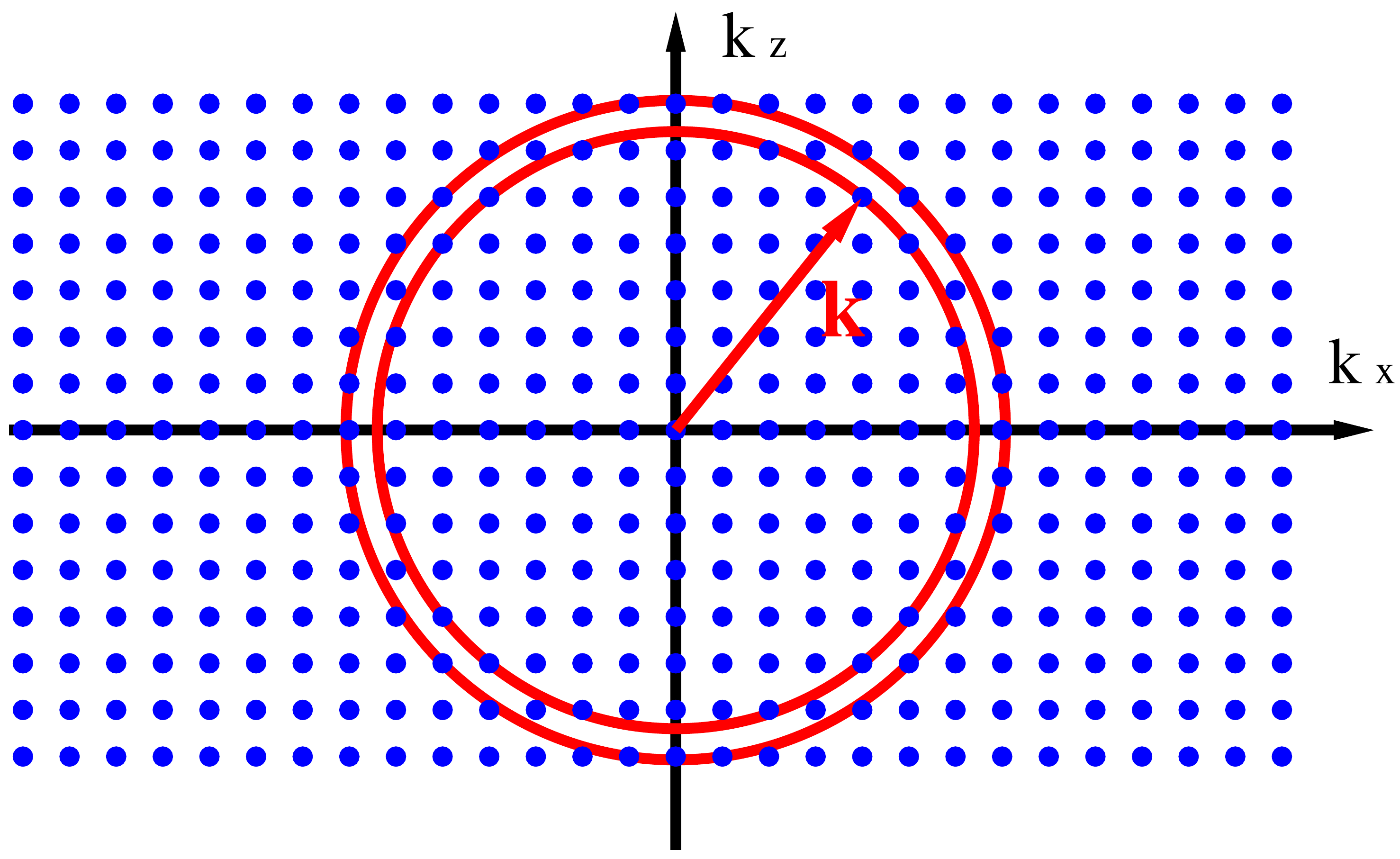} \quad                             %
\includegraphics[width=0.45\textwidth]{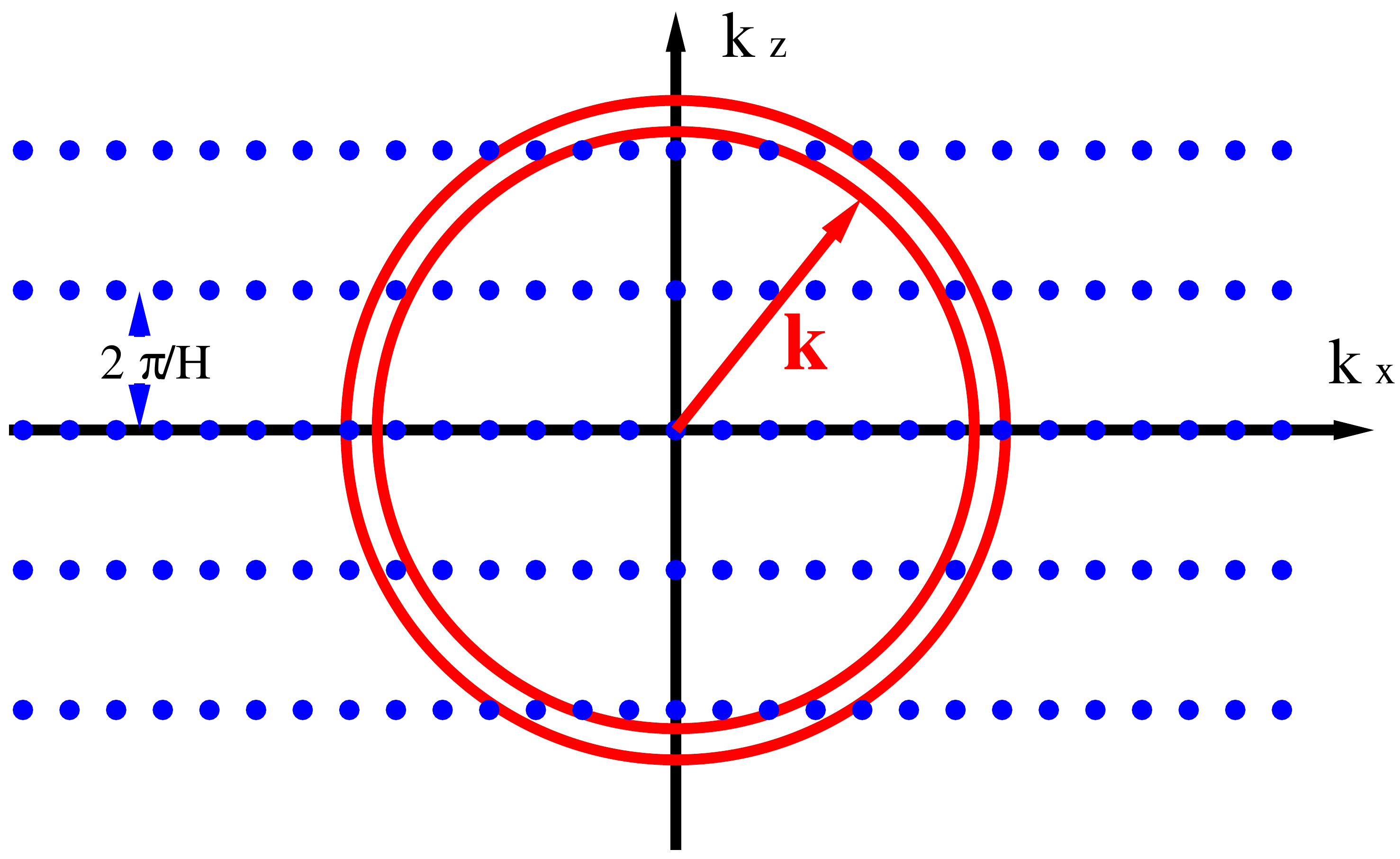}
\caption{                                                                                   %
         Distribution of discrete Fourier wavenumbers (blue dots)  in a cubic domain (left) %
         and in a thin layer (right). The distance between the wavenumbers  is proportional %
          to $2\pi/H$ in the vertical direction and $2\pi/L$ in the horizontal direction.   %
          The red circle is sketched to demonstrate the number of modes included in a       %
          spherical shell of radius $k$ and width $\Delta k$. if $k<2\pi/H$ only modes in   %
          the 2D3C manifold $k_z=0$ are included.                                       }  %
\label{fig:Fgrid}                                                                           %
\end{figure*}

 For simplicity the boundary conditions are assumed to be periodic in all directions.
 In experimental setups, however, we have no-slip boundary conditions and boundary-layer effects
 must be taken into account. As usual, we will assume the flow to be forced at some  particular length scale $\lin$. 
 An illustration of the flow set-up is shown in 
 Fig. (\ref{fig:box}). 
This configuration  has been examined in many numerical investigations
\cite{Celani2010turbulence,musacchio2017split,benavides2017critical} and it is, within  
a good approximation, a realistic description of flows in liquid metal embedded in a uniform magnetic field and forced by electromotive forces \cite{sommeria1982and,sommeria1986experimental,tabeling1987instability,Shats2010turbulence}. This set-up can also be important for the case
of Rayleigh-Benard convection at large aspect-ratio \cite{bailon2010aspect}. 
The compression of  the vertical direction to smaller and smaller  values leads to some anisotropic
distribution of Fourier modes, where  the vertical ones being wider apart than horizontal wavenumbers as shown in Figure \ref{fig:Fgrid}. This anisotropic distribution of wavenumbers
{ makes the $k_z=0$ modes with small horizontal wavenumbers 
$k_\perp=\sqrt{k_x^2+k_y^2}$ coupled to the remaining modes,  $k_z\ne0$,  only by highly non-local triads in Fourier space. In case of  full decoupling, we would remain with a pure 2D dynamics for the vertically averaged modes described by the so-called 2D3C manifold at $k_z=0$.}   \\
It is useful to start from the original 3D Navier-Stokes equations in the thick layer and
split it in the evolution of two coupled fields. 
The first is the  2D3C component, which evolves in the in-plane directions  $(x,y)$  only and it is
obtained by averaging over the vertical direction. The second is a full  3D field defined
as the remaining part once  the 2D3C component is subtracted.
The 2D3C field is defined as: 
${\bf u}_{_{2D}}(x,y,t) = (\overline{u}_x(x,y,t),\overline{u}_y(x,y,t),\overline{u}_z(x,y,t))$ where  with
$\overline{\bullet} = 1/H \int_0^H dz (\bullet)  $ we  indicate a $z$-averaged quantity, while  the 3D component is given by 
${\bf u}_{_{3D}}(x,y,z,t) = {\bf u}(x,y,z,t) -{\bf u}_{_{2D}}(x,y,t)$.
The original NSE rewritten in terms of the two
fields are: 
\begin{align}
\partial_t {\bf u}_{_{2D}} + {\bf u}_{_{2D}} \cdot \bnabla {\bf u}_{_{2D}}& = -\overline{ {\bf u}_{_{3D}} \cdot  \bnabla {\bf u}_{_{3D}} } - \overline {\bnabla P}  +  \nu \Delta {\bf u}_{_{2D}}  -  \alpha {\bf u}_{_{2D}}   +  {\bf f}_{_{2D}}, \label{thinlayera} \\
\partial_t {\bf u}_{_{3D}} + {\bf u}_{_{2D}} \cdot \bnabla {\bf u}_{_{3D}} &= -{\bf u}_{_{3D}} \cdot  \bnabla {\bf u}_{_{2D}}  +(\overline{ {\bf u}_{_{3D}} \cdot  \bnabla {\bf u}_{_{3D}} }
-{\bf u}_{_{3D}} \cdot  \bnabla {\bf u}_{_{3D}}) - (\bnabla P - \overline{\bnabla P})  -\alpha {\bf u}_{_{3D}} +  \nu \Delta {\bf u}_{_{3D}}   +  {\bf f}_{_{3D}}.
\label{thinlayerb}
\end{align}
It is important to realize that in the absence of the ${\bf u}_{_{3D}}$ components the evolution of ${\bf u}_{_{2D}}$ can be further split in two equations, one for the vertically averaged horizontal components, ${\bf u}_{_{2D}}^\perp= (\overline{u}_x,\overline{u}_y)$ which satisfies a purely 2D (and 2 components) Navier-Stokes equation:
\begin{align}
\partial_t {\bf u}^\perp_{_{2D}} + {\bf u}^\perp_{_{2D}} \cdot \bnabla {\bf u}^\perp_{_{2D}}& =- \overline {\nabla P}  +  \nu \Delta {\bf u}^\perp_{_{2D}}  -  \alpha {\bf u}^\perp_{_{2D}}   +  {\bf f}^\perp_{_{2D}}, \label{thinlayera2D}
\end{align}
and one for the third (out-of-plane) component that satisfy the equation of a passive scalar advected by the in-plane field:
\begin{align}
\partial_t \overline{u}_z + {\bf u}^\perp_{_{2D}} \cdot \bnabla \overline{u}_z& = +  \nu \Delta \overline{u}_z  -  \alpha \overline{u}_z   + \overline{f}_z. \label{thinlayera3C}
\end{align}
In many empirical studies, in order to make the energy injection almost independent of the thickness of the volume, it is customary to take the forcing invariant in the z-direction, with non vanishing horizontal components only, ${\bf f}_{_{3D}} =0$. Here, we prefer to keep the most general set-up and explicitly mention the restriction to the pure 2D forcing only when relevant.  \\
As said,  in the absence of ${\bf u}_{_{3D}}$ the system returns to the 2D Navier Stokes equations plus the evolution of a passive scalar. This system is often referred to as {\it two dimensional three component} (2D3C) flow.  The in-plane dynamics  conserves both energy and enstrophy and will be characterized by a dual cascade with energy flowing backward and enstrophy forward. The energy contained in the out-of-plane component will flow forward as in the case of a passive scalar in  incompressible flows (see also section \ref{sec:passivescalarcascades}). As a result,  if the ${\bf u}_{_{3D}}$ is non-zero but very small the system can still {\it quasi}-conserve enstrophy of the in-plane component and possibly lead to an inverse cascade while if ${\bf u}_{_{3D}}$ is big, no inverse cascade should be possible. \\
To quantify the role played by the different components in the global energy transfer we look at the energy balance among the 2D3C and 3D fields, by defining the total energy in the 2D plane:
$\mathcal{E}_{2D}=\frac{1}{2}\langle|{\bf u}_{_{2D}}|^2\rangle $  
and the one in the remaining 3D volume:
$\mathcal{E}_{3D}=\frac{1}{2}\langle|{\bf u}_{_{3D}}|^2\rangle $: 
\begin{align}
\partial_t \mathcal{E}_{_{2D}}(t) &= \ein^{{2D}}(t) - \mathcal{T}(t) - \enu^{{2D}}(t) - \emu^{{2D}}(t)\\
\partial_t \mathcal{E}_{_{3D}}(t) &= \ein^{{3D}}(t) + \mathcal{T}(t) - \enu^{{3D}}(t) - \emu^{{3D}}(t)
\end{align}
where $\ein^{2D},\ein^{3D}$ express the rate energy is injected in the 2D and 3D 
part of the flow respectively and similar for the dissipation terms 
$\enu^{{2D}} , \emu^{{2D}},\enu^{{3D}} , \emu^{{3D}}$. 
The term 
$\mathcal{T}=\langle  {\bf u}_{_{2D}} \cdot ({\bf u} \cdot \nabla{\bf u}_{_{3D}}) \rangle$ 
is responsible for the transfer of energy from the 2D plane to the 3D volume.\\
To fully describe the energy spectrum,
we need to control the distribution of energy $E(k_\perp,k_\|)$ where we indicate with $k_\perp=\sqrt{k_x^2+k_y^2}$ and $k_\| = |k_z|$ and we assumed isotropy only around the vertical axis.
It expresses the energy density in a ring of radius $k_\perp$ and width
$\Delta  k_\perp= 2\pi/L$ (in a box with finite horizontal width) located at height
$\pm k_\|$ and vertical width $\Delta k_\|= 2\pi/H$ (in the finite thickness case). 
Its relation with total energy is given by
\be
\cE =\Delta  k_\perp \Delta k_\|  \sum_{k_\perp,k_\|} E(k_\perp,k_\|) \label{eq:Aspec1}
\ee
while its relation with the spherically averaged energy spectrum is given by
\be
E(k) =  \frac{\Delta k_\perp \Delta k_\| }{\Delta k}
 \sum_{ \scriptsize
 \begin{array}{c}  k_\perp,k_\| \\  k \le \sqrt{ k_\perp^2+k_\|^2} < k+\Delta k  \end{array} } E(k_\perp,k_\|). \label{eq:Aspec2}
\ee
Note that the above  spherically averaged spectrum has units
of energy per wavenumber
while $E(k_\perp,k_\|)$ has units of energy per  wavenumber squared. To avoid further introduction of notation
we use the same symbol and distinguish the two by the number of their arguments. Because it is hard to draw
quantitative conclusions from a 2D spectrum, 
cylindrically averaged or plane averaged spectra
are often used:
\be
E_\perp(k_\perp) = \Delta  k_\|    \sum_{k_\|    } E(k_\perp,k_\|), \qquad 
E_\|   (k_\|   ) = \Delta  k_\perp \sum_{k_\perp } E(k_\perp,k_\|).  \label{eq:Aspec3}
\ee
Since in general $E(k_\perp,k_\|) $ depends on both wavenumbers, some  care needs to be taken when  comparing the spectra $E_\perp(k_\perp),E_\|   (k_\|   )$ with theoretical predictions. \\

\begin{figure*}[htbp]                                                                     
\centering                                                                                
\includegraphics*[height=0.30\textwidth]{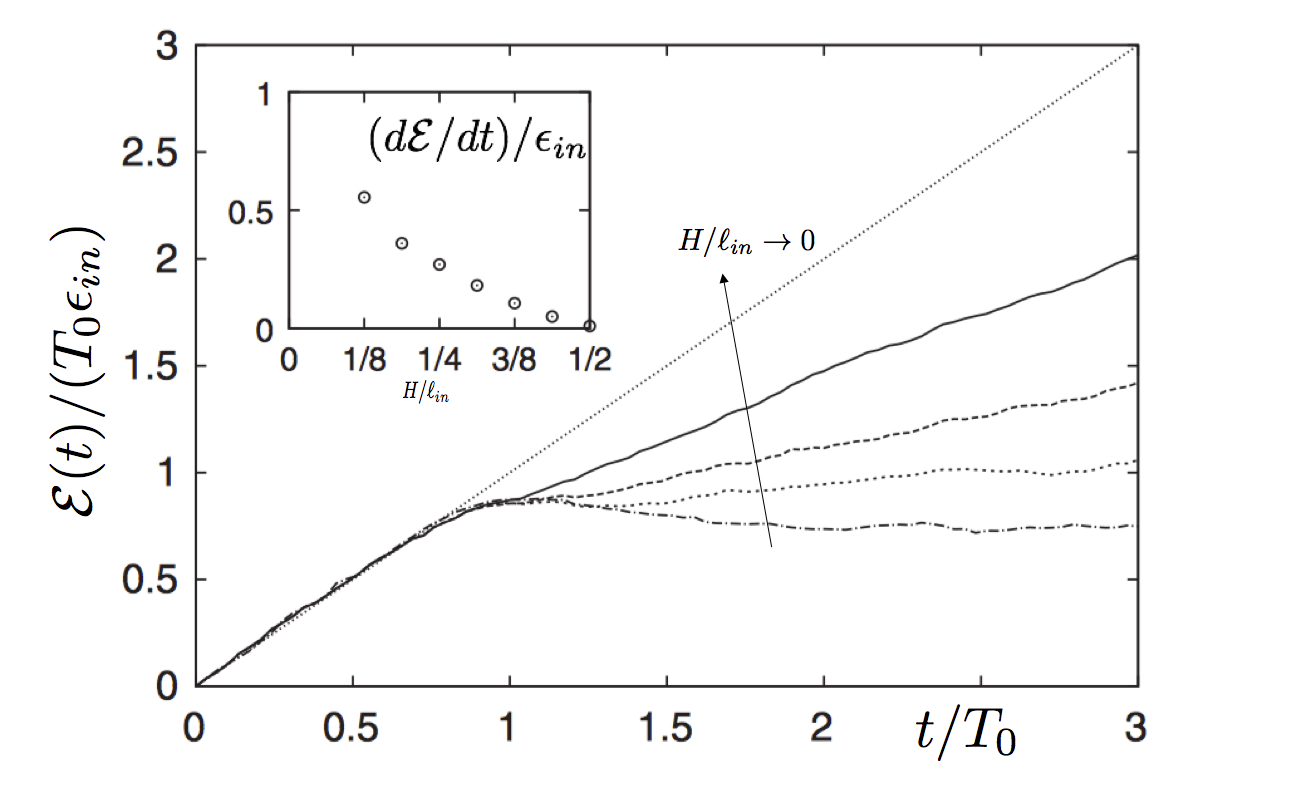}                                
\includegraphics*[height=0.30\textwidth]{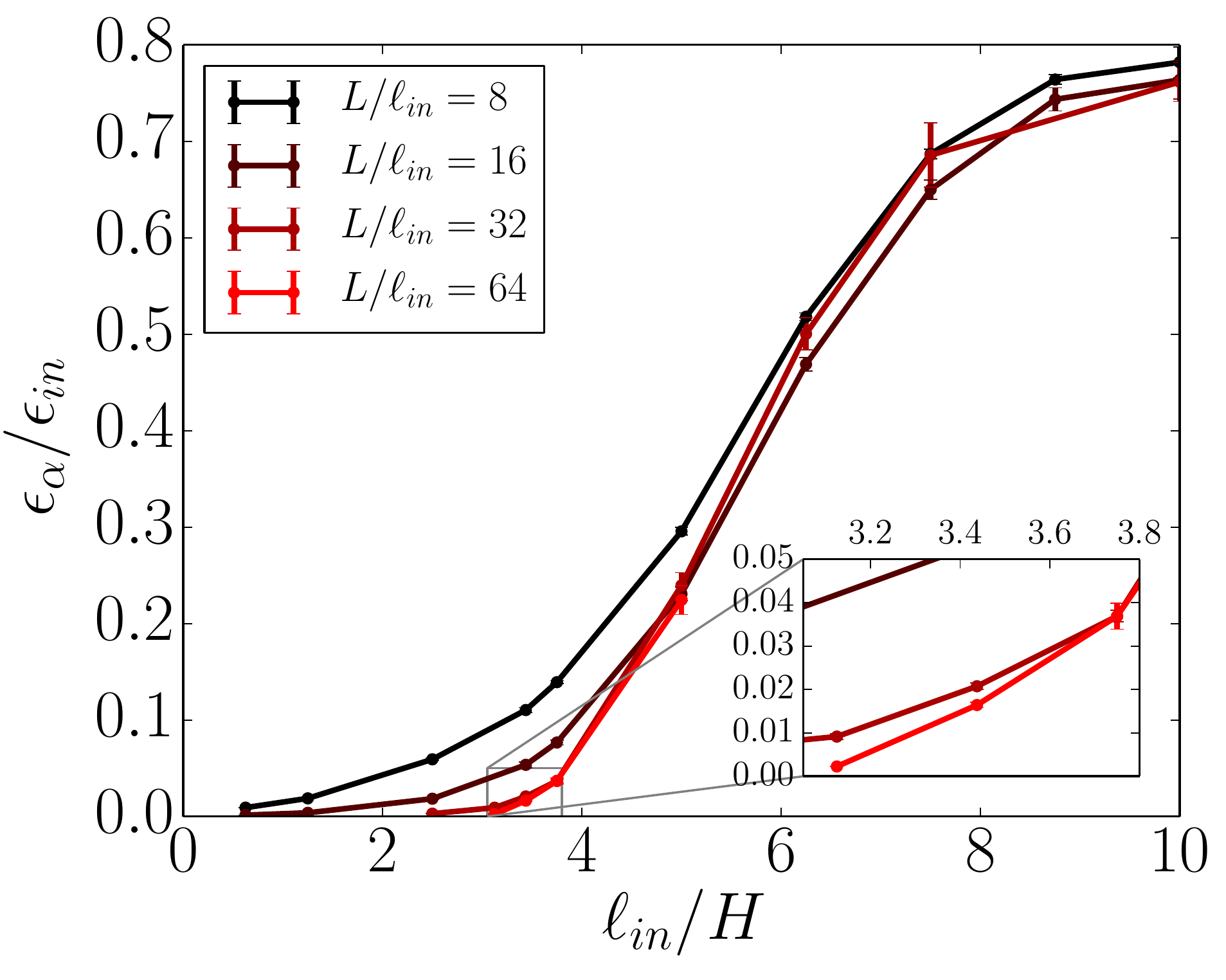}                           
\caption{Left panel: energy evolution for a turbulent flow in a layer of finite thickness 
for four different layer heights in the absence of  large-scale drag, $\alpha=0$.        
The linear behavior after $t\simeq T_0$ has a slope proportional to the                   
inverse energy flux, $d {\cal E}/dt$. Inset: the normalized inverse flux  {\it vs}        
the normalized height, $H/\lin$. Notice the apparent presence of a transition for         
$H/\lin\sim 1/2$. The figure is adapted from the results in \cite{Celani2010turbulence}.  
Right panel: $Q_\alpha=\emu/\ein$ as a function of $H/\lin$ for a model flow in a layer   
where only one vertical mode is allowed for four different values of the horizontal domain
size $L/\lin$. The transition converges to a critical (2nd order) for  $L/\lin\to \infty$.
         The figure is adapted from the results in \cite{benavides2017critical}. }        
\label{fig:converge}                                                                      
\end{figure*}

In terms of the evolution of the spectral fluxes the coupling among the 2D spectrum, $E_{_{2D}}(k,t) = E(k,k_\|=0,t)$,  and the 3D one, $E_{_{3D}}(k,t) = E(k,t) - E_{_{2D}}(k,t)$, is the following:
\begin{align} 
\sum_{k'<k} \partial_t  E_{_{2D}}(k',t)  &=
 -\langle {\bf u}_{_{2D}}^{<k} \cdot ({\bf u}_{_{2D}} \cdot \bnabla{\bf u}_{_{2D}}) \rangle
 -\langle {\bf u}_{_{2D}}^{<k} \cdot ({\bf u}         \cdot \bnabla{\bf u}_{_{3D}}) \rangle  
  +\langle \bbf \cdot \bu_{_{2D}}^{<k} \rangle  
 -\nu \langle |\bnabla {\bf u}_{_{2D}}^{<k}|^2\rangle
 -\alpha \langle |{\bf u}_{_{2D}}^{<k}|^2 \rangle
 \label{2DbalanceA}\\
 \sum_{k'<k} \partial_t E_{_{3D}}(k',t)  &=
 -\langle {\bf u}_{_{3D}}^{<k} \cdot ({\bf u} \cdot \bnabla{\bf u}_{_{3D}}) \rangle
 -\langle {\bf u}_{_{3D}}^{<k} \cdot ({\bf u} \cdot \bnabla{\bf u}_{_{2D}}) \rangle 
 +\langle \bbf\cdot \bu_{_{3D}}^{<k} \rangle 
 -\nu \langle |\bnabla {\bf u}_{_{3D}}^{<k}|^2\rangle
 -\alpha \langle |{\bf u}_{_{3D}}^{<k}|^2 \rangle
 \label{3DbalanceB}
\end{align}
where we have used the notation, ${\bf u}^{<k}(\bx,t) $ to denote a velocity field projected on all modes with $|\bk| < k$, as already introduced in (\ref{eq:lowpass}) . 
The first term 
$\Pi_{2D}(k) =  -\langle {\bf u}_{_{2D}}^{<k} \cdot ({\bf u}_{_{2D}} \cdot \bnabla{\bf u}_{_{2D}}) \rangle$ on the right hand side of (\ref{2DbalanceA})
is the flux due   to the  ${\bf u}_{_{2D}}$ components only and it is expected to have a negative contribution so that it transfers energy to the large scales. The terms 
$ -\langle {\bf u}_{_{2D}}^{<k} \cdot ({\bf u}         \cdot \bnabla{\bf u}_{_{3D}}) \rangle$ and  
$-\langle {\bf u}_{_{3D}}^{<k} \cdot ({\bf u} \cdot \bnabla{\bf u}_{_{2D}}) \rangle$
are terms that transfer energy from one field ${\bf u}_{_{2D}}$ to the other 
${\bf u}_{_{3D}}$, while the remaining term $-\langle {\bf u}_{_{3D}}^{<k} \cdot ({\bf u} \cdot \bnabla{\bf u}_{_{3D}}) \rangle$ is the term responsible for the transfer of the energy
of the ${\bf u}_{_{3D}}$ field to the small scales.  Adding these last three terms together
we obtain the contribution from  the 3D field only:
  \be
  \label{eq:2d3dlayer}
\Pi_{3D}(k)= \Pi(k)- \Pi_{2D}(k)
\ee
that is expected to lead to a positive flux. 
Just like we did  for homochiral and heterochiral contributions in   Sec. (\ref{sec:Helicity}), the  decomposition (\ref{eq:2d3dlayer})  is meant to separate  processes with a positive flux that favor the forward cascade from processes with a negative flux that favor the inverse cascade.
At varying the ratio $H/\ell_{in}$ and at changing the way we force, by switching on/off the three dimensional forcing components, ${\bf f}_{_{3D}}$, and the component of the forcing in the out-of-plane
direction, $\overline{f}_z$,  we can reinforce or deplete the relative role played by the 2D manifold in the global dynamics. As a result,
it is natural to expect that the system can make a transition from purely forward cascade to purely inverse energy cascade. In the following
we discuss numerical, experimental and theoretical evidences in favour of the existence of such a transition and concerning its properties.\\ 
\noindent If $H/\lin$ is small enough, the energy injected by the forcing is redistributed to smaller and larger scales forming  an inverse cascade at large scales with amplitude $\emu$ and a forward cascade at small scales with  amplitude $\enu$  such that $\emu+\enu=\ein$.  The value of the relative rate of inverse/forward cascade is a function of $H/\lin$.  The presence  of an inverse cascade in this system for different $H$ has been demonstrated in numerical simulations by measuring the energy flux \cite{smith1996crossover, ngan2005aspect, Celani2010turbulence, musacchio2017split}  and in experiments by measuring the third-order structure function \cite{Xia2011upscale}. In  \cite{Celani2010turbulence} the relative amplitude of the inverse cascade $Q_\alpha$  was measured as a function of $H$  and it was shown to drop to zero approximately when $H = H_c  \sim 1/2 \ell_{in}$ (see left panel of figure \ref{fig:converge}). Nevertheless, limits in the scale separation and in the value of the Reynolds number  did not allow  to conclude  if this transition is smooth or critical,  i.e. which  of the scenarios depicted in Fig.(\ref{fig:classII}) is realized. Moreover, one must not expect the value $H/\ell_{in}= 1/2$ to be universal, as it might depend on the details of the forcing, e.g. whether it is purely 2D or also the third vertical component is forced, on the temporal correlations etc... However, with the use of a different thin-layer modelling \cite{benavides2017critical}, where  only one  mode in the vertical direction is kept, it was shown  that the transition from direct to split energy cascade becomes increasingly close to a critical 2nd order transition like the one depicted in panel (b) of figure (\ref{fig:classification}) when we take the large box limit, $ L/\ell_{in} \to \infty$. This is shown in  the right panel of Fig. (\ref{fig:converge}).\\
\begin{figure*}[htbp]                                                                     
\centering                                                                                
\includegraphics*[width=0.45\textwidth]{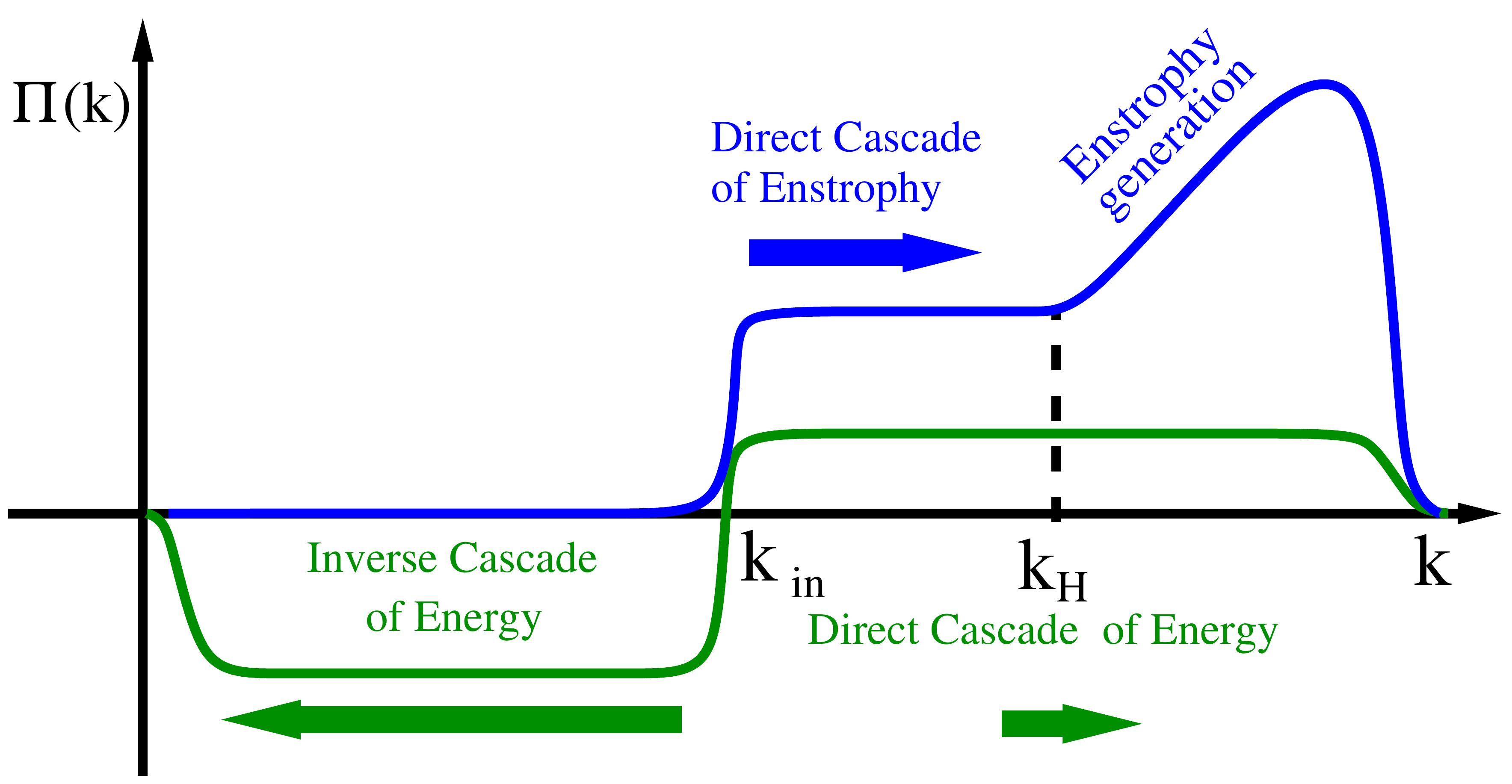}                                
\includegraphics*[width=0.45\textwidth]{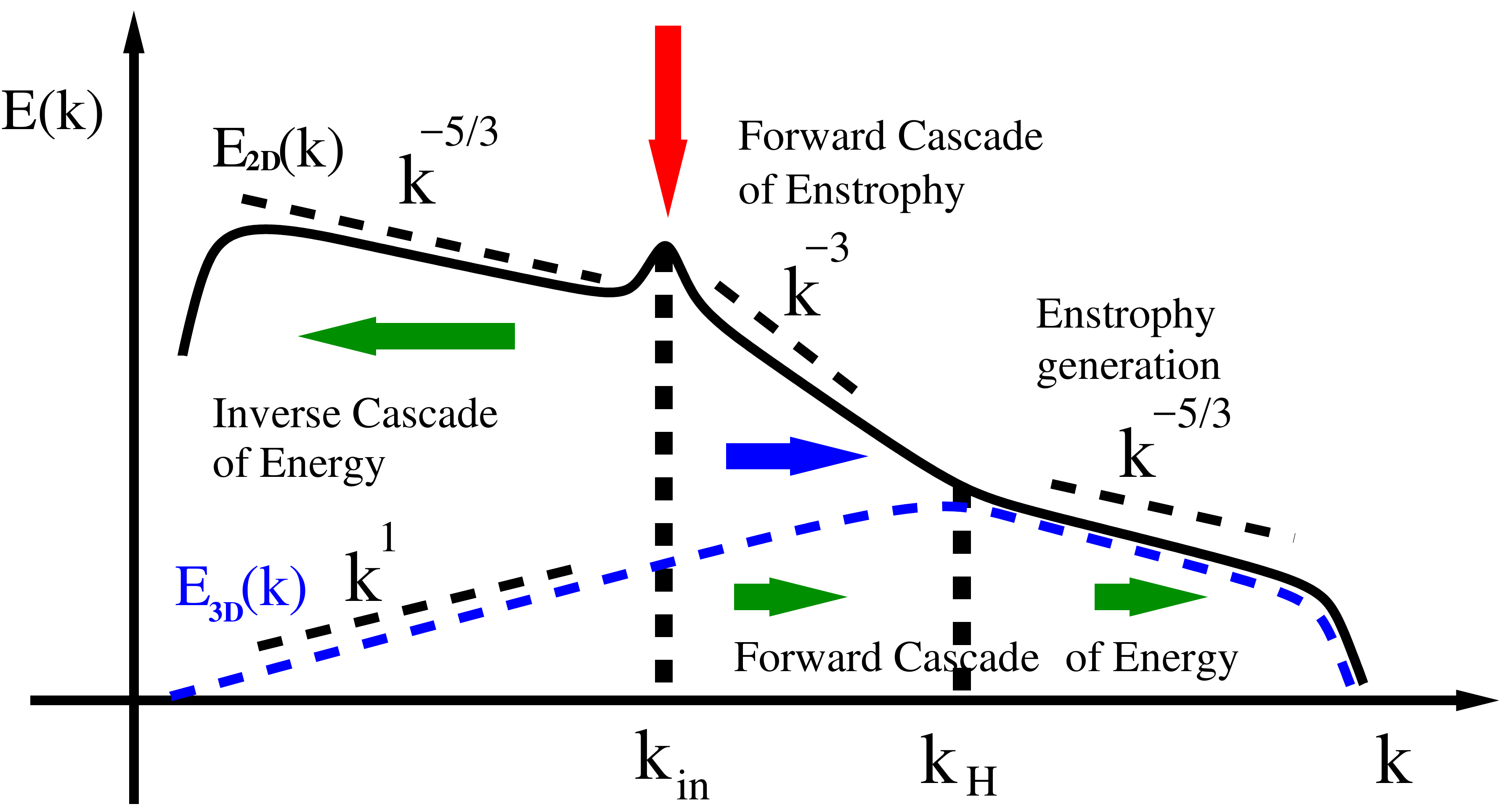}                             
\caption{ Energy and enstrophy flux (left panel) and energy spectra (right panel)         
for a thin layer. For wavenumbers $k$ smaller than $\kin$ the spectra are determined by   
the inverse cascade of energy, for intermediate wave numbers  $\kin < k < k_H$ the        
spectra are determined by the forward cascade of enstrophy while at the largest inertial  
wavenumbers the spectra are determined by the forward energy cascade. }                   
\label{fig:verythin}                                                                      
\end{figure*}                                                                             
\noindent To better understand the physics behind this split cascade it is useful to consider  the large and small $H$ limits separately.   \\
\subsubsection{Extremely thin layers, $H \ll H_c$}
\begin{figure*}[htbp]                                                                     
\centering                                                                                
\includegraphics*[width=0.9\textwidth]{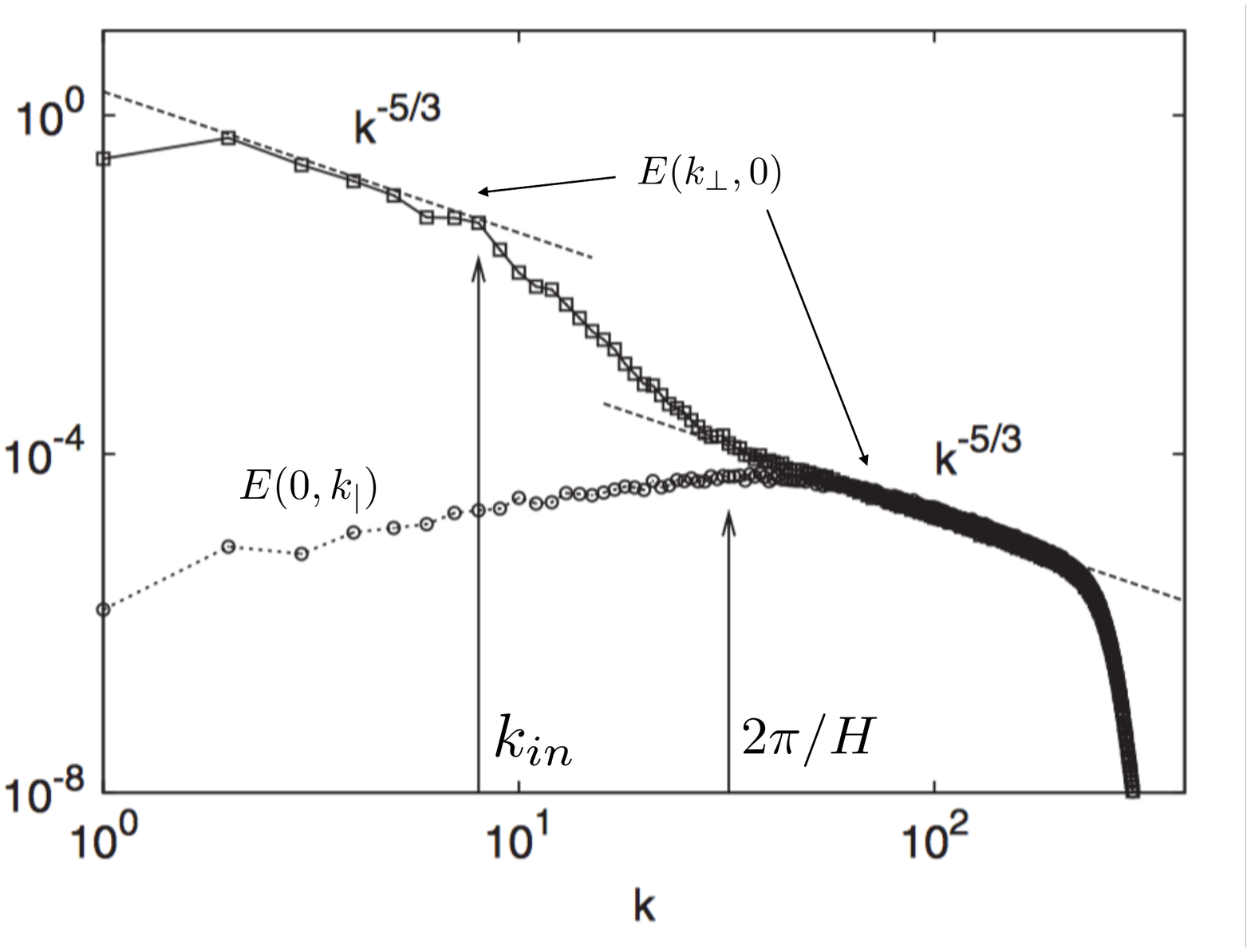}                               
\caption{ Energy spectra obtained from numerical simulations of a flow in a thin          
          layer. The figure is adapted from the results in \cite{Celani2010turbulence}. } 
\label{fig:celanispec}                                                                    
\end{figure*}

For very small values of $H\ll \lin$, the flow is strongly constrained to behave like a 2D flow
at $k_{in}$, since any variation along the $z$ direction is limited to very large vertical wave-numbers $k_z \ge k_{_{H}}= 2\pi/H$
(see right panel of Fig. \ref{fig:Fgrid}). 
Since $k_z \gg \kin$ it seems unlikely  that the $k_z$ modes affect the dynamics of the forcing modes $\kin$, because this would imply a sort of three-dimensional  backscatter of small scales to large scales.  Thus, the energy of the in-plane 2D components
of the flow ${\cal E}_{_{2D}}$  will cascade to the large scales while the out-of-plane 2D enstrophy will  cascade to the small scales, as in a purely 2D  flow.  
As the inverse cascade proceeds, energy arrives at scales $>\lin$ that satisfy 
better and better the 2D constraint and thus the inverse cascade is strengthened.
Asymptotically, at very large scales it is expected to reach a behavior identical
to 2D turbulence with a $k^{-5/3}$ energy spectrum. This argument is 
based on the assumption that as the scales $\ell$ and $H$ become further and further apart
they are more and more dynamically decoupled.  
For scales smaller than $\lin$ the dynamics are driven by the enstrophy cascade. As smaller and smaller scales are reached,
the constraint for 2D fluid motions  becomes weaker and weaker. At scales $\ell \sim H$ the 2D constraint breaks down and  enstrophy is no longer conserved, not even approximately, because of  enstrophy generation by 3D vortex stretching and a transition to a 3D forward energy cascade is expected.\\
This brings out the question how  a transition to a forward energy cascade at scales $\ell < H$ can take place if there is no forward cascade at scales $\ell > H$?.
To answer this question we need to remember that the 2D picture of no forward energy flux  and no inverse enstrophy flux holds in the infinite  $Re$ limit only. As the enstrophy cascade develops forward to smaller and smaller scales it always drags  some energy. However, the energy flux is decreasing and goes to zero for infinitely small scales. So, even in purely 2D flows  the small-scale energy flux is never exactly zero if the Reynolds number is finite as  quantitatively estimated by (\ref{eq:fwflxEn2D}).  
As a results for scales smaller than $H$, we expect the forward energy cascade spectrum $E(k) \sim \enu^{2/3} k^{-5/3}$
due to the residual flux, while for scales larger than $H$ one obtains the enstrophy cascade spectrum
$E(k) \sim \wein^{2/3} k^{-3} = \ein^{2/3}  \kin^{4/3} k^{-3} $. Equating the two expressions at $k \sim 1/H$ results in:
\be
\enu \propto \ein \left( \frac{H}{\lin} \right)^{2} .
\ee
These phenomenological arguments indicate that at scales $H < \ell < \lin$  there is a co-directional dual cascade of energy and enstrophy with the energy flux being sub-dominant.  The scaling above  have been  verified using a shell model in \cite{Boffette2011shell}. Figure \ref{fig:verythin} shows a sketch of the energy and enstrophy flux (on the left) and of the energy spectra (on the right) in the small  $H/\lin$ limit based on the aforementioned phenomenology, for the case that the layer thickness is much smaller than the forcing scale. The energy flux is negative and constant in the large scales while it is positive and constant (although weak) in the small scales. The enstrophy flux on the other hand is zero in the large scales, positive and constant in the range of wavenumbers $\kin \ll k\ll k_{_H}$ while it increases due to enstrophy production for wave numbers larger than $k_{_H}$. The spectrum accordingly forms a $k^{-5/3}$ power law behaviour in the large scales while a $k^{-3}$ power-law in the range of wave numbers $\kin \ll k\ll k_{_H}$, and beyond $ k_{_H}$ it recovers the $k^{-5/3}$ spectrum. The blue dashed line indicates the 3D part of the flow (composed from all the wavenumbers for which $k_z\ne0$) which  follows a thermal spectrum $\propto k$ for wavenumbers  smaller than $k_{_H}$ and a $k^{-5/3}$ for large wavenumbers. This phenomenological description has been nicely reproduced by the numerical work of \cite{Celani2010turbulence} as also reported in Fig. (\ref{fig:celanispec}).

\subsubsection{Critical layer thickness, $H \sim H_c$}       
\label{sec:CritThin}                                         
\begin{figure*}[htbp]                                                                      
\centering                                                                                 
\includegraphics*[width=0.62\textwidth,height=0.35\textwidth]{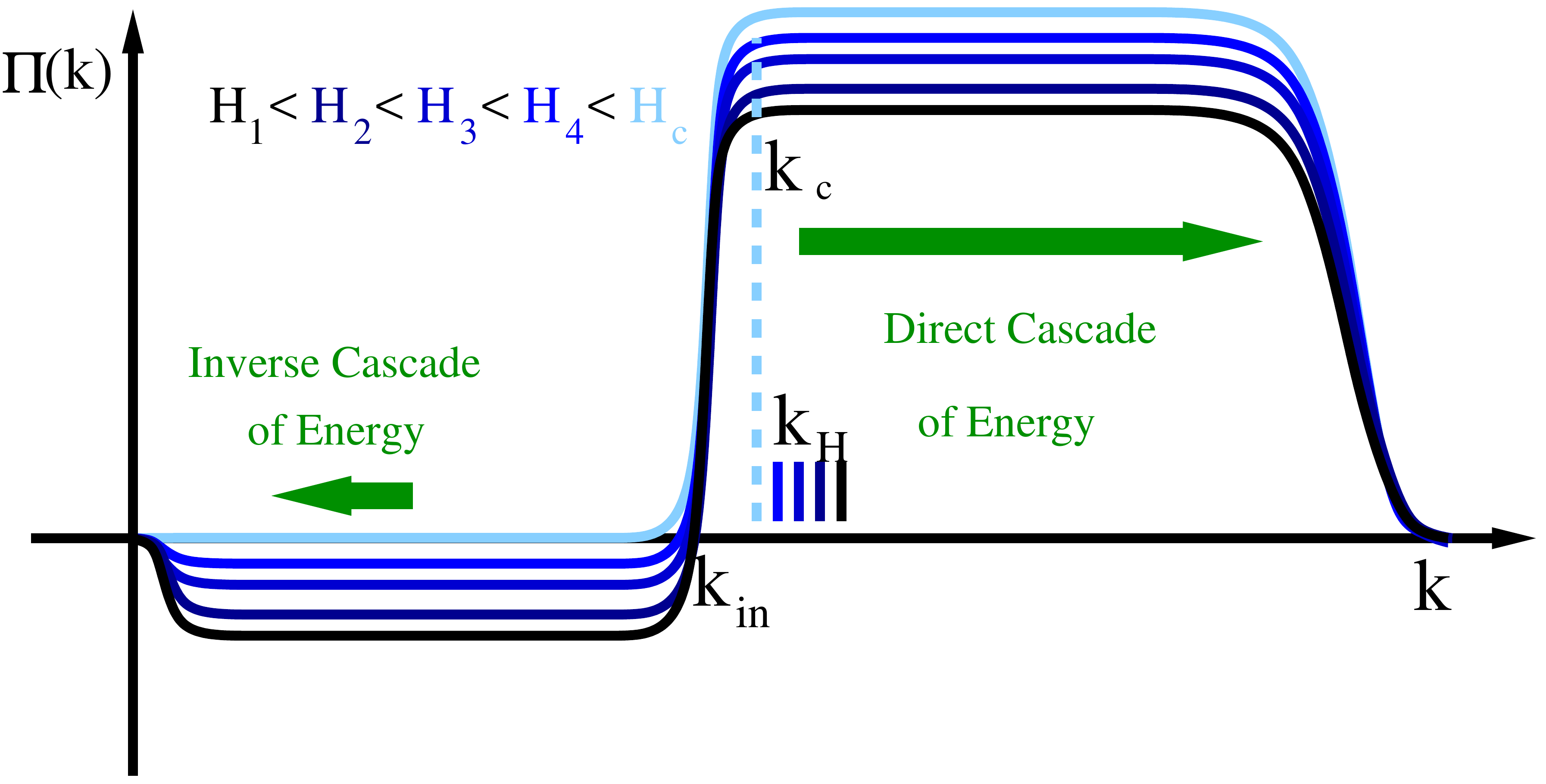}    \\     
\includegraphics*[width=0.42\textwidth]{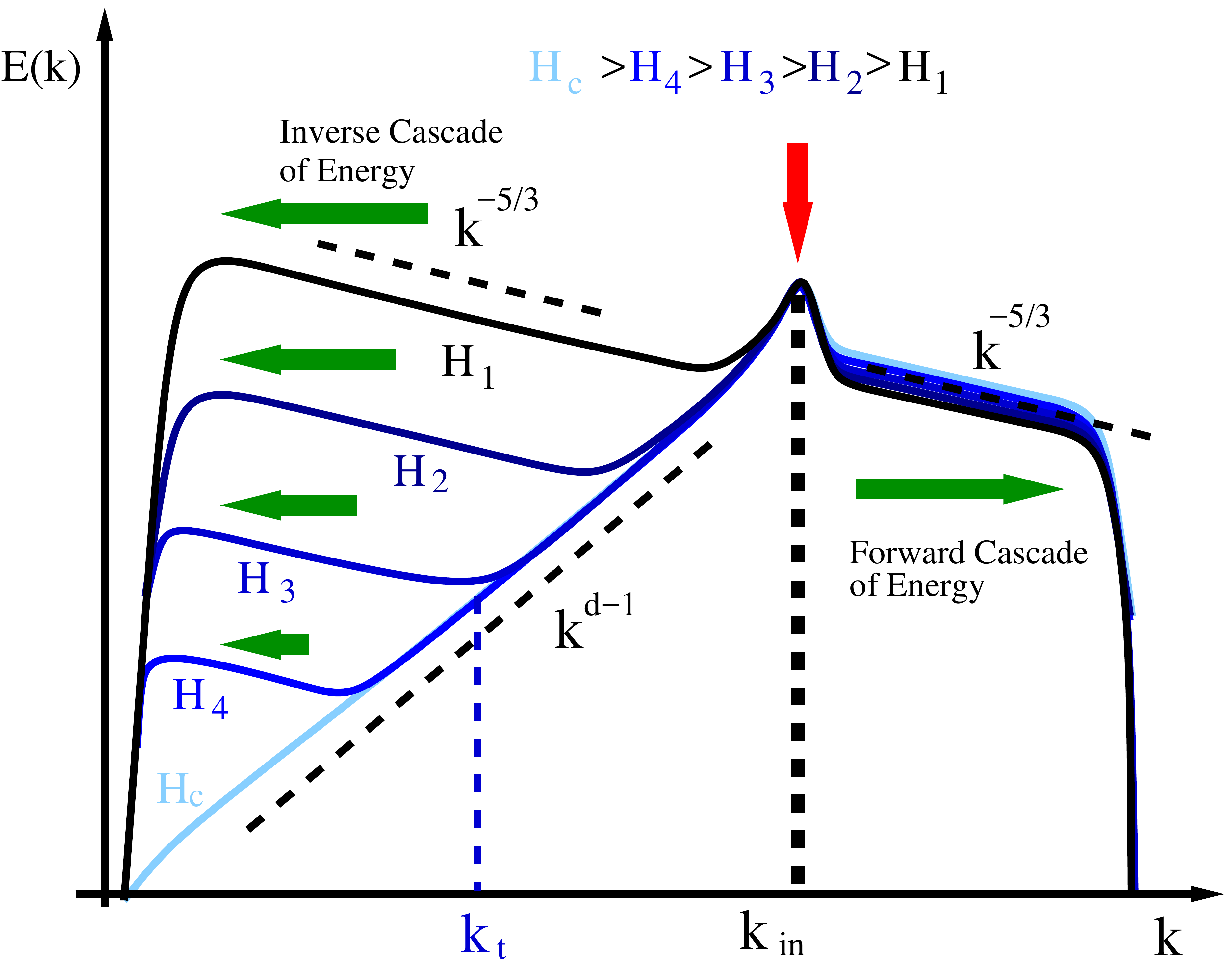}                              
\includegraphics*[width=0.42\textwidth]{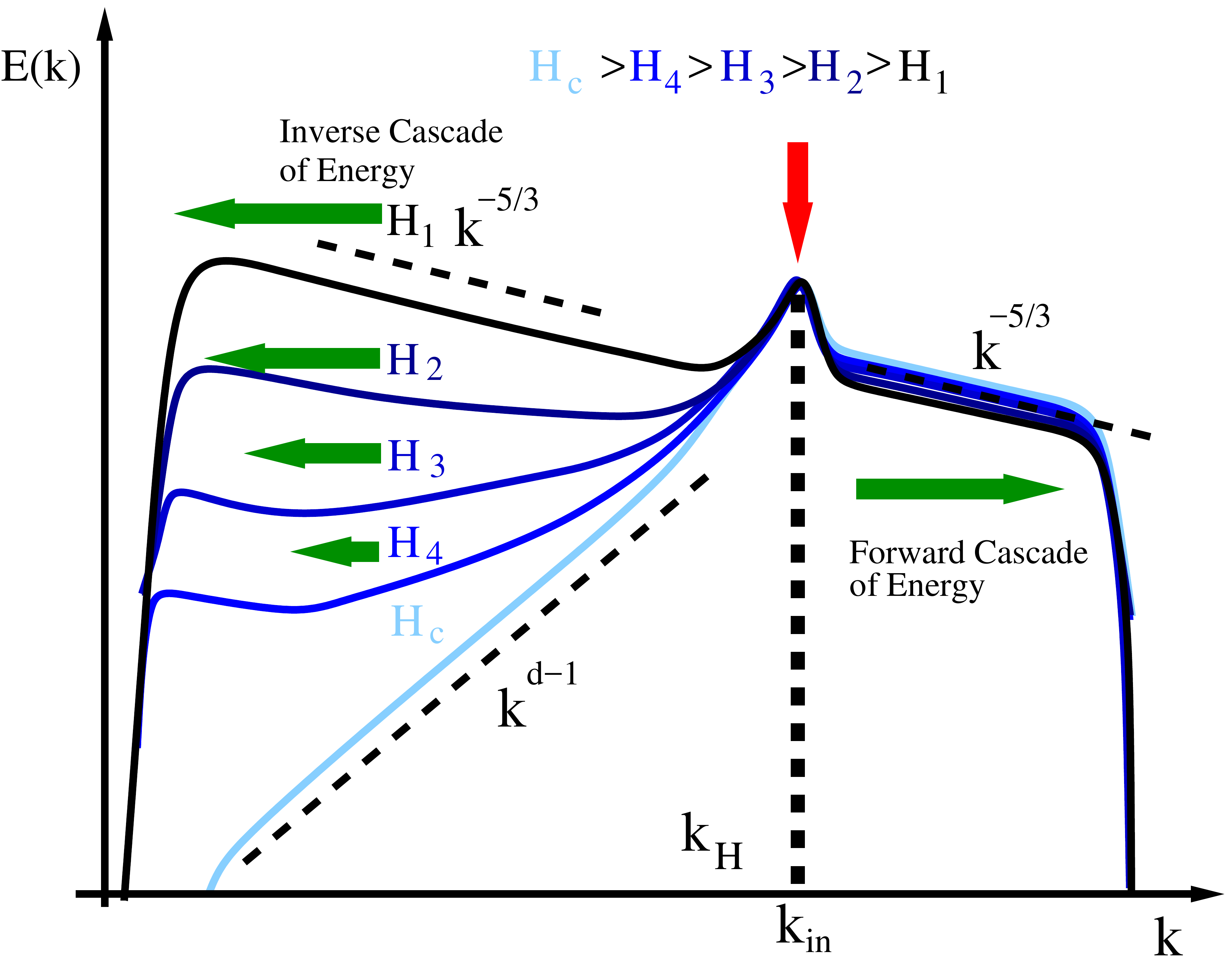}                             
\caption{                                                                                  
Top Panel:  Energy flux as for different $H$. The line colour becomes darker as the layer  
thickness $H$ is decreased. The location of the different $k_H=2\pi/H$ are marked by the vertical lines. Bottom panels: Two scenarios for the transition of the functional  
form of the energy spectra as $H$ approaches its critical value $H_c$ from below         
(from dark colours to bright). The left panel shows a transition from a thermal spectrum   
$\propto k^{d-1}$ to a $k^{-5/3}$ spectrum at a transition wavenumber $k_t$. In the right        
panel there is a continuous change of the exponent form 1 to -5/3 for a range of wave      
numbers.    }  \label{fig:closetocrit}                                                     
\end{figure*}

\begin{figure*}[htbp]                                                                      
\centering                                                                                 
\includegraphics*[width=0.42\textwidth,height=0.25\textwidth]{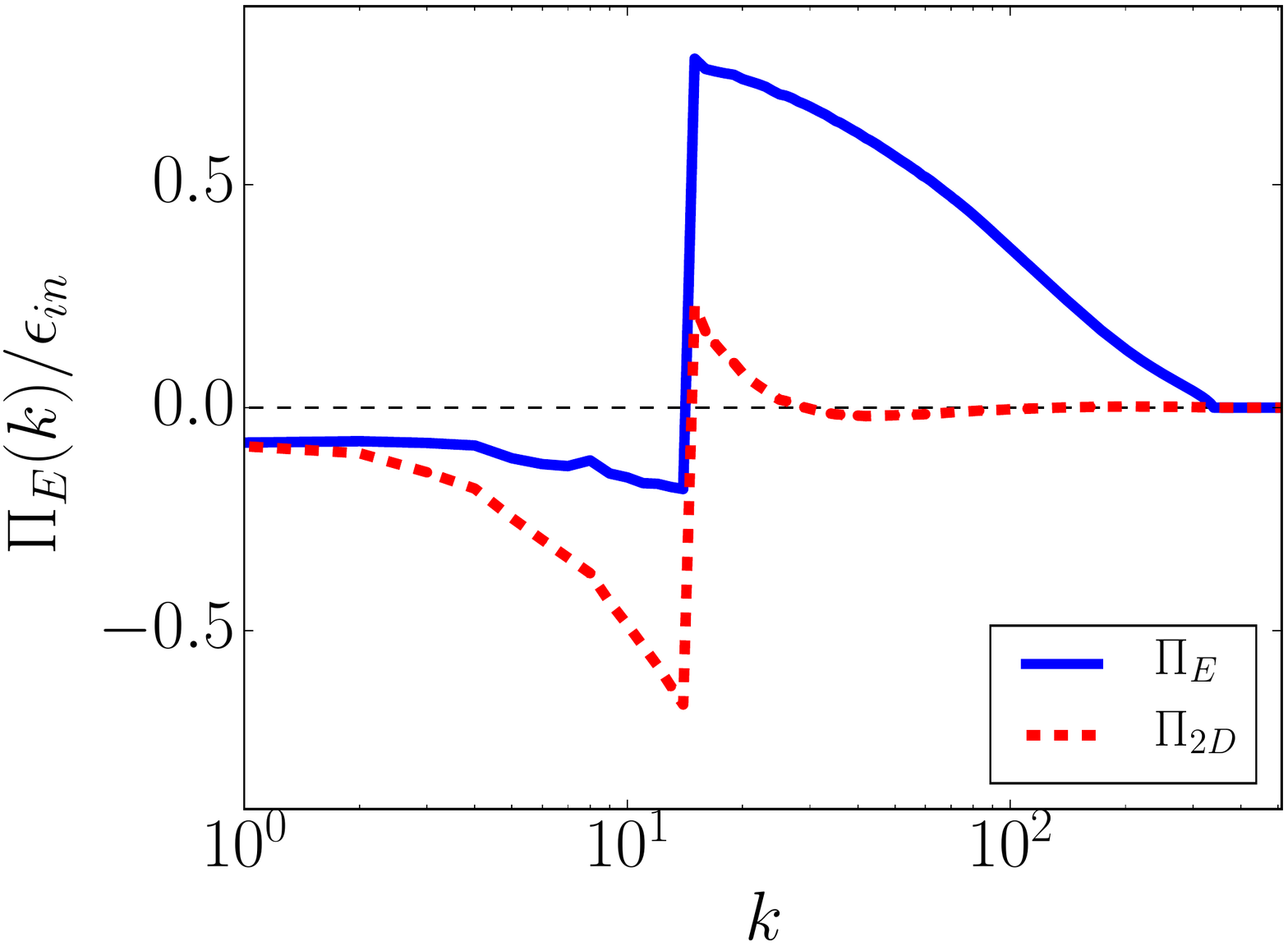} 
\qquad                                                                                     
\includegraphics*[width=0.42\textwidth,height=0.25\textwidth]{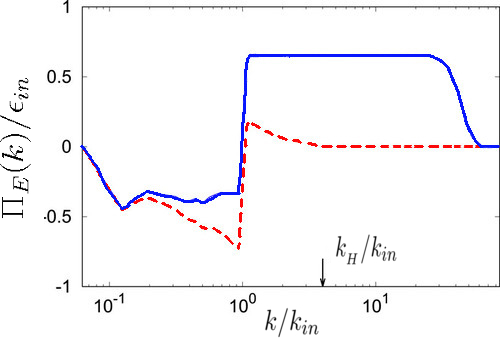}  
\caption{ Lin-log plot of the total energy  flux, $\Pi_E(k)$, (solid blue lines)  and 
the  $\Pi_{2D}(k)$ due to the 2D component (dashed red lines) as defined in                
(\ref{eq:2d3dlayer}).   The left panel is adapted from the results in                      
          \cite{benavides2017critical} while the right panel is adapted from the results   
          in \cite{musacchio2017split}.                 }                                  
\label{fig:BenavidesMusacchio}                                                             
\end{figure*}

\begin{figure*}[htbp]                                                                     
\centering                                                                                
\includegraphics*[width=0.62\textwidth]{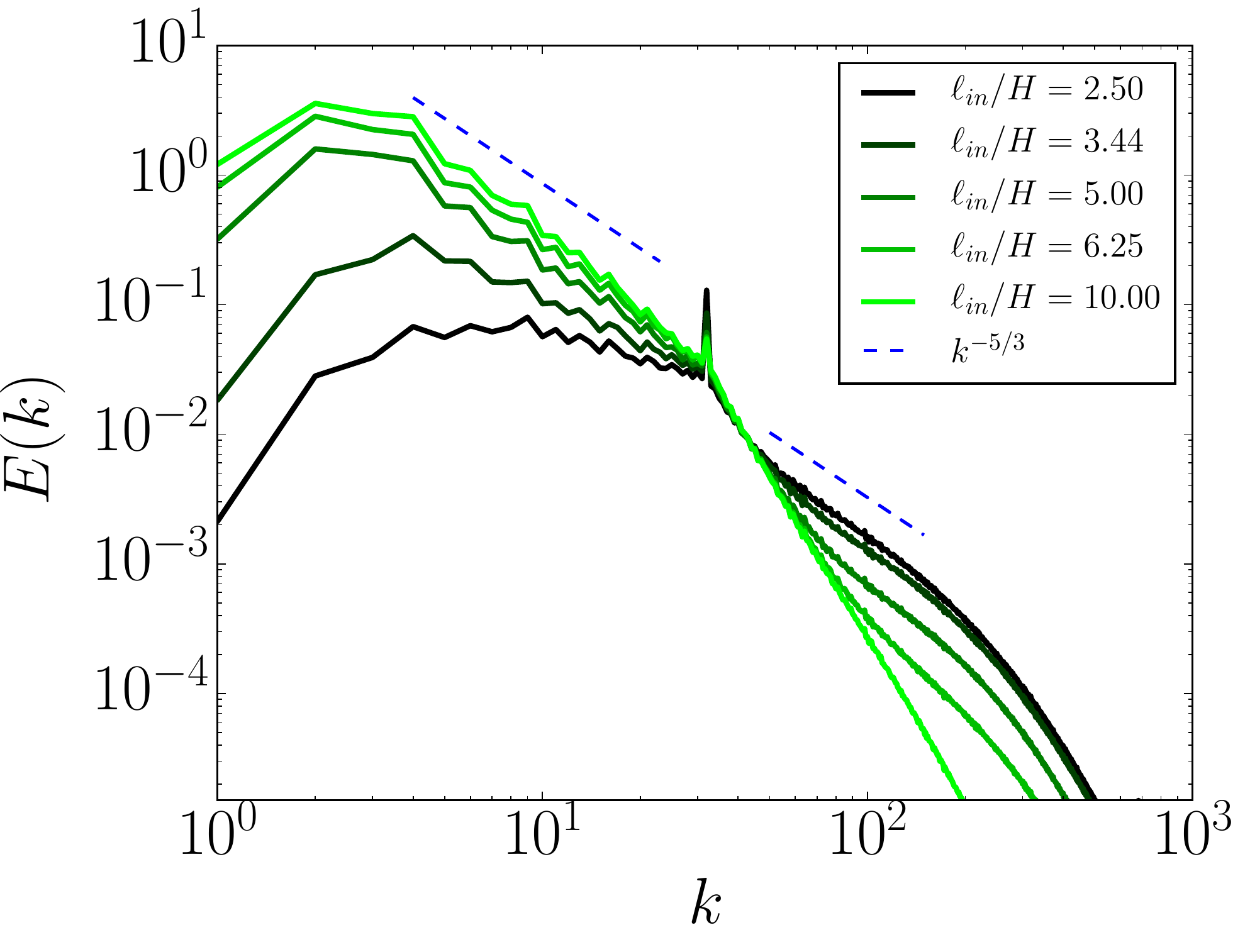}                            
\caption{ The energy spectrum from different values of $H$ for the thin-layer  model studied
  in \cite{benavides2017critical}.   }                          
\label{fig:Benavides2}                                                                    
\end{figure*}

For $H/\lin$ close to the critical value it is however  harder to have a simple phenomenological description. In fact it is not even certain that a critical transition exists, since we only have little numerical and qualitative evidence for it \cite{Celani2010turbulence,benavides2017critical}. One can not exclude that 
it is a smooth transition as for case (a) in Fig. (\ref{fig:classification}) and that there is always a weak inverse cascade even for very large $H$, i.e. $H_c = \infty$. For example, we can argue  that for large $H$ such that $k_{_{H}}\ll \kin$ a quasi thermal energy spectrum $E(k) \propto \ein^{2/3} \kin^{-5/3-(d-1)}\, k^{d-1}$ (where the flow dimension $d=3$ because we are in the thick-layer regime) builds up in the wavenumber range $ k_{_{H}}\ll k\ll \kin$, followed  in the range $k\ll k_{_{H}} $ by an inverse cascade with  $E(k) \propto \emu^{2/3} k^{-5/3}$. Equating the two spectra at $k=k_{_{H}}$ we get:  
\be
\emu=\ein \left(\frac{k_{_{H}}}{\kin}\right)^{11/2} =\ein \left(\frac{\lin}{H}\right)^{11/2}.  
\ee
In this scenario, we would always have a {\it weak} inverse cascade no matter how large $H$ is,  its amplitude decreasing  as $H^{11/2}$. 
The shape of the energy spectrum in such a case it is shown in the middle panel of Fig. (\ref{fig:closetocrit}), where  $H_c$ with zero  inverse flux corresponds to  $H_c=\infty$ and $k_{H_c} =0$.\\ 
These arguments assume that the Fourier  interactions among local wavenumbers for  $k\le k_{_H}$, that lead to the inverse cascade, dominate  over the non local interactions with the scales around the  forcing range or with the ones in the forward energy cascade range. In the $ k_{_{H}}\ll \kin$ case however the eddies of size $H$  have both energy and enstrophy smaller than the ones of the  forced scales.  As a result, it is more likely that  a nonlocal eddy viscosity \cite{kraichnan1976eddy, dubrulle1991eddy, cameron2016large, alexakis20183d} induced by the degrees of freedom in the forward cascade range  brings the energy back to the small scales and stops the inverse transfer. It is thus more likely that a finite critical height $H_c$ exists above which no inverse cascade is observed. This  issue is nevertheless open.
%
%
In the presence of a finite critical height, another important open question is  how  the energy spectrum changes as $H$ varies across $H_c$.
If the transition  is discontinuous (as in case (c) of figure \ref{fig:classification})
 one could have assumed that there is an abrupt change from the
thermal spectrum $E(k)\propto k^2$ to the inverse cascade spectrum $E(k)\propto k^{-5/3}$.
However all numerical investigations indicate that this transition is
continuous \cite{Celani2010turbulence,benavides2017critical,musacchio2017split}. 
There are two possible ways that the spectrum can change between the two exponents.
The first is by a change of slope from $k^{d-1}$ to a $k^{-5/3}$ at a 
particular wavenumber $k_t$   as depicted in the left lower panel of Fig.  (\ref{fig:closetocrit}),
where $k_t \to 0$  as $H \to H_c$ from below. The second is by a continuous
change of the spectral exponent $E(k)\propto k^{s(H)}$ where $s(H_c)=1$ and $s(0)=-5/3$. 
This is depicted in the right lower panel of Fig. (\ref{fig:closetocrit}). The first case is in
agreement with the notion that as the inverse cascade proceeds to large and larger scales 
it comes close to 2D turbulence decoupling from the smaller 3D velocity scales, while in
the second case the interactions with the 3D scales persist and alters the spectral exponent
from $-5/3$. The simulations done in \cite{benavides2017critical,musacchio2017split}
show that some interactions with the 3D modes reduce the cascade  intensity.  This was shown by comparing the total flux (see solid line in  Fig. \ref{fig:BenavidesMusacchio}) to the flux due to
the 2D modes (red dashed line in the same figure). The spectra shown in the right panel of figure \ref{fig:Benavides2} have a transition at large scales from a flat to a $ -5/3$ slope. However the scale separation is not sufficient to conclude  how the spectra change when $H \to H_c$.

As of now,   the exact behaviour close to the critical point lacks 
a precise theoretical description. The dependence of $Q_\alpha$ on the distance
from criticality $(H_c-H)$ has not been determined yet. 
Critical phenomena suggest  a power-law dependence:
\be 
Q_\alpha \propto (H_c-H)^\beta. 
\ee
The results of \cite{benavides2017critical} estimate $\beta \sim 1$. 
There is however no theoretical explanation for this exponent
nor verification from full DNS or from experiments.
In \cite{benavides2017critical} it has been proposed
that the transition could be viewed in configuration  space as the competition between 
the 2D vortices (prey) and 3D fluctuations (predators) that feed on 2D vortices
and thus following predator-prey dynamics. 
\begin{figure}                                                                                   
        \includegraphics[width=0.45\textwidth]{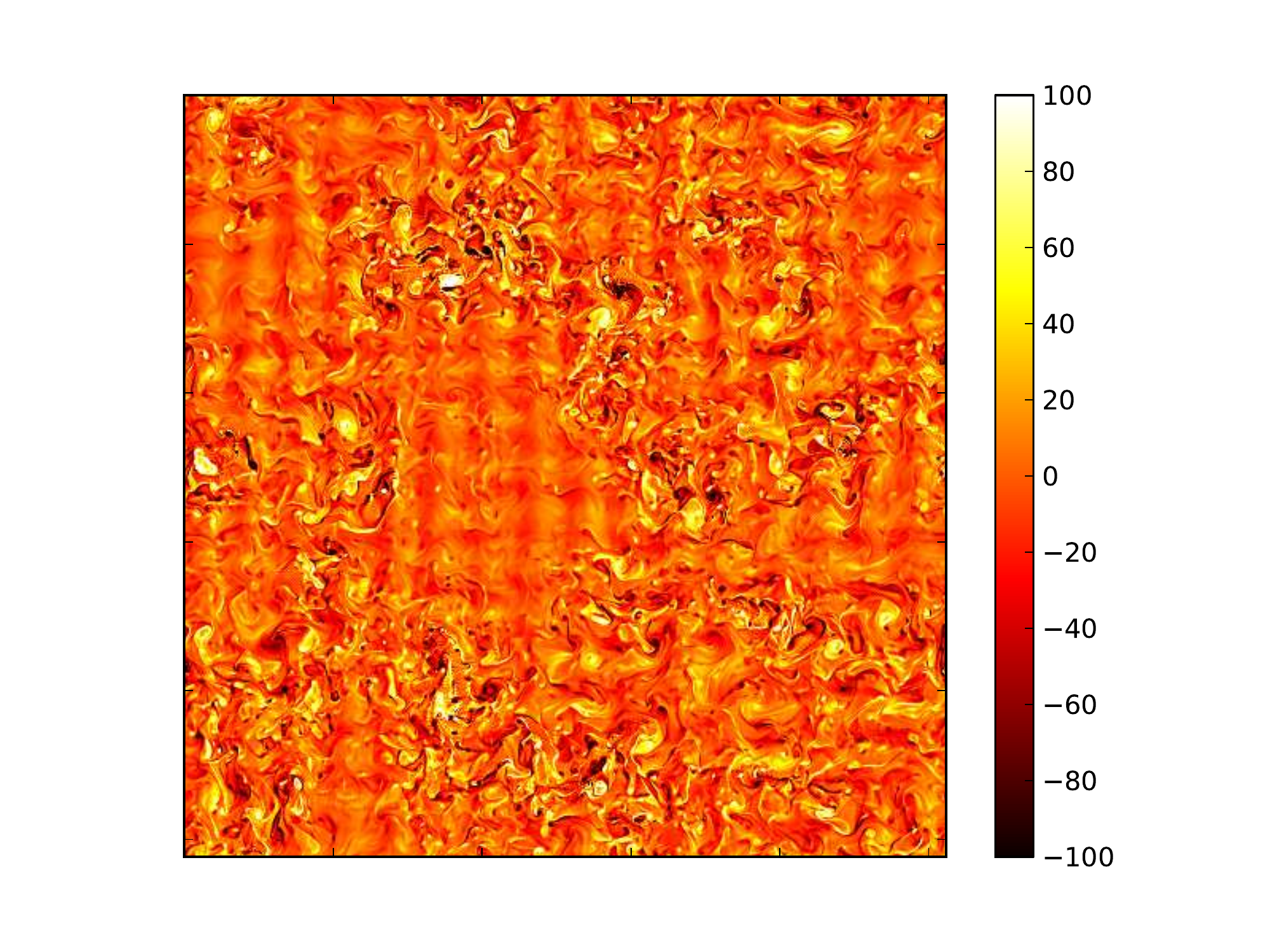}   \,\,\,                     
        \includegraphics[width=0.45\textwidth]{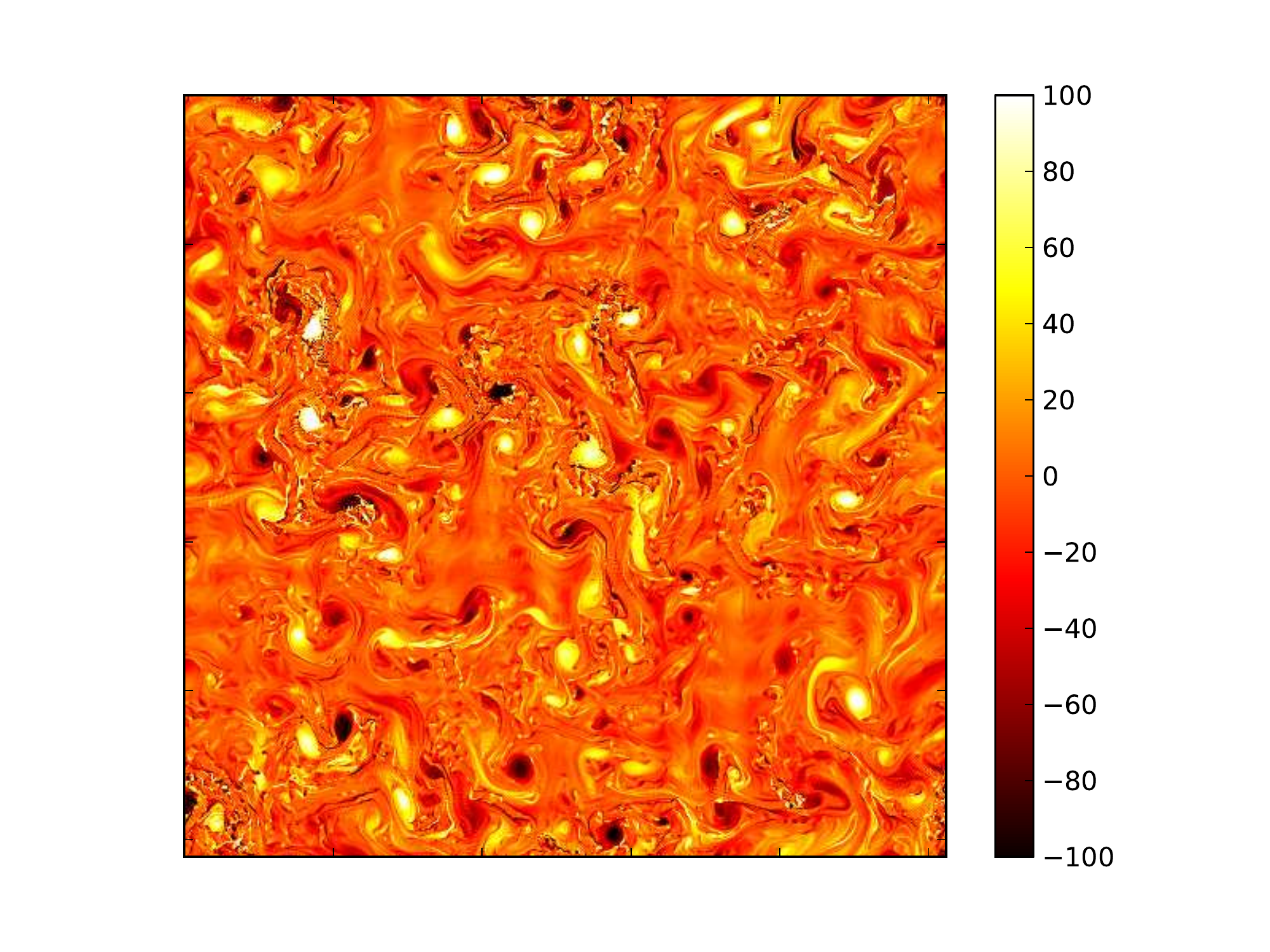}   \\                         
        \includegraphics[width=0.45\textwidth]{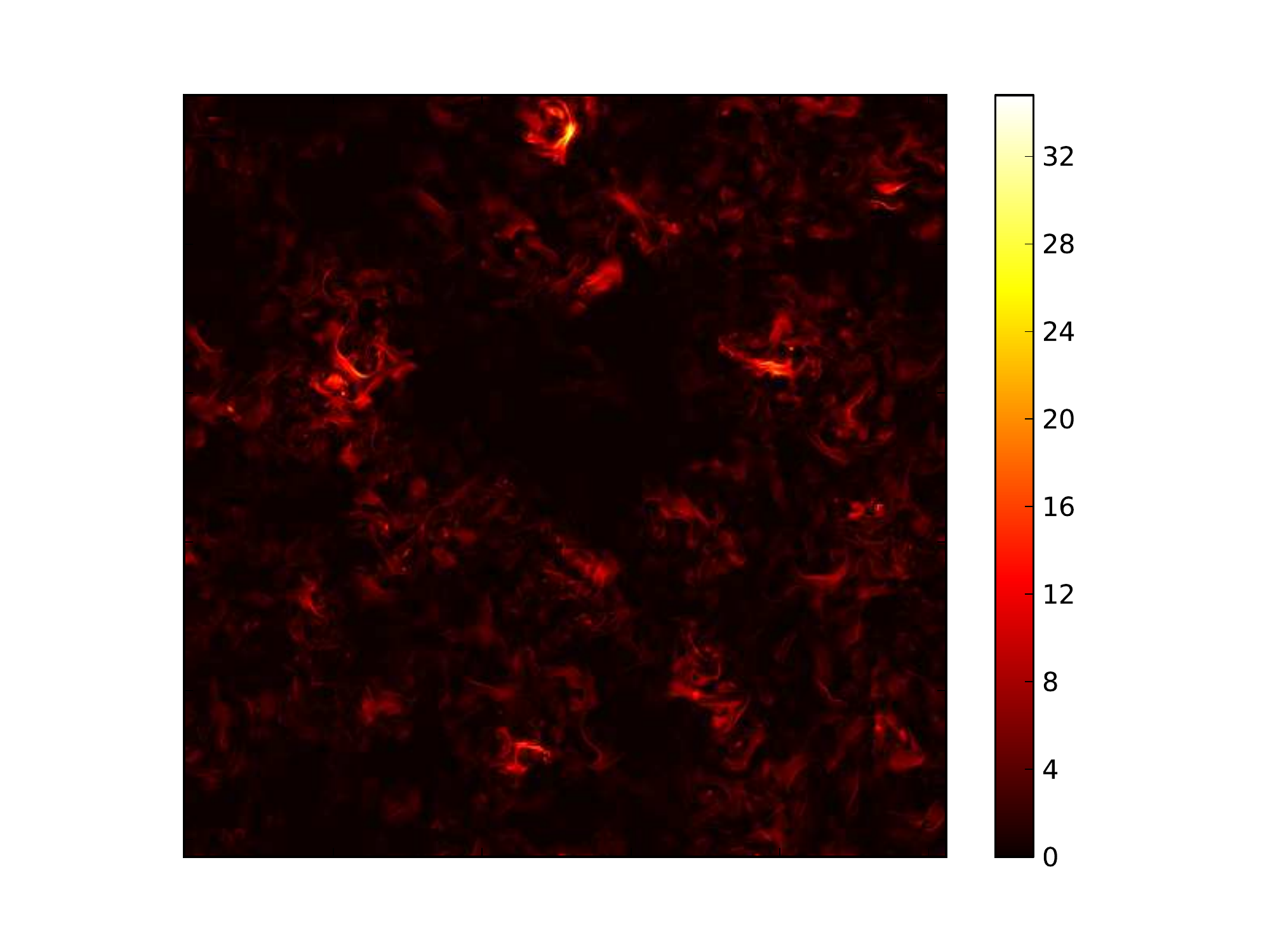}                              
        \includegraphics[width=0.45\textwidth]{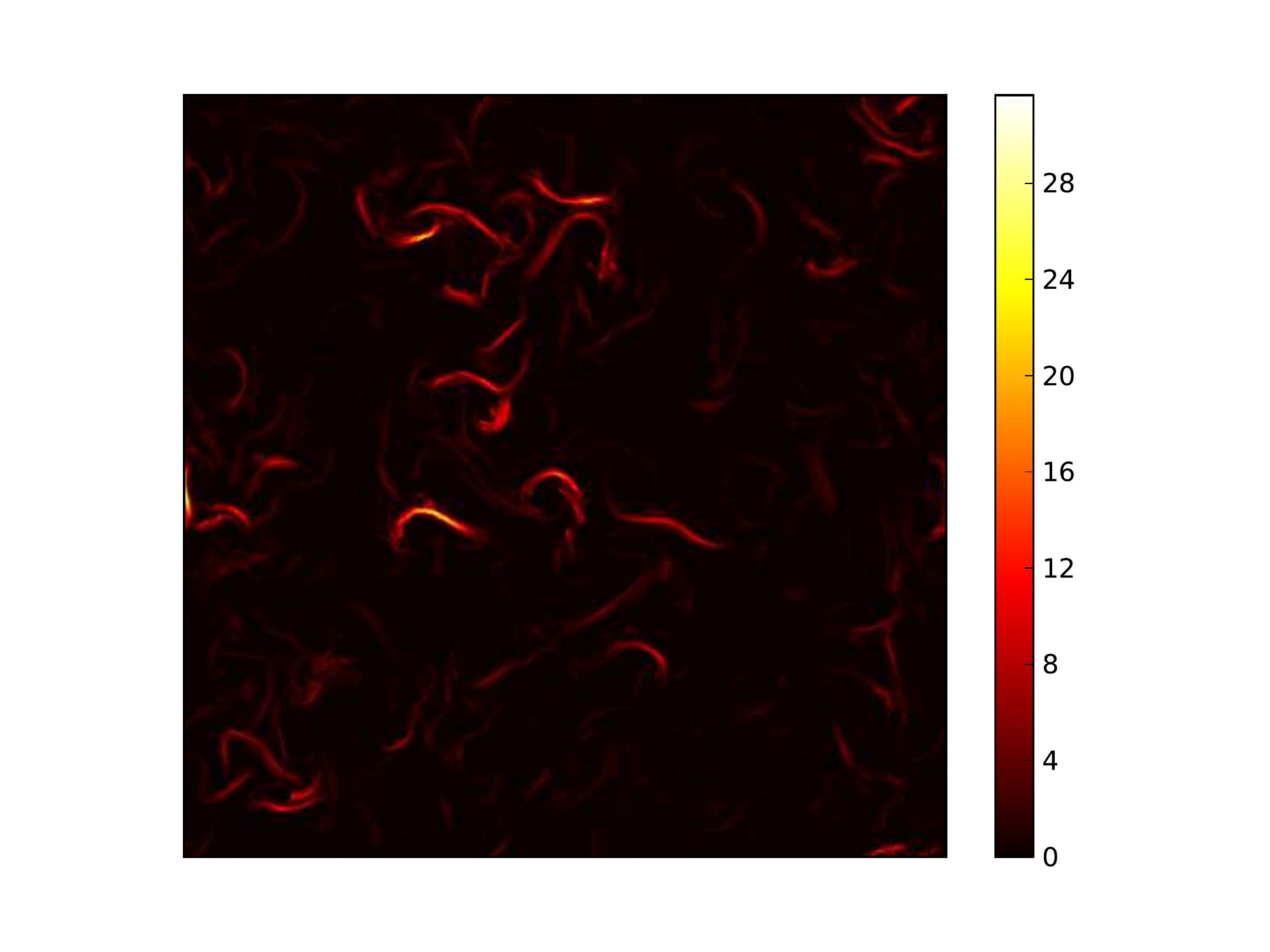}                              
\caption{The two-dimensional vorticity  (top) and 3D energy density                              
    (bottom), for large $H$ (left) small $H$ (right).                                            
   The large $H$ flow has more dense 3D active regions and less 2D vortexes while the  small     
   $H$ flow has less 3D active regions and more 2D vortexes.                                     
   The figures are adapted from the data in \cite{benavides2017critical}.                        
\label{fig:predator-vortices}     }                                                              
\end{figure}                                                                                     
This can be inferred from  Fig. (\ref{fig:predator-vortices}) produced from the thin layer model 
discussed in \cite{benavides2017critical} that shows characteristics both from 2D turbulent
coherent vortices clustering, leading to the  inverse cascade and from 3D phenomenology with small scale
vorticity filaments that are produced by the 3D motions and tend to  destroy the coherent vortices.
To what degree however such a description is plausible still needs to be explored.\\
\subsubsection{Finite horizontal size and finite Reynolds \label{sec:finitethin}}
So far we have discussed the cascades in the infinite box $L$, and infinite $Re$ limit. However, it is worth  discussing the case when these parameters are finite. In the finite $Re$ case it was shown that a second critical height exists below which the flow becomes exactly 2D and all 3D perturbations decay exponentially \cite{benavides2017critical}. This second critical height $H_\nu$ scales as $H_\nu \propto \lin Re^{-1/2}$ and corresponds to the height at which the viscous forces damp all variations along the $z$-direction making the 2D flow linearly stable. The region of stability of the turbulent 2D flow to 3D perturbations can been found analytically with the use of rigorous bounds. This was done in \cite{Gallet2015exact} for the more complicated  MHD case, from which
the hydrodynamic result can be simply obtained by setting the magnetic field to zero. It was further shown in \cite{benavides2017critical} that the amplitude of the 3D perturbations and thus also the dissipation rate of the 3D modes  scale like $(H-H_\nu)^\beta$ where $\beta$ is an exponent larger than 2. 
\begin{figure*}[htbp]                                                                        
\centering                                                                                   
\includegraphics*[width=0.82\textwidth]{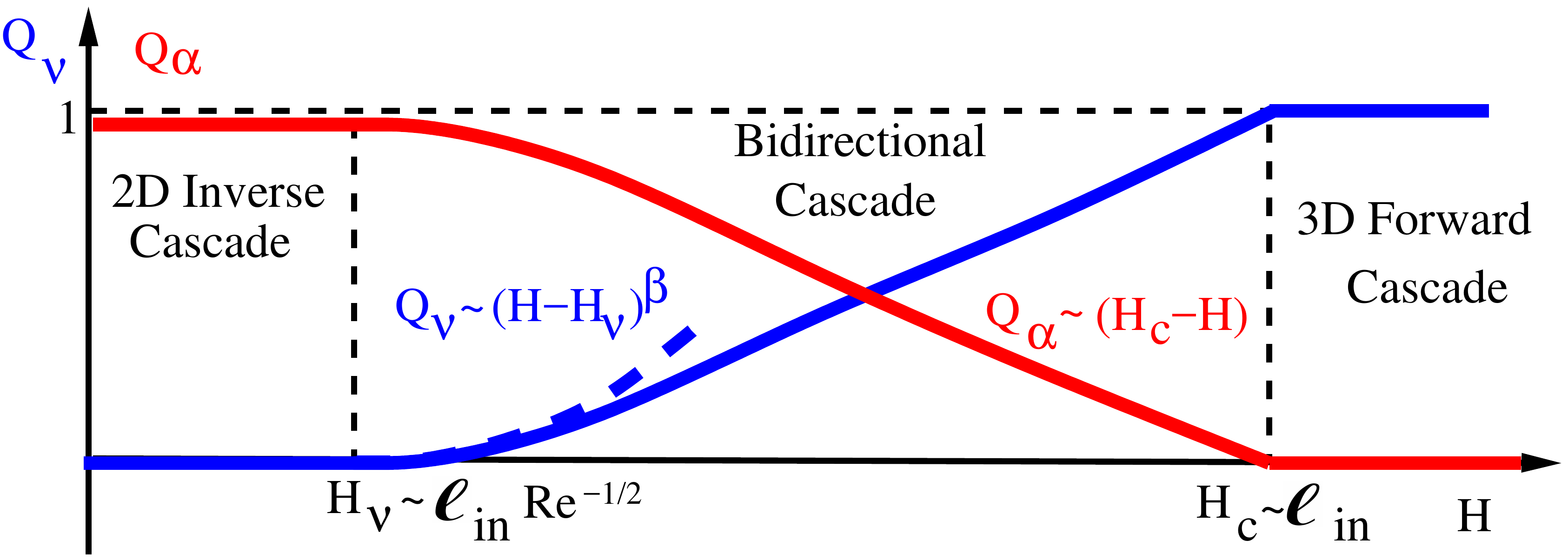}                                   
\caption{ Summary of the expected phase diagram for thin-layer flows.     }                  
\label{fig:thintran}                                                                         
\end{figure*}                                                                                
The flow close to this second critical point has an  intermittent behaviour 
both in space and in time. The transition diagram for finite $Re$ is shown in Fig. (\ref{fig:thintran}).

For finite $L$, if $H$ is such that an inverse cascade builds up,
a condensate will be formed in the absence of a linear drag, according to  Def. \ref{def13} in Sec. (\ref{sec:definitions}).
The formation of such condensate has also been realized in experiments \cite{Shats2010turbulence} and is shown in panel (c) of Fig. (\ref{fig:shats}). In 2D turbulence the condensate  reach very large velocity amplitudes such that the total energy scales inversely proportional to viscosity $\cE \propto \ein L^2/\nu$ so that the injection rate is balanced by dissipation (see Sec. \ref{sec:2DinvCsd}). In layers of finite thickness however this is not the only possibility. If 3D velocity fluctuations coexist one can assume that saturation is reached  when the eddy viscosity $\nu_{eddy}$ due to the 3D perturbations leads to saturation of the condensate by transferring energy back to the small scales:
\be  
\nu_{eddy} \frac{\cE}{L^2} \simeq \ein .
\label{eq:edvscThin}
\ee
The eddy viscosity for a 3D flow and at large $\Red$ is expected to scale like $\nu_{eddy} \propto u_{3D} H$ \cite{dubrulle1991eddy} where  $ u_{3D}\sim \ein^{1/3}H^{1/3} $ is the rms value of the 3D component of the flow. The balance (\ref{eq:edvscThin}) then leads to
\be 
\cE \propto  \frac{\ein  L^2}{ \nu_{eddy}} \propto \frac{ \ein^{2/3}  L^2 }{  H^{4/3} }
\ee
that is independent of viscosity. 
This case corresponds thus to one example of a  flux-loop condensate as given by  Def. (\ref{def15}) in Sec. (\ref{sec:definitions}) where the energy that cascades inversely due the 2D motions
is balanced by an eddy viscosity due to the 3D motions. This was demonstrated by numerical simulation \cite{vanKan2018condensates}. 

\subsubsection{Summary}
In this section, we have discussed flows in layers of finite thickness and the transitions that occur as the thickness is varied. The discussion has naturally introduced the notion of anisotropy in spectral space as well as the splitting of the flow in a 2D3C (vertically averaged) field and  the remaining 3D field that have different cascade properties. The first component drives -in some cases- an inverse cascade while the second one leads to a forward cascade. We examined different limiting configurations. First,  the very thin layer case where most of the energy cascades inversely while some sub-dominant forward cascade exists  affecting the energy spectrum at scales smaller than $H$ only. For a thickness close to the critical height where the inverse cascade vanishes, we discussed all possible transition scenarios. Numerical evidence suggests a second order behaviour (see Sec.\ref{sec:Classification}), but it is still not totally clarified  how the energy spectra and fluxes vary as a function of the distance from the  critical point. Different possibilities were discussed and open problems listed. Finally, we discussed the presence of a second critical point where the flow becomes exactly 2D for finite $Re$, and the effect of finite horizontal size when a flux loop condensate (see Def. \ref{def15} in Sec. \ref{sec:definitions}) can also be present.  

\subsection{ Rotation } \label{sec:Rotation}                                  

\begin{figure*}[h!]                                                                              %
\centering                                                                                       %
\includegraphics[width=1.0\textwidth]{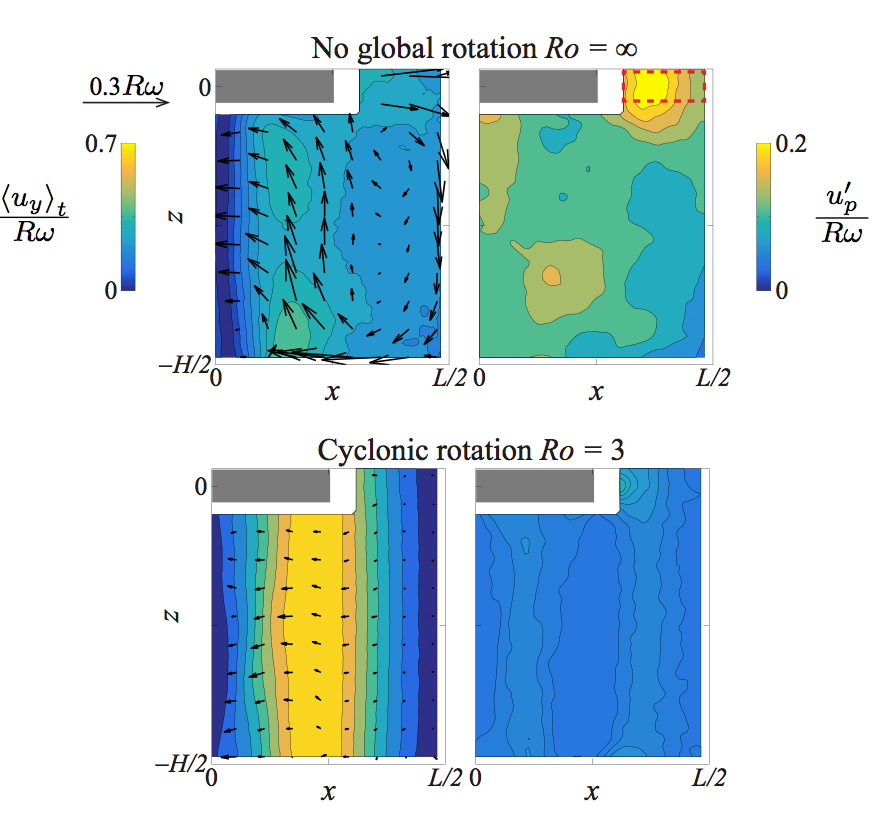}                                        %
\caption{    Particle Image velocimetry maps of the  mean flow (left) and rms fluctuations       %
for the poloidal flow (right) in a rotating turbulent flow driven by an impeller.                %
In the left panels, the colour codes the toroidal (out-of-plane) velocity for (top) no           %
background and (bottom)  cyclonic background rotation at $Ro=3$ (see \cite{campagne2016turbulent} for flow details). Data are normalized with        %
the mean frequency of the impeller here denoted with $\omega$ and the size of the container $R$. %
The two-dimensionalization results in a gradual weakening of the poloidal flow and a             %
strengthening of the toroidal flow.                                                              %
Reprinted by permission from \cite{campagne2016turbulent}}.                                      %
\label{fig:basile}                                                                               %
\end{figure*}

In this section, we consider the effect of rotation on the turbulent energy transfer.  Rotation plays an important role in most planetary and stellar flows making the flow anisotropic \cite{greenspan1968theory}. In some cases when rotation is strong it makes the flow quasi-two-dimensional (an effect known as the Taylor-Proudman theorem \citep{Hough1897,Proudman1916,Taylor1917}) and this can change the direction of the energy cascade. Rotation has been the focus of many studies in turbulence research and modeling \cite{cambon1999linear,hopfinger1993vortices}. The quasi-2D behaviour has been realized in experiments \citep{ibbetson1975experiments, hopfinger1982turbulence, dickinson1983oscillating, baroud2002anomalous, baroud2003scaling, Sugihara2005, ruppert2005extraction, morize2006energy, Staplehurst2008, Bokhoven2009experiments, Yoshimatsu2011, duran2013turbulence, Machicoane2016Two}  and numerical simulations \citep{yeung1998numerical, Smith1999transfer, godeferd1999direct, chen2005resonant, Thiele2009, mininni2009, Mininni2010Rotating, favier2010space, Sen2012anisotropy, Marino2013invers, alexakis2015rotatingTG, biferale2016coherent,valente2017spectral}. A visualization of a 2D cut from the
rotating experiment  \cite{campagne2016turbulent} is shown in Fig. (\ref{fig:basile}). 
Here, we focus on how rotation can affect the transfer of energy and helicity and review some recent results on how a change of cascade direction  can take place.

\subsubsection{Formulation} 
\begin{figure*}[h!]                                                                         %
\centering                                                                                  %
\includegraphics[width=0.70\textwidth]{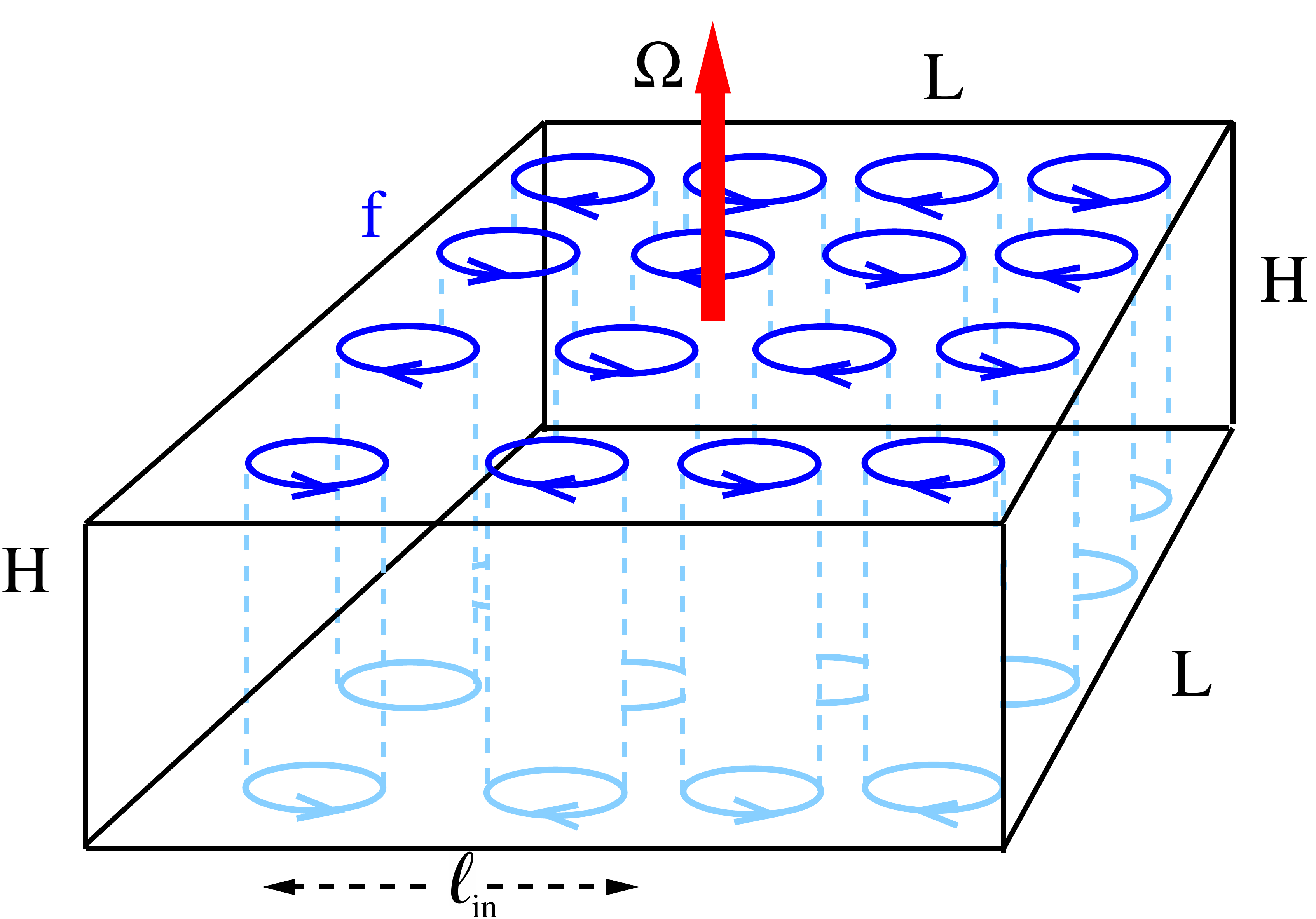}                                     %
\caption{A sketch of the domain considered for rotating turbulence. }                       %
\label{fig:Rbox}                                                                            %
\end{figure*}

To simplify our discussion, we  consider a layer of finite height $H$ with the rotation axis  being oriented in the vertical direction. This setup is demonstrated in Fig. (\ref{fig:Rbox}).  
In the rotating frame of reference the Navier-Stokes equations  are written as:
\be
\partial_t {\bf u} + {\bf u \cdot \bnabla u} + 2{ \Omega {\bf \hat{e}_z }\times {\bf u}  } =-\bnabla P + \nu \Delta {\bf u} -\alpha {\bf u}+ {\bf f}
\label{eq:NSR}
\ee
where $\Omega$ is the rotation rate and the linear term $2{ \Omega \bf\hat{e}_z \times u  }$ is the Coriolis force.  It is important to notice  that rotation does not do any work on the system, $\langle {\bf u} \cdot ( \bf {\hat{e}}_z \times {\bf u}  )\rangle =0 $ and therefore it does not alter the global energy balances (\ref{eq:globalbalance}-\ref{eq:globalbalance2}). Rotation  introduces a new non-dimensional control parameter  in the system, the Rossby number:
\be
\label{eq:Rossby}
Ro = \frac{u_f}{2\Omega \lin}
\ee
where (as for the Reynolds number) $u_f$ stands for the rms velocity at the forcing scale or $u_f=(\ein\lin)^{1/3}$ if the non-dimensionalization is based on the energy injection rate.\\
The existence of the new dimensionless control parameter $Ro$ allows for different possible limits to be examined. In particular, we will  first study what happens at changing the Reynolds and Rossby numbers and the thickness of the rotating volume, $H$. We will also comment on the role of having a finite horizontal  extension, $L$,  and how this is connected to the possibility of forming a quasi 2D condensate. \\
In the presence of rotation, one can study the existence of solutions of the linearized inviscid NSE:   
\be
\partial_t {\bf u} + 2{ \Omega \bf\hat{e}_z \times u  } =-\bnabla P.
\label{eq:LNSR}
\ee
It is easy to see that (\ref{eq:LNSR}) can be solved by using a superposition of the set of incompressible helical waves that have already been introduced in section (\ref{sec:Helicity}):
\be
   {\bf u}({\bf x},t) = \sum_{\bk}\sum_{s_k=\pm} \tilde{u}^{s_k}_\bk(t) {\bf h}_{\bk}^{s_k} e^{(i \bk \cdot \bx)}
   \ee
   with the wave amplitudes which exhibit fast harmonic oscillations:
   \be
   \label{eq:waverot}
   \tilde{u}^{s_k}_\bk(t) = A_{s_k} \exp{ (i \omega_{\bk}^{s_k} t)}; \qquad
   \omega_{\bk}^{s_k} = 2 s_k \Omega \frac{k_z}{k} 
\ee
where $\omega_{\bf k}^{s_k}$ is the frequency of the wave with positive or negative helicity,  $s_k = \pm $. 
It is important to notice that the modes with $k_z=0$ are not affected by rotation and cannot support any wave motion.  Indeed, by writing the evolution for the vorticity:
\be
\partial_t {\bf w} + 2{ \Omega \partial_z \bf u  } =  \curl({\bf u \times w}) +\nu \Delta {\bf w} + \curl{\bf f},
\label{VNSR}
\ee
it is easy to see that the vertically-averaged modes,  living in the $k_z=0$
plane, are not affected by $\Omega$. As a result, in the presence of strong rotation, one expects that the 2D modes can be amplified  independently of the Coriolis force, leading to a quasi-2D dynamics \cite{Hough1897,Proudman1916,Taylor1917}. 
Two-dimensionalization of the flow under rapid rotation has 
been observed in numerous simulations \cite{smith1996crossover,Smith1999transfer,bartello1995geostrophic,yeung1998numerical, godeferd1999direct, chen2005resonant, Thiele2009, mininni2009, Mininni2010Rotating, favier2010space, Sen2012anisotropy, Marino2013invers, alexakis2015rotatingTG, biferale2016coherent,valente2017spectral} and laboratory experiments \cite{ibbetson1975experiments, hopfinger1982turbulence, dickinson1983oscillating, baroud2002anomalous, baroud2003scaling, Sugihara2005, ruppert2005extraction, morize2006energy, Staplehurst2008, Bokhoven2009experiments, Yoshimatsu2011, duran2013turbulence, Machicoane2016Two}. The flow can be split in two components. The two-dimensional component  is made of  modes with $k_z=0$ that are not affected by the rotation and their are referred to as the {\it slow manifold}. The remaining 3D modes are potentially affected by waves with frequencies that increase with $\Omega$ and  they form the  {\it fast manifold}. As a result of the fast oscillation at large $\Omega$ the non-linear interactions becomes weaker because  the phases of the different modes decorrelate  with a rate $\propto 1/\Omega$ leading to a very small (asymptotically vanishing) contribution to the non-linear energy transfer in the limit $Ro \to 0$.\\  
This is true for any  generic Fourier-helical triad $(\tilde{u}^{s_p}_\bp,\tilde{u}^{s_k}_\bk,\tilde{u}^{s_q}_\bq)$  except for those that have  a combination of frequencies such that 
\be  
\omega_{\bq}^{s_q} + \omega_{\bk}^{s_k} + \omega_{\bp}^{s_p} = 0 \label{resrot2}. 
\ee
Such  triads are {\it resonant} and it can be shown that they continue to exchange energy even in the limit of very small Rossby number \cite{greenspan1968theory, waleffe1993inertial,newell2011wave,chen2005resonant}, leading to a network of interacting 3D waves in the fast manifold (see Sec. \ref{sec:WWT}). \\ 
Furthermore, there  exist quasi-resonant triads that  satisfy the condition (\ref{resrot2}) only up to order $O(Ro)$, which can play a role for energy transfer in some pre-asymptotic limit, when $ Ro \ll 1$. Finally all modes in the slow manifold have $\omega_{\bp}^{s_p} = 0$, and thus the 2D3C manifold is not affected by wave propagation.  
\begin{figure*}[htbp]                                                                           
\centering                                                                                      
\includegraphics*[width=0.52\textwidth]{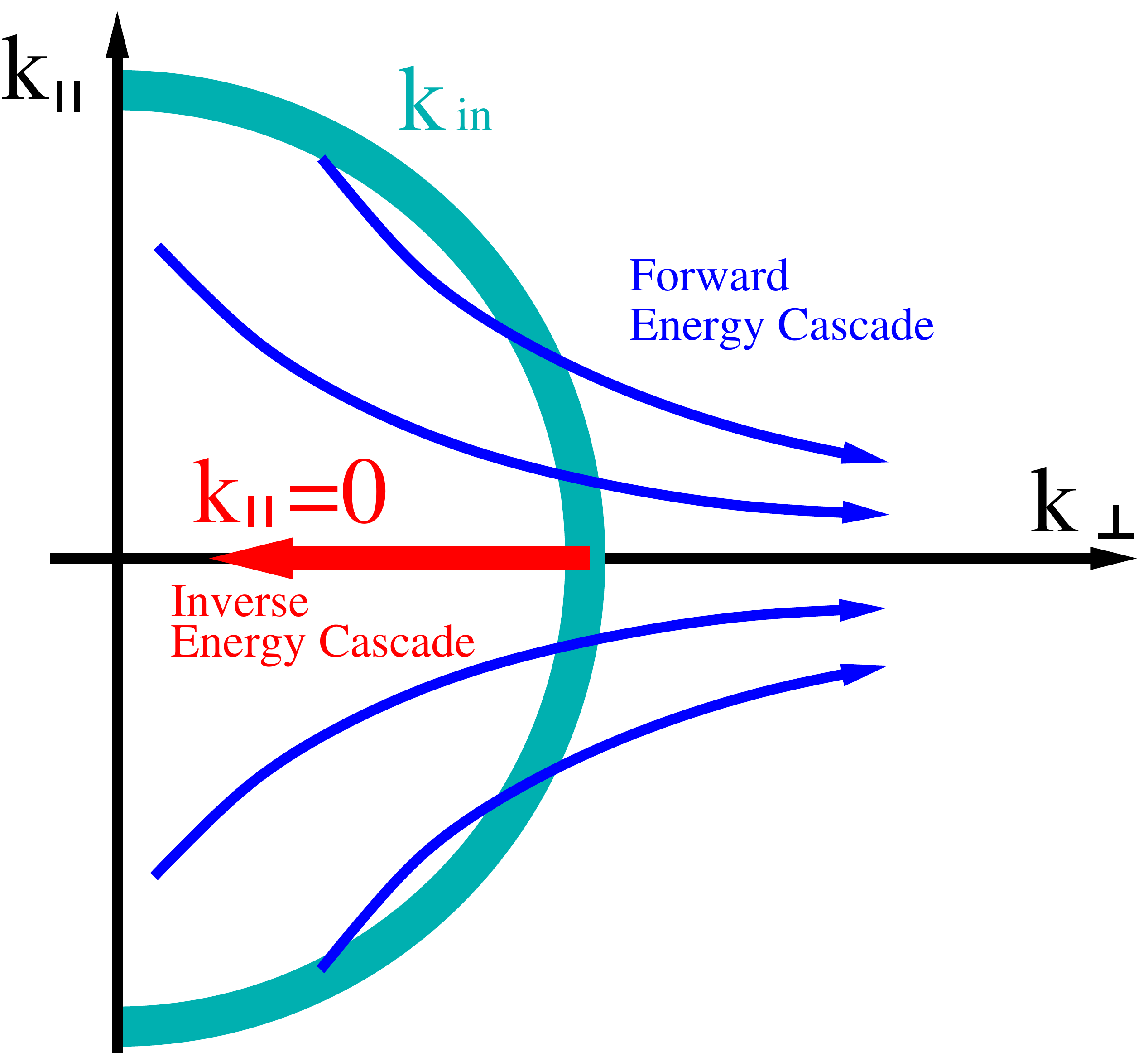}                                  
\caption{                                                                                       
          A sketch of the energy transfer  in the $(k_\perp,k_\|)$ plane in fast rotating              
          turbulence assuming that the 2D3C component is decoupled from the resonant waves.     
          The 3D waves tend to transfer energy to larger $k_\perp$ and smaller                  
          $k_\|$ without ever transferring energy to the $k_\|=0$ modes that act independently  
          and cascade energy inversely. }                                                       
\label{fig:Rotkpkp}                                                                             
\end{figure*}                                                                                   
\subsubsection{Decomposition in the 2D3C and 3D manifolds \label{sec:decomp}}
In order to understand the energy exchange and coupling among the 2D3C modes, the fully 3D resonant waves, the quasi-resonant triads and all the remaining 3D modes at changing the rotation rate it is useful,  as for the thin layer, to split the equations into the 2D3C vertical averaged component (slow modes)  and  the remaining 3D part (fast modes):
\begin{align}
\partial_t {\bf u}_{_{2D}} + {\bf u}_{_{2D}} \cdot \bnabla {\bf u}_{_{2D}}& = -\overline{ {\bf u}_{_{3D}} \cdot  \bnabla {\bf u}_{_{3D}} } - \overline {\bnabla P}  +  \nu \Delta {\bf u}_{_{2D}}  -  \alpha {\bf u}_{_{2D}}   +  {\bf f}_{_{2D}}, \label{rotlayera} \\
\partial_t {\bf u}_{_{3D}} + {\bf u}_{_{2D}} \cdot \bnabla {\bf u}_{_{3D}} + 2{ \Omega \bf\hat{e}_z \times u_{_{3D}}  }&= -{\bf u}_{_{3D}} \cdot  \bnabla {\bf u}_{_{2D}}  +(\overline{ {\bf u}_{_{3D}} \cdot  \bnabla {\bf u}_{_{3D}} }
-{\bf u}_{_{3D}} \cdot  \bnabla {\bf u}_{_{3D}}) \nonumber \\
& - (\bnabla P -\overline{\bnabla P}) - \alpha {\bf u}_{_{3D}}  +  \nu \Delta {\bf u}_{_{3D}}   +  {\bf f}_{_{3D}}, 
\label{rotlayerb}
\end{align}
where again ${\bf u}_{_{2D}}$ is a 2D3C velocity field obtained by vertically averaging the horizontal velocity components 
${\bf u}_{_{2D}} =(\overline{u}_x,\overline{u}_y,\overline{u}_z)$ and 
${\bf u}_{_{3D}}={\bf u}-{\bf u}_{_{2D}}$. 
The energy balance for the 2D3C and 3D  components is the same as the thin layer case
(\ref{2DbalanceA}-\ref{3DbalanceB}) because the rotation term does not transfer energy among scales. Nevertheless, rotation can decorrelate the 3D modes, and influence the 3D part of the  energy flux, $\Pi_{3D}$, changing its relative importance with respect to the 2D component, $\Pi_{2D}$, where the two are defined as in (\ref{eq:2d3dlayer}):
\be
\Pi_{2D}(k)=-\langle {\bf u}_{_{2D}}^{<k} \cdot ({\bf u}_{_{2D}} \cdot \bnabla{\bf u}_{_{2D}}) \rangle \qquad \mathrm{and} \qquad
\Pi_{3D}(k)=\Pi(k)-\Pi_{2D}(k).
\label{eq:2d3dflux}
\ee
\begin{figure*}[htbp]                                                                        
\centering                                                                                   
\includegraphics*[width=1.\textwidth]{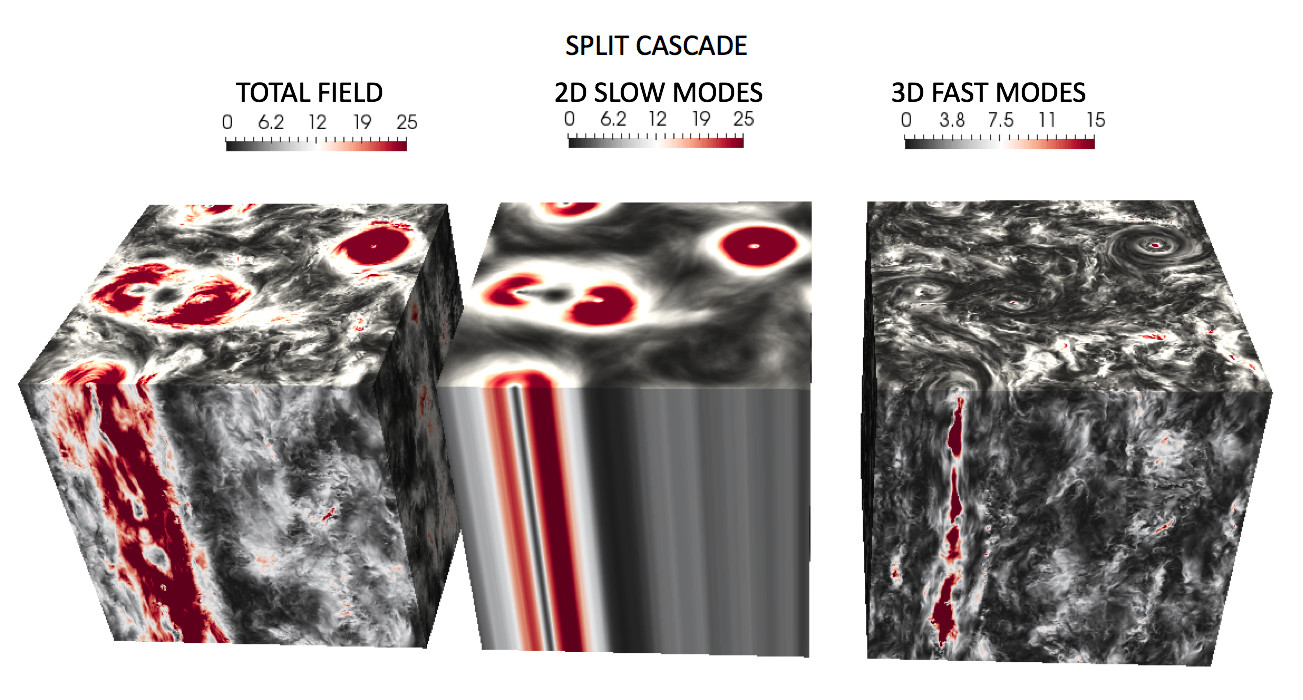}                              
\caption{     Color rendering of a 3D numerical simulation for rotating turbulence in the    
presence of a split energy cascade regime. We show the total field (left), the 2D3C          
slow manifold ${\bf u}_{_{2D}}$ (center) and the remaining 3D fast manifold field            
${\bf u}_{_{3D}}$. Colors are proportional to the velocity vector magnitude.                 
Figure adapted from \cite{buzzicotti2018energy}.   }                                         
\label{fig:splitcascade}                                                                     
\end{figure*}                                                                                
Indeed, it can be further shown that any two modes $\bp,\bq$ from the fast manifold  that belong to a triad including a mode in the slow manifold (e.g. with $p_{z}=-q_{z} \ne 0$ and  $k_{z}=0$) and which satisfy the resonant condition (\ref{resrot2})  lead to zero energy transfer \cite{greenspan1969on,waleffe1993inertial,chen2005resonant}. Thus, the slow 2D3C manifold becomes isolated from the remaining 3D part of the flow if the dynamics is restricted to resonant waves; i.e. the first term on the RHS of (\ref{rotlayera}) will be such that $\langle {\bf u}^{<k}_{_{2D}} (\overline{{\bf u}_{_{3D}} \cdot  \bnabla {\bf u}_{_{3D}}}) \rangle \to 0$ in the limit $Ro \to 0$. 
%
The 2D3C part of the flow can still advect the modes in the fast manifold, via the  $ {\bf u}_{_{2D}} \cdot \bnabla {\bf u}_{_{3D}}$ term  in (\ref{rotlayerb}),  but without
exchanging any energy or helicity with them. Thus, in the resonant limit  the $\bu_{2D}$ component evolves independently  of the  energy cascading to the large scales through the 2D in-plane  motions ${\bf u}^\perp_{_{2D}} = ( \overline{u}_x,\overline{u}_y)$ and to the small scales through the advection of the vertical component $\overline{u}_z$. If we further decompose the spectrum of the 2D3C flow in a contribution due to the 2D horizontal components and one for the vertical out-of-plane component, $E_{_{2D}}(k_\perp) = E^\perp_{_{2D}}(k_\perp) +E^\|_{_{2D}}(k_\perp)$ we have that the inverse cascade leads to  $E^\perp_{_{2D}}(k_\perp)\propto k_\perp^{-5/3}$ energy spectrum for $k_\perp \ll k_{in}$ limit and to $E^\perp_{_{2D}}(k_\perp)\propto k_\perp^{-3}$ for $k_\perp > k_{in}$ due to the 2D forward enstrophy cascade. The  process related to the  energy of the  $\overline{u}_z$ component  leads to a small-scale$E^\|_{_{2D}}(k_\perp) \propto k_\perp^{-1}$ spectrum as must be the case for a passive field in a smooth flow (see the discussion about (\ref{eq:batchelor}) in the section for passive scalars \ref{sec:passive} ).

On the other hand, the 3D part of the flow will cascade energy through resonant wave interactions, (\ref{resrot2}). Weak turbulence theory (see sec. \ref{sec:WWT}) predicts the existence of a forward anisotropic energy cascade to larger  $k_\perp$ and  smaller $k_\|$. Exact resonant interactions move energy to smaller and smaller $k_\|$ without however reaching $k_\|=0$. Nevertheless, for infinitely thick vertical domains, $H \to \infty$, there will be sufficiently small $k_\|\ne 0$ such that quasi-resonant triads, that satisfy the  condition (\ref{resrot2}) only  up  $O(Ro)$, will be able to transfer energy to/from the 2D3C manifold. As a result, the tendency of the 2D manifold to transfer energy backward can be suppressed.  Asymptotically, for infinite domains and for $k_\|/k_\perp\ll 1$, the wave theory leads to a forward energy cascade and \be E(k_\perp,k_\|) \propto \ein^{1/2} \Omega^{1/2} k_\perp^{-5/2}k_\|^{-1/2} \label{eq:RotWWT}\ee 
\cite{galtier2003weak,galtier2014theory}. Eq. (\ref{eq:RotWWT}) can be obtained phenomenologically from the expression (\ref{eq:wavebalance2}) discussed in section (\ref{sec:WWT}) assuming $k_\| \ll k_\perp$ that leads to $k\simeq k_\perp$. The validity of  weak turbulence theory breaks down when  sufficiently small perpendicular scales $r_\perp \sim k_\perp^{-1}$ are reached such that the amplitude of the non-linear term $\propto u_r^2/r_\perp$ is of the same order of the  Coriolis term $\Omega u_r r_\perp/ r_\|$, where  $u_r$ is a short-hand notation for the typical velocity fluctuation at scale $|\br| =r$, $u_r \sim \langle |\delta_\br \bu|^2\rangle^{1/2}$.  As a result, if the Reynolds number is large enough, there will always be a range of  small scales where  strong turbulence  is expected with $ u_r^3/r_\perp \propto \ein$. Combining these relations together, we can estimate that the transition between wave-turbulence and strong turbulence   occurs at scales where $r_\| \propto \Omega \ein^{-1/3} r_\perp^{5/3}$ or in spectral space:
\be
k_\| \propto \Omega^{-1} \ein^{1/3} k_\perp^{5/3}.
\label{eq:wavetransition}
\ee

According to the previous discussion, there are two mechanisms at play. One comes from the degrees of freedom in the slow manifold that bring energy to the large scales and the other comes from the 3D interactions that bring energy to the small scales. A sketch of the way energy flows in the $(k_\perp,k_\|)$-plane is shown in Fig. (\ref{fig:Rotkpkp}). Which of the two processes wins depends on $Ro$ and the value of $H/\lin$. For infinitely thick  layers and small $Ro$ the flow is described by weak wave turbulence theory \cite{galtier2003weak,galtier2014theory}. For $Ro \to 0$ and finite layer thickness, the dynamics is dominated by the decoupling of the 2D3C manifold from the 3D resonant wave-interactions. The former will have a strong tendency to an inverse energy cascade while the latter will preferentially move energy to small-scales (but anisotropically). In other words we are in the presence of a split energy cascade. This scenario can also be shown within the wave turbulence limit \cite{scott2014wave}. It has been supported by recent studies in rotating flows both in the lab \cite{Yarom2013experimental, yarom2014experimental, Bokhoven2009experiments, Moisy2014direct, campagne2016turbulent} and with high-resolution numerical simulations \cite{smith1996crossover, Smith1999transfer, baerenzung2011helical, Sen2012anisotropy, pouquet2013inverse, Deusebio2014dimentional, biferale2016coherent}. Recently,  it has been proven that for finite $H$ and $Re$ the flow becomes exactly 2D above a critical rotation rate \cite{gallet2015exact2}.

\begin{figure*}[htbp]                                                                 
\centering                                                                            
\includegraphics*[width=1.\textwidth]{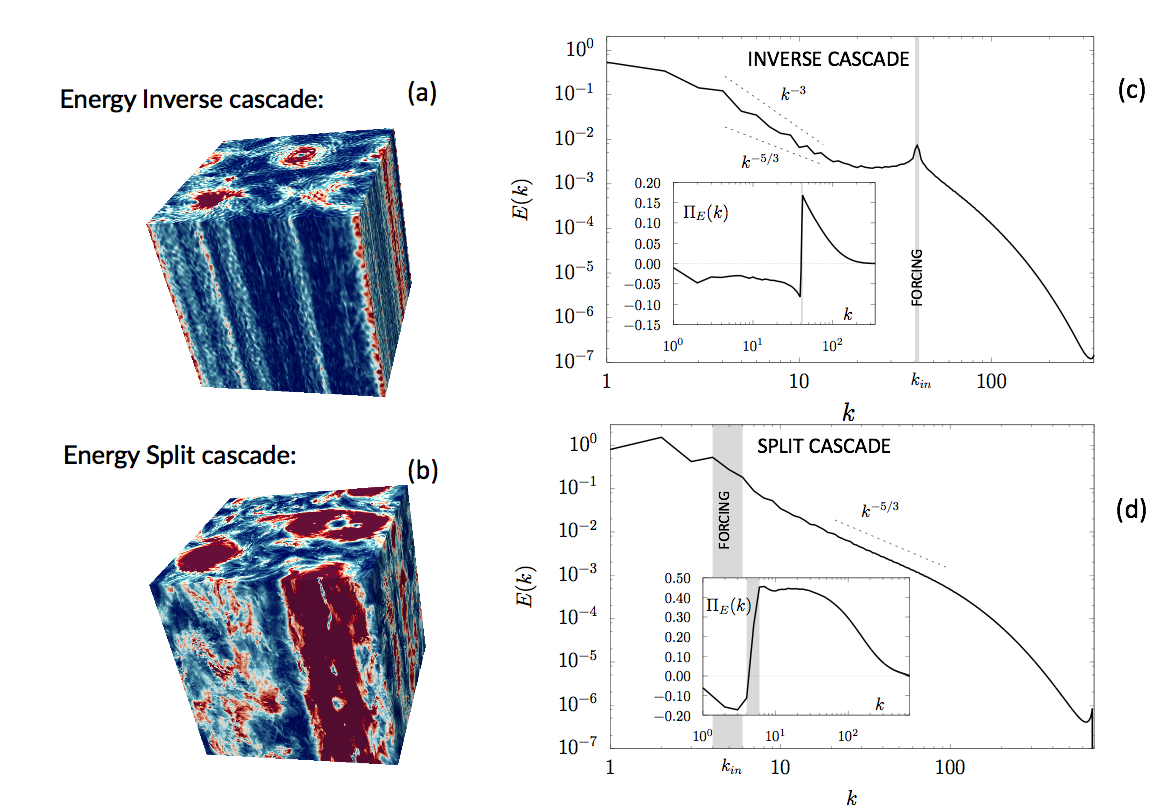}                
\caption{ Two different flow configurations for rotating turbulence at low Rossby and 
at changing the energy injection scale. Top: Spectrum for the case with small-scale  
forcing, $k_{in} \in [40:42]$ (grey band). Notice the presence of the inverse cascade 
given by the negative constant flux, $\Pi_E(k)$ (inset of panel c) and the poor       
scaling properties of $E(k)$. Bottom: the same flow configuration but forced at       
$k_{in} \in [4:6]$ (grey band)  with a clear $-5/3$ direct energy range and           
a split-cascade scenario for the flux (inset of panel d).  The 3D rendering also shows 
the complexity of the flow configurations with both 2D cyclonic and anticyclonic       
structures that coexist with a 3D turbulent background.                               
Data are adapted from \cite{buzzicotti2018energy}.}                                   
\label{fig:inversesplitcascade}                                                       
\end{figure*}

  It is important to stress that both the wave turbulence regime and the perfect decoupling among 2D and 3D modes might not be observable in many realistic situations, where Rossby number is never small enough. In such a case, the slow and fast manifolds interact and the flow evolution  is the result of a  mutual feedback among the two components. In Fig. (\ref{fig:splitcascade}) we show  a 3D rendering of a turbulent rotating flow side-by-side  with its  ${\bf u}_{_{3D}}$ and ${\bf u}_{_{2D}}$ components. From the figure it is evident that the two components are correlated. Furthermore, both in numerical and experimental realizations is very difficult to reach a resolution high enough or an experimental domain large enough to simultaneously resolve the inverse cascade regime and the two different scaling ranges in the direct energy cascade regime as depicted in Fig. (\ref{fig:Rotkpkp}) (see  \cite{biferale2016coherent} for a first attempt in this direction). Many studies are then limited to force either at large  or at small scales in order to highlight either the forward or the inverse energy transfer regime. In Fig. (\ref{fig:inversesplitcascade}) we show two flow realizations for rotating turbulence at small Rossby forced either at low $k$ (bottom panels) or at large $k$ (top panel) together with the corresponding spectra and fluxes. The flow forced at high $k$ develops only an inverse energy cascade range with a clear constant energy flux (see inset of panel c) but without any well developed power law properties for the  isotropic spectrum. The flow forced at low $k$ tends to develop a clean $-5/3$ spectrum at small-scales, while  the scaling for $k <k_{in}$ is not good due to the limited range of available wavenumbers. However, the flux shown in the inset of panel (d)  clearly shows  a split-cascade scenario with two ranges with opposite signs.

Finally, we remark that the phenomenology becomes more complicated as soon as one considers finite values of $Ro$, $Re$ and $H/\lin$. First, we need to consider that the statistics is always anisotropic and that one might imagine to recover exact scaling laws only in some asymptotic limit, e.g. for very large scales in the $k_\perp$ plane for inverse cascades or for very small scales  for the 3D fluctuations  (see \cite{biferale2005anisotropy} and Sec. \ref{sec:scalinganisotropic}). 
Isotropy is expected to be recovered  only at scales smaller that the Zeman scale,  $\ell_\Omega \sim (\ein/\Omega^3)^{1/2} $ where the eddy turnover time equals the rotation rate \cite{zeman1994note,mininni2012isotropization}.
 \begin{figure*}[htbp]                                                                         
 \centering                                                                                    
 \includegraphics*[width=0.45\textwidth]{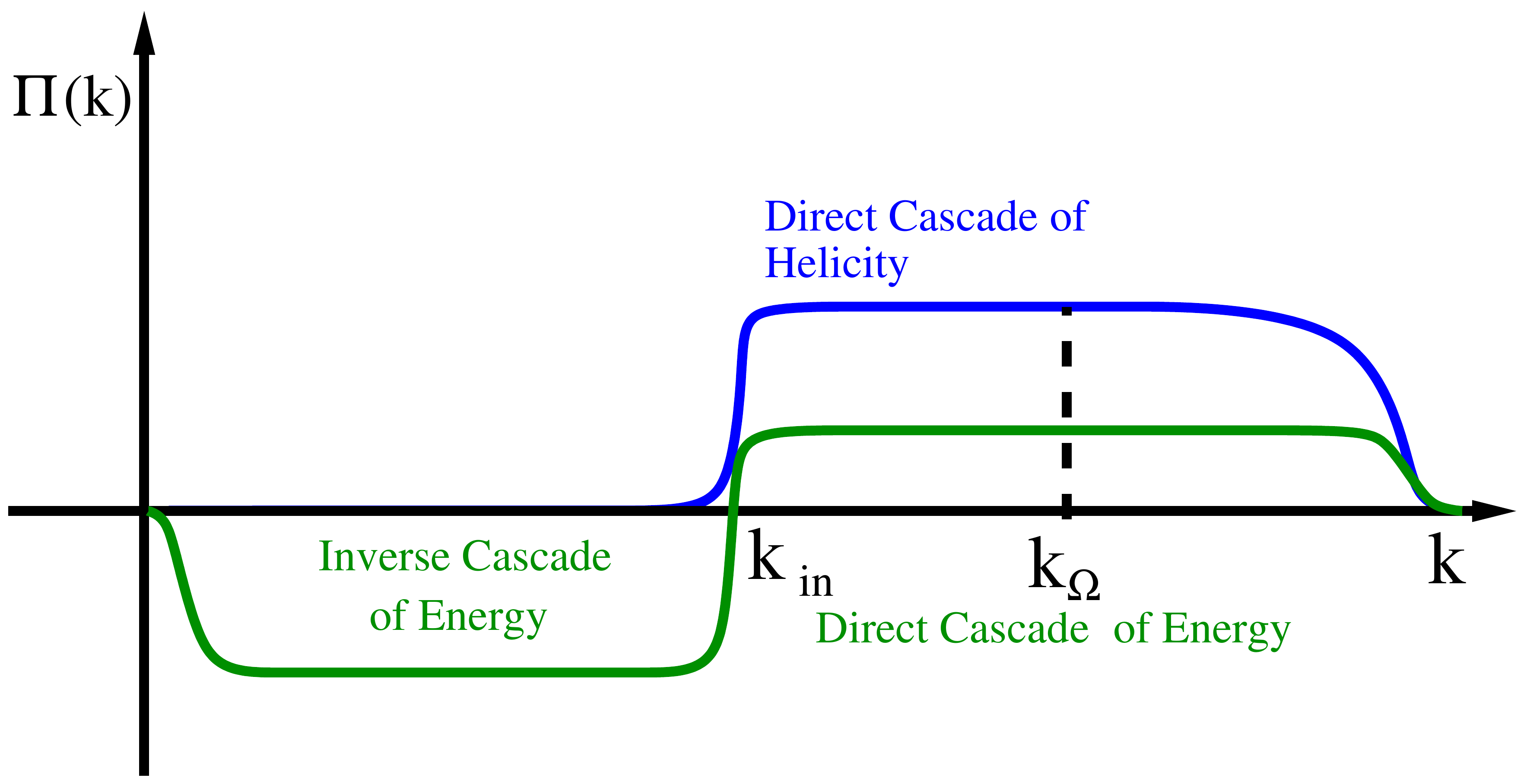}                                   
 \includegraphics*[width=0.45\textwidth]{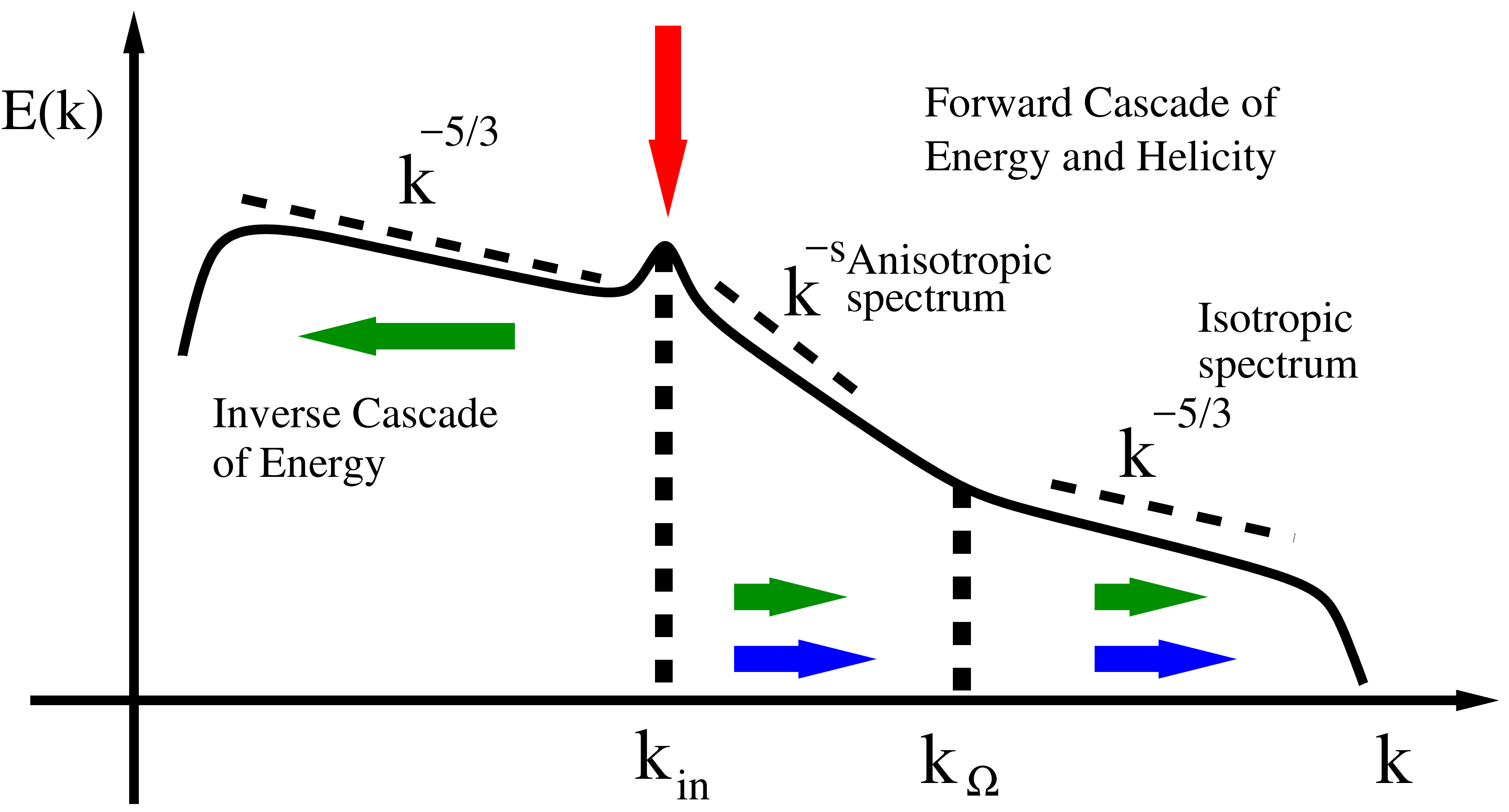}                               
 \caption{                                                                                     
   The fluxes and energy spectra  in rotating turbulence.                                      
           The different lines indicate the fluxes for the total energy $E$ (green) and        
           the helicity $H$ (blue). The spectral index $s$ is depends on  the process that     
           drives the cascade (weak turbulence, helicity cascade, enstrophy cascade) and can   
           also be a superposition of different power-laws.                                    
           $k_\Omega \sim 1/\ell_\Omega$ stands for the Zeman wavenumber \cite{zeman1994note}.} %
 \label{fig:RotFwd}                                                                            
 \end{figure*}                                                                                 
Furthermore, it was also suggested in \cite{mininni2009helicity,baerenzung2011helical,pouquet2010interplay} that for strongly rotating and strongly  helical flows, the small scales might be controlled by the forward cascade of helicity, forming a  spectrum $E(k)\propto h_{in}^{2/3} k^{-7/3}$ given by (\ref{eq:split}). Moreover, as we have seen in Sec.  (\ref{sec:Helicity}), there exist homochiral triadic interactions that have a pure 3D inverse energy transfer  and therefore one could argue that the global energy transfer in rotating turbulence is also affected by the combined effects of rotation and chirality.  As a result,   for the forward cascade in boxes of finite height and at fixed Rossby numbers there are different mechanisms that can determine the anisotropic scaling properties: the helicity transfer, the energy and the enstrophy transfers of the out-of-plane and in-plane components of the 2D3C field, weak wave turbulence or any combination of the above. Thus we can not give a unique  and general description for the functional form of the energy spectrum, except for the asymptotic regimes discussed above.  This scenario is summarized in Fig. (\ref{fig:RotFwd}). The different possibilities will be addressed below. \\

\subsubsection{Rotating turbulence at finite thickness, $H$, and at changing Re}
\begin{figure*}[h!]                                                                             
\includegraphics[width=0.45\textwidth]{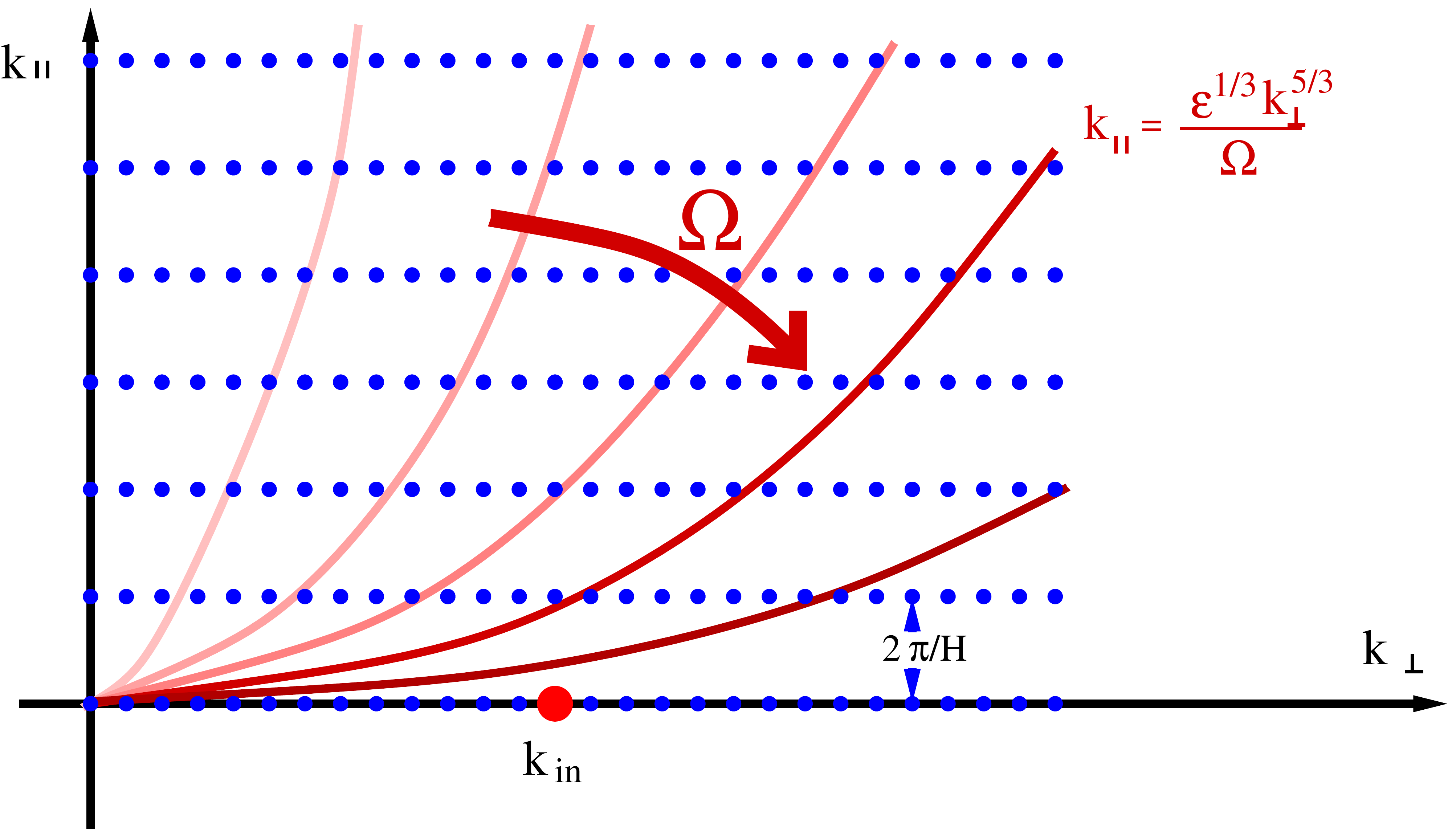}                                         
\includegraphics[width=0.45\textwidth]{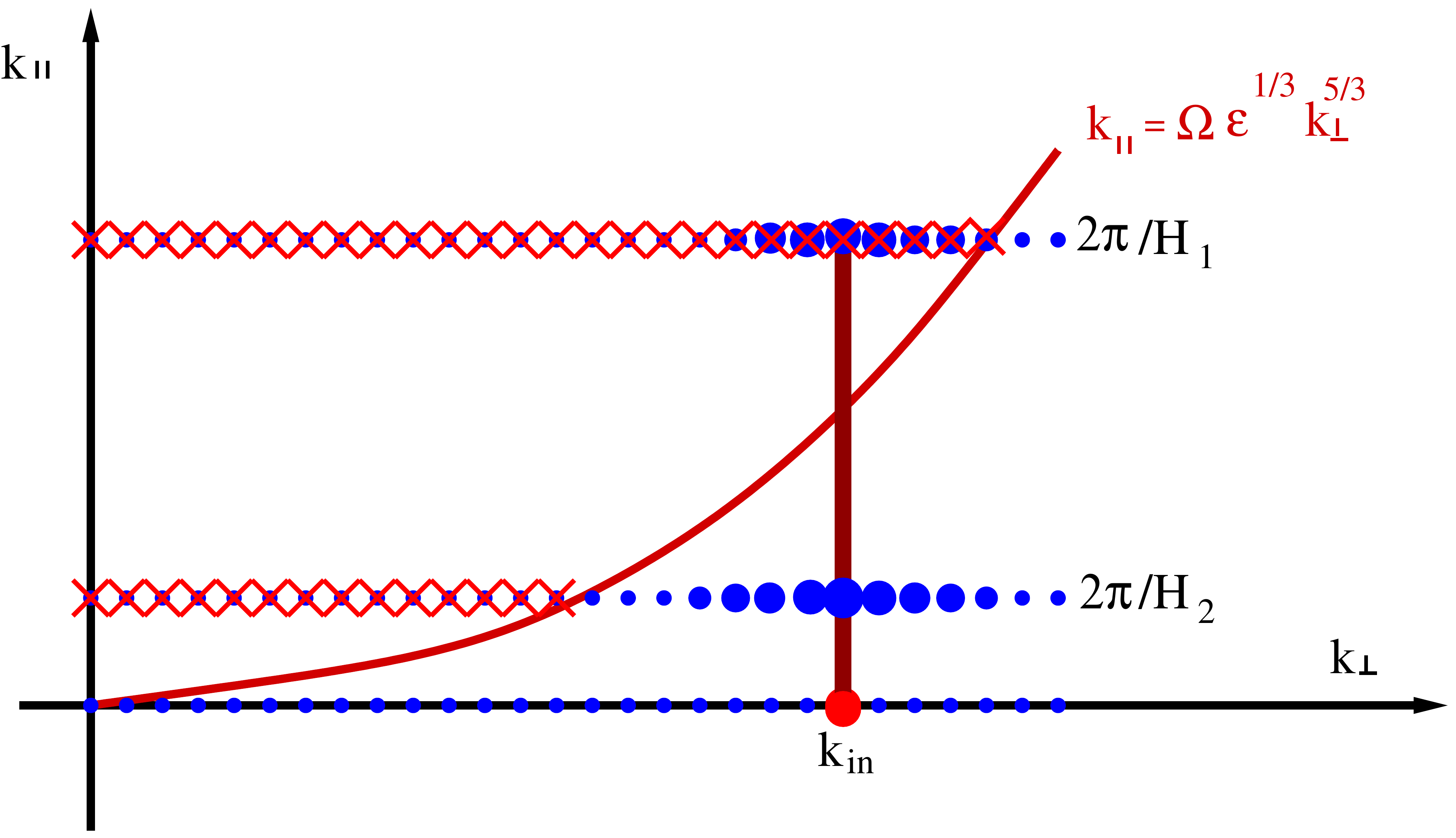}                                        
\caption{ 
  A demonstration of how the critical value of rotation and layer thickness                     
is determined. The left figure shows the discrete wave numbers in a layer of fixed height $H$   
(see also Fig. (\ref{fig:Fgrid}) in Sec. \ref{sec:Thin}). The 3D modes ($k_\|\ne0$) that        
interact with the 2D modes in the plane at $k_\|=0$ are the ones below the critical balance lines%
(displayed by brown lines for different values of $\Omega$). For weak $\Omega$ there are enough 
3D modes below the critical line, and close to the forcing wavenumbers $\kin$, that interact    
with the 2D slow modes                                                                          
and prevent an inverse cascade in the plane, due to the existence of non-linear 3D instabilities.%
If $\Omega$ is large enough, so that all 3D wavenumbers close to $\kin$ are above the critical  
balance line, then 2D modes act independently and cascade energy inversely to small $k_\perp$.  
Alternatively, the right panel shows the Fourier grid for two different layer thicknesses       
$H_1$ and $H_2$ with $H_1<H_2$. The modes which are closer to the forced mode $\kin$ are marked 
with larger symbols. Modes which can not interact with the forcing modes due to                 
the fast rotation are crossed out by the red x.                                                 
For $H_1$ the 3D modes close to the forcing are above the                                       
critical balance line and do not interact with the 2D modes that cascade energy inversely,      
while for $H_2$ the 3D modes are below the critical balance line and can interact with the 2D   
modes and prevent the inverse cascade. }                                                        
\label{fig:CBgrid}                                                                              
\end{figure*}

\begin{figure*}[htbp]                                                                     
\centering                                                                                
\includegraphics*[width=0.45\textwidth]{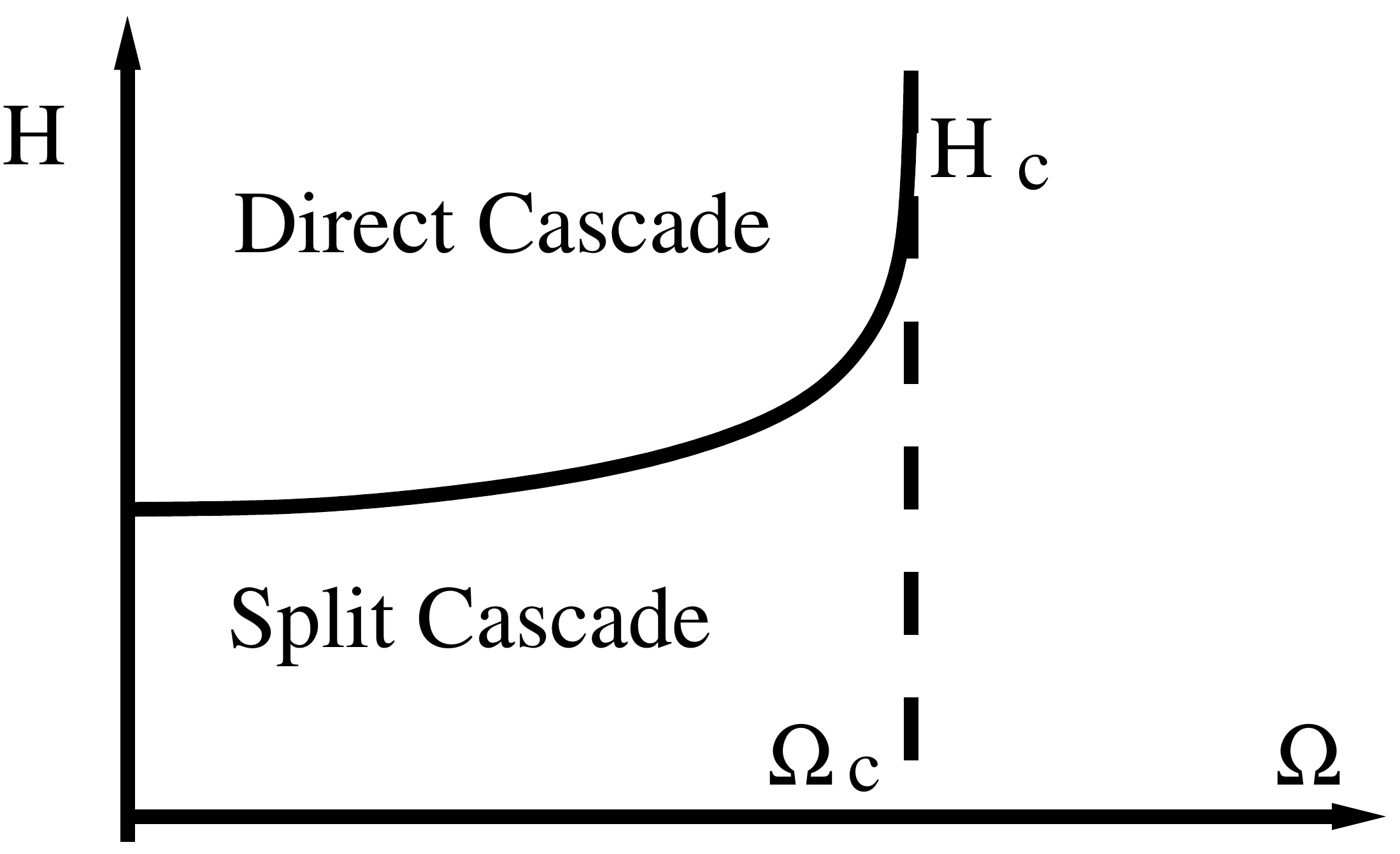}       \qquad                
\includegraphics*[width=0.45\textwidth]{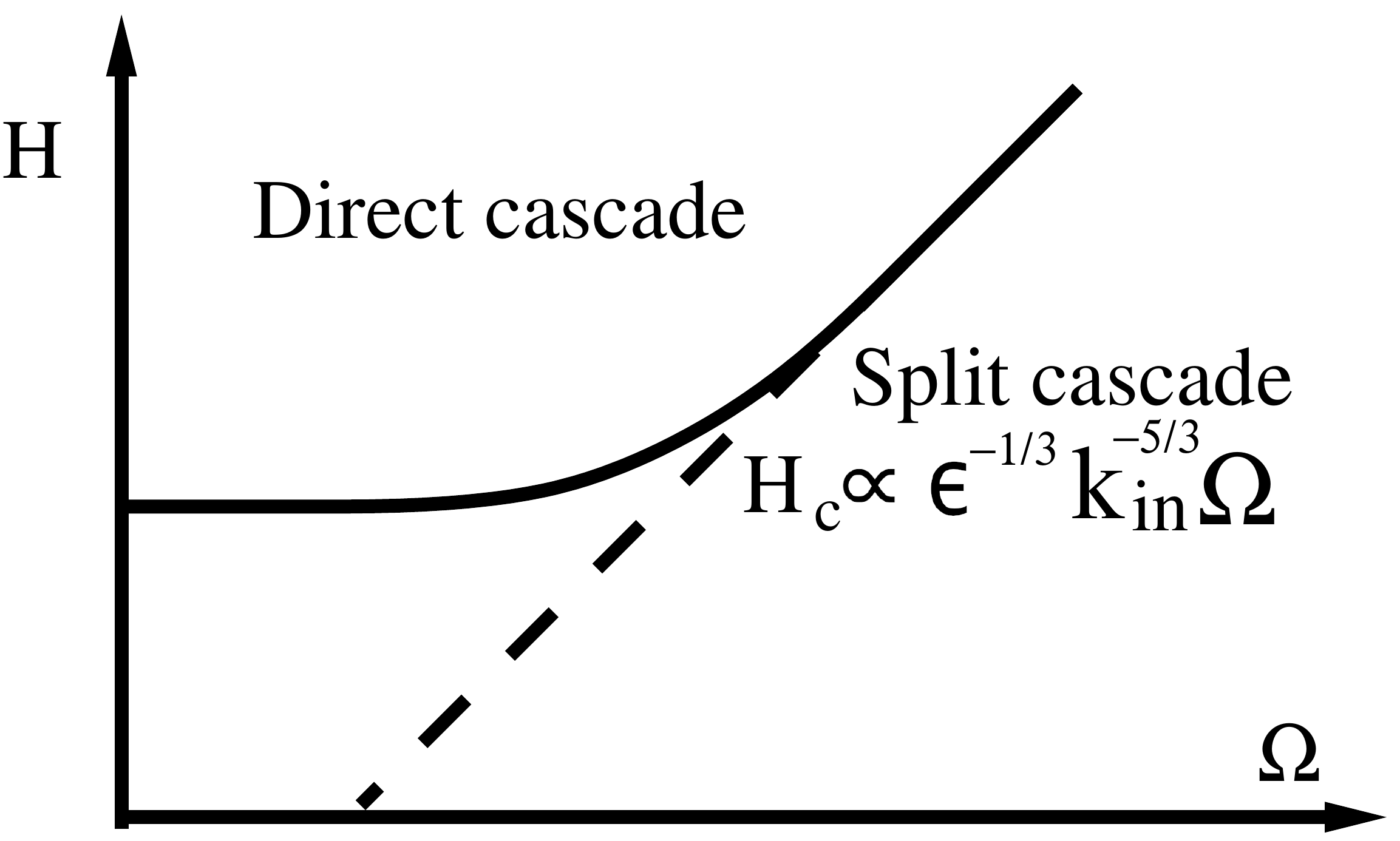}                             
\caption{ Two different phase-diagram scenarios  for rotating turbulence. The line        
separates the regions in $H-\Omega$ space for which a split cascade is present or not.    
In the  left  panel a critical rotation rate is assumed to exist above                    
which independently of how large $H$ is an inverse cascade exists. In the right 
panel the critical height depends on $\Omega$  as predicted by (\ref{eq:criticalH}) . }   
\label{fig:RotPhasespace}                                                                 
\end{figure*}

   As discussed in the previous section in fast rotating flows infinitely thick layers tend to cascade energy forward, while thin layers tend to cascade energy inversely. Here we explain how to  obtain  a prediction about the transition line in the $(\Omega,H)$ phase space. The main criterion is based on whether the 2D  modes are isolated from the 3D modes due to  fast rotation. The overall qualitative picture that emerges is the following. If the slow modes at the forcing scale  are isolated, they will start transferring energy to the larger scales. This process is strengthened as larger scales are reached because the eddy turnover time decreases compared to $\Omega^{-1}$, and thus the inverse cascade is reinforced. On the other hand,  if the slow modes close to the injection scale are interacting strongly with the  3D modes, then energy will cascade forward. More precisely, one might imagine that the 2D-3D coupling at the injection scale will be strong enough to stop the inverse cascade  if the  fast modes closest to the slow manifold (with $k_\|\sim 2\pi/H$ and $k_\perp\sim \kin$) fall in the strong turbulence regime determined by (\ref{eq:wavetransition}). This is demonstrated in Fig. (\ref{fig:CBgrid}) for  fixed $H$ and varying $\Omega$ (left panel) and for  fixed $\Omega$ and two different values of $H$ (right panel).  In the first case, for any finite $H$ and by increasing $\Omega$,    the relation  (\ref{eq:wavetransition}) predicts that the region where 2D3C and 3D modes can interact is pushed to larger and larger values of $k_\perp$. This scenario leads to  an inverse cascade due  to the isolation of the  2D dynamics close to $k_{in}$ (left panel of Fig. \ref{fig:CBgrid}).  
Vice-versa, for any fixed $\Omega$ and at increasing $H$,  one  predicts that the inverse cascade breaks down when 3D modes with $k \simeq k_{in}$ in the strong  wave-turbulence  region exist (right panel of Fig. \ref{fig:CBgrid}). In the latter case,  there are 3D interactions, outside the resonant sub-set, which couple the two manifolds and destroy the inverse cascade. This argument leads to the estimate:  
\be
\label{eq:criticalH}
H_c \propto \Omega \ein^{-1/3} k_{in}^{-5/3},
\ee
which  is  obtained from (\ref{eq:wavetransition}) by replacing $k_\| \sim H_c^{-1} $ and $k_\perp = k_{in}$. The same scaling can also be obtained by defining  $H_c$  from the balancing of the  eddy-turn-over frequency  with the  slowest wave frequency at the injection scale: $\ein^{-1/3} k_{in}^{2/3} = \Omega/(H_c k_{in})$.  As a result,  with respect to the thin layer case discussed in section (\ref{sec:Thin}), rotation is expected  to enhance the 2D inverse energy cascade, because pure resonant waves do not couple with the 2D3C manifold and tends to stabilize the 2D dynamics with respect to 3D perturbations. \\ 
\noindent In summary, in the $(\Omega,H)$ phase space, we expect that at  fixed $\Omega$  the {\it two-dimensionalization of the flow} starts to  lose validity when sufficiently tall boxes are considered such that there are always quasi-resonant modes arbitrarily close to the $k_z=0$ plane. These modes provide a link between the slow and the fast manifold that  according to  weak turbulence theory  will result in the energy cascade to be strictly forward in the infinite $H$ limit.\\
As rotation is increased it helps the {\it two dimensionalization} of the flow and thus one expects that the  the critical height, $H_c$, will increase with rotation according to (\ref{eq:criticalH}).
The increase of $H_c$ with $\Omega$ 
has been numerically demonstrated  in \cite{Deusebio2014dimentional}.
However, computational limitations did not allow 
 to investigate the asymptotic when  $H_c$ and  $\Omega$ tend to infinity. The right panel of Fig. (\ref{fig:RotPhasespace}) summarizes this  behaviour. In the same figure we also show
another potentially possible phase-diagram, where we imagine that  a critical $\Omega_c$ exists beyond which there will always be a split cascade regime, with a fully developed inverse range.
The latter  scenario would  be inconsistent with the wave-turbulence prediction but it cannot be excluded if there are 3D modes able to transfer part of the energy backwards as we have discussed in section  (\ref{sec:Helicity}) by analysing homochiral and heterochiral triads. \\
As for the case of the layer of finite height it is interesting to know how the spectrum  at the large scales changes from  the thermal distribution $E(k)\propto k^2$ to a spectrum  with a negative exponent in the presence of the inverse energy transfer. We note that like in thin layers 
as the inverse cascade develops to larger horizontal scales, 
the condition for {\it two-dimensionalization} 
becomes stronger both because $r_\perp$ becomes larger than $H$ but also because  the large-scale eddy-turn-over-time  becomes smaller than the rotation rate.  Thus, the inverse cascade is expected to reach asymptotically a pure 2D state and the spectrum  should  become  $E(k_\perp,0)\propto k_\perp^{-5/3}$ for $k_\perp \to 0$.
The transition from $k^2$ to $k^{-5/3}$ should then follow the path described in figure \ref{fig:closetocrit}.

Finally we note that  when the forcing is purely  2D there is another critical rotation rate at which the flow becomes exactly 2D.  Above this critical rotation rate all 3D fluctuations decay in time. The existence of this $Re$-dependent critical point was proven in \cite{gallet2015exact2}. The flow behaviour close to this second critical point has not yet been investigated  however.

\subsubsection{Dependency on the forcing}
\label{sec:depforcing}
\begin{figure*}[htbp]                                                                     
\centering                                                                                
\includegraphics*[width=0.85\textwidth]{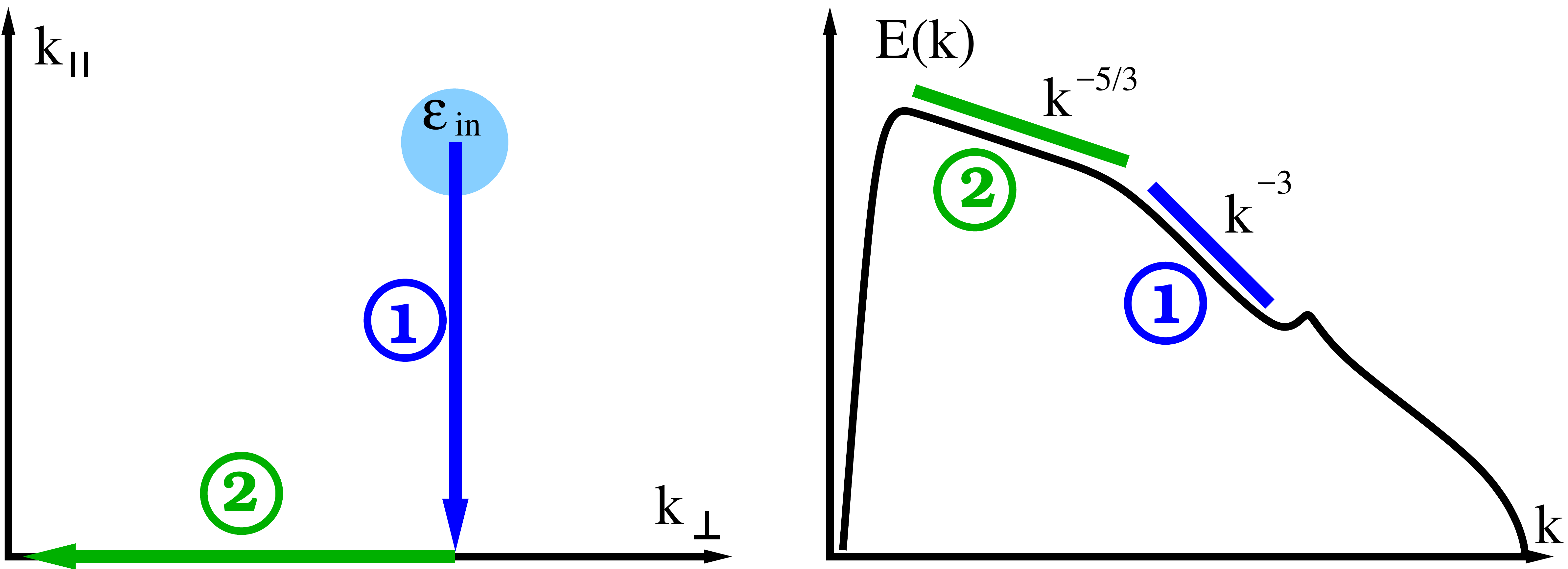}                                 
\caption{ The figure demonstrates the different processes that can occur when the    
forcing is located in the fast manifold $k_\|\ne0$ as indicated in the left panel.        
In this case there is a first process that transfers energy to the slow manifold $k_\|=0$ 
that can lead to steeper spectra (shown in the right panel) that is followed by a second  
process that transfers energy to the small $k_\perp$ wavenumbers by a 2D inverse cascade  
(see \cite{Sen2012anisotropy}).  }                                                        
\label{fig:kminus3}                                                                       
\end{figure*}

Another important issue that can modify the phenomenology is connected to the properties of the forcing. Due to the tendency of decoupling the 2D3C modes from the 3D wave-turbulence modes it is clear that the
relative intensities of $\emu$ and $\enu$ can depend on  whether the forcing is acting only on the 3D, only on the 2D3C or on both components \cite{Sen2012anisotropy}. Moreover,
the number, and efficiency, of resonant and quasi-resonant triads  depend on their helical properties,
suggesting the possibility that strong helicity  injection can alter the dynamics of the forward and inverse   cascades \cite{mininni2009helicity,baerenzung2011helical}.
For example,  recent results \cite{Sen2012anisotropy} have shown that a spectrum with a  $-3$ slope steeper than the $-5/3$ prediction appears in the inverse cascade range when the system is forced  also in  the fast manifold. This steeper power-law is attributed to a direct transfer of energy from the fast $k_\| \ne 0$  to the slow manifold, due to residual non-resonant interactions,  leading to a condensation at $k_\|=0$ after which the 2D inverse cascade starts. It is not know how much this scenario is stable at changing the control parameters in the flow.  The two processes that lead to the two spectra when the forcing is in the fast manifold are sketched in Fig. (\ref{fig:kminus3}). 

\begin{figure*}[htbp]                                                                      
\centering                                                                                 
\includegraphics*[width=1\textwidth]{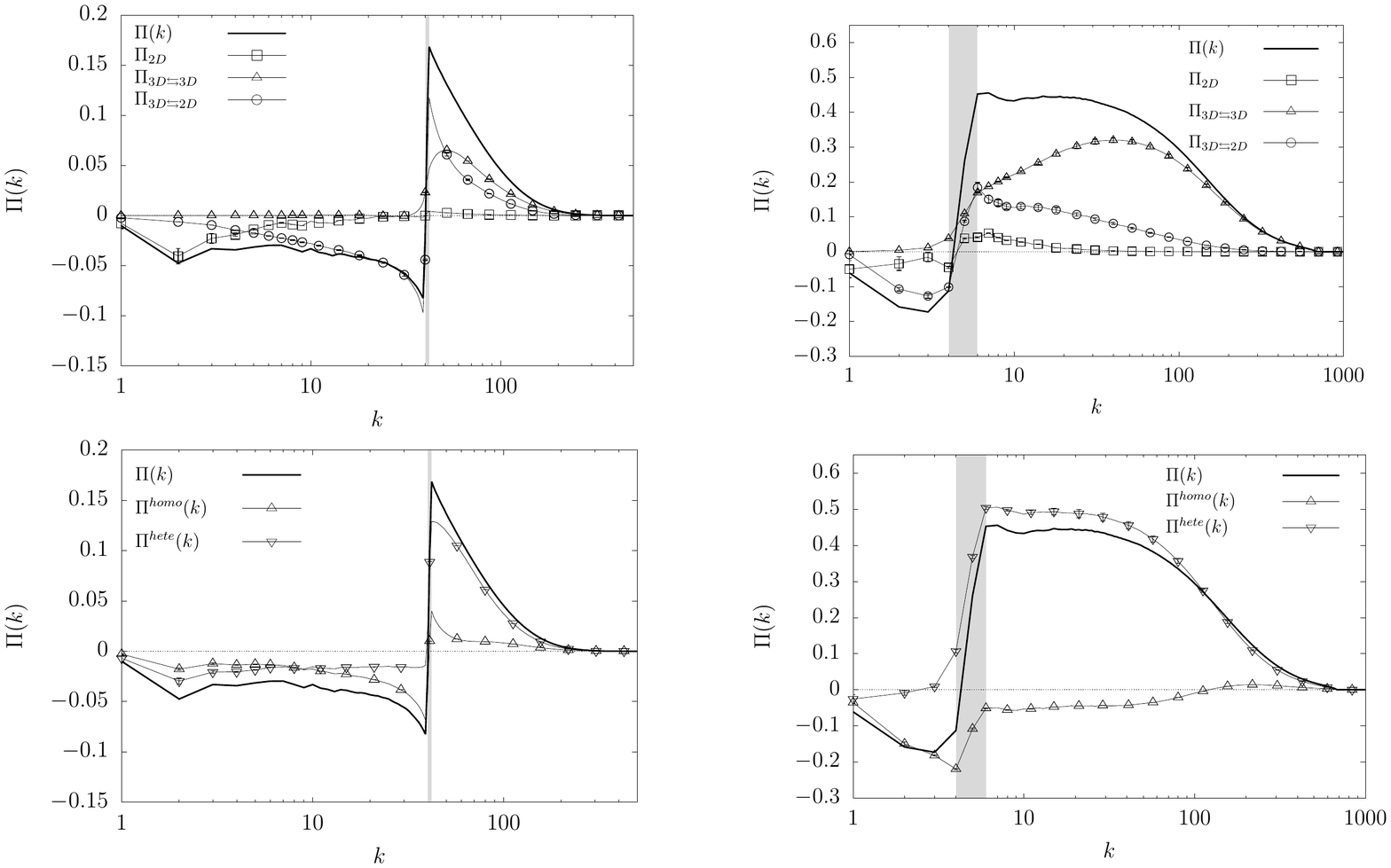}                           
\vskip -1.0 cm                                                                             
\caption{Decomposition of the total energy flux for turbulence under rotation in           
 in-plane and out-of-plane components (\ref{eq:2d3dflux}) (top) or  homochiral and         
heterochiral components (\ref{eq:decPi}) (bottom)  for two different forcing wavenumbers   
at small                                                                                   
scales (left) and large scales (right). In the top plots we have further decomposed the    
$\Pi_{3D}(k)$ flux given by (\ref{eq:2d3dflux})  into a component that transfer energy     
only among fast modes $\Pi_{3D \leftrightarrows  3D}(k)$ and in one that couples 2D and    
3D modes, $\Pi_{3D \leftrightarrows  2D}(k)$. In the bottom row we have added the two      
heterochiral components of the decomposition (\ref{eq:decPi}) to get to one total          
heterochiral contribution, $\Pi^{hete}(k) = \Pi^{hete,1}(k) + \Pi^{hete,2}(k)$.  Data are  
adapted from \cite{buzzicotti2018energy}. The two top panels show that the 3D-2D coupling  
is active for a large range of scales $k<k_{in}$. In the same range, the left panel of 
bottom row  shows that the homochiral channel is the one that transfers energy backward.   
Bottom right panel shows the chiral decomposition for a case forced at small $k$ from      
where we see that rotating turbulence falls in the split-cascade and bi-directional        
definitions (\ref{def4}) and (\ref{def6}) of Sec. (\ref{sec:definitions}) }                
\label{fig:homoheterorotation}                                                             
\end{figure*}

The above argument would indicate that  the 3D modes still play an important role and can not be neglected in the inverse cascade as also suggested by analysing the role of chirality in the triadic interactions. Indeed, as discussed in Sec. (\ref{sec:Helicity}), another exact way to decompose the total energy flux is based on the separation among homochiral and heterochiral components. In the latter case, in contrast to the in-plane and out-of-plane decomposition (\ref{rotlayera}-\ref{rotlayerb}), the distinction is made in terms of fields that evolve in a fully isotropic way  and fully 3D but still with a tendency to move energy either forward (heterochiral triads) or backward (homochiral triads). It is natural then to ask whether in turbulence under rotation, and in the presence of a split cascade, there is a tendency to enhance the homochiral channel in the range of scales where energy cascades backward \cite{buzzicotti2018energy}. In Fig. (\ref{fig:homoheterorotation}) we show that this is indeed the case by using two direct numerical simulations of rotating turbulence at high resolution and at changing the forcing scale, $k_{in}$, either close to the IR or to the UV limit in the energy spectrum. By decomposing the total inverse flux in homochiral and heterochiral components as described in (\ref{eq:decPi}), it is  demonstrated that for a large range of scales $k < k_{in}$, the main backward flux is brought by the homochiral interactions and that only in the limit of $k\to 0$, where the flow becomes fully 2D, the homochiral and heterochiral components have equal weight, as it must be in a 2D flow (see bottom left panel). In the same figure we also show the decomposition in homochiral  and heterochiral components   for a simulation forced in the IR, where most of the scales evolve in the direct energy cascade regime (bottom right panel). In the latter case, the bi-directional character of the direct energy cascade is clearly visible, where inverse homochiral and direct heterochiral transfers coexist in the same scaling range, $ k > k_{in}$,  following the definition (\ref{def6}) of Sec. (\ref{sec:definitions}). In the same figure (top row) we also show the alternative decomposition of the total flux in slow and fast components (\ref{eq:2d3dflux}) from where it is evident that the 2D plane is not isolated and receives energy from the 3D modes.  The above results show that if we are not in the limiting case where either $\Omega \to \infty$ or $H \to \infty$ the inverse energy transfer in rotating turbulence is more complicated than the simple uncoupled 2D3C-3D pictures would suggest, with both purely 2D (in-plane) and purely 3D (homo-heterochiral) channels that might compete or cooperate to send energy to large scales. Moreover, we finally remark that even in the presence of  a simple 2D3C flow, if the injection mechanism is fully helical, the third out-of-plane component becomes coupled to the in-plane 2D velocity field and does not behave any more as a passive scalar \cite{linkmann2018nonuniversal} (see also the discussion at the end of section \ref{sec:activescalarcascades}).

\subsubsection{Finite horizontal size}
\label{sec:finitehorizontalsize}
As in thin layers,  it is important to discuss also  the case when  the horizontal size, $L$  of the domain is finite and no drag force is present so that a condensate forms (in the split cascade regime). The difference with layers of finite height is that there is an additional path to saturate the inverse cascade. In square boxes the condensate takes the form of a vortex dipole. As the condensate increases in amplitude and is constrained not to exceed  the size of the box, its eddy-turn-over-time also increases. Thus, at sufficiently large amplitudes, the rotation rate of the counter-rotating vortex will locally cancel the effects of $\Omega$, the flow will no longer be  restricted to 2D dynamics and it will be  able to develop a forward energy cascade. This occurs when the amplitude of the condensate velocity $U$ is such that $U=\Omega L$. The system thus reaches a zero flux state that has a finite inverse flux in the slow manifold dynamics and a forward flux in the 3D fast manifold dynamics  as described by Definition (\ref{def7} and \ref{def15}) in Sec. (\ref{sec:definitions}). This flux loop mechanism was first proposed in \cite{bartello1995geostrophic} and later  demonstrated in \cite{alexakis2015rotatingTG, seshasayanan2018condensates}. It is worth noticing  that although there is zero flux at the large scales the system is far from being in thermal equilibrium. Similar flux loop dynamics have been observed in 2D stratified turbulence \cite{boffetta2011flux}, in compressible 2D turbulence \cite{falkovich2017vortices} and in numerical simulations with the NSE restricted to evolve on a subset of 2D3C modes \cite{biferale2017two}. It is also worth pointing out that the transition from a no-condensate state to a condensate state as the rotation is increased was found to be discontinuous when the forcing was acting on the fast manifold \cite{alexakis2015rotatingTG} (with $U$ displaying a hysteresis curve \cite{yokoyama2017hysteretic}) while it was found to be continuous when the forcing was acting on the slow manifold \cite{seshasayanan2018condensates}. 

\subsubsection{Summary}
In this section we have summarized the  phenomenology of turbulence under rotation, by analysing the energy transfer properties at changing the Rossby number, the Reynolds number and the height of the domain, $H$. We have discussed the tendency of the system to decouple the 2D3C plane from the resonant 3D wave dynamics in the limit $\Omega \to \infty$ at fixed $H$, leading to a transition from a forward energy cascade to a split energy cascade as of our DEf. (\ref{def4}) in Sec. (\ref{sec:definitions}). Concerning the {\it order} of the transition, it is an open problem to understand  which of the routes depicted in Fig. (\ref{fig:classification}) is realized. For finite vertical domains, the formation of a condensate with a rich and still not fully understood dynamical behaviour has been discussed. We have also analysed the role of quasi-resonant waves in breaking  the exact decoupling and the existence of a region in the $(k_\perp,k_\|)$ plane where wave-theory cannot be applied and we must expect a fully 3D turbulent energy transfer if the Reynolds number
is large enough for any finite Rossby number  and $H$. Such intermediate asymptotics necessarily introduces interactions between the 2D and 3D dynamics, leading  to a series of open questions. First, it is not known if the wave-turbulence prediction of the  inverse cascade absence for $H \to \infty$, and for any fixed rotation, is correct or not. Second, it is not known whether purely 3D homochiral interactions are efficient enough to maintain some inverse cascade independently of the role of the 2D3C triads. Third, the dependency  on the forcing mechanism has not been clarified yet, and the different asymptotic in the presence of only 3D, only 2D3C forcing and of helical forcing are not fully under control yet.   

 \subsection{Stratification} \label{sec:stratification}        

\begin{figure*}[h!]                                                                          
\centering                                                                                   
\includegraphics*[width=1.0\textwidth,angle=0]{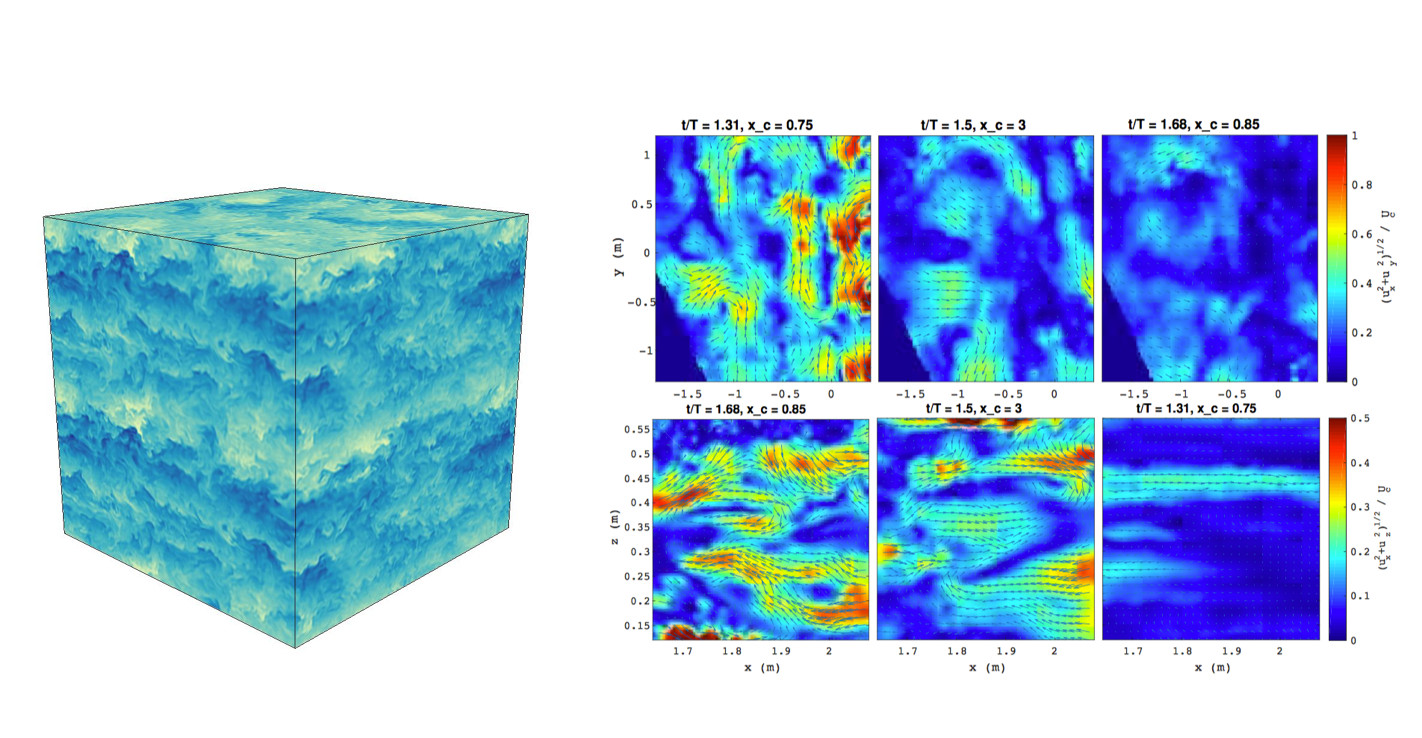}                         
\caption{Left: 3D rendering of a turbulent simulations with stable stratification.           
Color represent the fluctuating density field $\phi$  (from white to blue) from a direct     
numerical simulation of (\ref{eq:NSS}) at a$512^3$ resolution with $Fr=0.26$, $Pe=1$. The    
presence of horizontal layers is clearly visible,  confined in the vertical direction,       
especially at large scales  \cite{sozza2018inertial}. Courtesy of A. Sozza. Right:           
Instantaneous horizontal  fields (top row)  and vertical fields (bottom) for three different 
times  in an experimental realization of a stratified flow with $Fr = 0.1$ and  $Re= 450$ in 
the Coriolis platform \cite{campagne2016archive}. Background colors indicate the norm of the 
2D components normalized by the mean velocity of the oscillating energy injecting carriage.  
Different times correspond to three different instances after the passage of the carriage.   
The horizontal field exhibits strong persistence vortices while vertical fields show strongly %
horizontally elongated structures, which is a robust feature of turbulence in stratified     
fluids.                                                                                      
{Reprinted under permission of  P. Augier and                                                
A. Campagne \cite{campagne2016archive}.}}                                                    
\label{fig:Strt}                                                                             
\end{figure*}

 For non uniform density fluids in the presence of gravity additional forces need to be added in  the NSE to account for  the feedback of the advected density field on the velocity. For stable stratification where the mean density decreases with height, gravitational forces tend to suppress vertical motions and form layers of almost constant density as the one shown in Fig.  (\ref{fig:Strt}). For unstable stratifications, gravity acts as an energy source transforming the stored potential energy into kinetic energy. Both  cases have attracted the interest of turbulence research due to the applicability of stratified turbulence (both stable and unstable) to atmospheric dynamics, planetary and stellar interiors \cite{davidson2015turbulence}. Recently a large number of new results have been revealed due to extensive numerical simulations \cite{lilly1983stratified, herring1989numerical, brethouwer2007scaling, chung2012direct, rorai2015stably, rorai2013helicity, waite2006stratified, rorai2014turbulence, Sozza2015dimensional,feraco2018vertical} and experiments \cite{campagne2016archive, billant1998experimental, augier2014experimental, smyth2001efficiency, billant2000experimental, praud2005decaying}. Here we
 review some of these results and describe their implications for the turbulent cascades. More extensive reviews of stratified flows can be found in \cite{thorpe1973turbulence, hopfinger1987turbulence, riley2000fluid, peltier2003mixing, lohse2010small, chilla2012new, ahlers2009heat, boffetta2017incompressible}. 

\subsubsection{Stable stratification} 
Concerning turbulent flows in the presence of stable stratification we consider the case of  a layer with horizontal size $L$, height $H$ and  gravity ${\bf g}$ in the $z$-direction. Indicating with $\sigma$ the mean stratification, the density is given by $\rho = \rho_0 + \sigma z + \delta \rho({\bf x},t)$. The flow is forced by a mechanical forcing function ${\bf f}$. We also assume a general heat source $S$ that  locally modifies the density.  
\begin{figure*}[h!]                                            
\centering                                                     
\includegraphics[width=0.40\textwidth]{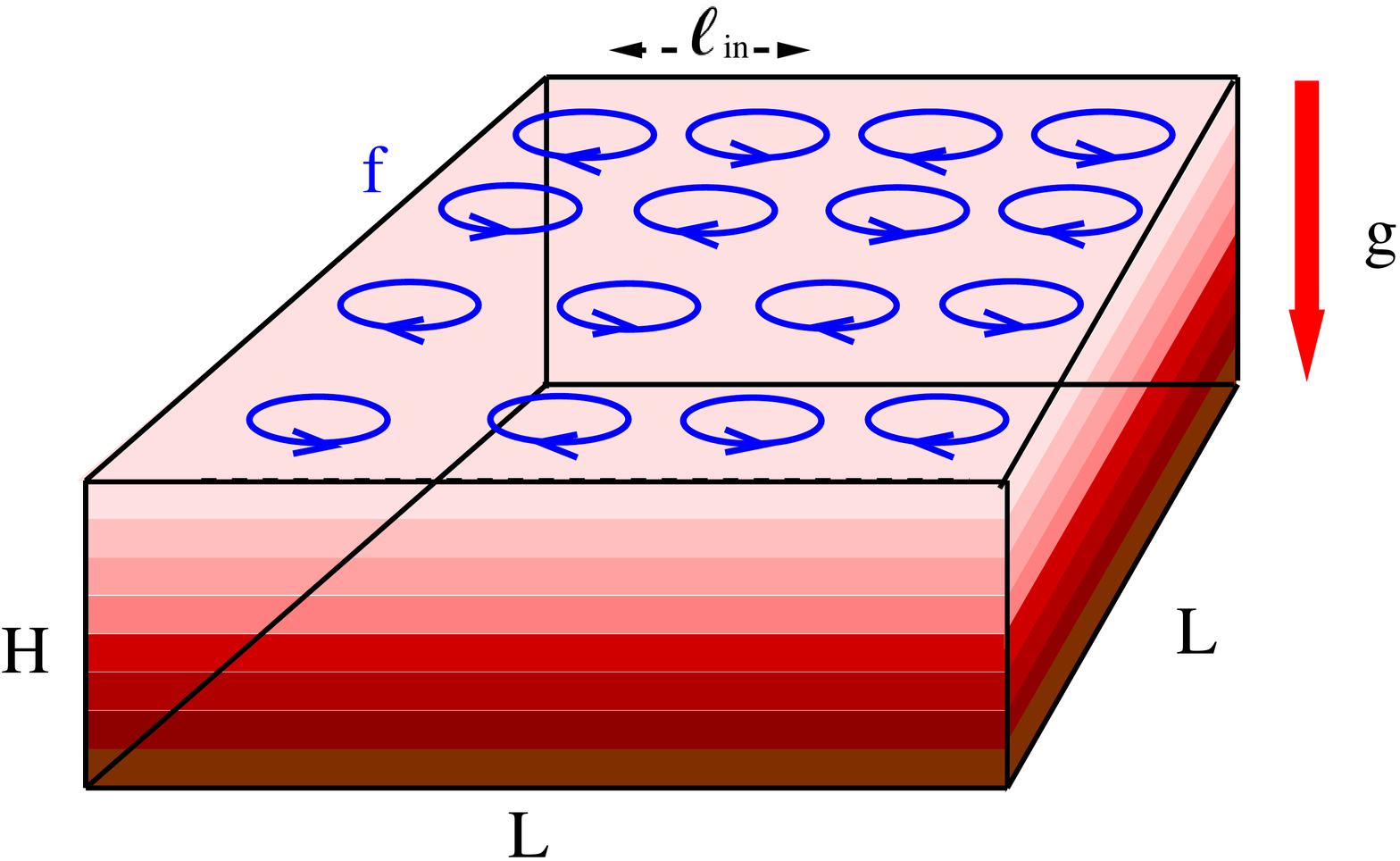}        
\caption{Sketch of a flow set-up in a stratified  layer,       
where density variation is depicted with different intensity   
of red colour. Circles represent forcing effects.}             
\label{fig:Sbox}                                               
\end{figure*}                                                  
To simplify our discussion we limit ourselves to the Boussinesq equations
obtained in the limit  $|\rho-\rho_0| \ll \rho_0$:
\begin{align}
\partial_t {\bf u} + {\bf u \cdot \bnabla u}       =& -N {\bf e}_z \phi          -\bnabla P      + \nu    \Delta {\bf u} - \alpha {\bf u}+ {\bf f}, \\
\partial_t \phi    + {\bf u \cdot \bnabla  } \phi  =& +N {\bf e}_z \cdot {\bf u}                + \kappa \Delta \phi                    + S.
\label{eq:NSS}
\end{align}
Here $\kappa$ is the diffusivity, $N$ is  the Brunt-Vais\"al\"a frequency given by $N=\sqrt{g \sigma /\rho_0}$ and $\phi(\bx,t) =N \delta \rho(\bx,t)$.
Two new non-dimensional control parameters must be considered. First, the 
Froude number that measures the strength of the stratification with respect to inertia:
\be Fr = \frac{u_f}{N\lin}. 
\label{eq:Froude}
\ee
Second,  the relative strength of advection and diffusive terms is  measured by the P\'eclet number:
\be Pe=  \frac{u_f \lin}{ \kappa}, \ee
where, as usual, we have indicated the rms velocity at the injection scale with
$u_f=(\ein\lin)^{1/3}$.  A sketch of the setup considered is shown in Fig. (\ref{fig:Sbox}).
The total energy, $\cE(t)$  can be split into the kinetic energy, $\cE_K(t)= \frac{1}{2}\langle {\bf u}^2 \rangle$ and
the potential energy, $\cE_P(t)=\frac{1}{2}\langle \phi^2 \rangle$.
In the absence of forcing and dissipation only $\cE$ is conserved
while kinetic and potential energy  can be exchanged through the gravitational force.  The energy balance equations thus read 
\begin{eqnarray}
&\frac{d}{dt} \cE_K(t) = \ein(t) -\enu(t) -\emu(t) - \iepsilon_{_{T}}(t) \\
&\frac{d}{dt} \cE_P(t) = \iepsilon_{_{S}}(t) + \iepsilon_{_{T}}(t) - \iepsilon_{\kappa}(t)
\label{eq:stratified}
\end{eqnarray}
where $\iepsilon_{_{S}}=\langle \phi S\rangle $ is the  potential 
energy injection rate,
$\iepsilon_{\kappa} = \kappa \langle |\bnabla \phi|^2\rangle $ is the dissipation rate due to  diffusion and  
$ \iepsilon_{_{T}} = N \langle u_z \phi \rangle $ is the conversion rate of 
kinetic energy  to potential energy.
The latter can be either positive or negative,
its time averaged value however has to be positive if $S=0$ and ${\bf f} \ne 0$,
while it is negative if $S\ne 0$ and ${\bf f}=0$.   
Helicity is not conserved in this system. It can be generated or removed by gravity and the helicity balance reads
\be
\partial_t \cH(t) = -\hnu(t) -  \halpha(t) + \hin(t) + h_{_{N}}(t)
\ee
where $h_{_{N}}(t) = N \langle w_z \phi\rangle $ is the helicity injection rate by
the gravitational forces that can be of either signs.  Potential vorticity is an additional inviscid invariant
conserved along each  particle trajectory:
\be
V_p = -N w_z + {\bf w} \cdot \bnabla \phi 
\ee
Its evolution in the unforced inviscid and non-diffusive limit satisfies
\be
\partial_t V_p + {\bf u} \cdot \bnabla  V_p =0
\ee
thus any moment $m$ of the potential vorticity $\langle V_p^m \rangle$ is conserved by the flow.
The first nontrivial moment is the second (the first being identically zero) and reads
\be
\mathcal{V}_2 =\langle V_p^2\rangle = 
N^2 \langle w_z^2 \rangle - 2 N  \langle w_z ({\bf w} \cdot \bnabla \phi)\rangle + \langle ({\bf w} \cdot \bnabla \phi)^2 \rangle,
\ee
which has terms of third and quartic order too. This implies that $\mathcal{V}_2$ is not equal to the sum of the squared potential vorticity of each Fourier mode as it involves terms that couple different wavenumbers.   This limits the interpretation of the transfer of this quantity in terms of a cascade in Fourier space \cite{aluie2011joint}. When $N$ is large, the first term dominates and the squared potential vorticity can be approximated as  $\mathcal{V}_2 \approx N^2\langle w_z^2 \rangle$ which is quadratic and equal to the vertical enstrophy. We  further discuss the implications of this invariant in the next section.  In Fourier space we can write the evolution of the  kinetic  energy,  
$\cE_K^{<k} =\frac{1}{2}\langle |{\bf u}^{<k}|^2\rangle $, and  potential energy, 
$\cE_P^{<k} =\frac{1}{2}\langle |{\phi }^{<k}|^2\rangle $, in a sphere of radius $k$   as
\begin{align*}
\partial_t \cE_K^{<k} =& - \Pi_K(k) - \Pi_{KP}(k)  -\alpha \langle |{\bf u}^{<k}|^2 \rangle - \nu \langle |{\bf w}^{<k}|^2 \rangle 
+ \langle {\bf u}^{<k}\cdot { \bf f} \rangle,\\
\partial_t \cE_P^{<k} =& - \Pi_P(k) + \Pi_{KP}(k) - \kappa \langle |\bnabla {\phi}^{<k}|^2 \rangle + \langle {\phi}^{<k} S \rangle,
\end{align*}
where $\Pi_K(k)=\langle {\bf u}^{<k} \cdot ({\bf u}\cdot \bnabla {\bf u})  \rangle$ is the inertial  hydrodynamic (un-stratified) 
energy flux, $\Pi_P(k)=\langle {\phi}^{<k} \cdot ({\bf u}\cdot \bnabla {\phi})  \rangle$ is the potential energy flux while
$\Pi_{KP}(k)=N \langle  u_z^{<k} \phi^{<k} \rangle$ is the rate kinetic energy
is transferred to potential energy within the examined sphere in Fourier space. It is important to stress  that the gravitational force is linear and homogeneous, it can result 
in an energy exchange among kinetic and potential  terms  but does not produce any transfer across scales. Its presence however
modifies the properties of the flow so that the flux of energy through the non-linear terms can be altered.
This leads to an anisotropic version of the  von K{\'a}rm{\'a}n-Howarth-Monin relation \cite{augier2012kolmogorov} that reads 
\be 
\bnabla_{\bf r} \cdot \la [ |\delta_{\bf r} \bu|^2 + \delta_{\bf r} \phi^2 ] \delta_{\bf r} \bu  \ra = -4\ein. \label{eq:HKMstr}
\ee \\
%
 \noindent It is worth examining the limits of weak and strong stratification to elucidate the processes involved.\\ 
\paragraph{Weak stratification.}
For very weak stratification, $Fr\gg 1$, the gravitational force does not modify the flow and the density fluctuations $\phi$ are advected as a passive scalar. The energy transfer  is thus dominated  by the hydrodynamic part of the flux $\Pi_K$ and  the kinetic energy cascades either forward or inversely depending on the dimensionality of the system. The potential energy  cascades forward as for any passive scalar advected by an  incompressible flows, $\Pi_P = \iepsilon_\kappa$,  with a sub-dominant energy exchange with the kinetic component, $\Pi_K \gg \Pi_{KP}$ (see Sec. \ref{sec:passive} for the scaling of a passive scalar advected by a turbulent  flow). The flow statistics is close to HIT  and (\ref{eq:HKMstr}) leads in 3D to the relations $\delta_r u \propto   \enu^{1/3} r^{1/3}$ and $ \delta_r \phi \propto \enu^{-1/6} \iepsilon_\kappa^{1/2} r^{1/3}$  that for the kinetic and potential energy spectra read:
\be 
E_K(k) \propto  \enu^{2/3} k^{-5/3},\qquad  E_P(k) \propto \iepsilon_\kappa \enu^{-1/3} k^{-5/3}.    
\label{eq:kolmostra}
\ee
\paragraph{Strong  stratification and strong turbulence.}
For strong stratification another balance has been suggested, assuming that  the 
total energy cascade is driven by the potential energy flux $\Pi_P$ and
the kinetic energy cascade is balanced by the potential energy transfer,
  $ \Pi_K \sim \Pi_{KP}$  leading to  $\delta_r u \propto \iepsilon_\kappa^{1/5} N^{2/5} r^{3/5}$ and $\delta_r \phi \propto \iepsilon_\kappa^{2/5} N^{-1/5}r^{1/5}$ 
that is referred as the Bolgiano-Obukhov scaling \cite{bolgiano1959turbulent,obukhov1959effect}, with the corresponding   spectra:                                   
\be 
\label{eq:spestrat}
E_K(k) \propto  \iepsilon_\kappa^{2/5} N^{4/5} k^{-11/5},\qquad  E_P(k) \propto \iepsilon_\kappa^{4/5} N^{-2/5} k^{-7/5}.  
\ee
The two predictions meet at the scale that is given by $\ell_B =\enu^{5/4}\iepsilon_\kappa^{-3/4} N^{-3/2} $.
The  Bolgiano-Obukhov scaling is expected to dominate at scales $\ell \gg \ell_B$ while at smaller scales the Kolmogorov spectrum should be recovered.
Evidence for this scaling  has been recently observed in numerical simulations \cite{Rosenberg2015evidence,kumar2014energy}.

\paragraph{Strong stratification and Wave turbulence.}
 The above arguments are based on the dominance of potential over kinetic energy flux or vice versa and did not take into account neither anisotropy nor the presence of waves.  
Stratified flows, however, sustain internal gravity waves that have the following dispersion relation: 
\be
\label{eq:wavestra}
\omega^2 = (k_\perp^2/k^2) N^2 
\ee
where $k_\perp^2 =k_x^2 +k_y^2$. 
The frequency $\omega$ becomes zero for modes with  zero vertical velocity $\tilde u_z(k)=0$, that define  the slow manifold for stratified turbulence \cite{bartello1995geostrophic,herbert2014restrictede}.  A particular role  is played by the subset with $k_\perp=0$, that due to incompressibility satisfy  $\tilde u_z(k)=0$. For $Fr\ll 1$  the internal gravity waves are expected to play a dominant role. Wave turbulence theory  leads to the estimate: $$ E(k) \sim \ein^{1/2} N^{1/2} k^{-2}$$ if isotropy is assumed \cite{garrett1972space}, see also (\ref{eq:wavebalance}). If we take into account anisotropy, the spectrum becomes: 
\be 
E(k_\perp,k_\|)\sim \ein^{1/2} N^{1/2} k_\perp^{-1/2} k_\|^{-5/2}
\label{eq:wavestra2}
\ee
as one can verify  using  (\ref{eq:wavebalance2}) where $k_\|\gg k_\perp$ has also been assumed.  More detailed analyses have led to different predictions for the gravity wave turbulence spectra \cite{pelinovsky1977weak,caillol2000kinetic,lvov2001hamiltonian}. The wave  regime ends at scales small enough  such that turbulent eddies can efficiently overturn and 3D isotropic turbulence sets in. The scale where  this transition takes place is known as the Ozmidov scale, $\ell_o$,  determined  by the  balance between  the inverse of the eddy turnover time $\delta_r u/r \sim \enu^{1/3} r^{-2/3}$ and the  wave frequency (\ref{eq:wavestra}), $N$:  $\ell_o \propto \enu^{1/2}N^{-3/2}$ \cite{ozmidov1965turbulent}. We note that the ratio between the Ozmidov and Bolgiano scales is given by $\ell_o/\ell_B \propto (\iepsilon_\kappa/\enu)^{3/4}$. For weakly stratified systems, $\iepsilon_\kappa\ll \enu$, and the two scales can be considerably apart.  Taking into account the anisotropy of the dispersion relation we arrive at  an anisotropic relation for the Ozmidov scale, or at a critical balance relation that reads 
$ k_\perp \propto \enu k_\|^{3}/N^3 $ \cite{nazarenko2011critical}. This last scale marks where wave turbulence theory ceases to be valid. Let us remark that wave turbulence (\ref{eq:wavestra2}) and the Bolgiano scaling (\ref{eq:spestrat}) describe different mechanisms for the anisotropic energy transfer in the  strong stratification limit  and it might well be that there exists a wide range of scales where both transfers coexist. Even the definition of $\ell_o$ would change if the scaling (\ref{eq:spestrat}) is used to estimate the eddy-turn-over-time. The different theories are not necessarily compatible with each other and they may be realized in different limits or with different forcing mechanisms \cite{dewan1997saturated}.

From the empirical side we know that in the presence of stratification,  vertical motions are suppressed leading to a strongly anisotropic flow forming  shear layers with large velocity gradients in the vertical direction and sharp jumps of density variations across them as shown by  Fig. (\ref{fig:Sbox}). Although there is early evidence of a weak inverse energy transfer in stratified turbulence \cite{herring1989numerical}, most numerical simulations have found a direct cascade of energy \cite{lindborg2006energy}. More recent calculations have clearly demonstrated that stratification aids the forward cascade  due to  the formation of layers \cite{Sozza2015dimensional,rorai2015stably}. The flow within these layers acts as a quasi 2D flow but interacts with other layers and transfers energy to small scales by the formation of sharp gradients along the $z$ direction. Furthermore, numerical simulations indicate the presence of energy spectra with a very steep power-law behaviour along the direction of gravity and a Kolmogorov like behaviour in the horizontal direction, i.e. with a scaling  that does not coincide with any of the previous arguments
\cite{lindborg2006energy,kitamura2006kh,brethouwer2007scaling,billant2001self}. Based on these results and using similarity arguments the following energy spectra were proposed in \cite{lindborg2006energy}
\be \int E_K(k_\|,k_\perp) dk _\perp \propto N^2 k_\|^{-3} \quad \mathrm{and} \quad \int E_K(k_\|,k_\perp) dk_\| \propto \ein^{2/3} k_\perp^{-5/3}.\ee

\begin{figure*}[htbp]                                                                     
\centering                                                                                
\includegraphics*[width=0.45\textwidth]{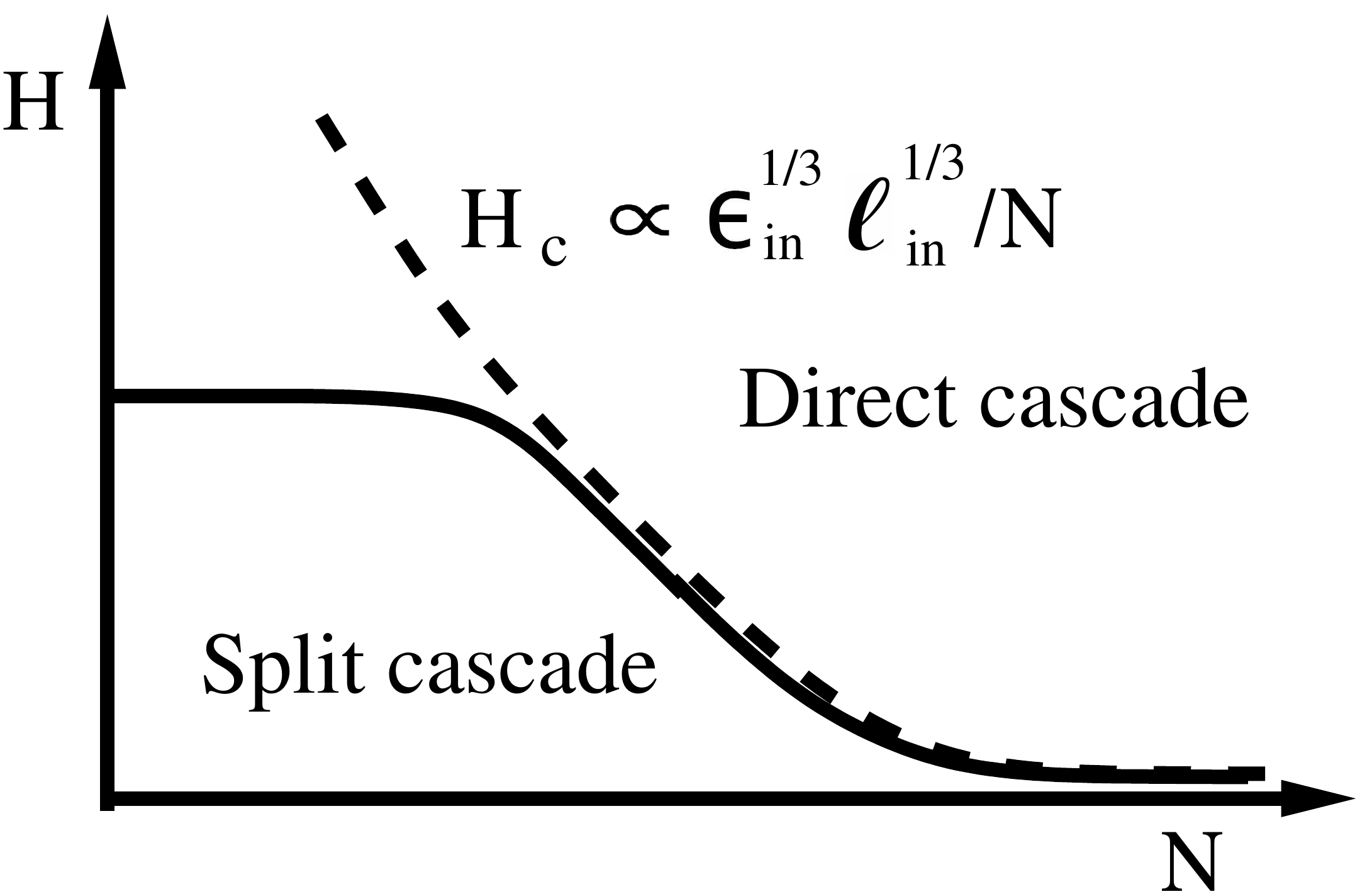}       \qquad              
\caption{ Phase diagram for stratified flows. The line                                    
separates the regions in the $H-N$ space for which a split energy  cascade is present or  
not, according to the prediction  (\ref{eq:strcrt}). }                                    
\label{fig:StrPhasespace}                                                                 
\end{figure*}                                                                             
\subsubsection{Stratified flows at finite thickness, $H$}
 It is worth looking at what happens for stratified flows in a  thin box, where the vertical height $H$ is much smaller than the horizontal domain, forcing the   flow to have a  quasi 2D behaviour. As discussed earlier, stratification aids the forward cascade by opening a new channel for the energy transfer,  converting kinetic energy to potential energy, which always cascades forward. Then, one  might ask in the limit of a thin box and strongly stratified flow which process dominates and whether the system displays an inverse cascade. This question was addressed in \cite{Sozza2015dimensional}.
It was shown that for weak stratification the transition to an inverse cascade remained unaltered and the critical height
$H_c$  where  a transition from a forward to a split energy cascade happens is close to the injection scale. 
However as stratification increases it suppresses the inverse cascade decreasing the critical height $H_c$.
The critical height dependence on the stratification is then given by: 
 \be 
 H_c \propto \lin \quad \mathrm{for} \quad Fr \gtrsim 1 \quad \mathrm{and} \quad
 H_c \propto  \ein^{1/3} \lin^{1/3}/N  \quad \mathrm{for} \quad Fr \ll 1       \label{eq:strcrt} ,
 \ee 
where, for  $H\ll\lin$ and $Fr\ll1$ the transition (\ref{eq:strcrt}) is  obtained by equating the gravity wave 
frequency $H/\lin N$ with the local eddy-turn-over-time $\ein^{1/3}\lin^{-2/3}$.
This is a sort of anisotropic expression of the Ozmidov scale in the limit of very thin layers where $ k \sim k_z$. This transition can be understood by noting that stratified turbulence leads to the formation of layers of width $\ell_z\simeq  \ein^{1/3} \lin^{1/3}/N $ \cite{billant2000experimental, lindborg2006energy}. The transition occurs when the domain is thick enough to allow the formation of these layers.
The resulting  phase diagram for stratified flows in a layer of finite thickness is  shown in figure  \ref{fig:StrPhasespace}. 
\subsubsection{2D vertically stratified flows}
\label{sec:2Dstratification}
\begin{figure*}[htbp]                                                                     
\centering                                                                                
\includegraphics*[width=0.45\textwidth]{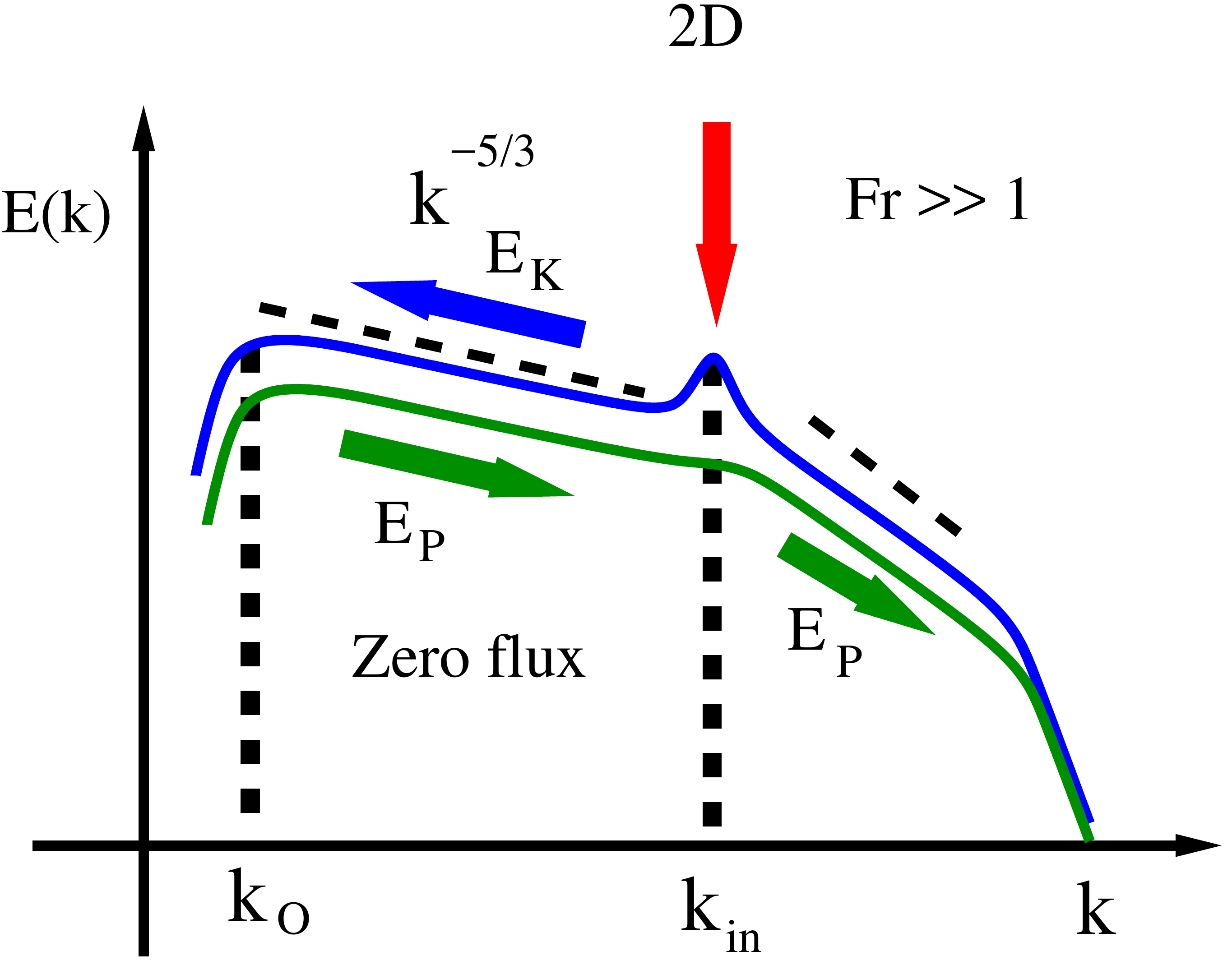}       \qquad               
\includegraphics*[width=0.45\textwidth]{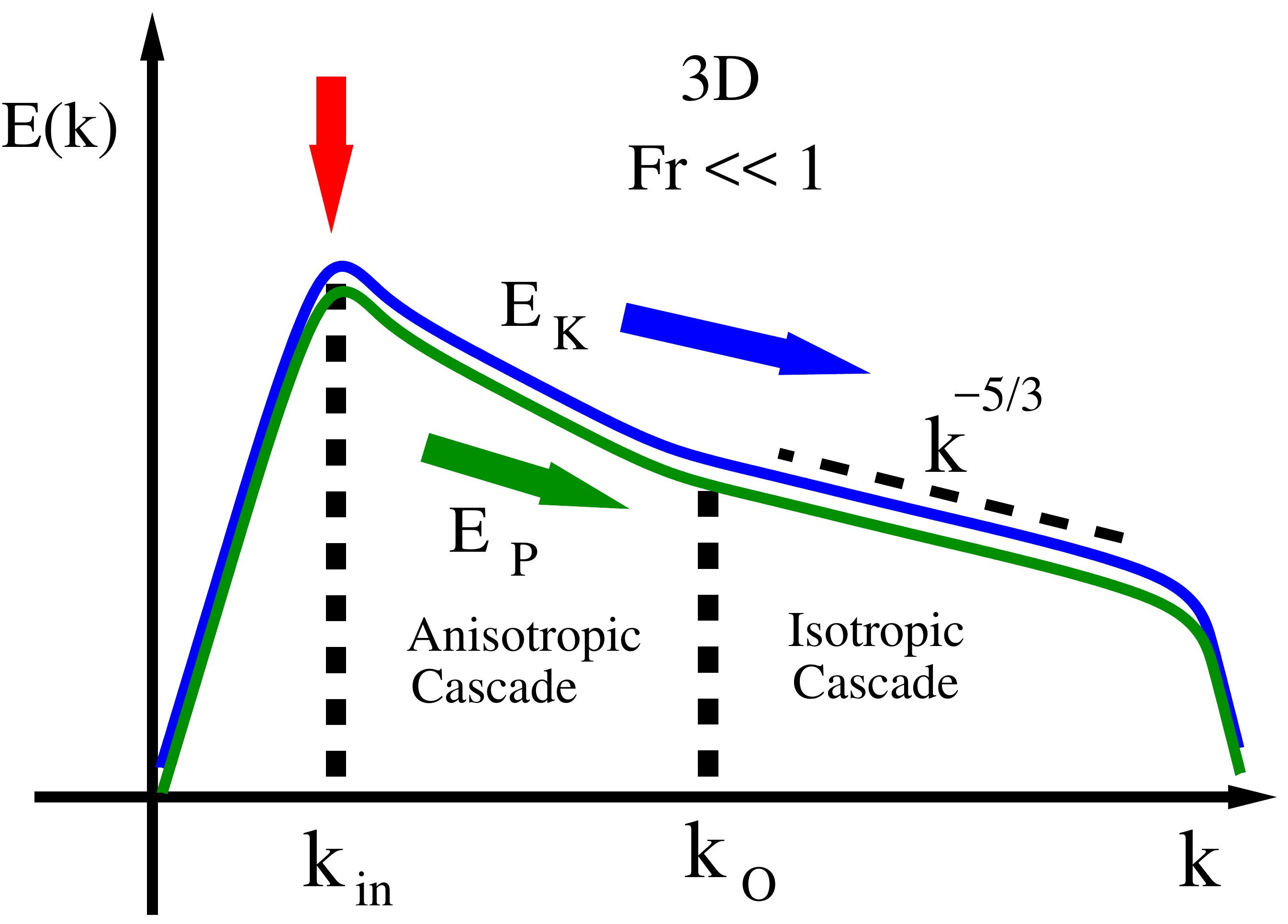}                            
\caption{ Energy spectra for stably stratified flows in 2D (left panel) and 3D            
(right panel).  }                                                                         
\label{fig:Strat2D3D}                                                                     
\end{figure*}

An interesting different  transition happens in 2D stratified flows (e.g. a soap flow in the $(x,z)$-plane). 
For weak stratification and for flows forced at small scales, the  gravitational
force can be initially neglected and a regular 2D inverse cascade develops.
However as structures of larger and larger scales are formed, the eddy turnover time decreases
until the eddies reach a size comparable with  the Ozmidov scale $\ell_o$ where it becomes  comparable 
to the wave frequency. At this scale gravity effects are important and can move energy back to the
small scales through the potential energy flux $\Pi_P$. The system reaches a steady
state with zero total inverse flux which is the result of a perfect balance between a  finite  negative kinetic energy flux 
and a finite  positive potential energy flux. 
This mechanism was demonstrated in \cite{boffetta2011flux}. 
 Figure \ref{fig:Strat2D3D} summarizes the two scenarios for the spectral properties  in 3D and 2D.

\subsubsection{Unstable stratification}
Unstable stratification refers to the case that the background density gradient is inverted so that denser fluid lies on top. 
This is a very common situation that is met in Rayleigh-B{\'e}nard convection and for the Rayleigh-Taylor instability. Turbulent convection drives most atmospheric and planetary flows,
and has been the subject of numerous investigations that we can not possibly cover in this review. We limit ourselves to issues related to cascades in a periodic box with a mean background stratification, see Sec. (\ref{sec:bounded}) for a short discussion about the case with solid boundaries.
We refer the reader to the recent works \cite{lohse2010small, chilla2012new, ahlers2009heat, verma2017phenomenology, boffetta2017incompressible} for extensive reviews of convection and unstably stratified flows. 
In unstably stratified flows the inversion of the density gradient results in a change of sign in 
the buoyancy term in (\ref{eq:NSS}):
\begin{align}
\partial_t {\bf u} + {\bf u \cdot \bnabla u}       =& N {\bf e}_z \phi          -\bnabla P      + \nu    \Delta {\bf u} - \alpha {\bf u} \\
\partial_t \phi    + {\bf u \cdot \bnabla  } \phi  =& N {\bf e}_z \cdot {\bf u}                + \kappa \Delta \phi           
\label{eq:NSUS}
\end{align}
where we have also set $\bbf=0$ and $S=0$ because energy is now injected in the system by the buoyancy term. The potential and kinetic energy balance now reads,
\begin{eqnarray}
&\frac{d}{dt} \cE_K(t) =  \iepsilon_{_{T}}(t) -\enu(t) -\emu(t)   \\
&\frac{d}{dt} \cE_P(t) =  \iepsilon_{_{T}}(t) - \iepsilon_{\kappa}(t).
\label{eq:ustratified}
\end{eqnarray}  
The total energy balance thus becomes
\begin{eqnarray}
 \frac{d}{dt} \cE(t) = 2\iepsilon_{_{T}}(t) -\enu(t) -\emu(t) - \iepsilon_{\kappa}(t).
\label{eq:unstratified}
\end{eqnarray} 
The term $2\iepsilon_{_{T}}=N \la u_z \phi \ra$ is now the only source of energy in the system and acts at all scales. We note that  in the presence of periodic boundary conditions in the vertical direction,  unphysical  exponentially growing configurations with   $u_x=u_y=0$ and $u_z=\phi=e^{\gamma t} \sin(k_x x+\theta_x)\sin(k_y y+\theta_y)$ exist
that are exact solutions of (\ref{eq:NSUS}) \cite{calzavarini2006exponentially}. These solutions tend to be unstable and a steady state is nonetheless reached.  \\
\noindent As for the stably stratified case both  Bolgiano-Obukhov (\ref{eq:spestrat}) and Kolmogorov scaling are possible (\ref{eq:kolmostra}).
Most numerical investigations of 3D convection point to the Kolmogorov prediction \cite{calzavarini2002evidences,mishra2010energy,kumar2014energy} (see \cite{lohse2010small}  for a discussion).
In contrast,  in 2D where kinetic energy cascades inversely and potential energy forward, the Bolgiano-Obukhov scaling is observed \cite{celani2001thermal,mazzino2017two}.
A transition between the two scaling has been observed in unstably stratified thin layers \cite{boffetta2012bolgiano}.

\subsubsection{Summary}
In this section we have summarized the  phenomenology of stratified  turbulence for stable and unstable stratifications.
We have discussed the cascade in the two  kinetic and potential energy channels.
Different predictions exist for the energy spectra depending if  weak or strong stratification is considered in the stable regime. In particular, for strong turbulence in the strongly stratified regime, the Bolgiano scaling  (\ref{eq:spestrat}) is predicted, while in the weak wave turbulence regime, different predictions can be made depending whether isotropic or anisotropic spectra are assumed. Empirical data show mixed results and it is still not clear whether this is due to numerical and experimental limitations, to deficiencies in the above theories, to the difficulties to achieve asymptotic values of small Froude and large Reynolds numbers or to a combination of all the above. For 2D turbulence it was shown that a flux-less cascade (see Def. (\ref{def7}) in Sec. \ref{sec:definitions}) can exist where kinetic energy cascades inversely in balance with potential energy that cascades forward. Transitions to split cascades can also exist in stratified turbulence with the critical height being a decreasing function of stratification.

\subsection{ Rotating and stratified flows }  
\label{sec:RotationStratification}            

\begin{figure*}[h!]                                                                         %
\centering                                                                                  %
\includegraphics[height=0.30\textwidth]{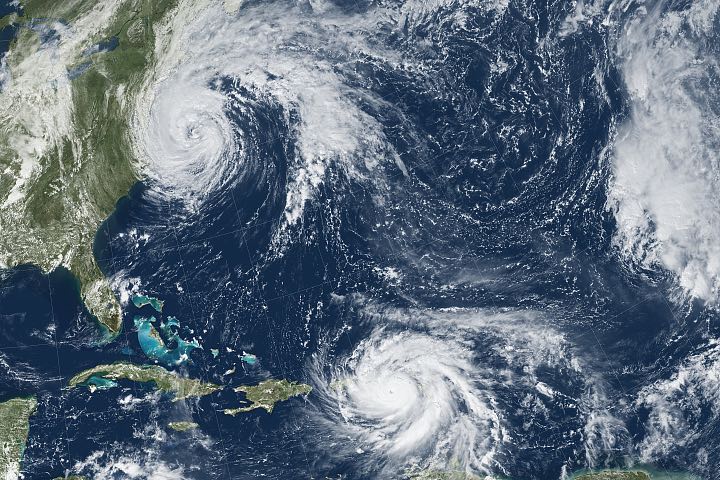} \quad                         %
\includegraphics[height=0.30\textwidth]{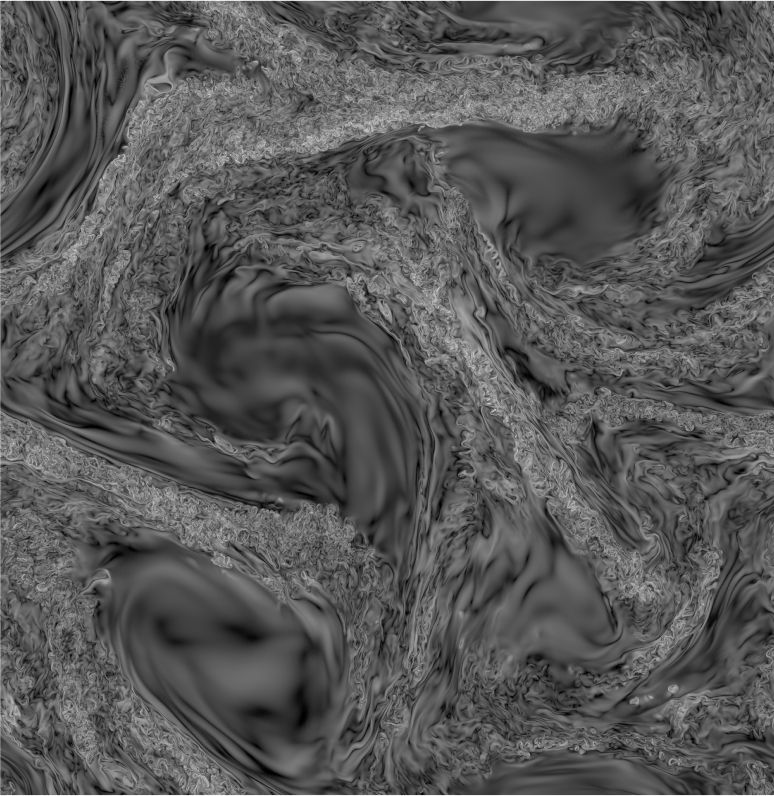}                              %
\caption{Left: Satellite image of the atmosphere                                            %
(Courtesy of the NASA Earth Observatory/NOAA)                                               %
It is an example of rotating and stratified flow where both small turbulent scales and      %
large coherent structures can be seen. The figure on the right is a visualization from      %
a $4096^3$ grid points simulation of a rotating and stratified turbulent flow where         %
a similarly large-scale  and small-scale structures coexist.  The figure is constructed     %
from the data in \cite{Rosenberg2015evidence}. Courtesy of D. Rosenberg. }                  %
\label{fig:AtmOce}                                                                          %
\end{figure*}
\begin{figure*}[h!]                                                                         %
\centering                                                                                  %
\includegraphics[width=0.70\textwidth]{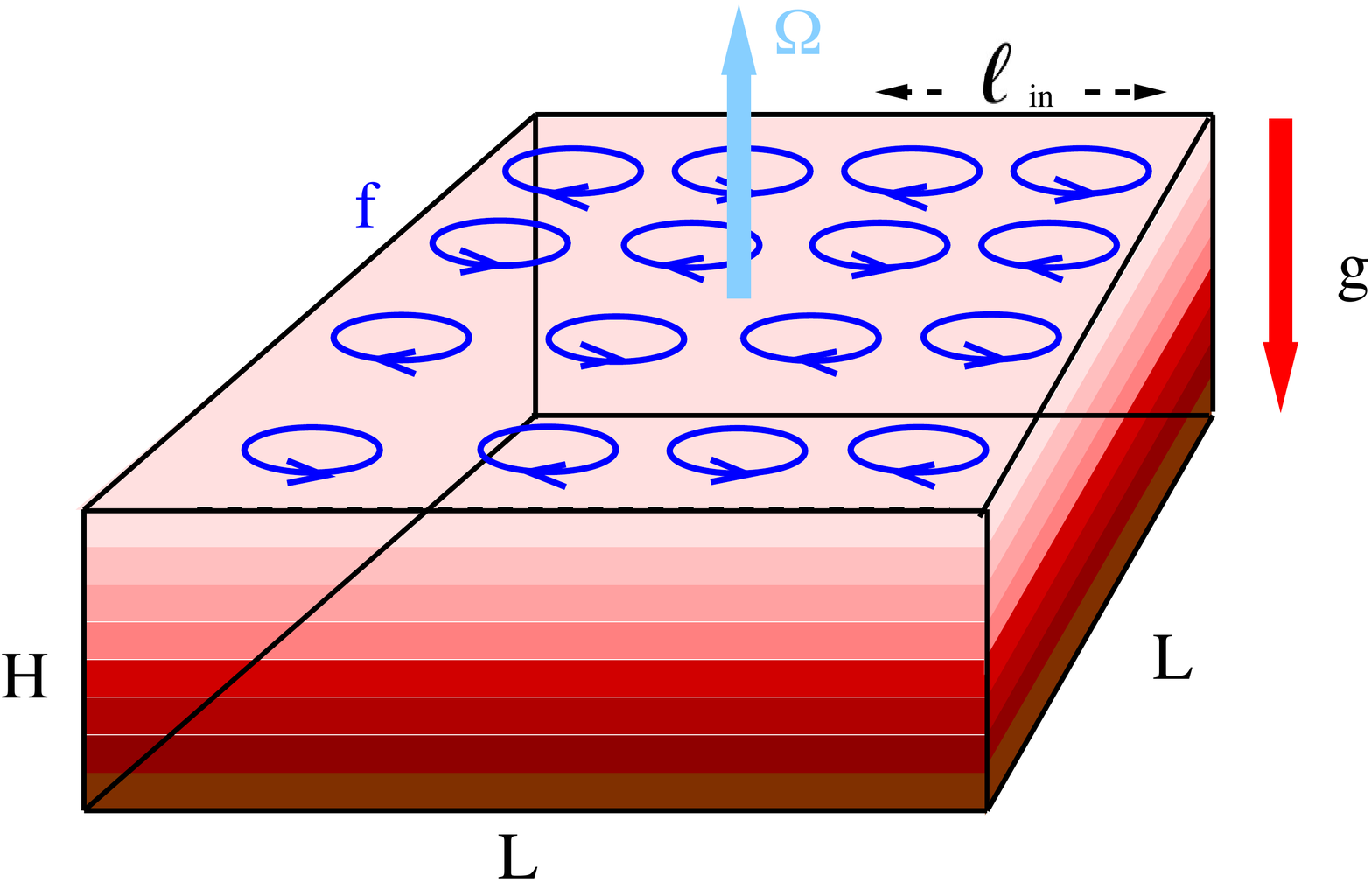}                                  %
\caption{Domain of the flow in a rotating and stratified  layer. }                          %
\label{fig:RSbox}                                                                           %
\end{figure*}

                                                                               %
Rotating and stratified flows  represent the simplest model that  describe to some accuracy the motion of planetary atmospheres and have attracted the  attention of
many different disciplines (see Fig. \ref{fig:AtmOce}) \cite{maxworthy1975experiments,gibson1991laboratory,davidson2013turbulence}. 
Having two mechanisms that break isotropy (gravity and rotation) and two independent dynamical control parameters ($Ro$, $Fr$) there are  numerous different set-ups.  A recent review on rotating and stratified flows can be found in \cite{pouquet2017dual}. Here we limit the discussion to the direction of the cascades  for the case where gravity and rotation are aligned as shown in
Fig. (\ref{fig:RSbox}).

To its simplest form the governing equations can be written as
\begin{align}
\partial_t {\bf u} + {\bf u \cdot \bnabla u} + 2\Omega\times {\bf u}       =& -N {\bf e}_z \phi          -\bnabla P      + \nu    \Delta {\bf u} - \alpha {\bf u}+ {\bf f}, \\
\partial_t \phi    + {\bf u \cdot \bnabla  } \phi  =& +N {\bf e}_z \cdot {\bf u}                + \kappa \Delta \phi                    + S.
\label{eq:NSSR}
\end{align}
Because rotation does not do  any work on the system, the global energy balance is the same of that  discussed for the zero-rotation case in a stratified system (\ref{eq:stratified}).

The potential vorticity is also conserved,  but it now depends explicitly on $\Omega$:
\be
V_p = 2\Omega \partial_z \phi -N w_z + {\bf w} \cdot \bnabla \phi. 
\ee
Again all moments of the potential vorticity are conserved, and as for the non rotating case,  the lowest non-trivial order  is the second one:
\be
\mathcal{V}_2 =\langle V_p^2\rangle = 
        \langle (2\Omega \partial_z \phi - N w_z)^2  \rangle
 - 2    \langle (2\Omega \partial_z \phi - N w_z) ({\bf w} \cdot \bnabla  \phi)\rangle 
 +      \langle ({\bf w} \cdot \bnabla \phi)^2 \rangle
\ee
The combination of  rotation and stratification leads to inertia-gravity waves with the following dispersion relation: 
\be
\label{eq:waveB}
\omega^2 = (4\Omega^2 k_\|^2 + N^2 k_\perp^2)/k^2.
\ee
One recovers the dispersion relation of inertial waves (\ref{eq:waverot}) for $N=0$ and that of gravity waves (\ref{eq:wavestra})  for $\Omega=0$ with the corresponding slow manifolds.
For non-zero values of $N$ and $\Omega$  the two inertial and gravity modes are mixed and the slow manifolds are not isolated in Fourier space. 
An important subset of the slow modes is given by the pure 2D modes: $k_\|=0$ and $\tilde{u}_z=0$. A detailed description of the linear modes for this system can be found  in  \cite{leith1980nonlinear,bartello1995geostrophic,herbert2014restrictede,sukhatme2008vortical}. There are numerous investigations of rotating and stratified turbulent flows for which a split cascade is observed \cite{smith2002generation,Rosenberg2015evidence,marino2014large,Marino2015resolving,Marino2013invers,kurien2008anisotropic,oks2017inverse}.
For unstable stratification in the presence of rotation an inverse cascade of energy has been reported leading to the formation of large-scale condensates \cite{julien2018impact, favier2014inverse, guervilly2017jets, guervilly2014large, rubio2014upscale}. 

\begin{figure*}[h!]                                                                              %
\centering                                                                                       %
\includegraphics[width=0.70\textwidth]{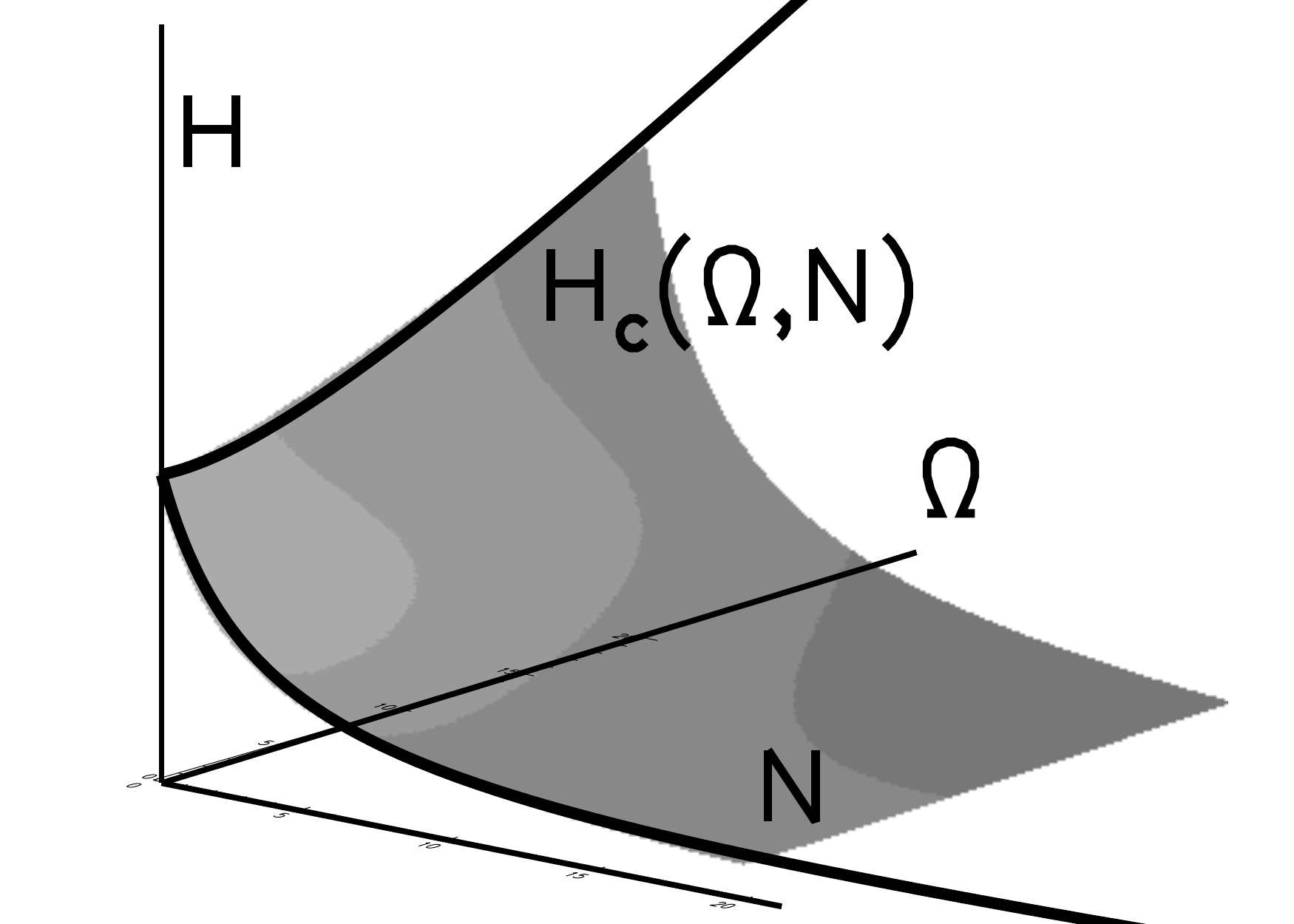}                                            %
\caption{ 
  Grey surface: critical  height $H_c(\Omega,N)$ as a function of Rotation and Stratification.}  %
  For values of $H<H_c$ there is a split cascade and for $H>H_c$ only a forward cascade.         %
\label{fig:RSphase}                                                                              %
\end{figure*}                                                                                    %
It is fair to say that due to the large number of parameters  it is difficult to obtain a precise prediction for  the transfer properties in the whole phase space. In the limits  $N=0$ or $\Omega=0$ one expects to recover the results from  un-stratified or non rotating cases but it is unclear what happens for intermediate values. It is  possible that for  given  $\Omega$  and $N$ there is  a critical height $H_c$ below which a split cascade exists. Such a possible phase-diagram for rotating and stratified flows is shown in figure \ref{fig:RSphase} where the limits $N=0$ and $\Omega=0$ recover the predictions of pure rotation and pure stratification discussed in the previous sections (see Figs. \ref{fig:RotPhasespace},\ref{fig:StrPhasespace}). 

\begin{figure*}[h!]                                                                         %
\centering                                                                                  %
\includegraphics[width=0.70\textwidth]{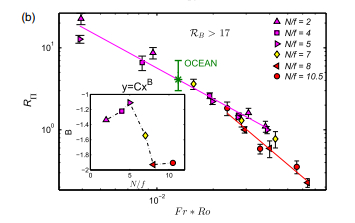}                              %
\caption{ The ratio $R_\Pi=\emu/\enu$ as a function of $RoFr$ for different values of       %
$N/f$ (where $f=2\Omega$ in our notation) measured from  high-resolution DNS together with   %
data from observation in the ocean. 
The inset gives the slope of the                        %
variation of $R_\Pi$ with $FrRo$ for various $N/f$. The figure is taken from                %
\cite{Marino2015resolving}  and used after permission.        }                             %
\label{fig:RSphase2}                                                                        %
\end{figure*}

We must note  that the existence of $H_c$ on the planes where   $\Omega=0$ or  $N=0$   does not necessarily implies criticality also for the cases  $N\ne0,\Omega\ne0$. In fact, it has been suggested  that the fraction of energy that cascades to the large scales follows a power-law behaviour in terms of  $Ro\cdot Fr$ as 
\be \frac{\emu}{\enu} \propto (Ro  Fr)^{-1} \label{eq:marino},
\ee with the power-law becoming steeper for larger values of the ratio $N/\Omega$. This has been demonstrated based on the results of numerical simulations and by plotting the ratio of the rate that energy that cascades to the small scales to the rate that energy cascades to the large scales
$$R_\Pi=\frac{\emu}{\enu}$$ \cite{Marino2015resolving}, see  Fig. (\ref{fig:RSphase2}). 
\begin{figure*}[h!]                                                                         %
\centering                                                                                  %
\includegraphics[width=0.48\textwidth]{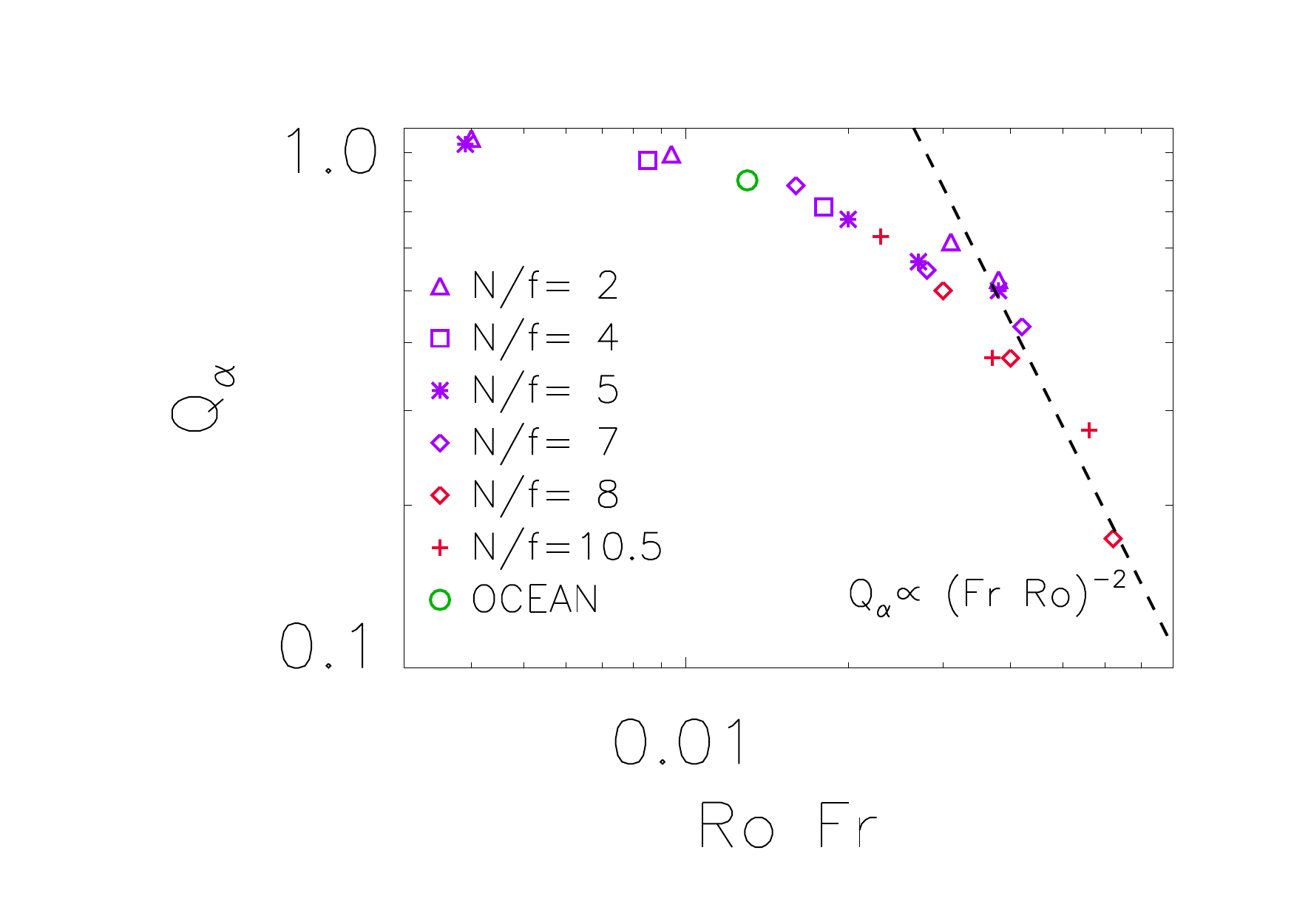}                               %
\includegraphics[width=0.48\textwidth]{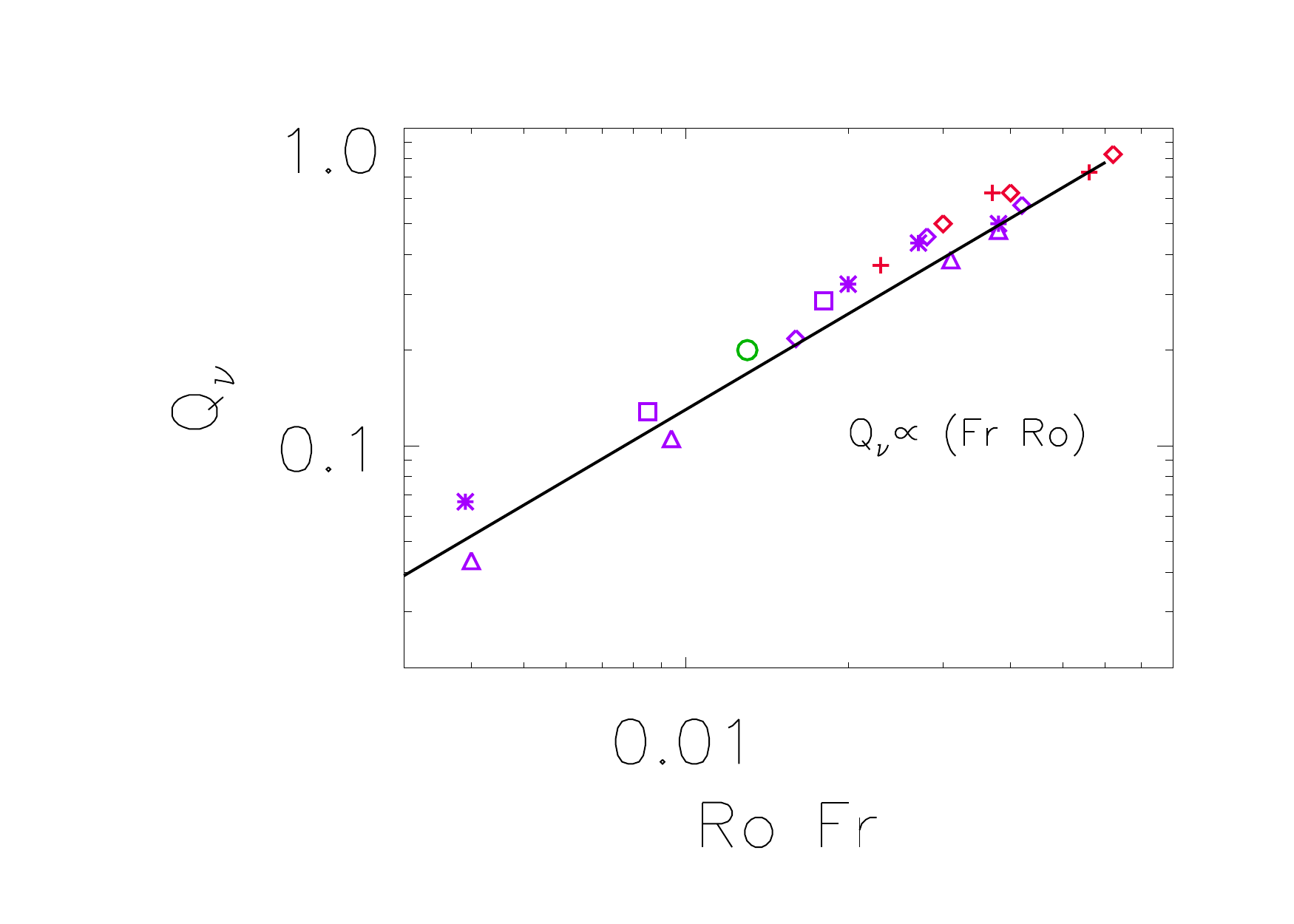}                                %
\caption{ The plot shows $Q_\alpha$ and $Q_\nu$ as a function of $RoFr$ based on the data   %
extracted from  Fig.(\ref{fig:RSphase2}) of \cite{Marino2015resolving}.                     %
As suggested in \cite{Marino2015resolving}                                                  %
$Q_\nu$ follows the scaling $Q_\nu \propto RoFr$,                                           %
while $Q_\alpha$ displays a sharp decrease that could be a steep power law                  %
$Q_\alpha \propto (RoFr)^{-2} $ or the indication of a critical 2nd order transition at     %
$FrRo \simeq 0.08$.} \label{fig:RSphase3}                                                   %
\end{figure*}                                                                               %
 A power-law dependence implies that the transition is smooth and thus different  from  the critical  scenario of Fig. (\ref{fig:RSphase}). In Fig. (\ref{fig:RSphase3}) we also plot 
\be
    Q_\alpha=\frac{\emu}{\ein}=\frac{R_\Pi}{1+R_\Pi} \quad \mathrm{and} \quad
    Q_\nu   =\frac{\enu}{\ein}=\frac{1          }{1+R_\Pi}
\ee
(see Def. \ref{def3} in Sec. (\ref{sec:CascadeDef}) ).  {
The data here have been extracted directly from fig. \ref{fig:RSphase2}.} The left panel of Fig. (\ref{fig:RSphase3}) indicates that $Q_\nu$  follows the scaling $Q_\nu\propto (Ro  Fr)$. Therefore the amplitude of the forward cascade decreases as a power-law of the product $RoFr$. For  $Q_\alpha$ the data follow a steep decrease with $RoFr$. This can be a power-law $Q_\alpha \propto (RoFr)^{-2} $ but we can not also exclude the possibility that this steepening is due to a critical transition with $Q_\alpha\to 0$ at $RoFr\simeq 0.08$. 

Rotating and stratified turbulence has a large number of control parameters and it is thus not easy to draw precise  conclusions from the limited amount of data presently available. It is thus possible that the phase diagram for this flow configuration  is  much more complex than previously believed. A first systematic  analysis of many different realizations at varying all parameters in simulations and experiments has  been attempted only recently \cite{pouquet2018scaling,rosenberg2016variations}.  Further  numerical, experimental and  theoretical studies are needed to address this issue.

\subsubsection{Summary}
We have briefly summarized recent results in rotating and stratified turbulence whose complexity is higher than what encountered for only stratified or only rotating flows.  
Flows in the presence of rotation and stratification can also display split cascades. However the precise location, properties or even existence of the critical surface is not clear yet. It is thus an open problem to determine how the turbulent cascades behave in the various different limits that can be considered.
 \subsection{ MHD (3D and 2D) \label{sec:MHD}}  

\begin{figure*}[h!]                                                                         %
\centering                                                                                  %
\includegraphics[width=0.30\textwidth]{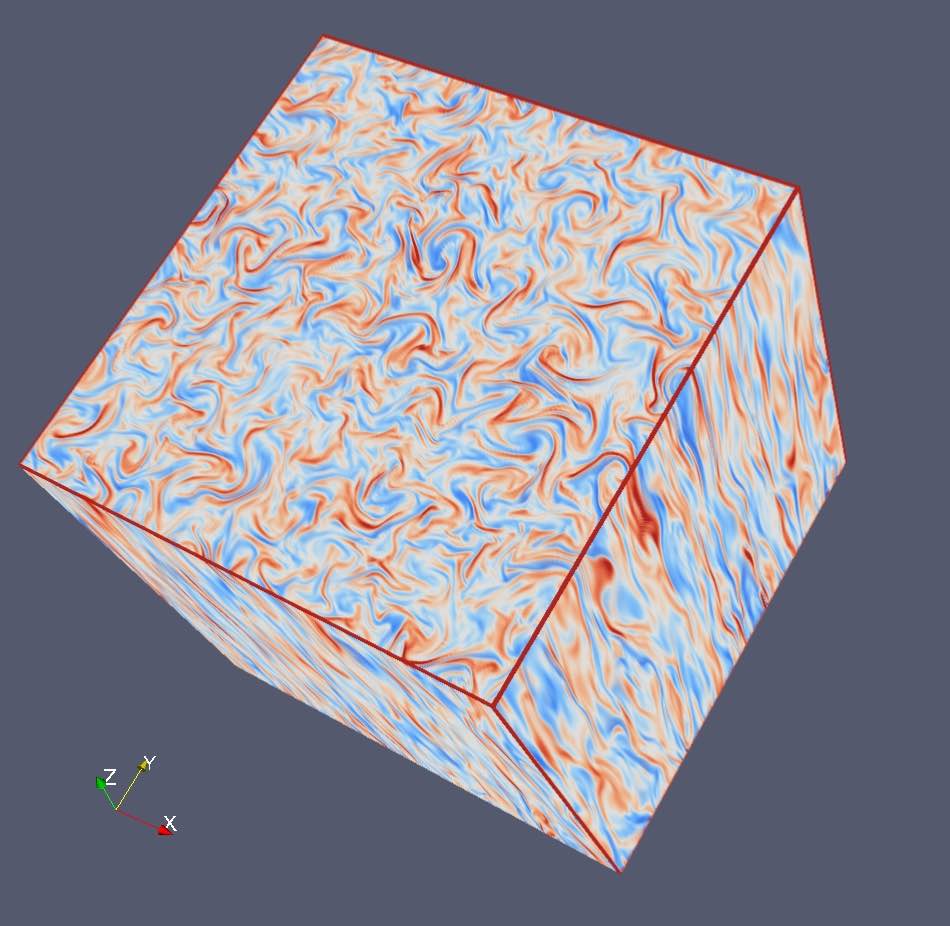}                                 %
\includegraphics[width=0.30\textwidth]{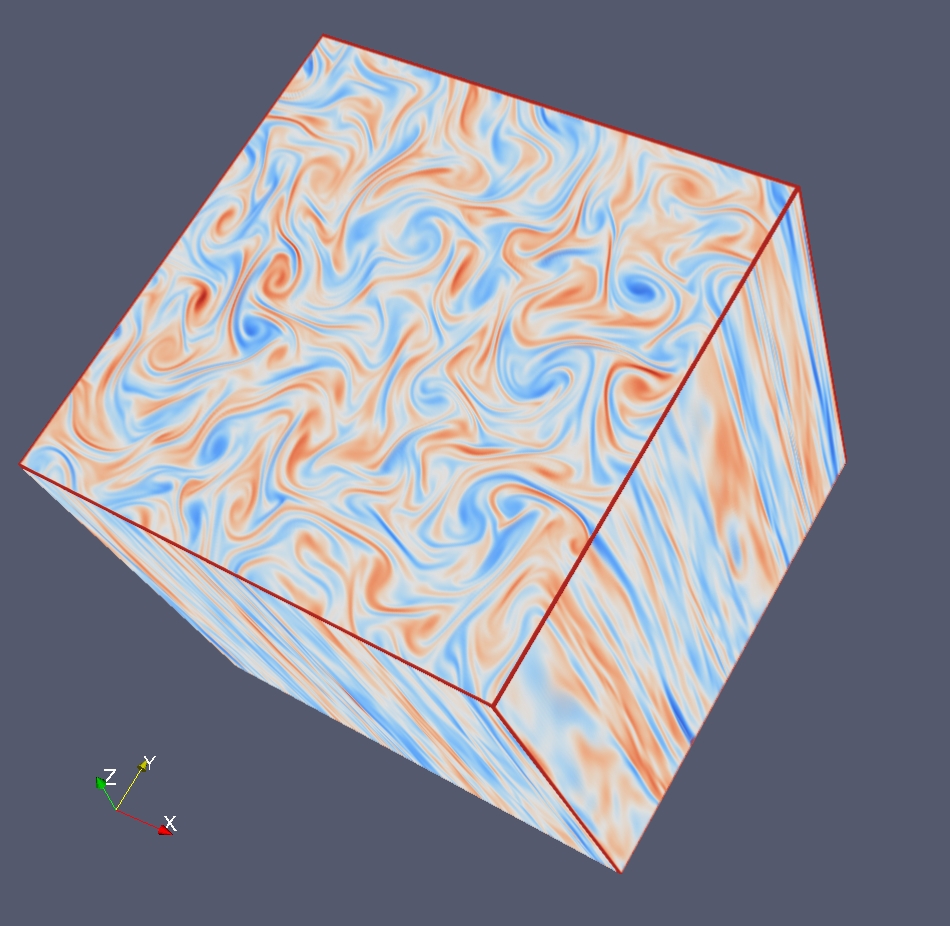}                                %
\includegraphics[width=0.30\textwidth]{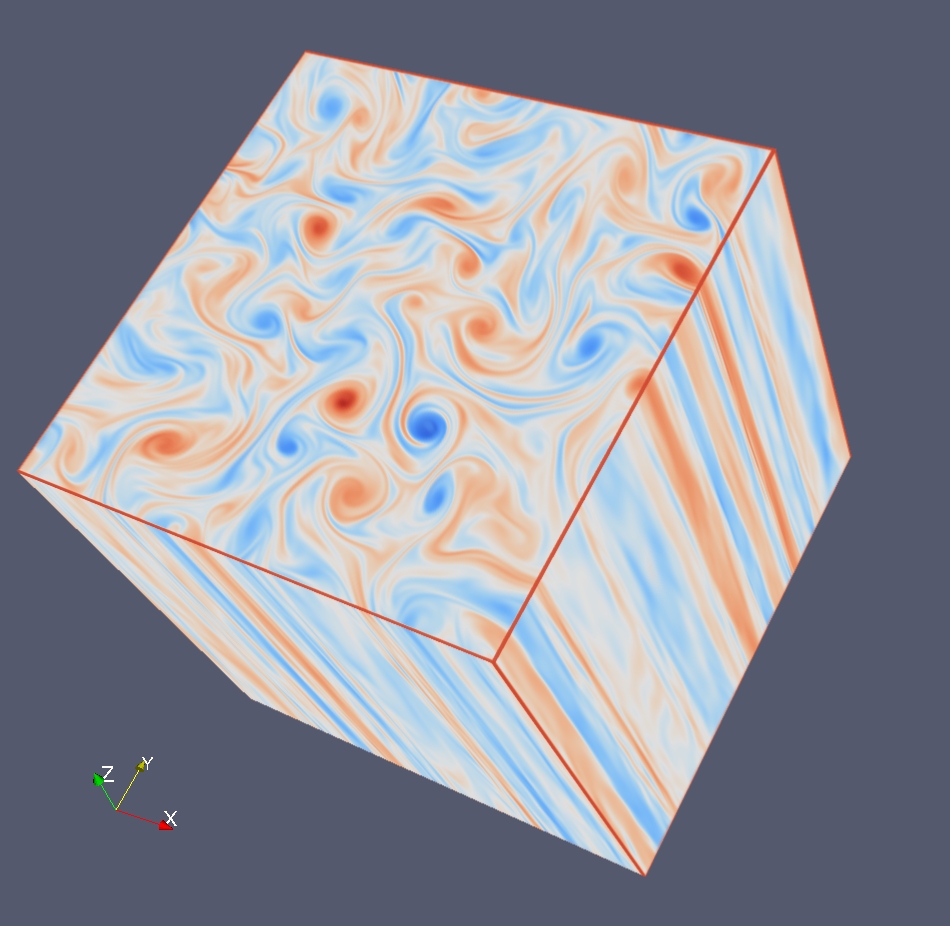}                                 %
\includegraphics[width=0.50\textwidth,height=0.3\textwidth]{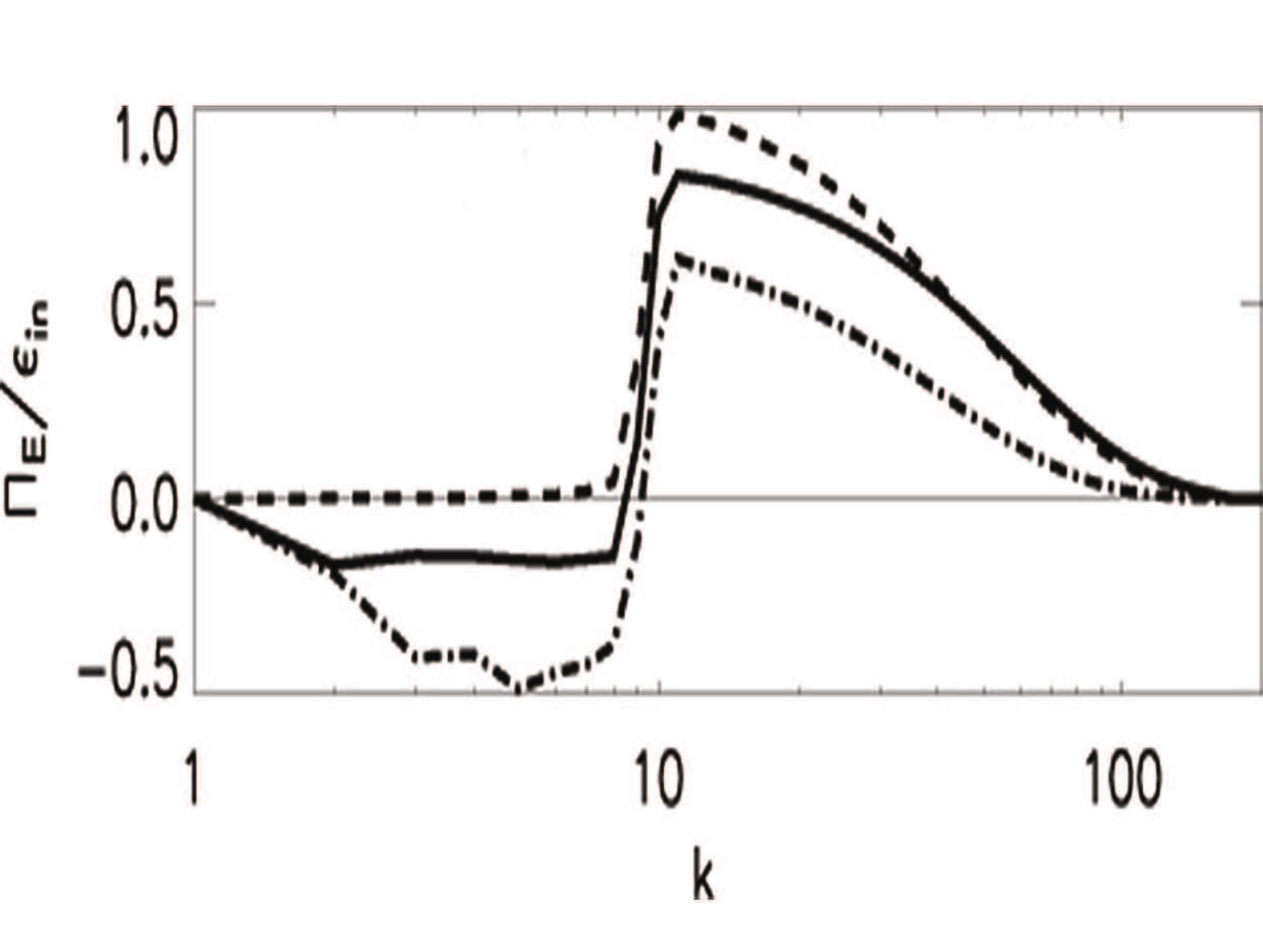}            %
\caption{ Top row:  vorticity component aligned along the mean magnetic field, $\bB_0 = B_0 \hat {\bf z}$ from a MHD simulation at increasing intensity $B_0$ from left to right. The visualizations are based on the runs in            %
\cite{Alexakis2011two}. The lower panel shows the energy flux for the three cases with      %
dashed line for the smallest value of $B_0$ solid line for the medium value and dashed dot  %
for the largest value. }                                                           %
\label{fig:MHDB0}                                                                           %
\end{figure*}

 In this section we discuss the properties of turbulence in conducting fluids coupled to magnetic fields known as Magnetohydrodynamic (MHD) turbulence.  It is met in a variety of systems such as in the stellar interiors \cite{priest1991advances}, the solar wind \cite{bruno2013solar,goldstein1995magnetohydrodynamic},  accretion disks \cite{pringle1981accretion} and the galaxies \cite{falgarone2003turbulence,zeldovich1983magnetic} but also in industrial applications \cite{Diamond2005plasma,white2017theory}. Even in its simplest form, MHD turbulence possesses a  vast richness of phenomena and different regimes that is not possible to fully cover  in this review. The interested reader can refer to many reviews and books on this subject \cite{biskamp2003magnetohydrodynamic,zhou2004colloquium,davidson2016introduction,galtier2016introduction} 
 and most recently \cite{pouquet2018helicity}. Here, we  focus on the properties of the cascades of the different invariants in the following five different limiting flow configurations: isotropic and  anisotropic 3D-MHD turbulence, 2D MHD, quasi-static MHD  and helical MHD. It is shown that in MHD turbulence  a change of cascade direction can be observed and, most importantly, it can be realized in the laboratory too. Fig. \ref{fig:MHDB0} shows the visualization of three different flow realization with increasing uniform magnetic field (from left to right) that demonstrates the  quasi-two-dimensionalization  as the magnetic field increases.

\subsubsection{Formulation and invariants for 3D MHD}  
%
The simplest model for flows of a conducting fluid is given by the incompressible MHD equations:
\begin{eqnarray}
\partial_t {\bf u} + {\bu \cdot \bnabla \bu} & =& {\bB \cdot \bnabla \bB} -\bnabla P +\nu  \Delta {\bu} -\alpha \bu    +  {\bbf_u}     \label{MHD1}\\
\partial_t {\bf B} + {\bu \cdot \bnabla \bB} & =& {\bB \cdot \bnabla \bu}            +\eta \Delta {\bB} -\alpha \bB    +  {\bbf_b}     \label{MHD2}
\end{eqnarray}
where  $\bB$ is the magnetic field,
the two fields are divergence-free $\bnabla \cdot \bu =\bnabla \cdot \bB =0$,
the coefficients $\nu$ and $\eta$ are the viscosity and Ohmic diffusivity respectively and 
to maintain a high level of generality we have included a large scale damping term for both fields.
Here we have allowed the possibility that energy is injected in the system by both mechanical, $\bbf_u$, and electromotive, $\bbf_b$, forces.
For simplicity the domain is assumed again to be a periodic box. It will be convenient to split the magnetic field $\bB$ as $\bB= \bB_0 +\bb$ where $\bB_0=\langle \bB \rangle$ is the uniform component of $\bB$ and $\bb$ is a spatially varying component. Unless forced, the uniform component $\bB_0$ is 
constant time. If $\bB_0=0$  and $\bbf_b=0$ then $\bb=0$ is always a solution, however this solution is in general unstable. 
A weak initial  magnetic field $\bb$ can be amplified and sustained by a dynamo process \cite{moffatt1978field}.

The new control parameter for this system other than the kinetic Reynolds number is the magnetic Reynolds number $Rm$
\be
Rm = \frac{u_f\lin}{\eta} 
\ee
where as usual $u_f$ stands for the rms velocity at the injection scale or $u_f=(\ein\lin)^{1/3}$ if the non-dimensionalization  is based on the energy injection rate as defined in (\ref{eq:Re}-\ref{eq:Red}).
The ratio of the two Reynolds numbers is referred to as the magnetic Prandtl number $P_M=\nu/\eta$.
In the presence of a non-zero mean magnetic field,  $\bB_0$, also 
the Alfv\'en Mach number must be introduced to express the relative amplitude of the velocity fluctuations to the amplitude of the mean field $B_0$:
\be
\label{eq:Alfven}
M = \frac{u_f}{B_0}.
\ee
An alternative and useful way to write the MHD equations is using the Els\"asser variables $\bz^\pm = \bu
\pm \bb$: 
\begin{eqnarray}
\partial_t {\bz^\pm } \pm {\bB_0 \cdot \bnabla \bz^\pm} + {\bz^\mp \cdot \bnabla \bz^\pm} &=& -\bnabla P +\nu  \Delta {\bz}^\pm -\alpha \bz^\pm +  {\bbf^\pm}     \label{MHD3}
\end{eqnarray}
where $ {\bbf^\pm = \bbf_u \pm \bbf_b}$ and we have assumed $\nu=\eta$ for simplicity.  Written in
this way, it can be seen that a uniform magnetic field $\bB_0$ advects $\bz^+$ fluctuations along its direction while the $\bz^-$ fluctuations are advected in the opposite direction. This transport of the $\bz^\pm$ fluctuations corresponds to the so called Alfv\'en waves \cite{alfven1942existence} that follow the dispersion relation: 
\be
\omega_{\bf k} = \bB_0\cdot \bk .
\ee
Furthermore,  $\bz^+$ advects  $\bz^-$ and vice versa without having any self-interactions ($\bf z^+$ with $\bf z^+$ or $\bf z^-$ with $\bz^-$ ). 
As a direct consequence if one of the two Els\"asser variables is zero the other variable advected by $\bB_0$
is a solution of the inviscid MHD equations because the non-linearity is zero, no matter how complex  the remaining field is.  

There are three ideal quadratic invariants for the 3D MHD equations. The total energy
\be
\mathcal{E} = \frac{1}{2}\langle |{\bf u}|^2 + |{\bf b}|^2 \rangle, 
\ee
the cross helicity
\be
\mathcal{H}_c =  \langle {\bf u\cdot b} \rangle, 
\ee
and the magnetic helicity (if ${\bB_0}=0$)
\be
\mathcal{H}_\cM = \frac{1}{2}\langle {\bf a \cdot b} \rangle 
\ee
where $\ba$ is the  magnetic field vector potential, $\bb= \bnabla \times \ba$.
Note that the kinetic helicity (\ref{eq:inviscid3D}) is not conserved by (\ref{MHD1}-\ref{MHD2}) for a non-zero magnetic field.
Concerning  the above quantities, only the energy is positive definite, while the two helicities
can take either sign. The total energy however and the cross helicity can be combined to give
two positive definite invariants that in terms of the Els\"asser variables are written as
\be
\mathcal{E}^\pm = \frac{1}{2}\langle |{\bf z}^\pm|^2  \rangle = \mathcal{E} \pm  \mathcal{H}_c
\ee
and we will refer to them as the Els\"asser energies.

The balance for the total energy then reads 
\be
\partial_t \mathcal{E}(t) = \ein(t) - \enu(t) -\iepsilon_\eta(t) -\emu(t)
\ee
where now $\ein(t)=\la \bu\cdot {\bf f}_u \ra + \la \bb \cdot {\bf f}_b \ra$, $ \iepsilon_\eta(t) = \eta \la |\bnabla \bb|^2  \ra $
and $\emu(t)=2\alpha \cE$. In terms of 
Els\"asser energies the balance reads
\be
\partial_t \mathcal{E}^\pm(t) = \ein^\pm(t) -\iepsilon^\pm(t) -\emu^\pm(t)
\ee
where $\ein^\pm(t)=\la \bz^\pm \cdot {\bf f}^\pm \ra$ and $\iepsilon^\pm(t)= \frac{\nu+\eta}{2} \la |\nabla \bz^\pm|^2\ra +  \frac{\nu-\eta}{2} \la \nabla_i z_j^\pm \nabla_i z_j^\mp \ra  $ is the viscous and Ohmic dissipation rate of the two energies and $\emu^\pm(t)=2\alpha \cE^\pm$.
The second term in the definition of $\iepsilon^\pm$ is not sign definite and is related to the dissipation of the cross-helicity that is given by $\iepsilon_c(t)= ( \nu + \eta )\la \nabla_i u_j \nabla_i b_j \ra$.
Note that $\iepsilon^\pm =(\iepsilon_\nu+\iepsilon_\eta) \pm \iepsilon_c$. 

Finally the magnetic helicity satisfies the balance: 
\be
\label{eq:magnetichelicity}
\partial_t \cH_\cM(t) = \mu_{in}(t) - \mu_\eta (t) - \mu_\alpha(t)   
\ee
where     $\mu_{in}(t)=\la {\bf f}_b \cdot {\bf a}\ra$ is  the magnetic helicity injection rate, 
 $\mu_\eta(t)=\eta \la {\bf b} \cdot \bnabla \times {\bf b} \ra $ is the magnetic helicity Ohmic dissipation rate
and $\mu_\alpha(t) =2 \alpha \cH_\cM(t)$ is the large scale drag. Note that the mechanical forcing ${\bf f}_u$ cannot inject magnetic helicity in the system
but nonetheless magnetic helicity can be generated by $\mu_\eta$ or $\mu_\alpha$ that are not sign definite. This is discussed in Sec. (\ref{sec:mhelicity}). 

\subsubsection{Isotropic 3D MHD turbulence}  
\label{sec:IsoMHD} 
\begin{figure*}[h!]                                                                         %
\centering                                                                                  %
\includegraphics[width=0.70\textwidth]{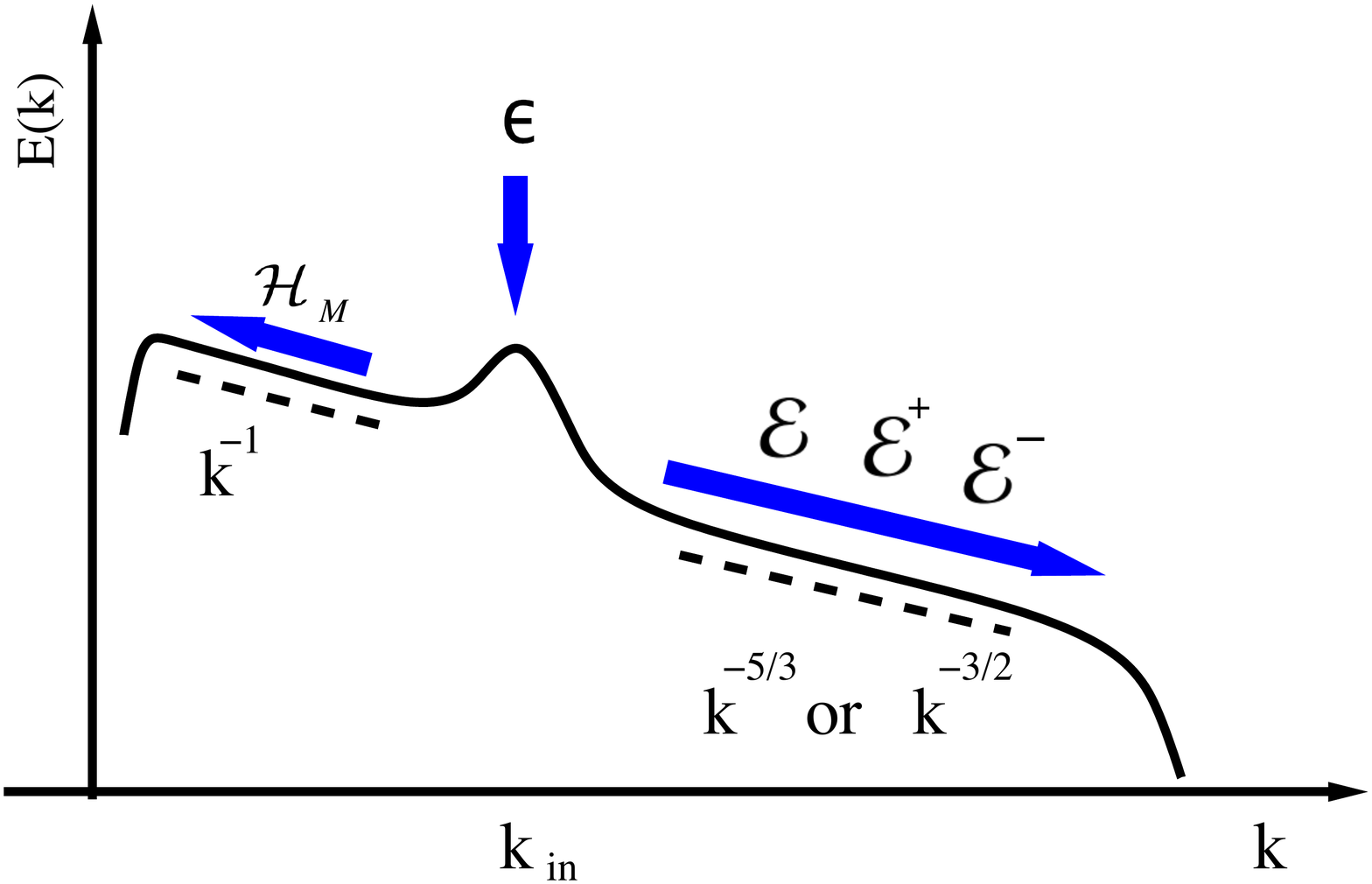}                                   %
\caption{A general picture of energy cascades in isotropic MHD turbulence. }                %
\label{fig:MHDspec1}                                                                        %
\end{figure*}

As a first step we consider the ${\bf B}_0 =0$ case and assume the flow to be isotropic.
It is an empirical fact \cite{mininni2009finite,biskamp2003magnetohydrodynamic}  that for ${\bB}_0=0$ the energies $\mathcal{E},\mathcal{E}^\pm$ and the cross helicity $\mathcal{H}_c$ cascade forward to the small scales while the magnetic helicity $\mathcal{H}_\cM $ cascades inversely to the large scales. The general picture of the developed cascades
is shown in Fig. (\ref{fig:MHDspec1}). 
The helical case will be treated separately in  Sec. (\ref{sec:mhelicity}), here we proceed by assuming
$\mathcal{H}_\cM=0$. The same phenomenology that we used for 3D hydrodynamic turbulence can be used for MHD turbulence and it predicts a K41 energy spectrum,  $E(k)\propto \ein^{2/3} k^{-5/3}$. Here $E(k)$ stands for the total energy spectrum that is composed of the kinetic energy spectrum $E_u(k)$  and the magnetic energy spectrum $$E_b(k) =  \frac{1}{2\Dk} \sum_{k \le |\bk| < k +\Dk} |\bbt(\bk)|^2$$ that measures the energy in the magnetic modes so that $E(k)=E_u(k)+E_b(k)$.  It can be argued that the magnetic field fluctuations, $\langle \bb^2\rangle^{1/2}$, can directly influence the small scales by acting locally as a guiding field that suppresses the nonlinear interactions. This suppression then is expected to modify the energy spectrum (see Eq. \ref{eq:wavebalance} in Sec. \ref{sec:WWT}) to a less steep power-law $E(k)\propto\ein^{2/3} \langle \bb^2\rangle^{1/2}  k^{-3/2}$ referred to as the Iroshnikov-Kraichnan (IK) theory \cite{Iroshnikov64, kraichnan1965inertial}.  In configuration space, the existence of  inertial cascades lead to a series of exact relations for third order quantities similar to the ones derived in (\ref{eq:KHMstationary}) for hydrodynamical turbulence \cite{chandrasekhar1951invariant,politano1998dynamical}. Using longitudinal velocity and magnetic field increments, for the total energy one obtains
\be
\langle \delta_\br u ^3 - 6 b_\br ^2 \delta_\br u  \rangle = -\frac{4}{5} (\iepsilon_\nu +\iepsilon_\eta)  r.
\label{eq:mhd45}
\ee
Anisotropic versions have been examined in
\cite{podesta2008laws, podesta2007anisotropic, galtier2009exact} and have been extended to
high Prandtl number regimes in \cite{yousef2007exact} and to Hall MHD in \cite{galtier2008karman} (see \cite{pouquet2015review} for a review). It is important to note that in (\ref{eq:mhd45}) both field increments and single point quantities, $b_\br = \bb \cdot \hat \br$, appear. In fact, the latter  term was shown to be dominant  \cite{yoshimatsu2012examination}. In terms of the Els\"asser variables,  a different law, involving only field increments, can be derived  \cite{politano1998dynamical, politano1998karman,politano2003karman}:
\be
\langle \delta_\br z^\mp |\delta_\br {\bz}^\pm|^2 \rangle  = -\frac{4}{3} \iepsilon^\pm r.
\label{43mhd1}
\ee
where  $\iepsilon^\pm$ are the dissipation rates of the $\cE^\pm$ energies,  and both  longitudinal and perpendicular components appear. Returning to the $\bu,\bb$ variables, adding and subtracting the relations (\ref{43mhd1}) and assuming isotropy, we have  \cite{podesta2008laws, podesta2007anisotropic}:
\be
  \langle \delta_\br u ( |\delta_\br \bu|^2+ |\delta_\br \bb|^2 )\rangle  
-2\langle \delta_\br b ( \delta_\br \bu \cdot \delta_\br \bb ) \rangle
= -\frac{4}{3} (\iepsilon_\nu+\iepsilon_\eta) r
\ee
and 
\be
  \langle \delta_\br b ( |\delta_\br \bu|^2+ |\delta_\br \bb|^2 )\rangle  
+2\langle \delta_\br u (\delta_\br \bu \cdot  \delta_\br \bb) \rangle
= -\frac{4}{3} \iepsilon_c r
\ee
where $\iepsilon_c$ is the dissipation rate of cross helicity.
The scaling relations obtained for the third order structure functions suggest the Kolmogorov
phenomenology, $\delta_\br \bz^\pm \propto r^{1/3}$; however IK scaling: $\delta_\br \bz^\pm \propto r^{1/4}$ can not be excluded due to possible scale dependent correlations between the two fields.

Numerical investigations for isotropic MHD \cite{verma1996numerical, biskamp2000scaling, muller2000scaling, muller2005spectral, alexakis2013large} also tend to favour the Kolmogorov $E(k)\propto \ein^{2/3} k^{-5/3}$ spectrum contrary to the anisotropic case that is examined in the next section and seems to favour the $k^{-3/2}$ prediction.  The discrepancy between the two predictions is due to the assumption made in the IK theory that large-scale magnetic fluctuations might play the role of a uniform  --locally anisotropic-- magnetic field.  Furthermore another possibility exists that the cascade in the presence of large scale but not-uniform fields is restricted in regions of space where amplitude of the magnetic field is small \cite{alexakis2013large}. Due to the limited scale separation achieved in numerical simulations, it is difficult to distinguish among the different predictions. We discuss the issue of the two different spectral predictions further in the next section where the anisotropic case is examined.
\subsubsection{ Anisotropic MHD } 
\label{sec:AnisMHD}
\begin{figure*}[h!]                                                                        
\centering                                                                                 
\includegraphics[width=0.70\textwidth]{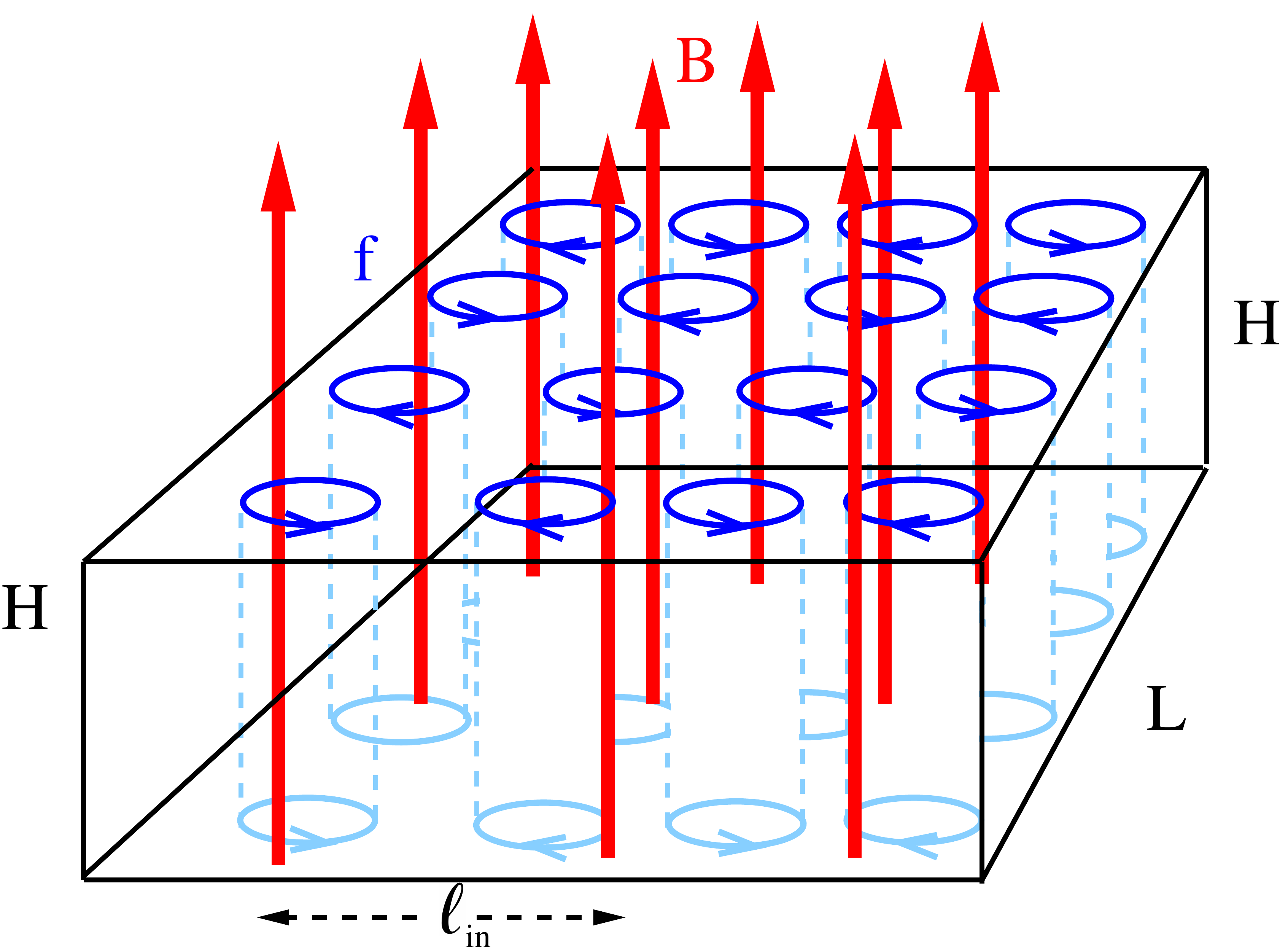}                                    
\caption{Domain of the flow embedded in uniform magnetic field  layer. }                
\label{fig:MHDbox}                                                                          
\end{figure*}                                                                              
We next treat the case of 3D turbulence in the presence of a uniform field $\bB_0$ as shown in figure \ref{fig:MHDbox}. We consider a conducting fluid embedded in a fluid layer of finite height $H$ that is penetrated by a uniform magnetic field of intensity $B_0$ across the layer.  A strong guiding field results in  the Alfv\'en waves having a large  frequency (\ref{eq:waveB})  except  for those modes with wavevectors  perpendicular to $\bB_0$, i.e.  ($k_\|=0$), that have zero frequency and compose the slow manifold in MHD.  We need to distinguish three different regimes, depending on the intensity of the guiding field and on the domain thickness. \\
\noindent For infinite vertical domains and strong $B_0$, so that the Alfv\'en  wave period, $\tau_{wave}(k)=1/B_0k$, is much smaller than the typical non-linear time-scale, $\tau_{nonlin}(k)=E(k)^{-1/2} k^{-3/2}$, the non-linearity is weak and the system falls in the weak turbulence regime where it can be treated by an asymptotic expansion (see sec.
\ref{sec:WWT}).  If $\tau_{wave} > \tau_{nonlin}$,   then the system is in the strong turbulence regime. Finally, for finite domains and strong $B_0$ there is a third regime where quasi-2D behaviour together with an inverse cascade occurs.

\paragraph{Weak turbulence regime  } 
For infinite domains, the slow manifold is not isolated due to quasi-resonances and the 2D-2D interactions  play a sub-dominant role.

\begin{figure*}[h!]                                                                        
\centering                                                                                 
\includegraphics[width=0.70\textwidth]{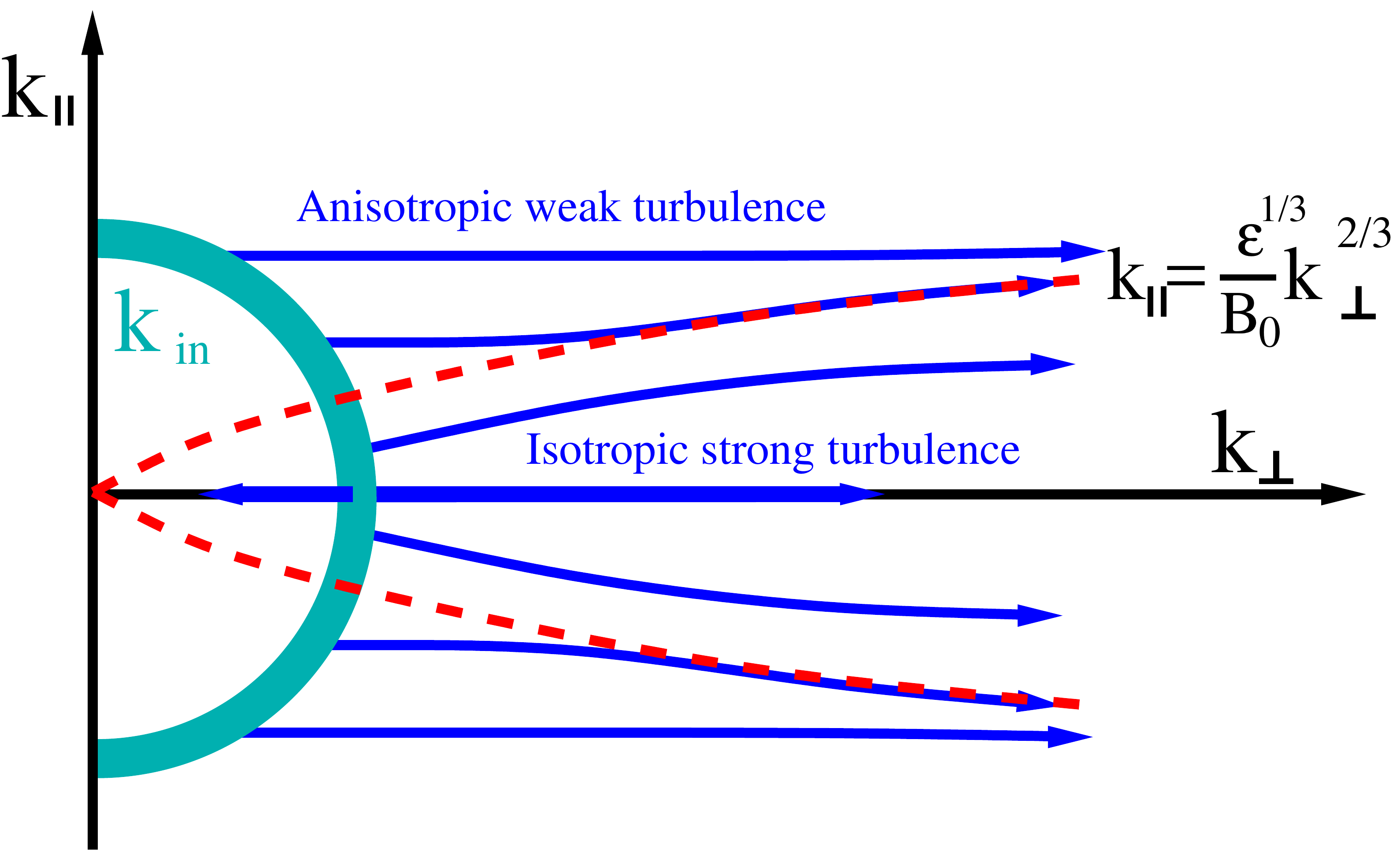}                              
\caption{ Anisotropic Cascade in MHD turbulence.                                           
The dashed lines indicate the critical balance relation                                    
$B_0 k_\|\propto \ein^{1/3} k_\perp^{2/3}$. If the $k_\|=0$ modes become isolated          
an inverse cascade can develop also. Blue-sky circle represents the modes were energy is
injected (here imagined isotropically at $k=k_{in}$). }                                                      
\label{fig:MHDAnis}                                                                        
\end{figure*}

In this case the cascade is driven by weak wave interactions and can be treated perturbatively \cite{galtier2000weak}. The cascade is strongly
anisotropic with the magnetic field reducing the rate of energy transfer across scales by a factor $\tau_{wave}(k)/\tau_{nonlin}(k)$ as will be discussed in Sec. (\ref{sec:WWT}). Assuming isotropy  and using (\ref{eq:wavebalance}), one obtains the IK spectrum $E(k) \propto \iepsilon^{2/3} B_0^{1/2} k^{-3/2}$ \cite{Iroshnikov64,kraichnan1965inertial}. However the assumption of isotropy is far from being realized. An anisotropic version of the IK argument can be made by noting that for strong magnetic fields the cascade occurs dominantly in the direction perpendicular to $B_0$ and time scales must be corrected as  $\tau_{wave}(k_\perp,k_\|)=1/B_0k_\|$ and $\tau_{nonlin}=[E(k_\perp,k_\|)k_\| k_\perp]^{-1/2} k_\perp^{-1}$ where we have assumed $k_\| \ll k_\perp$ and used the  energy spectrum defined
in (\ref{eq:Aspec1}). Wave turbulence theory prediction for the flux (\ref{eq:wavebalance2}) leads to \cite{galtier2000weak}: 
\be 
E(k_\perp,k_\|) \propto f(k_\|) k_\perp^{-2}k_\|^{-1/2} 
\label{eq:WWTMHD}
\ee
where the spectrum is defined up to an arbitrary non-dimensional function $f(k_\|)$ that might depend on the forcing mechanism
because the cascade proceeds only along the $k_\perp$ direction. 
\paragraph{Strong turbulence}  
For strong turbulence further assumptions are needed to derive an expression for the energy spectrum.  
The most popular point of view is given by the {\it critical balance theory} \cite{goldreich1995toward,nazarenko2011critical} as already discussed for rotating and stratified flows.
In the {\it critical-balance} description, energy is concentrated around wavenumbers that satisfy a `{\it balance}' between the wave time scale and the nonlinear time scale $\tau_{wave} \sim \tau_{nonlin}$. Substituting the latter equality in  equation for the wave turbulence energy balance (\ref{eq:wavebalance1}) we obtain:
\be 
\ein \sim (E(k_\perp,k_\|)k_\| k_\perp)^{3/2}k_\perp; \qquad B_0 k_\|\propto \ein^{1/3} k_\perp^{2/3}
\label{eq:critical} 
\ee
where the second relation is obtained by the equality of the two time scales. For the spectrum along the critical line we then obtain:
\be
E(k_\perp,k_\|) \sim \ein^{2/3}k_\perp^{-5/3} k_\|^{-1} .
\ee
In this description, one assumes that  energy is mainly transferred along the critical line in $k_\|,k_\perp$ space. This is graphically summarized  in figure \ref{fig:MHDAnis} where energy flows  along the dashed lines. However, numerical simulations show that energy is concentrated in the entire strong turbulence region below the critical balance line 
where $\tau_{nonlin}(k_\perp,k_\|) <\tau_{wave}(k_\perp,k_\|)$ and strong interactions take place. Alternative descriptions that take some of these issues in to account can be found in  \cite{alexakis2007nonlocal,galtier2005spectral}. 
Based on the previous assumptions of critical balance the anisotropic cylindrically averaged energy spectrum (see definition \ref{eq:Aspec3}) follows the scaling $E_\perp(k_\perp)\propto k_\perp^{-5/3}$. Although this spectrum has been observed in some recent simulations \cite{beresnyak2011spectral}, most numerical studies  have shown a spectrum closer to $E(k_\perp)\propto k_\perp^{-3/2}$ \cite{perez2012energy}. This was interpreted as an effect of a scale-dependent alignment of the two fields $\bu,\bb$ that reduced the non-linear interactions \cite{boldyrev2005spectrum}. The two spectra are debated in the literature and a recent review of the results on the issue can be found in \cite{schekochihin2017mhd,galtier2016introduction}.  As of now,  it is fair to say that we do not have an accurate and commonly-agreed description of the anisotropic energy spectrum $E(k_\perp,k_\|)$ in MHD even from a phenomenological point of view.

\paragraph{Layers of finite thickness}
For layers of finite height, $H$, further complications arise since the slow modes can have isolated dynamics. In this case the slow modes at $k_\|=0$ can dominate and drive an inverse cascade. This is similar to the strongly rotating flows as a strong magnetic field also tends to suppress fluctuations and bi-dimensionalize the flow. This result has been proven in \cite{Gallet2015exact}. Using the same arguments as  in Sec. (\ref{sec:Rotation}) for rotating turbulence and summarized in Fig. (\ref{fig:CBgrid})  the flow will become 2D  when the smallest non-zero $k_\|= 2\pi/H$ has a time period $\tau_{wave}= H/B_0$ smaller than the eddy turnover time  evaluated at the forcing scale \cite{Alexakis2011two,nazarenko20072d}, i.e. when $ 2\pi/H$ is above the critical balance line:
\be 
\frac{H}{B_0} \le \ein^{-1/3} \kin^{-2/3} .
\ee 
In 2D-MHD flows,  energy cascades forward (see next section),  so a bi-dimensionalization does not necessarily imply  an inverse cascade of energy. Nonetheless, if only a mechanical force is used, the hydrodynamic 2D modes can not excite a magnetic field and a genuine 2D inverse energy cascade develops. This was indeed observed in \cite{Alexakis2011two, sujovolsky2016tridimensional}. The 2D MHD case is discussed in more detail in the next section. 

\subsubsection{2D MHD}  
\label{sec:MHD2D}
We now consider MHD turbulence in 2D for which the equations simplify significantly  when expressed in terms
of the vector potential ${\bf a}= a(x,y,t) {\bf e}_z$ and of the vertical
vorticity $\bw = w(x,y,t) {\bf e}_z$.
In this case we have: 
\begin{eqnarray}
\partial_t {w} + {\bf u} \cdot \bnabla w = {\bf b} \cdot \bnabla j + \nu  \Delta { w} 
-\alpha w +  {f_w}\label{MHD2Da}\\
\partial_t {a} + {\bf u} \cdot \bnabla a  =  {\bf e}_z \cdot ( {\bf u} \times {\bf B}_0)    + \eta \Delta a   
-\alpha a +  {f_a}\label{MHD2Db},
\end{eqnarray}
where $ j=-\Delta a$ is the current density in the $z$ direction and ${\bB_0}$ is
assumed  to be on the $x,y$ plane. Setting $a=\bB_0=f_a=0$ one recovers the 2D hydrodynamic NSE (\ref{eq:2DNS}).  

In the absence of forcing and dissipation this system conserves the total energy  $\cE=\cE_u+\cE_b$ written as the sum of the kinetic energy $\cE_u =\frac{1}{2}\langle {\bf u^2} \rangle$ plus the magnetic energy  $\cE_b =\frac{1}{2}\langle {\bb^2} \rangle$  and, if $B_0=0$ , the square vector potential  $\cA =\frac{1}{2} \langle a^2 \rangle$.  The latter satisfies the following equation: 
\begin{eqnarray}
  \partial_t \cA = \langle {a {\bf e_z} \cdot (\bu \times \bB_0)}
  \rangle -\alpha \cA -\eta \cE_b + \langle a f_a \rangle.
\end{eqnarray}
The square vector potential is a positive-definite quantity that is restricted by the evolution of the energy cascade like the enstrophy in hydrodynamic 2D turbulence. In particular the square vector potential spectrum satisfies $E_\cA(k)= E_b(k)k^{-2} \le [E_b(k)+E_u(k)]k^{-2}=  E(k)k^{-2}$. Thus, the vector potential spectrum falls as $k^{-2}$ with respect to the energy spectrum. Following the arguments given  in (\ref{eq:fluxConstraintAB1}) and (\ref{eq:fluxConstraintAB2}) we can deduct that $\cA$ cascades inversely \cite{fyfe1976high, pouquet1978two, biskamp2003magnetohydrodynamic, banerjee2014statistics}. Furthermore, it is empirically known  that the energy cascade in 2D MHD is forward. Therefore, in 2D MHD turbulence a dual counter-directional cascade is expected with a forward cascade of energy and an inverse cascade  of the squared vector potential. Assuming that the dominant time scale is that of the eddy turnover time $\tau_r=r/\langle |\delta_\br \bu|^2 \rangle^{1/2} $ and further assuming that the two fields have similar amplitudes
\be
  \delta_\br \bu \sim \delta_\br \bb
\label{eq:similar}
\ee
one  recovers the Kolmogorov prediction for the forward energy cascade 
\be E_b(k) \simeq E_u(k) \propto \ein^{2/3} k^{-5/3}, \quad
\mathrm{for} \quad k>\kin \label{eq:2dmhdSpec1} \ee 
and the spectrum 
\be E_b(k) \simeq E_u(k) \propto (\ein/\kin^2)^{2/3} k^{-1/3}, \quad \mathrm{for} \quad k<\kin\label{eq:2dmhdSpec2} \ee 
for the inverse cascade of the square vector potential \cite{pouquet1978two,fyfe1976high}. However the  assumption (\ref{eq:similar})  is hard to justify in the case of the inverse cascade and indeed different spectral exponents have been measured in the literature \cite{banerjee2014statistics, pandit2017overview}.

We need to further notice that in the absence of an external magnetic field or magnetic forcing, the vector potential decays and therefore any magnetic fluctuations that exist at $t=0$ will die out.  This property is referred to as the anti-dynamo theorem of 2D flows \cite{zeldovich1957magnetic}. In this case, the system reduces to 2D fluid turbulence with an inverse cascade for energy. Furthermore, if the vector potential is non-zero but too weak to feedback on the flow,  it acts passively and its variance $\cA$ cascades forward (see next section \ref{sec:passivescalarcascades}). We have therefore two different limits. For $f_a\kin^2 \gtrsim f_w  $ there is a forward cascade of $\cE$ and an inverse cascade of $\cA$, for ${f_a} \kin^2 \ll f_w $ there is an inverse cascade of $\cE$ and a forward cascade of $\cA$. It is thus interesting to examine how the system transitions from one case to the other when the magnetic forcing is varied.  

For very weak magnetic forcing, the kinetic energy dominates $E_u(k) \gg E_b(k)$ and the flow does not feel the effect of the magnetic field (Lorentz force). Thus, the flow behaves as a hydrodynamic system cascading energy inversely with the kinetic energy spectrum following (\ref{eq:2dkraichnan}) and (\ref{eq:2denstrophy}) at scales smaller and larger than the forcing scale, respectively. At the same time, the vector potential $a$  is advected passively and $\cA$  cascades forward like the variance of a passive scalar (see section \ref{sec:passive})  forming a spectrum  
\[E_\cA(k) = k^{-2}E_b(k) \propto \iepsilon_b \iepsilon_u^{-1/3}  \kin^{-8/3}  k^{-1}\] where here $\iepsilon_b=\la \bb \cdot \bbf_b\ra$ and $\iepsilon_u=\la \bu \cdot \bbf_u\ra$. This implies that the kinetic and magnetic energy spectra read:
\[ E_b(k) \propto \iepsilon_b \iepsilon_u^{-1/3}  \kin^{-8/3} k, \quad
   E_u(k) \propto            \iepsilon_u^{ 2/3}  \kin^{4/3} k^{-3} .\] 
For moderate values of $Re,Rm$ (here we will assume $Re=Rm$ for simplicity) the magnetic field will always be sub-dominant and the flow will be at this hydrodynamic state with passive $a$ at all scales. However, for large $Re,Rm$ since $E_u(k)$ decreases with $k$ and $E_b(k)$ increases with $k$, the condition $E_u(k) \gg E_b(k)$ will necessarily break down at some wavenumber $k_c=\kin (\iepsilon_u/\iepsilon_b)^{1/4}$. This will happen if $\Red$ is large enough so that $k_c < \knu$. 
For 2D turbulence $\knu \propto \kin Re^{1/2}$  (see eq. \ref{eq:knu2D}) therefore the magnetic field becomes non-linear when $\iepsilon_u/\iepsilon_b > Re^2 $.
 For wavenumbers larger than $k_c$ a forward cascade of energy will take place similar to the forward cascade of energy observed in thin layers when $k>1/H$ (see Sec. \ref{sec:Thin}). The spectra for $k>k_c$ will thus follow (\ref{eq:2dmhdSpec1}): $ E_b(k) \sim E_u(k) \propto \enu^{2/3} k^{-5/3}$.   

\begin{figure*}[h!]                                                                         %
\centering                                                                                  %
\includegraphics[width=0.32\textwidth]{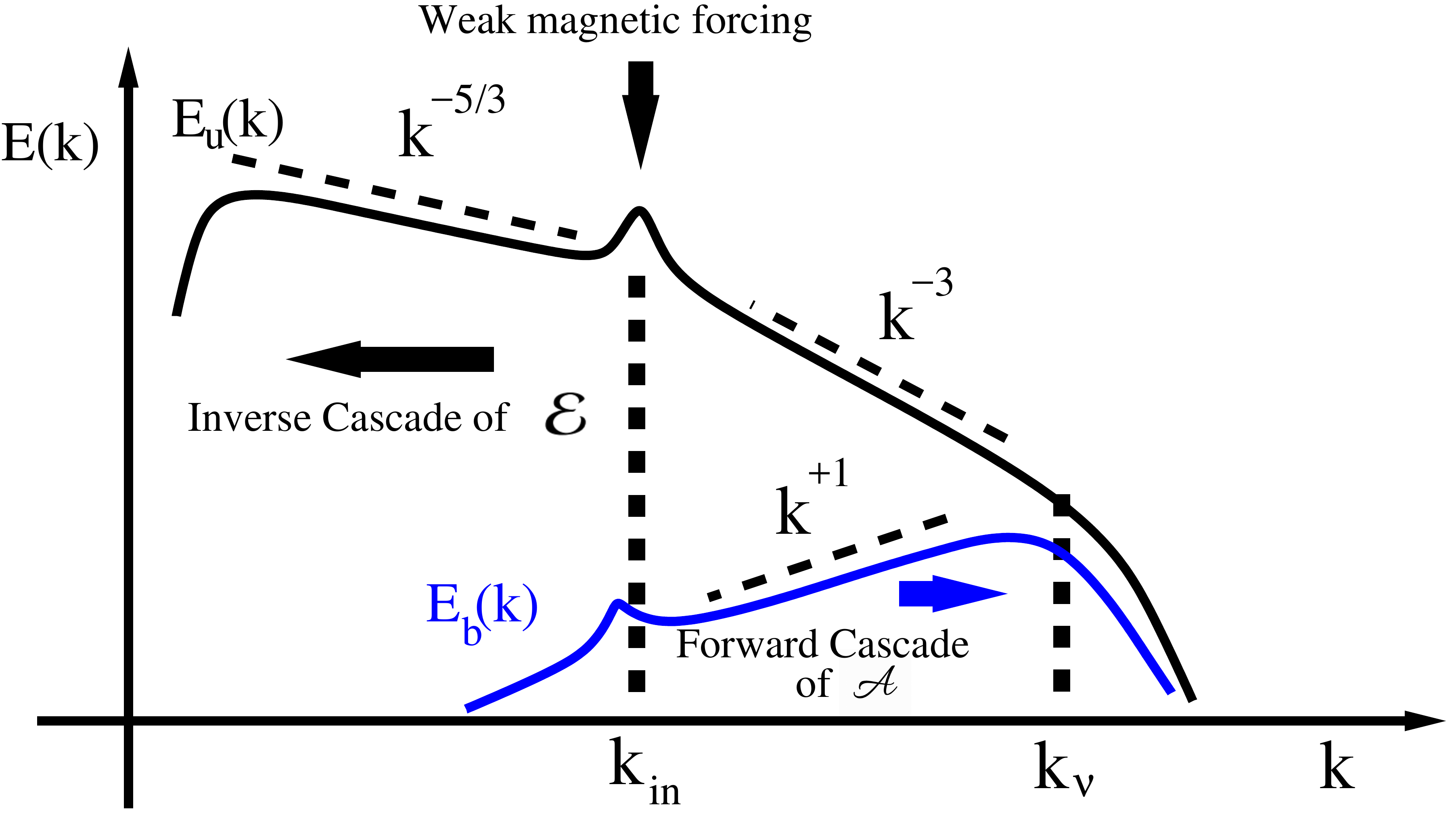}                                     %
\includegraphics[width=0.32\textwidth]{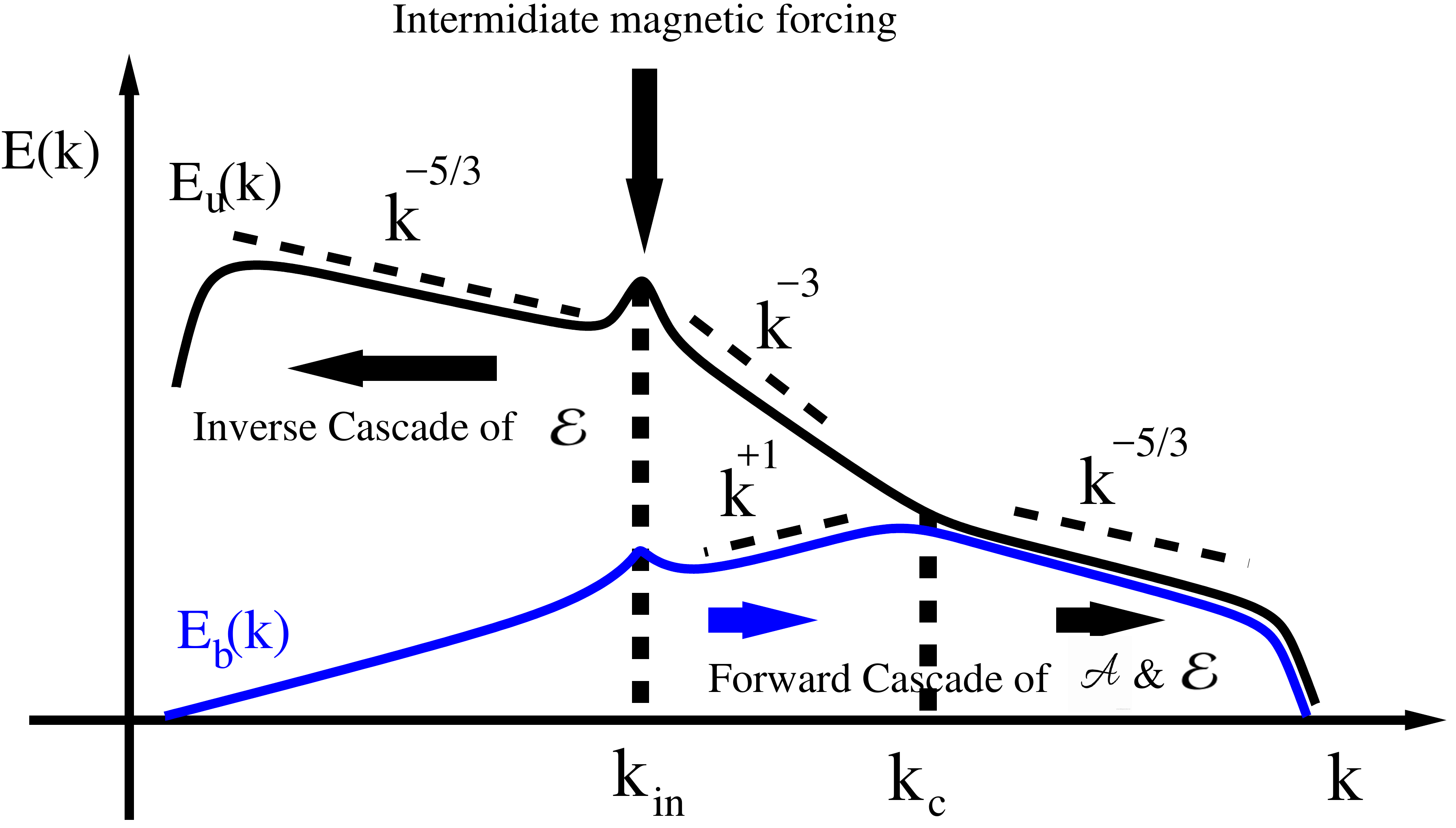}                                     %
\includegraphics[width=0.32\textwidth]{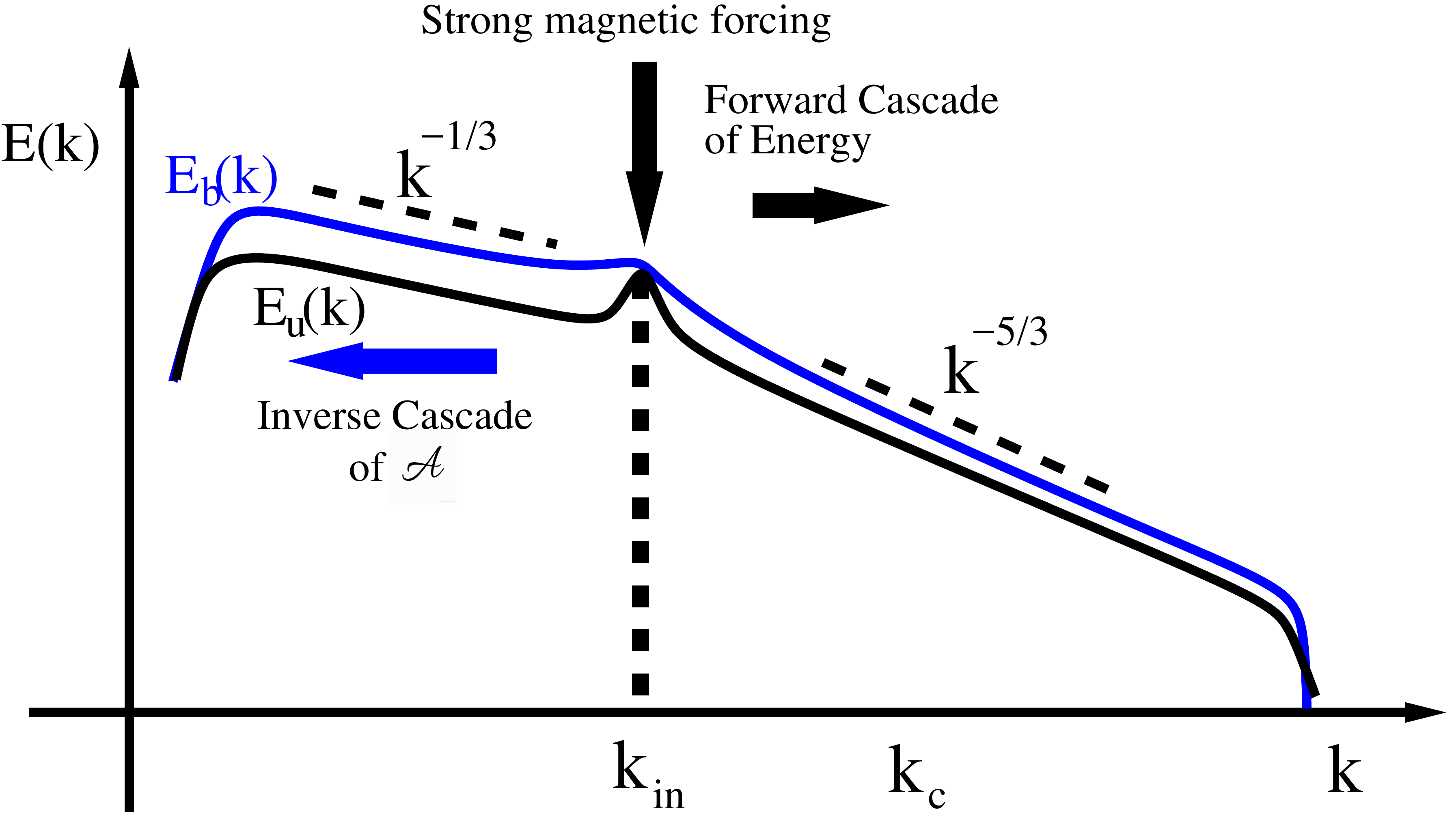}                                     %
\caption{ The kinetic and magnetic energy spectra for weak (left), intermediate (center)    %
and strong (right) magnetic forcing.                                                        %
\label{fig:2DMHDSpectra}  }                                                                 %
\end{figure*}

As the magnetic forcing is further increased, the magnetic field becomes active at larger and larger scales until the flow becomes magnetically dominant with a forward cascade of $\cE$ and an inverse cascade of $\cA$ and the spectra (\ref{eq:2dmhdSpec1},\ref{eq:2dmhdSpec2}) are valid for the two cascading ranges. There are thus three regimes for 2D-MHD flows, 
(i) for very weak magnetic forcing it behaves like 2D hydrodynamics  at all scales.
(ii) for intermediate magnetic  forcing the system behaves like 2D hydrodynamics up to scales $k_c^{-1}$ and like a 2D-MHD at smaller scales, and 
(iii) for strong magnetic forcing  the system behaves as 2D-MHD at all scales. Spectral properties for all
three regimes are summarised in Fig. (\ref{fig:2DMHDSpectra}). 

\begin{figure*}[h!]                                                                         %
\centering                                                                                  %
\includegraphics[width=0.5\textwidth]{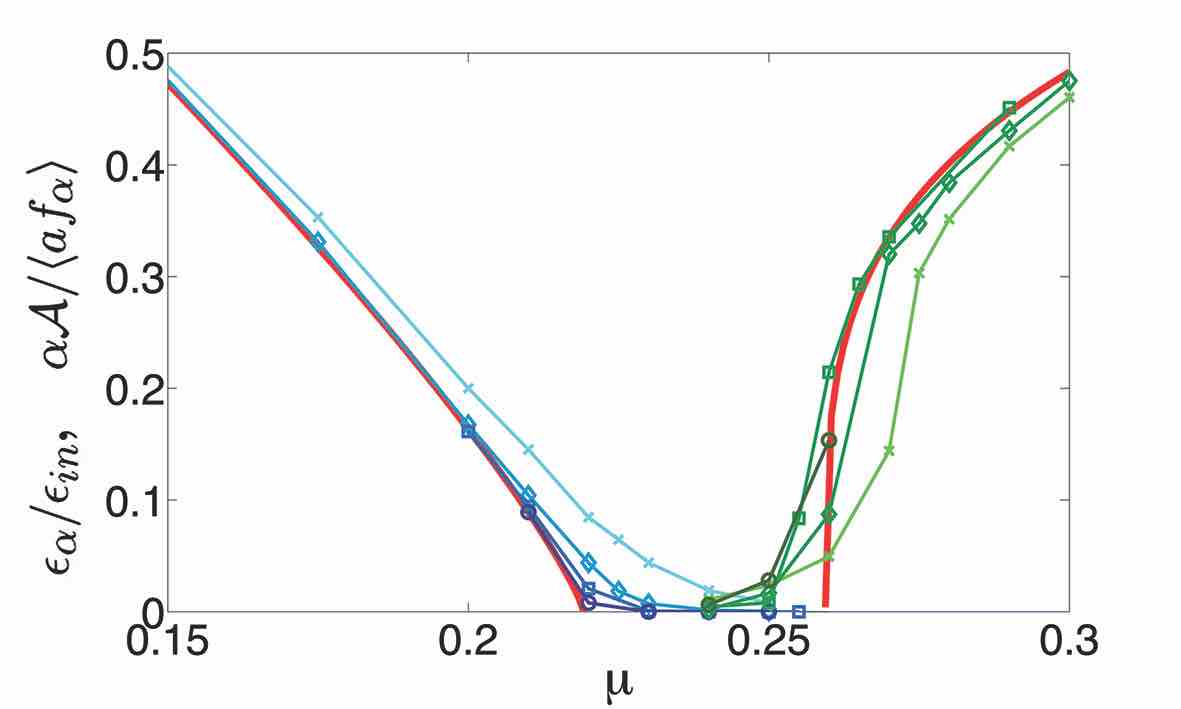}                               %
\includegraphics[width=0.4\textwidth]{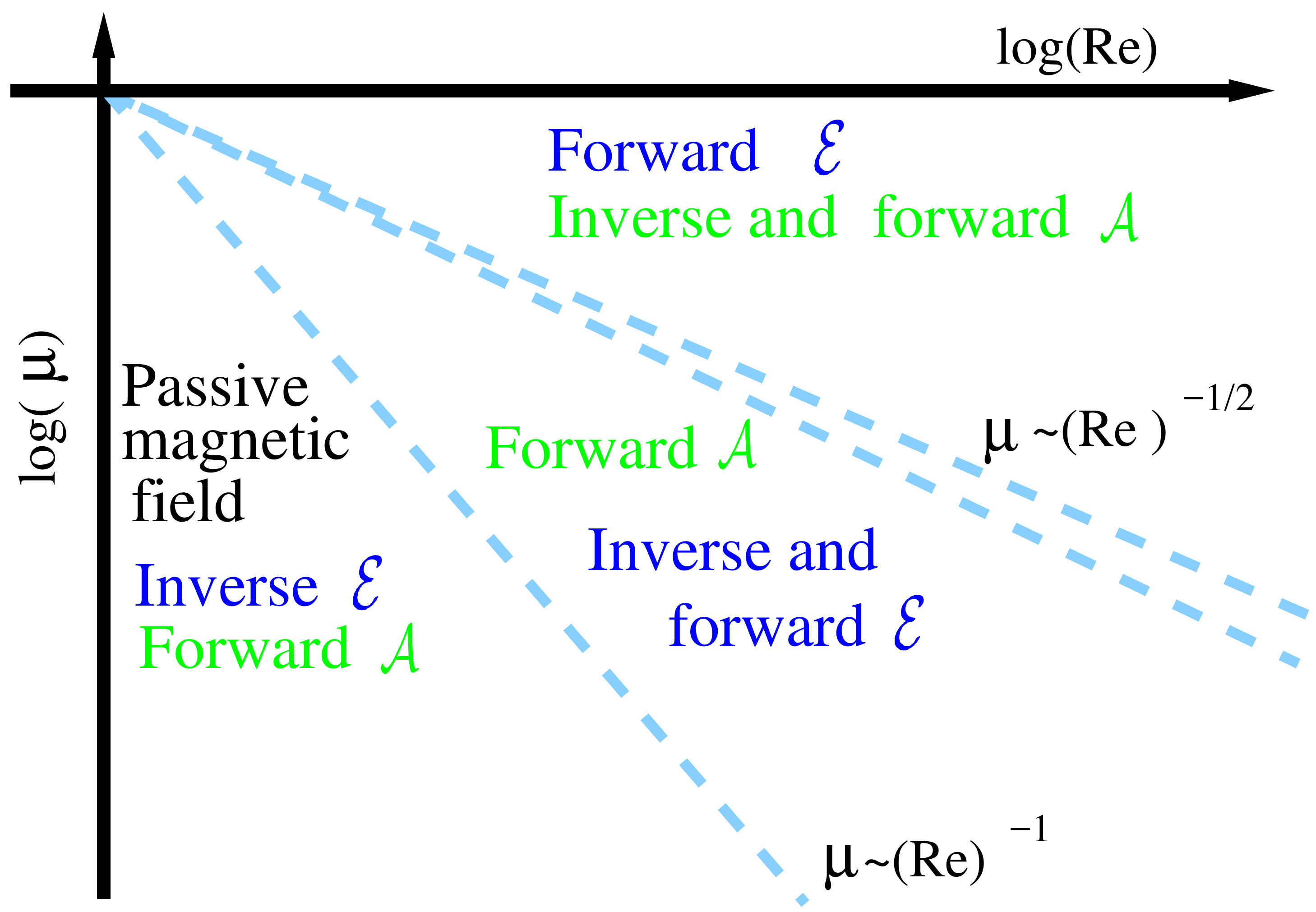}                                 %
\caption{                                                                                   %
Left Panel: Amplitude of the relative large-scale drag energy dissipation rate (blue lines) and square vector potential (green lines) as a function of    %
$\mu=\|{\bbf_b}\|/\|{\bbf_u}\| $. Dark lines imply larger box sizes.                        %
The red solid line corresponds to a fit for the infinite box size                           %
limit extrapolation, supporting the existence of two distinct critical values               %
$\nu_c$ one for the transition of the energy cascade and one for the transition             %
of the square vector potential cascade. Right Panel: Summary of the phase space at changing %
the injection rate ratios and the Reynolds number (with $Re=Rm$). 
The figures are based on the                                                                %
results of \cite{Seshasayanan2014edge, Seshasayanan2016critical}. }                         %
\label{fig:2DMHDTran}                                                                       %
\end{figure*}                                                                               %
The transition from a forward to inverse cascade was investigated in \cite{Seshasayanan2014edge, Seshasayanan2016critical} using the  relative amplitude of the forcing  $\mu=\|{\bbf_b}\|/\|{\bbf_u}\|$ as a control parameter.
 The left panel of Fig. (\ref{fig:2DMHDTran}) shows how the amplitude of the inverse cascade changes both for the energy (blue lines) and for the square vector potential (green lines).  In agreement with the previous discussion, it was shown  that  for $\mu<Re^{-1}$ the magnetic field acts like a passive scalar, while for larger values a split cascade of energy is formed. The transition from split to forward cascade of $\cE$ and from a forward to a split cascade of $\cA$ was found empirically to occur when 
$$
\mu_c \propto Rm^{-1/2}.
$$  
This result implies that in the limit $\Red,Rm\to \infty$ there is a forward energy cascade and inverse vector potential cascade for any non-zero value of $f_a$
which is consistent with the results in (\ref{eq:fluxConstraintAB1}) for the inverse cascade of the square vector potential. The phase diagram of the transition is shown in the right panel of Fig. (\ref{fig:2DMHDTran}).

\subsubsection{Quasi-static MHD flows}                                   
\label{sec:quasistatic}
\begin{figure*}[h!]                                                                         %
\centering                                                                                  %
\includegraphics[height=0.4\textwidth,width=0.45\textwidth ]{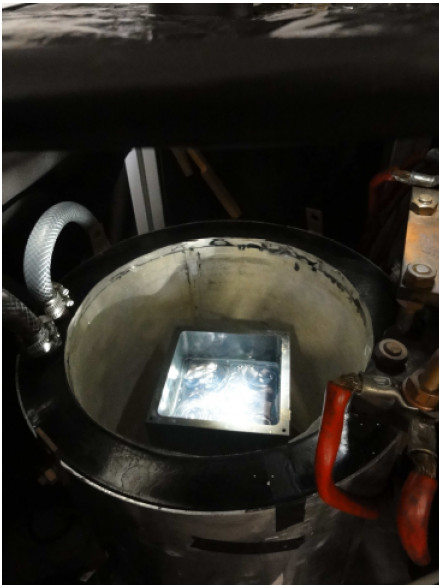} \quad %
\includegraphics[height=0.4\textwidth,width=0.45\textwidth]{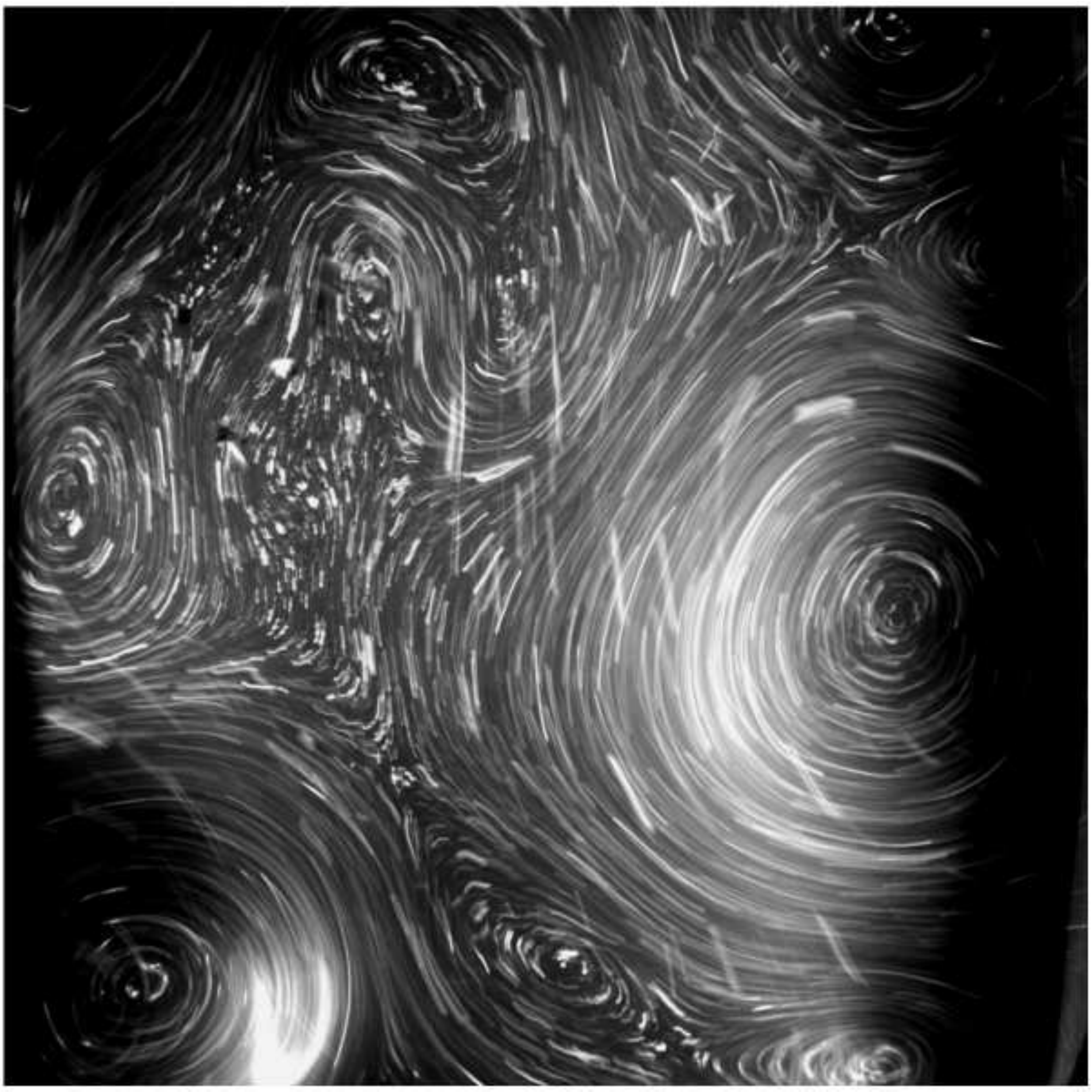}         %
\caption{ Left panel: View of an experimental set-up to study liquid metal flows in the     %
quasi-static limit. Right panel: Visualization by tracer particles of the resulting large-scale vortex in the presence of an inverse cascade. The images provided by J. Herault       %
display the experimental apparatus used in the works                                        %
\cite{gallet2012reversals,herault2015experimental,michel2016bifurcations}.   }              %
\label{fig:labJH}                                                                           %
\end{figure*}

Quasi-static MHD flows refers to a special limiting case of MHD for which the magnetic Reynolds number of the flow is very small but there is a strong magnetic field that sustains magnetic fluctuations. This limit is applicable to flows of liquid metals that have very small magnetic Prandtl numbers $P_M=Rm/Re=\mathcal{O}(10^{-5})$ and has received considerable attention due to its industrial applications. Furthermore, because liquid metals are more easily accessible in the lab than plasmas, the quasi static limit of MHD is easily investigated  in the lab \cite{alemany1979influence, sommeria1982and, sommeria1986experimental, tabeling1987instability, potherat2014why,gallet2012reversals, herault2015experimental, michel2016bifurcations, baker2018inverse, Shats2010turbulence} as can be seen in Fig. (\ref{fig:labJH}). 

In limit of $Rm \ll 1$ but with a uniform magnetic field  strong enough (so that $B_0\lin/\eta$ is of order one) the induction equation reduces to a balance between the diffusive and the magnetic shearing terms of the uniform component due to velocity fluctuations. We thus obtain  $ {\bf B_0 \cdot \bnabla u}   = - \eta \Delta {\bf b}$. Solving for the magnetic fluctuations and substituting this balance in the NSE we obtain 
\be
\partial_t {\bf u} + {\bf u \cdot \bnabla u} = -
{ \frac{B_0^2}{\eta} \nabla^{-2} \partial_z^2 \bf u } 
-\bnabla P +\nu  {\Delta} {\bf u} +  {{\bbf_u}}.   \label{eq:MHDlowPm} 
\ee
We have thus reduced to a single equation where the effect of the magnetic field is limited to the first term on the right-hand side of (\ref{eq:MHDlowPm}). This is a dissipative term that removes energy from velocity fluctuations that vary along the direction of the magnetic field. The energy balance equation therefore reads: 
\be
\partial_t \cE = \ein -\enu -\emu -  \frac{B_0^2}{\eta}  \langle ( {\nabla^{-1} \partial_z \bf u})^2  \rangle. 
\ee
Note that the Ohmic dissipation  affects  3D modes that vary in the direction of the field $\bB_0$ only. If the latter are not forced, and if a sufficiently large magnetic field is used, the oOmic dissipation will render the flow 2D. As in the previously discussed cases varying the strength of the magnetic field the flow can transition from a 2D state that cascades energy inversely to a 3D state that cascades energy forward. For a system where  only the 2D modes are forced, the transition from  forward  to  inverse cascade happens at the onset of the instability of a 3D mode, when the dissipation time scale due to the Lorentz force $\eta H^2/(B_0^2 \lin^2)$ becomes comparable to the eddy turnover time 
$\lin/u_f \sim \ein^{-1/3} \lin^{2/3}$ both evaluated at the forcing scale. Equating these two
time-scales we can predict that the transition occurs when: 
\be 
B_0^2 = \frac{ \eta \ein^{1/3} H^2 }{ \lin^{8/3} } ,
\label{eq:B2cr} 
\ee
where $H$ is the height of the layer.
These different states have been demonstrated in numerous numerical simulations \cite{reddy2014strong, verma2015modeling, reddy2014anisotropic, verma2017anisotropy, burattini2008spectral, burattini2008anisotropy, knaepen2008magnetohydrodynamic, favier2011quasi}.
We should note that unlike for the cases of fast rotating flows and  MHD at large $Rm$, where 
the tendency toward a 2D dynamics is the result of fast-decorrelating waves, in this case the 3D modes are directly damped. This renders the equations analytically more tractable  and one can rigorously prove
the  two-dimensionalization \cite{Gallet2015exact}. 
Besides the analytical treatment, there exist many  experiments that   test both the MHD predictions and   2D turbulence properties 
\cite{sommeria1982and, sommeria1986experimental, tabeling1987instability, potherat2014why, gallet2012reversals, herault2015experimental, michel2016bifurcations, baker2018inverse, Shats2010turbulence}. 
In these experiments, a layer of a liquid metal is embedded in a uniform magnetic field while smaller magnets are placed under the layer. Dividing a current either across the layer or through an array of contacts one controls the
injection  mechanism  via the Lorentz force. Depending on the strength of the magnetic field  this system can give rise to an inverse energy cascade. 
This setup was one of the first that demonstrated an inverse cascade whose strength can be varied with a control parameter.

\subsubsection{ Magnetic helicity }  
\label{sec:mhelicity}
In this section, we discuss the effects of  magnetic helicity for 3D MHD.
Like kinetic helicity, the magnetic helicity is a sign-indefinite quantity and does not provide a restriction on the direction of the energy cascade. The cascade of the magnetic helicity itself can not be forward in the presence of a forward energy cascade. Using the same arguments as in (\ref{eq:fluxConstraintAB1}) one can show that the absolute value of the magnetic helicity flux has to decrease at large $k$ \cite{berger1984rigorous,taylor1974relaxation}. A recent rigorous proof of the absence of a forward cascade of magnetic helicity can be found in \cite{aluie2017coarse}, see also the discussion provided in \cite{linkmann2017triad}. 
The inverse cascade of helicity was first shown within turbulent EDQNM models \cite{frisch1975possibility,pouquet1976strong} leading to a power-law behaviour for the  magnetic helicity,  $H_M(k)$, and energy $E(k)$ spectra for $k < k_{in}$:
\be
\label{spectrahelicity}
H_{\cM}(k) \propto   \mu_\alpha^{2/3} k^{-2} \quad \mathrm{and} \quad E(k)  \propto \mu_\alpha^{2/3} k^{-1}.
\ee
These spectra can be obtained using the same arguments discussed for  the kinetic helicity
case in Sec. (\ref{sec:Helicity}) and further assuming that velocity and magnetic field have similar
amplitudes $\delta_\br \bu \sim \delta_\br \bb$. The inverse cascade of magnetic helicity has been demonstrated in various simulations \cite{pouquet1978numerical, meneguzzi1981helical, kida1991statistical, brandenburg2001inverse, gomez2004direct, Alexakis2006mhelicity}. Nonetheless, recent simulations at large scale separations have shown strong deviations from this spectral exponent and find $H_M(k) \propto k^{-3.3}$
\cite{muller2012inverse, malapaka2013large}. The difference was attributed to the presence of kinetic helicity and different modeling schemes have been proposed  to account for the  deviations from the
dimensional prediction (\ref{spectrahelicity})  \cite{muller2012inverse}.

In configuration space, one can derive a series of inertial relations for third-order moments that are linked to the cascade of magnetic helicity. These have been examined by \cite{politano2003karman} and  more recently for anisotropic turbulence in \cite{podesta2008laws, podesta2007anisotropic, galtier2008karman}:
\be
\la { \bb \cdot (\delta_\br \bb \times \delta_\br \bu } )\ra   =  \iepsilon_\cM. 
\ee
As in (\ref{eq:mhd45}), the presence of single-point quantities together with  velocity and magnetic field increments makes  the derivation of a scaling law difficult, if not impossible.
\begin{figure*}[h!]                                                                        
\centering                                                                                 
\includegraphics[width=0.70\textwidth]{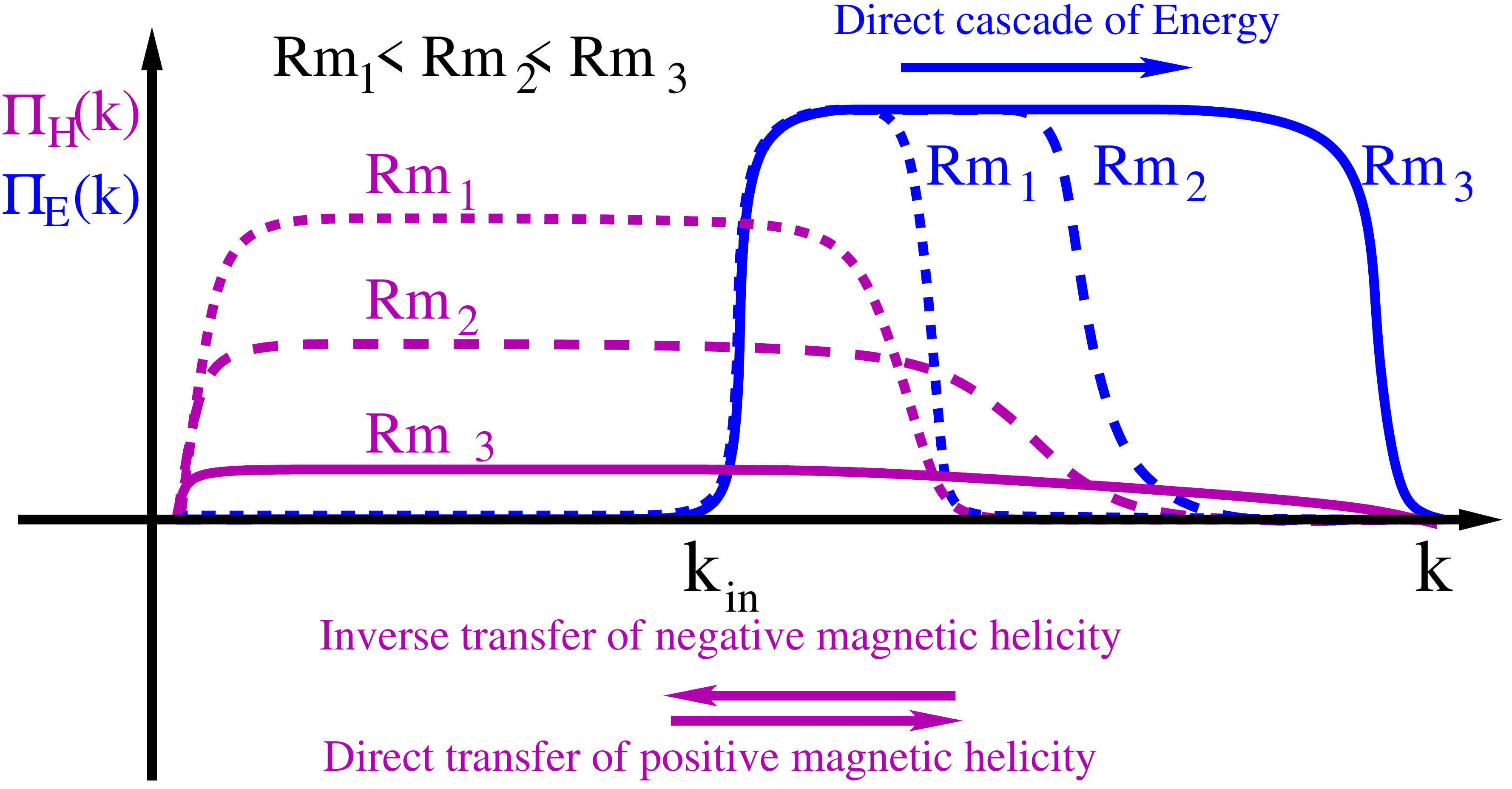}                                    
\caption{The flux of energy (in blue) and of magnetic helicity (in purple)
  in a dynamo (no magnetic forcing) driven by a       
         positively helical flow for three                                                 
         different magnetic Reynolds numbers $Rm_1<Rm_2<Rm_3$. The amplitude of the        
         inverse/direct flux of magnetic helicity retains the same sign across all scales  
         but is decreased as $Rm\to\infty$. }                                              
\label{fig:Mhelly}                                                                         
\end{figure*}                                                                              
It is also worth examining the case of dynamo flows for which there is no magnetic forcing in (\ref{MHD2}) and
the magnetic field is sustained by the dynamo of the fluid motions alone. In this case,  magnetic helicity is not externally supplied, $\mu_{in}=0$,  but it can be generated by the Ohmic forces. This will lead to the stationary balance
\be
\mu_\alpha=-\mu_\eta
\label{eq:balancemhd}
\ee
from (\ref{eq:magnetichelicity}).
For finite values of  $Rm$, one can imagine that Ohmic forces generate helicity at small scales, which then
cascades up to the large scales where it is dissipated by the drag term.  
This is indeed what happens in dynamo flows at finite $Rm$ when the flow has kinetic  helicity, as predicted by the {\it alpha dynamo} theory \cite{parker1955hydromagnetic, steenbeck1966berechnung, moffatt1978field} that provides the large-scale magnetic field evolution in  the kinematic regime when the magnetic field is advected passively. 
In a dynamo, positive kinetic helicity generates large-scale negative magnetic helicity
and small-scale  positive magnetic helicity, keeping the averaged helicity generation due to non-linear interactions
zero. The positive helicity that arrives at the small scales is dissipated by Ohmic dissipation while the negative magnetic helicity by the drag term such that the balance (\ref{eq:balancemhd}) is satisfied . Thus, although there is no injection of magnetic helicity by the non-linearity, the flow  self-organizes such that positive helicity is preferentially dissipated at small scales. This process can be viewed as an inverse transfer of negative magnetic helicity (or equivalently a forward transfer of positive helicity). The opposite happens when the  dynamo flows is of  negative kinetic helicity. Figure (\ref{fig:Mhelly}) demonstrates this mechanism.  In the absence of a drag term, $\mu_\alpha=0$, the negative magnetic helicity will pile up at large scales forming a condensate, see Def. (\ref{def13}) of Sec. (\ref{sec:definitions}).

However, the magnetic helicity transfer can not continue for arbitrary large $Rm$ because at small scales the flux of helicity is constrained in absolute value by the forward energy cascade (see \ref{eq:fluxConstraintAB1}) so eventually the amplitude of the flux of helicity must  go to zero as $Rm \to \infty$. Thus, although a constant flux of negative/positive magnetic helicity can be built from the smallest diffusive scales to the largest scales of the system this process is not a cascade according to the definitions that we have given in Sec. (\ref{sec:CascadeDef}) because in the large  $Rm$ limit  the amplitude of the magnetic helicity flux vanishes. 

Before concluding this section it is worth noticing  that the combined effects of kinetic and magnetic helicity can be studied by applying the same chiral decomposition (\ref{upm})  studied in Sec. (\ref{sec:Helicity}) to both velocity and magnetic fields. In the MHD case, the number of possible different Fourier-helical triads becomes larger and the analysis more involved. A first series of attempts in this direction can be found in \cite{linkmann2016helical,linkmann2017effects,linkmann2017triad}.

\subsubsection{Summary}
In this section we discussed the different limits of MHD turbulence  and reviewed the different phenomenological predictions and recent numerical and experimental results.
For isotropic flows we summarized the different exact relations for third-order quantities connected to the flux of all ideal invariants. For the anisotropic case, we discussed three different limiting configurations: strong, weak and quasi-2D that appear in layers of finite thickness and which can lead to an inverse cascade of energy. In 2D we also discussed the transitions from a forward to an inverse cascade that can occur due to the presence of the additional conservation of the square vector potential. A particular limit that deserves further attention is that of quasi-static MHD where most liquid metal experiments operate. Finally, we discussed the subtle case of magnetic helicity transfer that despite the agreement of phenomenological approaches with numerical simulations for the direction of the cascades, the resulting spectra are still far from  all predictions.


\subsection{{Passive and active scalars} \label{sec:passive}}  

In this section we review the main findings concerning the transfer properties of passive and active scalars in turbulence by changing the dimensionality of the
embedding space and the statistical properties of the advecting velocity field. 
Active/passive advection/transport/diffusion equations describe a huge set of natural phenomena, ranging from atmospheric physics \cite{pasquill1977atmospheric}, combustion \cite{peters2000turbulent} transport and amplification of magnetic fields \cite{moffatt1978field} or transport of density in stratified fluids  as already discussed in Sec. (\ref{sec:stratification})  and (\ref{sec:MHD}). Here we limit ourselves to review what is known for some of the most generic and important configurations, as for the case of the
advection of a dye or contaminant in 2D and 3D compressible and incompressible velocity fields, for the passive  case, and for  2D magnetic potential, 2D  vorticity  and potential temperature in surface flows, for a few paradigmatic active cases.  In particular, most of the physics concerning the active cases has already been discussed in the previous sections, here we revisit some of the previous results by putting them {\it face to face} with the evolution  of a passive field advected by the same velocity configuration in order to highlights the main differences/similarities. \\Detailed results about the most rigorous aspects can be found in \cite{crisanti1991lagrangian, falkovich2001particles, celani2004active, haynes2005controls,vanneste2006intermittency, drivas2017lagrangian}, experimental and numerical results concerning passive scalar advection can be found in \cite{warhaft2000passive, sreenivasan1991local, donzis2010batchelor, gotoh2015power, gotoh2011universality, donzis2010resolution}.
\begin{figure*}[htbp]                                                                        %
\centering                                                                                   %
\includegraphics*[width=0.95\textwidth]{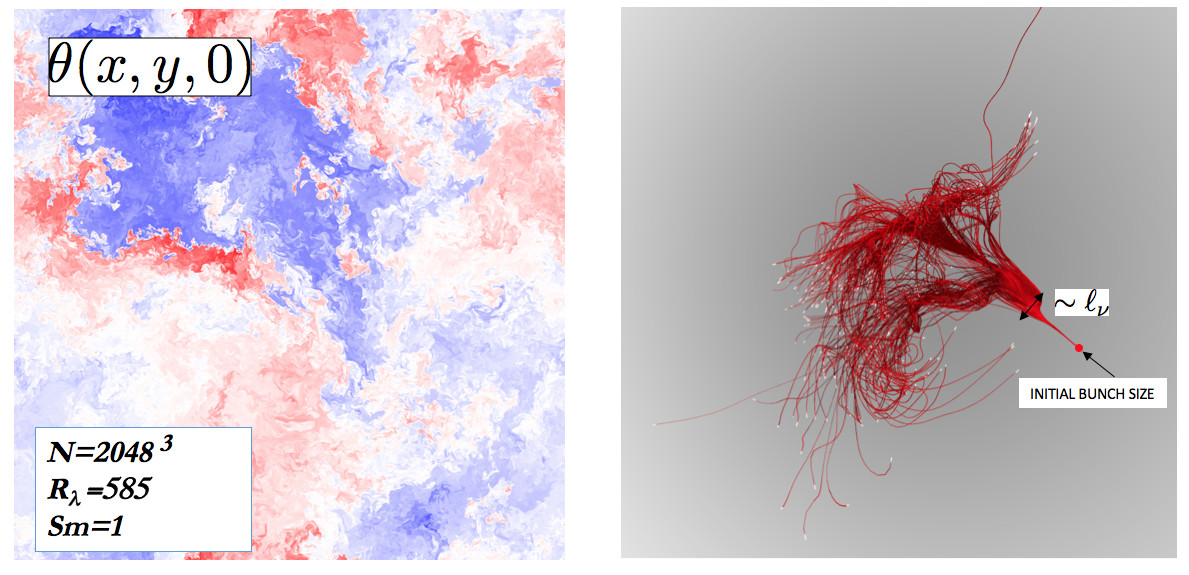}                             %
\caption{Left: A 2D Eulerian                                                                 %
snapshot from a  3D DNS of a passive scalar in a highly turbulent flow, with the characteristic  %
front/cliff structures where scalar dissipation is concentrated and leading to a direct      %
cascade for the passive scalar variance. Courtesy of T. Gotoh and T. Watanabe, adapted from \cite{watanabe2007scalar}.%
Here $R_\lambda  = u_{f} \lambda/\nu$ where $\lambda$ is the Taylor scale defined in terms of 
the rms velocity gradient \cite{Frisch}. Right: Lagrangian evolution of a bunch of tracer particles injected in a  tiny %
region of size $\ll \ell_\nu$. In the first stage of the evolution, until the bunch remains  %
smaller then the dissipative scale $\ell_\nu$, the dispersion is deterministic and           %
exponential. At a late time, the typical distance among particles is $> \ell_\nu$, the       %
underlying flow increments become rough and  the bunch blows up, with trajectories of nearby %
particles that reach distances $r \gg \ell_\nu$ in a finite time and with a                  %
{\it  spontaneous} stochasticity behaviour (Courtesy of R. Scatamacchia).} \label{fig:Gotoh} %
\end{figure*}                                                                                %
 The linearity of the equations in the passive case allowed considerable analytical progress and many important theoretical achievements have been obtained, in the last two decades.  The problem is considered solved in some idealized cases. In particular, this is the case for the advection of a passive scalar in a  stochastic, self-similar,  Gaussian and delta-correlated in time velocity field, the so-called Kraichnan model \cite{kraichnan1994anomalous}. \\ Analytical and rigorous progress has been made both using Eulerian and Lagrangian approaches, i.e. following the evolution of tracers particles along fluid streamlines. In Lagrangian coordinates, the existence of {\it spontaneous stochasticity}, i.e. the explosive separation  of two nearby trajectories in the presence of rough advecting velocity field is connected to the existence of the dissipative anomaly for the passive field  and to a direct cascade (see Fig. \ref{fig:Gotoh} for a visualization of  a passive scalar Eulerian configuration \cite{watanabe2004statistics} and for a Lagrangian evolution of a particle bunch \cite{scatamacchia2012extreme}). Phase transitions from direct to inverse cascades are then possible as a function of the degree of compressibility of the advecting field and its dimensionality \cite{gawedzki2000phase,Celani2010turbulence}. 
 In Eulerian coordinates intermittent corrections and their universality can be analytically approached as it will be briefly discussed in Sec. (\ref{sec:intermittency}).\\
The presence of an active feedback on the flow leads to a plethora of new phenomena that can change the transfer direction depending on the coupling with the velocity field and/or the degree of correlation among the external forcing field and the advecting velocity.

 \subsubsection{Background definitions}
 The typical setup describing the  combined evolution of a turbulent flow  and advected scalar field $\theta(\bx,t)$ is described by the following set  of equations:
\begin{align}
  \partial_t {\bf u} + {\bf u \cdot \bnabla u}      &= -\bnabla P + \nu  \Delta {\bf u} -  \alpha  \bu + {\bf f}  \\ 
  \partial_t \theta + {\bf u \cdot \bnabla } \theta &=  \kappa  \Delta \theta  -  \alpha_\theta  \theta + f_\theta  \label{eq:activepassive}
\end{align}
where we have introduced the scalar diffusion coefficient, $\kappa$ and the large-scale scalar sink, $\alpha_\theta$ needed to reach stationarity in the presence of inverse cascades (if any). It is important to remark that in the presence of a compressible velocity field, a passive  concentration, i.e. the density of a pollutant, evolves in a different way from $\theta(\bx,t)$ in (\ref{eq:activepassive}) because of the presence of the extra term $({\bf \nabla \cdot u} ) \theta$ on the LHS of the evolution equation.  
 The field $\theta(\bx,t)$ in (\ref{eq:activepassive}) is considered as active and not passive if at least one of the following three conditions is satisfied:\\   
\noindent (i) it enters in the definition of the velocity forcing mechanism:
\be
\label{eq:convection}
 \bfo \to \bfo[\theta, {\bf \nabla}\theta, \dots],   
\ee
 \noindent (ii)  there exists a functional relationship that gives $\bu$ in terms of $\theta$:  
\be
\label{eq:2d}
  \bu (\bx,t) = \int d^dy {\bf K}(\bx-\by) \theta(\by,t)  
\ee
\noindent (iii)  scalar and velocity injections are correlated, i.e. their PDFs are such that:
\be
\label{eq:2d3c}
P[f_\theta,\bfo] \ne P[f_\theta] P[\bfo].
\ee
Convection is a paradigmatic case belonging to  (\ref{eq:convection}), while
(\ref{eq:2d}) is realized  for  the relation among vorticity and velocity in 2D or  for the evolution of density fluctuations in the
quasi-geostrophic approximation of stably stratified fluids, where the scalar is coupled to the stream function, $(u_x,u_y) =  (\partial_y \psi,-\partial_x \psi)$,  of the flow by the relation:
\be
\label{eq:SGS}
 \theta(\bx,t) = (-\Delta)^{1/2} \psi(\bx,t).  
\ee
The  case (\ref{eq:2d3c}) occurs when the velocity field and the scalar injection are strongly correlated.
One example is the advection of a scalar $\theta$ in a 2D flow for which $f_\theta=f_w=\hat{{\bf z}} \cdot
\bnabla \times \bbf$ \cite{linkmann2018nonuniversal}. In the latter case, the field $\theta(\bx,t)$ cannot be considered {\it active} strictly speaking, because it does not influence directly the evolution of $\bu(\bx,t)$. Still because the two fields $\theta$ and $w=  \hat{{ \bf z}} \cdot \bnabla \times \bu$
follow the same equation, the correlation among the two inputs make the problem different from the one of a passive advection where the forcing term must be independent  of the velocity field. \\ 
{For incompressible flows and in} the absence of sinks and inputs, the equation
(\ref{eq:activepassive}) conserves all moments, $ \partial_t \langle \theta^n\rangle =0$, as one can easily verify considering that the scalar field is conserved along  velocity streamlines:
\be
\label{eq:advection}
\frac{d \theta(\bX(t),t)}{d t} = 0; \qquad 
\ee
 when $f_\theta,\kappa, \alpha$ vanishes and we have denoted with
\be
\label{eq:tracer}
\frac{ d \bX (t)}{dt} = \bu(\bX(t),t)
\ee 
the evolution of a  tracer  in the flow. Of course, the same  happens for the active field under similar conditions. \\
Beside Reynolds number, there exists a dimensionless number that characterizes the evolution of the active/passive system (\ref{eq:activepassive}), which weighs the relative importance of the viscous and molecular diffusion, 
 the so-called Schmidt number: $Sm = \nu/\kappa$ {(also often referred as the Prandtl number)}. { Alternatively we can define the P\'eclet number: $ Pe = (\langle |\bu|^2\rangle)^{1/2} \ell_{in}^\theta/\kappa $ where $\ell_{in}^\theta$ is the typical length scale where the scalar fluctuations are injected by the forcing $f_\theta$. In this work we only consider cases for which $\ell_{in}^\theta=\lin$ 
in which case $Pe=Re Sm$. A discussion of cases for which $\lin \ll \ell_{in}^\theta$ (homogenization limit \cite{majda1999simplified,keating2010homogenization}) or $\lin \gg \ell_{in}^\theta$ (Townsend Batchelor limit \cite{townsend1951diffusion,batchelor1959small}) can be found in \cite{ottinokinematics,alexakis2011bounding,doering2006multiscale,thiffeault2012using}.
The P\'eclet number is a measure of the relative importance of velocity advection and scalar diffusivity in  (\ref{eq:activepassive}).  }\\
\subsubsection{Passive scalar cascades \label{sec:passivescalarcascades}} 
In the presence of a stationary  evolution, with an incompressible velocity field,  a balance among input and output of scalar variance must establish similarly to (\ref{eq:globalbalance2}) for the kinetic energy:
\begin{align}
\label{eq:balanceactive}
&  \thetain = \thetachi + \thetaalpha, 
\end{align}
where we have introduced the mean  passive scalar variance injection $\thetain = \langle \theta f_\theta \rangle $, mean variance dissipation $\thetachi = \kappa \langle  (\nabla \theta )^2 \rangle$, and large-scale sink, $\thetaalpha = \alpha \langle \theta^2 \rangle$. 
For the two-point correlation function in a homogeneous statistics,  $C_2^\theta(\br) = \langle \theta(\br)\theta(0)\rangle$, one can  write the equivalent of (\ref{eq:KHMstationary}) to get:
\be
\label{eq:passiveKHM}
-\frac{1}{4} {\bf \bnabla}_{\br} \cdot \la {(\delta_{\br} \theta) ^2 \delta_{\br} \bu} \ra =  \kappa_\theta \Delta_r {C}^\theta_2(\br) +   F_\theta(\br) 
-\alpha_\theta  {C}^\theta_2(\br) 
\ee
where we have introduced the injection correlation $F_\theta(\br) =   \langle \theta(\br) f_\theta(0) \rangle$. 
To control the existence of inverse/direct scalar cascades we need to distinguish two different ways to perform the limit of vanishingly small diffusivity.  Either we send first $Re \to \infty$ (i.e. $\nu \to 0$ for fixed flow configuration) and then $Pe \to \infty$ ($\kappa \to 0$) or we send first  $Pe\to \infty$ and then $Re\to \infty$.  Let us first consider sending $Re \to \infty$ and let us suppose for the sake of simplicity that the velocity is described by a 3D turbulent homogeneous and isotropic forward cascade at all scales with a $k^{-5/3}$ spectrum. By repeating the same arguments used for the direct cascade of homogeneous and isotropic kinetic energy, one expects that if passive fluctuations are only transferred to small scales the large-scale dissipation must vanish, $\thetaalpha=0$,  and a dissipative anomaly of scalar fluctuations must exist  for  $Pe \to \infty$.  This is a direct consequence of  the fact that stationarity implies $\thetain = \thetachi$: 
\be
\label{eq:anomalyscalar}
\lim_{Pe \to \infty } \kappa \langle |\nabla \theta|^2\rangle  = \thetachi = Const.   
\ee
As a result, for any finite $Pe$, we can define  a typical length scale $\ell_\kappa \sim k_\kappa^{-1}$ where the dissipative term starts to counterbalance the advection. In the inertial range of scales for the scalar field, $ \ell_{in}^\theta >  r > \ell_\kappa$, we can further simplify the exact relation (\ref{eq:passiveKHM}) by noticing that the diffusive term, $ \kappa \Delta_r {C}^\theta_2(\br) \to 0 $
for fixed $r$ and $\kappa \to 0$, and that the forcing-scalar correlation is dominated by its value for $r=0$, $F_\theta(\br) \sim F_\theta(0) + O(r^2) \sim \thetain$.
As a result, from (\ref{eq:passiveKHM})  one gets the celebrated isotropic inertial-range  Yaglom relation concerning the third order correlation made of   longitudinal velocity increments and  scalar increments:
\be
\label{eq:yaglom}
\langle \delta_\br u (\delta_\br \theta)^2 \rangle = -\frac{4}{3} \thetain  r,
\ee
where $\delta_\br u$ refers to the longitudinal field increment as introduced in Sec.  (\ref{sec:HIHT}). For a pure  inverse scalar cascade, one proceeds by repeating exactly the same balancing used to lead to (\ref{eq:45inverse}), now we assume $\thetachi=0$ and the balancing in (\ref{eq:passiveKHM}) for $r > \ell_{in}^\theta$ will be among the LHS and the large-scale scalar sink, leading to 
\be
\label{eq:yaglominverse}
\langle \delta_\br u (\delta_\br \theta)^2 \rangle = \frac{4}{3} \thetaalpha r .
\ee
The exact scaling relations (\ref{eq:yaglom}) and (\ref{eq:yaglominverse}) are not closed in terms of the scalar fluctuations only, {since they depend on the velocity field as well}. Hence,  it is not possible to derive an explicit expression for the scalar spectrum (or for high-order scalar correlations). Standard dimensional and phenomenological arguments, based on the idea that there exists the equivalent of the Richardson cascade also for the scalar field, lead
to prediction that $ \langle \delta_\br u (\delta_\br \theta)^2 \rangle \sim \langle (\delta_\br u)^2 \rangle ^{1/2}
\langle (\delta_r \theta)^2 \rangle $ and  to:
\be
\label{eq:k41scalar}
\langle (\delta_\br \theta)^2 \rangle  \sim \ein^{-1/3} \thetain r^{2/3}
\ee
where we have assumed the inertial range  K41 scaling for the velocity fluctuations, $  \langle (\delta_\br u)^2 \rangle \sim \ein^{2/3} r^{2/3}$. It is important to notice that the conceptual steps leading to (\ref{eq:k41scalar}) are   {\it complicated} by the need to control the correlation among the velocity and scalar cascades, as also evident from  the fact that the RHS of (\ref{eq:k41scalar}) is expressed in terms of a negative power of $\ein$. As a matter of fact, the entanglement among the two transfers has defeated, up to now,  any systematic attempt to go beyond the
above rough estimates, including cases when joint multi-fractal cascades  have been proposed \cite{meneveau1990joint,jensen1991intermittency}. The spectral counterpart of (\ref{eq:k41scalar}) is
given by a $-5/3$ law similar to the forward energy cascade:
\be
\label{eq:53passive}
E_\theta(k) \sim  \ein^{-1/3} \thetain k^{-5/3}.
\ee
It is instructive to double check the self-consistency of the derivation, by estimating the scalar diffusive scale $\ell_\kappa$ in terms of the P\'eclet number by assuming the $-5/3$ spectrum for the scalar inertial range of scales. It is easy to see that balancing the scalar diffusive term in (\ref{eq:passiveKHM}) with the inertial LHS one gets $\ell_\kappa \sim Pe^{-3/4}$. As a result, by taking first the limit $Re \to \infty $  we  found that  the infinite P\'eclet number  limit discussed above is consistent with the existence of a range of scales where $ \ell_\kappa < r < \ell_{in}^\theta$ because, $\ell_\kappa \to 0$. For $ \ell_\nu < r < \ell_\kappa$ we have for the scalar a balance between diffusion and advection by a rough velocity field,  a case discussed in  \cite{batchelor1959smallb} for which the spectrum $E_\theta(k) \sim k^{-17/3}$ is predicted.

If we change the order of the limits and take first $Pe \to \infty$, we need to consider that the viscous  Kolmogorov scale is not vanishingly small,  $\ell_\nu > 0$.  For $Pe$ larger and larger, the scalar diffusivity becomes negligible and we need to study the advection of a scalar in a smooth velocity field, $ r < \ell_\nu$. The advection of a (passive) scalar by a smooth incompressible stochastic velocity field is another important application, first studied by Batchelor in the 50's \cite{batchelor1959small,monin2013statistical}. In this case,  the spectrum of the scalar is characterized by a $E_\theta(k) \sim k^{-1}$ scaling range, corresponding to logarithmic scaling properties for the two-point correlation function in  configuration space.   

We can summarize the phenomenological prediction for the advection of a passive scalar advected by an homogeneous and isotropic incompressible  turbulent flow in the phase space (Re, Pe)  as follows.\\
If $\ell_\kappa \gg \ell_\nu$  (small Schmidt number regime) we  have :
\begin{equation}
\begin{cases}
   r  \gg \ell_\kappa \gg \ell_\nu        \to  C_2^\theta(r)   \sim r^{2/3} \qquad    &\mathrm{(convective-inertial)} \\
       \ell_\kappa \gg r   \gg \ell_\nu   \to  C_2^\theta(r)   \sim r^2     \qquad    &\mathrm{(diffusive -inertial)}\\
     \ell_\kappa \gg \ell_\nu \gg r       \to  C_2^\theta(r)   \sim r^2     \qquad    &\mathrm{(diffusive -dissipative)}
\end{cases}
\end{equation}
If $\ell_\kappa \ll \ell_\nu$ (large Schmidt number) we  have :
\begin{equation}
  \begin{cases}
   r  \gg \ell_\nu \gg \ell_\kappa        \to  C_2^\theta(r) \sim r^{2/3}                \qquad  &\mathrm{(convective-inertial)}\\
   \ell_\nu \gg  r   \gg \ell_\kappa      \to  C_2^\theta(r) \sim log(r/\ell_{in}^\chi)  \qquad  &\mathrm{(convective-dissipative)}   \label{eq:batchelor} \\ 
   \ell_\nu         \gg  \ell_\kappa > r  \to  C_2^\theta(r)   \sim r^2                  \qquad  &\mathrm{(diffusive-dissipative)}.
\end{cases}
  \end{equation}
 The scaling behaviour for the scalar spectrum is also summarized in Fig. (\ref{fig:scalarspectra}) below.
\begin{figure*}[htbp]                                               %
\centering                                                          %
\includegraphics*[width=0.45\textwidth]{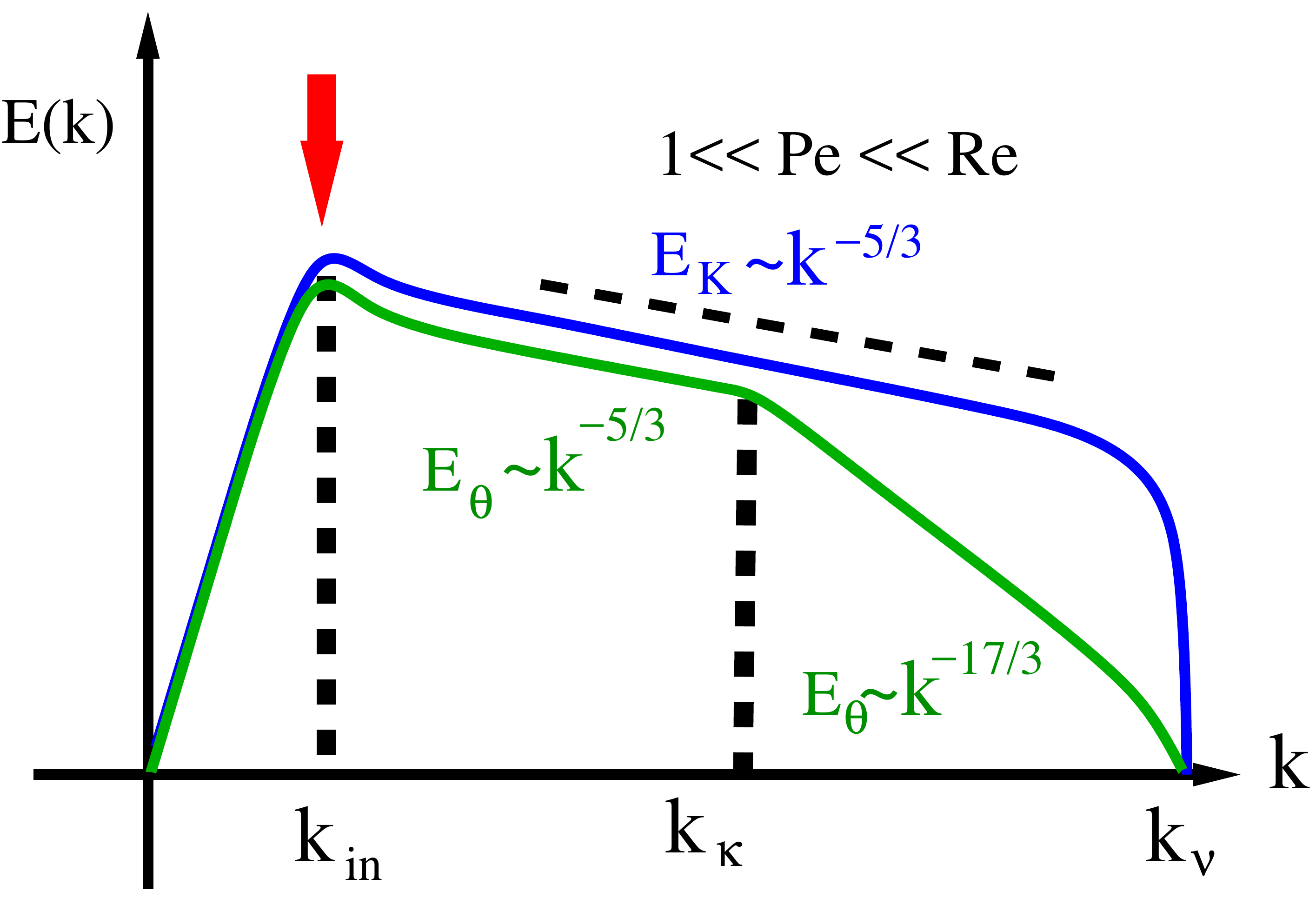}           %
\includegraphics*[width=0.45\textwidth]{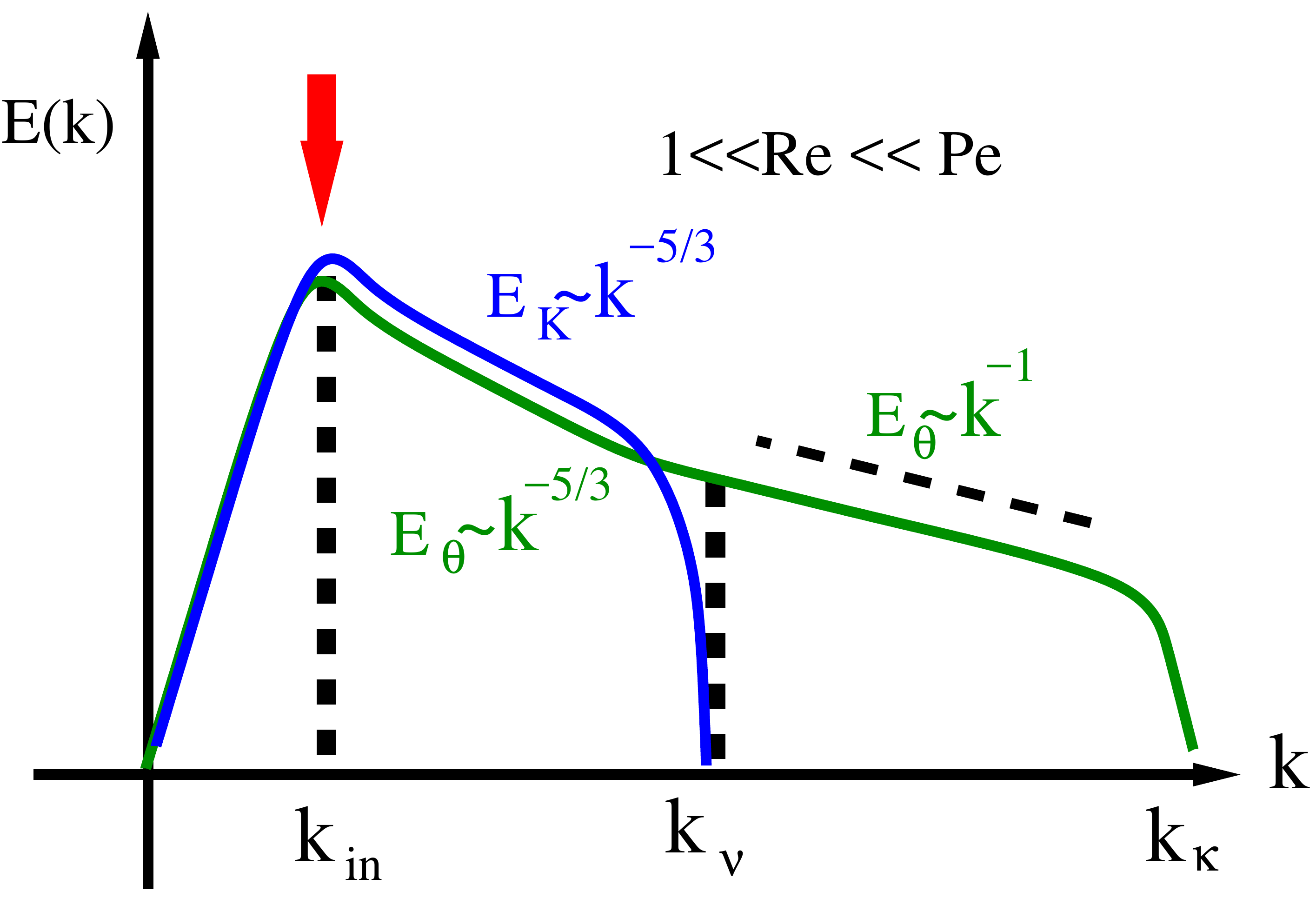}           %
\caption{ {Kinetic energy spectra, $E_K(k)$,  and scalar variance spectra, $E_\theta(k)$, for   %
small $\Sm$ (left) and large $\Sm$ (right) numbers.}}                 %
\label{fig:scalarspectra}                                           %
\end{figure*}                                                       %
\subsubsection{Active scalar cascades \label{sec:activescalarcascades}}
If the advected scalar is active, it is not possible
to make  any general conclusions about its cascading  properties. 
This is evident when the scalar is active in the sense of (\ref{eq:convection}) as for the case of convection. In the latter case,  the feedback on the velocity  is given by
the buoyancy   $\bfo(\bx,t) = - g \beta \hat {\bf z} \theta(\bx,t) $, expressed in terms of the gravity, $g$, the thermal expansion coefficient, $\beta$ and the active temperature (or density)  field $\theta(\bx,t)$,  
 (see section \ref{sec:stratification}).
The velocity field as well as the correlation between the velocity and the advected field might change scaling properties depending on the  dimensions of the embedding space and on whether the system is stably or unstably stratified, as discussed in Sec. (\ref{sec:stratification}). 
This is  a paradigmatic example of what was discussed after (\ref{eq:yaglom}): the existence of an exact scaling law for the mixed third order correlation function is not enough to
make a prediction on the scaling of velocity and temperature/density separately.
Even  the direction of the cascades of  active and passive scalars, transported by the same velocity field, can be different, as discussed later in  Sec. (\ref{sec:MHD2D}), where the magnetic potential plays the role of an active field. 

Similarly, no universality among passive and active quantities are expected when  a functional relation among the advecting velocity and the advected scalar fields (\ref{eq:2d}) exist. 
This is well exemplified in the two paradigmatic cases of vorticity evolution in 2D turbulence and potential temperature in the surface quasi-geostrophic (SQG) approximation  \cite{pedlosky2013geophysical}. 
The dynamic equations for both cases can be written by generalizing the Navier-Stokes equation in 2D by introducing 
a relation that connects the Fourier-space velocity field, $\but$  with its vorticity $\tilde w$ such that
\be \label{eq:2DNSG}  \partial_t w + {\bf u \cdot \bnabla } w            =  \nu  \Delta w  -  \alpha  w  + f_w  \ee where    
\be
\label{eq:2dactive}
(\ut_x(\bk),\ut_y(\bk)) = - i k^{-z}(k_y,-k_x) \tilde w(\bk,t). 
\ee
The value of $z$ corresponds to different  degrees of {\it locality} in the functional relation \cite{pierrehumbert1994spectra,held1995surface,sukhatme2002surface,sukhatme2009local}. 
Indeed,  for $z=2$ we have the equivalent of the  relation connecting vorticity and velocity  $ \bw = \bnabla \times \bu$, while for $z=1$ we recover the relation among velocity and potential temperature
(\ref{eq:SGS}). The system (\ref{eq:2DNSG},\ref{eq:2dactive}) conserves the variance of the advected scalar, 
$\cA=\la w^2\ra$, and an energy-like quantity, $\cE=\la w (-\Delta)^{-z/2} w  \ra$.  The quantity $\cA$ corresponds to the enstrophy for $z=2$. 
  Note that $\cE$ has units of velocity square only for $z=2$ when the 2D Navier-Stokes is recovered. For $z=1$, $\cA$ has units of velocity square.
The spectra of the two quantities $E_\cA(k)$ and $E_\cE(k)$ are related by $E_\cA(k)= k^{z} E_\cE(k)$ and thus according to the discussion leading to (\ref{eq:fluxConstraintAB1}-\ref{eq:fluxConstraintAB2}) we can conclude that
$\cA$ cascades forward and  $\cE$ cascades inversely for $z>0$, while the opposite occurs for $z<0$. Thus, we observe a discontinuous (1st order) transition as $z$ is varied across $0$, as discussed in Sec. (\ref{sec:Classification}). Note, however, that for $z=0$ the non-linearity vanishes.  The standard dimensional arguments for the two spectra lead for $z>0$  to the predictions:
\begin{align}
E_\cE(k) \propto \iepsilon_\cE^{2/3} k^{-7/3+z/3},\qquad  E_\cA(k) \propto \iepsilon_\cE^{2/3} k^{-7/3+4z/3}, \qquad for \,\, k\ll \kin \\
E_\cE(k) \propto \iepsilon_\cA^{2/3} k^{-7/3-z/3},\qquad  E_\cA(k)  \propto \iepsilon_\cA^{2/3} k^{-7/3+2z/3}, \qquad for \,\, k\gg \kin 
\end{align}
where $\iepsilon_\cE$ is the rate of injection of $\cE$ and $\iepsilon_\cA \sim \kin^z \iepsilon_\cE$ is the rate of injection of $\cA$. The spectra reduce to the 2D turbulence spectra (\ref{eq:2dkraichnan}-\ref{eq:2denstrophy}) for $z=2$. Note that for the 2D turbulence case the spectrum of the advected quantity (the enstrophy spectrum) follows up to logarithmic corrections a $k^{-1}$ scaling as for the cascade of the variance of a passively advected field by a smooth flow. For vorticity in 2D turbulence, as already discussed in Sec. (\ref{sec:2DinvCsd}),  phenomenological and numerical results predict that in the absence of a large scale sink (Ekman friction), the enstrophy develops a direct cascade with a $-1$ spectrum corresponding to a smooth  velocity field with a  spectrum given by the $-3$ Batchelor-Kraichnan law (\ref{eq:2denstrophy}). As a result, enstrophy evolves similar to a passive quantity in a differentiable flow, developing the same Batchelor scaling as predicted by the $log(r)$ law discussed by (\ref{eq:batchelor}). Extensive numerical simulations \cite{boffetta2002intermittency} have shown that also in the presence of the Ekman friction, $\alpha \neq 0$,  the enstrophy behaves as a  passively advected scalar, at least for what concerns the scaling properties in the forward cascade range, despite  the fact that the enstrophy spectrum changes to a $E(k) \sim k^{-1-y}$ and the exponent $y$ depends on the intensity of $\alpha$, due to non-local effects induced by the large-scale drag on the entire direct enstrophy cascade range. The equivalence of the scaling properties of enstrophy and a passive scalar advected by a  2D turbulent flow cannot be proved analytically. However, phenomenological arguments  based on the observation that both fields are stretched exponentially along Lagrangian trajectories is provided in \cite{nam2000lagrangian,celani2004active}. For the SQG case ($z=1$ in Eq. \ref{eq:2dactive}) the spectrum of the advected field variance follows the spectrum of a scalar advected by a 3D turbulent flow. Numerical evidence exists where the passive and the active (the SQG density) fields  have a direct cascade but with different scaling properties \cite{celani2004active}.  

Finally, we comment on the case examined in \cite{linkmann2018nonuniversal} where the simultaneous advection of vorticity $w$ and  passive scalar $\theta$ by a 2D flow was investigated with $\nu=\kappa$. This case corresponds to 2D3C-flow  where the scalar is the vertical component of the velocity that is advected passively. It was shown that the spectrum of the scalar quantity can be strongly affected by varying the degree of correlation between the scalar and vorticity injections (\ref{eq:2d3c}).    When the passive scalar forcing $f_\theta$ is fully correlated with vorticity forcing $f_w \propto f_\theta$,  the two scalars follow exactly the same equation and it is straightforward to show that at late times $\theta \|f_w\| =w \| f_\theta\| $. Such a case corresponds to a  3D  forcing injecting maximal helicity into the system. By varying the degree of correlation of the forcing it was shown that the spectrum at large scales of the passive field changed from the thermal spectrum $k$ when the forcing were uncorrelated to that of the enstrophy, $\propto k^2 E(k) \propto k^{1/3}$. We note that the scalar field $\theta$ does not fed back to the flow. However, its persistent correlation with the  vorticity field induced by the helical forcing leads to a change in the spectral exponent. 

\subsubsection{Active {\it vs} Passive scalar cascades in Lagrangian coordinates} %
It is important to revisit the existence of direct and inverse cascades for passive and active scalars advected by a turbulent flows by using Lagrangian coordinates, i.e. by solving the advection equations (\ref{eq:activepassive})
for the scalar field by the method of characteristics. Let us for the moment put the large-scale sinks to zero $\alpha_a=\alpha_\theta=0$. In the absence of external forcing and scalar diffusivity the solution along a fluid characteristic is given by (\ref{eq:advection}), which simply states that the scalar is constant along fluid-particle trajectories. In order to take into account of the diffusivity we need to generalize (\ref{eq:tracer}) considering the stochastic evolution of all particle paths that are at point $\bx$ at time $t$:
\be
\label{eq:sto}
 \frac{d \bX(s|\bx,t)}{ds}  = \bu(\bX(s|\bx,t),s) + \sqrt{2\kappa} \dot {\bm \eta}(s);\qquad \bX(t|\bx,t) = \bx. 
 \ee
 where  $\dot {\bm \eta}$ is a Wiener vector process with  components given by
 a Gaussian variable of zero mean and delta-correlated in time. It is possible  to show \cite{frisch1999lagrangian} that
 for any realization of the external forcing $f_\theta(\bx,t)$ and of the advecting velocity $\bu(\bx,t)$ the value of the passive scalar at the position $\bx$ at time $t$,   can be expressed as
 \be
 \label{eq:lagrangian}
 \theta(\bx,t)  = \left\la \int_0^t ds f_\theta(\bX(s|\bx,t),s)\right\ra_{\bm \eta},
 \ee
 where we have assumed without any loss of generality that at time $t=0$ the scalar is identically zero, $\theta(\bx,0)=0$. Equation  (\ref{eq:lagrangian}) is telling us that the value of the Eulerian scalar field
 evolved under  (\ref{eq:activepassive})   can be reconstructed by summing the contributions of the forcing estimated along all stochastic Lagrangian paths that end
 on  $\bx$ at  time $t$. The statistical properties of the trajectories of a given velocity configuration can be summarized by the   particle propagator
 $ P(\by, s|\bx,t) = \langle \delta(\by - \bX(s|\bx,t))\rangle_{\bm \eta}$ which is the probability to find a particle that evolves according to (\ref{eq:sto})  in $\bx$ at the final time $t$ provided that it
 is at the  earlier time $s$ in $\by$.  By writing the backward-in-time Kolmogorov equations, $ \partial_t P + {\bm \partial}_x (\bu P) = \kappa \Delta_x P $ with final condition $P(\by,t|\bx,t) = \delta(\by-\bx)$
 we can express the value of the scalar field without  making anymore reference to the trajectories as in \cite{celani2004active}:
 \be
  \label{eq:lagrangianP}
 \theta(\bx,t)  = \int_o^t ds \int d^dy  f_\theta(\by,s)P(\by, s|\bx,t).
 \ee
An  active field evolving in the same velocity configuration can be described  by the same stochastic Lagrangian evolution and the  propagator would  also  be the same. The difference is that for  the active case 
the forcing, $f_a$ and the propagator in the RHS of (\ref{eq:lagrangianP}) will not be uncorrelated because $a(\bx,t)$ influences the evolution of $\bu$ and therefore the expression of $P$ itself.  \\
Starting from the Lagrangian stochastic expressions (\ref{eq:lagrangian}) or (\ref{eq:lagrangianP}) one can understand the meaning of the dissipative anomaly and connect the existence of forward or inverse cascade
with the properties of the propagator $P$ in the limit for $Pe \to \infty$, i.e. in the zero-noise case $\kappa \to 0$ in  (\ref{eq:sto}). Before doing that, let us remember that for vanishing large-scale sink, the exact balancing for the passive fluctuations at all times must satisfy:
\be
\label{eq:balance}
\partial_t \langle \theta^2\rangle = \thetain-\thetachi.
\ee
As a result, in the absence of the dissipative anomaly, $\thetachi \to 0$ for $\kappa \to 0$,  the passive variance would growth indefinitely in time, being always possible to choose $\thetain >0$. Let us now go back to (\ref{eq:lagrangianP}), by  averaging over the forcing realization and over $\bu$ we can write for the two-point scalar correlation function:
\be
\label{eq:c2}
C_2^\theta(\bx_1,\bx_2|t) = \int_0^t ds \int d^dy_1 d^dy_2 \langle P_2(\by_1,\by_2,s|\bx_1,\bx_2,t)\rangle_{\bu}F(|\by_1-\by_2|/\ell_{in}^\theta)  
\ee
where we have introduced the two point propagator $P_2$ which gives the probability for two particles being in $\by_1$ and $\by_2$ at time $s$ to arrive at points $\bx_1$ and $\bx_2$ at time $t$. In the above expression we have  disconnected  the average of $P_2$  over the velocity realization from the  two-point forcing correlation, $\langle f_\theta(\by_1,t) f_\theta(\by_2,t')\rangle = \delta(t-t') F(|\by_1-\by_2|/\ell_{in}^\theta)$, because for the passive case the velocity field does not depend on the injection of the passive scalar. For simplicity we have also considered the forcing to be delta correlated in time. 
If we evaluate the above expression at coinciding points, $\bx_1=\bx_2=\bx$,  we get the passive variance $\langle \theta^2(\bx,t)  \rangle = C_2^\theta(0,t)$.  From (\ref{eq:balance}), one derives that the existence of a dissipative anomaly
is equivalent to the requirement  that the passive variance saturates to a constant (does not grows in time) even in the limit  $\kappa \to 0$. Translated to (\ref{eq:c2}), it implies that the probability to have two trajectories that end at the same point $\bx$ at time $t$ is not proportional to a  delta function of the initial distance, i.e. that $ \langle P_2(\by_1,\by_2,s|\bx,\bx,t)\rangle_{\bu}$ is not proportional to $\delta(\by_1-\by_2)$. Indeed, in the latter case,
the RHS of (\ref{eq:c2}) would grow linearly in time and the passive variance cannot be bounded. On the other hand, if $P_2$ is not a delta-function of $\by_1-\by_2$, since  $F(r)$
is vanishingly small for separation larger than the forcing length, $\ell_{in}^\theta$, the time integral in (\ref{eq:c2}) will be cut-off at the -finite- time
needed for two coinciding particles to separate  a distance of the order of the forcing correlation length. This is the mechanism leading to the existence of the dissipative anomaly.   \\The fact that even in the limit of zero diffusivity, $\kappa \to 0$,  the 2-particles propagator with coinciding final positions does not collapse on
$\delta(\by_1-\by_2)$ is the mathematical equivalent of the so-called {\it spontaneous stochasticity}, i.e. the possibility for two particles  to separate for any, arbitrarily small, noise intensity in (\ref{eq:sto}) even in the limit of coinciding initial points.
In other terms, the existence of a forward passive cascade and the presence of a dissipative anomaly for the scalar goes together with the requirement that the process obtained by sending the noise to zero in (\ref{eq:sto}) is still a stochastic process with a non-trivial 2-point and multi-point kernel. This is possible only when the advecting flow is rough (i.e. by sending first $Re \to \infty$ for the case of a turbulent scaling flow with Kolmogorov scaling at all scales).
It can be  understood considering that  if the advecting flow is not Lipshitz continuous,  trajectories starting from the same point are not unique. Two coinciding particles might separate forward or two initially different trajectories might cross at the final point. Applied to passive scalars this is one of the assumptions at the basis of the  well-known Richardson turbulent dispersion, where two particles with initial distance $r_0 = |\by_1-\by_2|$ are advected by a rough velocity field and they  separate with a law:
$ \langle r(t)^2 \rangle \propto \ein t^3 $
which is asymptotically independent of $r_0$, i.e. there exists an explosive separation of trajectories. \\
For any realizable flows in Nature, it is impossible to have an infinitely extended inertial range and the advecting flow must be smooth for separations smaller then the Kolmogorov scale, $\ell_\nu$.
If two tracer particles  start with an initial separation $r_0 <\ell_\nu$ we will observe a transient initial  exponential stretching due to the smoothness of the  underlying flow, followed by an explosive separation for large times (see, e.g., the right panel of  Fig. (\ref{fig:Gotoh}) for an example of the whole process).  As a result, the spontaneous stochasticity will
emerge as soon as the separation is larger then $\ell_\nu$, i.e. for any separation if $Re \to \infty$.
Standard phenomenology predicts that for a long enough time the
 Richardson  law  is recovered independently of $r_0$ \cite{bitane2013geometry,scatamacchia2012extreme,salazar2009two}. \\
For advection-diffusion in an everywhere smooth velocity field (Batchelor limit), two nearby trajectories will only separate exponentially in time, and
two coinciding trajectories will lead to a unique evolution if $\kappa=0$, as a result there is not dissipative anomaly and the total passive energy $\langle \theta^2\rangle$
will not reach stationarity as is also reflected by the existence of the logarithmic singularity for $r \to 0$ in (\ref{eq:batchelor}). \\
\noindent We have now the possibility to understand  also the opposite case, i.e.  when we are in the presence of an inverse passive cascade. Whenever all Lagrangian trajectories collapse into a unique trajectory in the limit $\kappa \to 0$ the $P_2$ propagator tends to a delta-function and the scalar variance will grow linearly in time, as it must be the case for an inverse cascade in the absence of a large scale sink. Spontaneous stochasticity is absent and  cannot have a dissipative anomaly.\\

For active fields, nothing can be derived using the previous arguments  because the particle propagator $P_2$ is correlated with the active forcing. Let us indeed re-interpret in terms of Lagrangian evolution the instructive case of  2D MHD turbulence (see Sec. \ref{sec:MHD2D}), where the active (magnetic potential) field develops an inverse cascade, while the  passive scalar has  a direct cascade. It is clear that the inverse cascade for the active field cannot be due to the collapse of the trajectories of the active tracer: both passive and active fields  are advected by the same rough velocity field and we must have spontaneous stochasticity for both cases. To understand what happens we need to go back to the Lagrangian stochastic evolution by writing the equivalent of (\ref{eq:lagrangian}) for two-point observables in a form that explicitly considers the cumulative effect of the forcing along  trajectories that end on points $\bx_1$ and $\bx_2$ at time $t$. For the active case we have:
\be
\label{eq:c2a}
 C_2^\theta(\bx_1,\bx_2|t) = \langle \theta (\bx_1,t) \theta(\bx_2,t) \rangle= \int_0^t ds_1 \int_0^t ds_2  \langle f_\theta(\bX(s_1|\bx_1,t),s_1) f_\theta(\bX(s_2|\bx_2,t),s_2)\rangle_{\bu, f, {\bm \eta}}.
\ee
It is  key to understand that even in the limit of coinciding points $\bx_1\to \bx_2$, the stochastic trajectories that must be used to evaluate the effect of the forcing are independent. As a result, for the direct cascade of the passive scalar the mean on the RHS of (\ref{eq:c2a})  will suffer from strong cancellations and does not grow in time. For an active field, it might be  dominated by a systematic correlation among the trajectories and the value of the forcing, if the latter is correlated with the advecting velocity. For 2D MHD (see section \ref{sec:MHD2D}) this is exactly what happens: despite  the separation of nearby trajectories, the contribution of the forcing for the active field is biased and leads to a value for (\ref{eq:c2a}) that grows in time when the two final points coincide \cite{celani2002active}.
\subsubsection{Compressible flows and transitions to inverse cascades} \label{compadvection}        
 A transition from a forward to an inverse cascade of a passively advected  scalar can be observed when the flow is compressible.  
 We will refer to a direct cascade when a dissipation anomaly exists at the small scales, Lagrangian trajectories separate explosively,  thus generating a small scale dissipation of scalar variance. Similarly we refer to an inverse cascade if Lagrangian trajectories  tend to collapse in time, and thus transfer
scalar variance to large-scale structures as discussed in the previous subsections.

\begin{figure*}[htbp]                                                                 %
\centering                                                                            %
\includegraphics*[width=0.75\textwidth,angle=0]{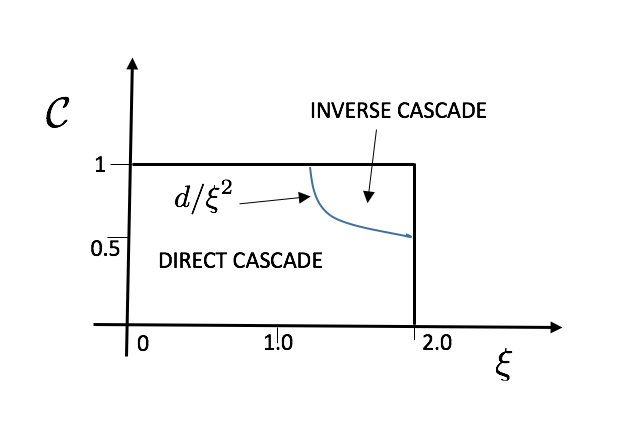}            %
\caption{Summary of the transition among  direct/inverse cascades                     %
  for passive scalars in $d$-dimensions in a compressible flow within the Kraichnan %
  model as a function of the roughness exponent $\xi$ and compressibility ${\cal C}$  %
  \cite{gawedzki2000phase}. }\label{fig:compressiblepassive}                          %
\end{figure*}                                                                         %
Transition from direct to inverse cascade can be rigorously analyzed \cite{gawedzki2000phase}  in the $d$-dimensional compressible Kraichnan
model where the advecting velocity field is Gaussian, self-similar and delta-correlated in time, with two-point variance given by 
$$
\langle \delta_\br u_i(t) \delta_\br u_j(t')\rangle =\delta(t-t') D_{ij}(\br) 
$$
with $ D_{ij}(\br) = A(d,\xi)\delta_{ij}r^\xi +B(d,\xi) r_ir_j r^{\xi-2}$, where $ 0 \le \xi \le 2$ is the roughness exponent of the advecting field and $A,B$ are two functions which depend  on the degree of compressibility of the field  defined as the ratio: ${\cal C} = \langle (\partial_i u_i)^2\rangle/\langle \sum_{ij} \partial_i u_j^2\rangle  $. 
In the compressible Kraichnan model, a  phase transition from a direct to inverse cascade develops as a function of the dimensionality, $d$, of the roughness exponent $\xi$  and of the degree of compressibility, ${\cal C} $ as shown for the first time in \cite{gawedzki2000phase} and graphically summarized in Fig. (\ref{fig:compressiblepassive}).  {The transition from a forward cascade to an inverse cascade 
is discontinuous and thus it falls in case (c) of the transition scenario discussed in Sec. (\ref{sec:Classification}). } \\

Another case that displays a transition of cascade direction  is passive scalar advection when the domain in one of two dimensions is severely constrained similar to turbulence in thin layers.  
This case was examined also within the compressible Kraichnan model in \cite{Celani2010turbulence}.
When the 2D layer is very thin the flow behaves like 1D with the variance cascading inversely while when the thickness of the layer becomes very large the flow behaves like 2D and the scalar
variance cascades forward. It was shown that in this system, small-scale dissipation and transfer to large scales coexist. The asymptotic analysis done in  \cite{Celani2010turbulence} revealed that the transition 
from one extreme case to the other as the layer thickness is varied is smooth and the amplitude of the forward cascades scales like a power-law and thus falls in case (a) of the transition discussed in Sec.  (\ref{sec:Classification}). 
{The dispersion of fluid particles on the surface of a $d$-dimensional cylinder  was further investigated in \cite{celani2010dispersion}, where it was shown that the compactification results in to an invariant measure for the two-particle separation transition probability.} 
\subsubsection{Summary}
In this section, we have summarized the most important phenomenological and theoretical
advancements on the transfer of scalar quantities advected by turbulent flows at changing
the properties of the advecting field and the forcing mechanisms.
We have also discussed the different scaling regimes expected at changing the relative intensity of Reynolds and P\'eclet numbers.\\
We have analyzed the main difference among the passive and active case. For the former, we have connected the existence of a forward cascade and of the dissipative anomaly in the Eulerian framework to the phenomenon of {\it spontaneous stochasticity} for the equivalent Lagrangian description in terms of tracer particles. As a result, the passive scalar is supposed to cascade forward unless the advecting field is compressible enough to
collapse two (or more) trajectories on a unique path  when their starting points tend to coincide.
In the latter case the transition shown in Fig. (\ref{fig:compressiblepassive}) from forward to backward cascade is predicted to happen sharply, following the scenario  illustrated in panel (c) of Fig. (\ref{fig:classification}). \\ For ideal situations when the advecting field is self-similar, Gaussian and delta-correlated in time (Kraichnan model), the above results can be proven rigorously. For the most general case, the linearity of the problem is typically a sufficient condition  to advocate for similar results to hold, even if no rigorous calculations can be made \cite{celani2001statistical,arad2001statistical}. \\
Active fields might be transferred either forward or backward even in the presence of {\it spontaneous stochasticity}, i.e. even when passive scalar fluctuations cascade to small scales. This is due to the fact that the advecting velocity and the scalar injections are in general correlated, opening the possibility for a preferential sampling of the effects of the forcing along the Lagrangian trajectories of the active scalar advection. \\
Paradigmatic applications to 2D turbulence, 2D MHD, surface-quasi-geostrophic  flows and convection have been briefly discussed. 
Recently, the signatures of direct and inverse energy cascade properties have also been connected to the breaking of time-reflection symmetry along single-particle trajectories in 2D and 3D turbulence, as shown by experimental and numerical data in  \cite{pumir2014redistribution,cencini2017time}. 

\section{Further topics about cascades \label{sec:further}}  
%
\subsection{Wave turbulence \label{sec:WWT}}

Wave turbulence refers to systems for which the main interactions are dominated by waves. We have already met wave turbulence in the sections regarding rotating flows (\ref{sec:Rotation}), stratified flows (\ref{sec:stratification}), and in MHD (\ref{sec:MHD}) where inertial waves, gravity waves, or Alfv\'en waves  play a dominant role.  Wave turbulence however is not limited to incompressible flows, as it is also met in elastic plates, quantum turbulence, acoustic waves to mention just a few.  In the limit of very fast waves, referred to as the {\it weak wave turbulence} (WWT) regime, the nonlinear equations can be closed  and an analytic expression for the energy spectrum can be derived. In fact, not only the energy spectrum but also the  probability distribution function of the wave amplitudes can be solved for. A  comprehensive and extensive coverage  of WWT can be found in the book \citep{nazarenko2011wave} in the review \cite{newell2011wave} 
as well as 
{
in the book \cite{zakharov2012kolmogorov} that is based on the the classical works by Zakharov in
collaboration with other authors. }
%
A review on wave turbulence interactions in the stable atmospheric boundary layer can also be found in
\cite{sun2015review}. Here, after shortly describing the main ideas we discuss the implications and limitations of WWT for the systems examined so far.

Weak wave turbulence is realized at scales where the linear wave frequency is much larger than the typical inverse nonlinear time scale $\tau_{nonlin}^{-1}$ of the system. In this limit, we can define a small parameter $\sepsilon$ given by the ratio of the  wave period, $\twv$, and the eddy-turn-over-time,  $\tnl$: 
\be  
\label{eq:wavecond}
\sepsilon=\twv/\tnl \ll 1.\ee
If measured at the forcing scale $\lin$ this parameter coincides with the Rossby number for rotating turbulence (\ref{eq:Rossby}), with the Froude number in stratified turbulence (\ref{eq:Froude}) and with  the Alfv\'en Mach number in MHD turbulence (\ref{eq:Alfven}). To demonstrate how WWT proceeds we consider the NSE in Fourier space (\ref{NSH}) with an arbitrary dispersion relation $\omega_\bk$ for the helical modes $\tilde{u}^{s_k}_\bk$:
\be
\label{WNSH}
\partial_t \tilde{u}^{s_k}_\bk  - \frac{i}{\sepsilon} \omega^{s_k}_\bk \tilde{u}^{s_k}_\bk =  \sum_{\bf p+q+k=0}  \sum_{s_p,s_q} \mathcal{C}^{s_k,s_q,s_p}_{\bp,\bq,\bk} (\tilde{u}^{s_p}_\bp)^* (\tilde{u}^{s_q}_\bq)^* 
 - \nu k^2  \tilde{u}^{s_k}_\bk  - \alpha   \tilde{u}^{s_k}_\bk + \tilde{f}^{s_k}.
\ee
where we have assumed that the wave term, $\omega^{s_k}_\bk \tilde{u}^{s_k}_\bk$, is  order $O(1/\sepsilon)$ with respect to the non-linear contributions. 
For example, in rotating turbulence the inertial wave frequency would be $\omega^{s_k}_\bk=s_k Ro^{-1} k_z/k$ (see section \ref{sec:Rotation}).
To filter out the fast waves we can use the substitution 
\[\tilde{u}^{s_k}_\bk(t) = a^{s_k}_\bk(t) e^{\frac{i}{\sepsilon}  \omega_\bk t}\] 
to obtain an equation for the slow amplitude $a^{s_k}_\bk$: 
\be
\label{WNSH2}
\partial_t \tilde{a}^{s_k}_\bk  =   \sum_{\bf p+q+k=0}  \sum_{s_p,s_q} e^{-\frac{i}{\sepsilon} (\omega^{s_k}_\bk+\omega^{s_q}_\bq+\omega^{s_p}_\bp) t }    \mathcal{C}^{s_k,s_q,s_p}_{\bp,\bq,\bk} (\tilde{a}^{s_p}_\bp)^* (\tilde{a}^{s_q}_\bq)^* 
 - \nu k^2  \tilde{a}^{s_k}_\bk  - \alpha   \tilde{a}^{s_k}_\bk + \tilde{f}^{s_k}e^{-\frac{i}{\sepsilon}  \omega_\bk t} .
 \ee
We can  treat this equation perturbatively by expanding $\tilde{a}^{s_k}_\bk$ as
\[ \tilde{a}^{s_k}_\bk = \tilde{a}^{s_k\,(0)}_\bk+ \sepsilon \tilde{a}^{s_k\,(1)}_\bk  + \sepsilon^2 \tilde{a}^{s_k\,(2)}_\bk + \dots  \]
Before proceeding we need to distinguish two cases that correspond to which order the limits $\sepsilon\to 0$ and $L\to \infty$ are taken.
\paragraph{Discrete wave turbulence:}
If the limit $\sepsilon\to 0$ is taken first, then the fast oscillating term in (\ref{WNSH2}) time-averages to zero for any finite time window unless the resonance condition:
\be
\omega^{s_k}_\bk+\omega^{s_q}_\bq+\omega^{s_p}_\bp =0
\label{eq:reso}
\ee
is satisfied. Therefore, in the limit of  $\sepsilon\to 0$ only Fourier modes that satisfy both $\bk+\bp+\bq=0$ and (\ref{eq:reso}) can interact. If these exact resonances exist they lead to a network (that could be finite or infinite) of interacting wavevectors \cite{kartashova2009discrete,Kartashova2007exact}. The  convolution term in
(\ref{WNSH2}) is thus significantly reduced.   Finding  waves that satisfy the resonance condition (\ref{eq:reso})  on a discrete lattice is not an easy task and only a few cases have been treated \cite{bustamante2013complete}.  If no exact resonances exist, then higher order terms in the expansion need to be considered and the system evolves on a slower time scale. 
\paragraph{Continuous  wave turbulence:} 
In the case that the large box size limit is considered first, the sums in (\ref{WNSH2}) need to be replaced by integrals over $dq^3,dp^3$ and the condition $\bk+\bp+\bq=0$
by a Dirac delta function $\delta(\bk+\bp+\bq)$. A time average of the oscillating term becomes 
\be \la e^{-\frac{i}{\sepsilon} (\omega^{s_\bk}_\bk+\omega^{s_\bq}_\bq+\omega^{s_\bp}_\bp) t }\ra_T \simeq \sepsilon \delta(\omega^{s_k}_\bk+\omega^{s_q}_\bq+\omega^{s_p}_\bp). \label{eq:res2}\ee
where we have defined \[ \la f(t) \ra_T = \frac{1}{2T} \int_{-T}^T f(t')dt'. \]
For any small but finite value of $\sepsilon$ the  resonance condition does not need to be exact but can be violated up to order $\sepsilon$,  i.e. $\omega^{s_k}_\bk +\omega^{s_q}_\bq +\omega^{s_p}_\bp=O(\sepsilon)$. These interactions are referred to as quasi-resonances.  We further note that  (\ref{eq:res2}) reduces the intensity of the nonlinearity by $\sepsilon$,  thus 
  the expansion needs to be carried out until order $\sepsilon^2$.
Wave turbulence proceeds by calculating the evolution of the statistical average of different moments like  $n_\bk =\la |  \tilde{a}^{s_k,(0)}_\bk|^2 \ra_S$
(where the subindex $S$ stands for an average over an ensemble of initial conditions). The energy spectrum $E(k)$ is obtained by spherically averaging 
$n_\bk$ over the surface of a sphere of radius $k$. This procedure leads to a kinetic equation that takes the form 
\begin{align}
\partial_t n_\bk &= \int (V^1_{\bk,\bq,\bp} n_\bq n_\bp + V^2_{\bk,\bq,\bp} n_\bk n_\bp + V^3_{\bk,\bq,\bp} n_\bq n_\bk ) \delta(\bk+\bp+\bq) \delta(\omega^{s_k}_\bk+\omega^{s_q}_\bq+\omega^{s_p}_\bp) dq^3 dp^3 \nonumber \\
                 & - \nu' k^2 n_\bk -\alpha n_\bk + F_\bk
\end{align}
where the potentials $V^i_{\bk,\bq,\bp}$ are obtained from the non-linear coupling term, $\nu'$ is a rescaled viscosity $\nu'= \nu \sepsilon^{-1}$, and $F_\bk$ is the rate the forcing injects energy at the wavenumber $\bk$. If there are no modes that satisfy the resonance conditions then the expansion needs to be carried over  to the next order
where 4-wave interactions take place.\\

There are a few remarks that need to be made. 
The  weakening of the non-linearities  implies that the LHS of   balance (\ref{eq:flux}) is multiplied by a factor $\sepsilon$. This  leads to the following estimate for the drag coefficient:
\be
 \lim_{\Red\to\infty }  
\frac{ \oein } 
{{\cE}_{in}^{2/3} \kin } = O(\sepsilon).
\ee
The above expression  has been often  used as a good proxy for the system to be in a
wave turbulent state \cite{alexakis2013large,alexakis2015rotatingTG,campagne2016turbulent,seshasayanan2018condensates}. 
It also provides a phenomenological way to predict scaling relations for the structure functions or energy spectra by assuming constancy of the energy flux. 
The idea is to assume that in the wave-regime, the wave-wave interaction time-scale is shorter than the eddy-turn-over-time, $\tau_{wave}(k) < \tau_{nonlin}(k)$, hence the energy is transferred via a less efficient mechanism due to the quasi-Gaussian wave background and the total flux is depleted by a factor $\tau_{wave}(k)/\tau_{nonlin}(k)$:
\be
\label{eq:wavebalance}
\ein \sim \frac{(\delta_r u)^3}{r_\perp} 
\frac{\tau_{wave}(r)}
{\tau_{nonlin}(r)}
\quad or \quad
\ein \sim {E(k)}^{3/2}k^{5/2} 
\frac{\tau_{wave}(k)}
{\tau_{nonlin}(k)}
\ee
where $\tau_{nonlin}(k) \sim k^{3/2} E(k)^{1/2}$ is the eddy turnover time and $\tau_{wave}(k)$ is evaluated from the dispersion relation $\tau_{wave}(k)=\omega_\bk^{-1}$. The isotropic result can be extended to higher wave interactions assuming isotropy \cite{nazarenko2011wave}. 
For anisotropic systems we need to use the 2D energy spectrum $E(k_\perp,k_\|)$ with $\delta_r u^2\sim E(k_\perp,k_\|)k_\perp k_\|$ and 
$1/\tau_{nonlin}(k) \sim [E(k_\perp,k_\|)k_\perp k_\|]^{1/2} k$. The same relation then reads:
\be
\ein \sim [ E(k_\perp,k_\|)k_\perp k_\| ]^{3/2} k \, 
\frac{\tau_{wave}(k_\perp,k_\|)}
{\tau_{nonlin}(k_\perp,k_\|)}
={E^2(k_\perp,k_\|)} k_\|^2 k_\perp^2  k^2 \omega_\bk^{-1}.
\label{eq:wavebalance1}
\ee
The relations (\ref{eq:wavebalance},\ref{eq:wavebalance1}) can be used to predict 
   the energy spectra in the wave turbulence regime as
\be
 E(k_\perp,k_\|) \sim \ein^{1/2} \omega_\bk^{1/2}  (k_\| k_\perp  k)^{-1}.
\label{eq:wavebalance2}
\ee
We need  to warn the reader that the anisotropic case (\ref{eq:wavebalance2}) does not take into account the direction of the cascade in the $(k_\perp,k_\|)$ plane, and sometimes further assumptions are required. In fact, we have already used these relations to derive the energy spectra in rotating, and MHD turbulence.
We also note that since the non-linearity is depleted by a factor of $\sepsilon$, in order for the viscous effects to be neglected in the inertial range we need to require that $\sepsilon \Red \gg 1$ and not
just $\Red \gg 1$.\\

Concluding this short section about  wave turbulence we need to express  a few words of caution.
First we note that the condition (\ref{eq:wavecond}) is not uniform in  wave-number space because both $\twv$ and $\tnl$ depend on  $k$. It is thus common that wave turbulence theory breaks down at some scale where $\twv/\tnl =O(1)$ after which turbulence becomes strong.
We also stress  that changing the ordering of the limits $\sepsilon\to 0$ and $ L,H \to \infty$  can drastically impact the dynamics of the system.
Not only the nonlinearity can appear at different order but as seen in rotating  and in MHD flows (see secs. \ref{sec:Rotation} and \ref{sec:MHD} ) confinement can change the direction of the cascade. Finally we stress the difficulty to realize a weak turbulent state in experiments or numerical simulations.
The problem  originates from the different limits that need to be considered. First the large box size limit needs to be taken $H,L\to \infty$, 
then the fast wave $\sepsilon\to 0$ and the large rescaled Reynolds limit $\sepsilon Re \to \infty$, and finally the system needs to run long enough, of the order $O(\tnl/\sepsilon)$, for the cascade to build.

\subsection{{Intermittency and multi-scaling} \label{sec:intermittency}}  
\begin{figure*}[htbp]                                                          %
\centering                                                                     %
\includegraphics*[width=0.92\textwidth,angle=0]{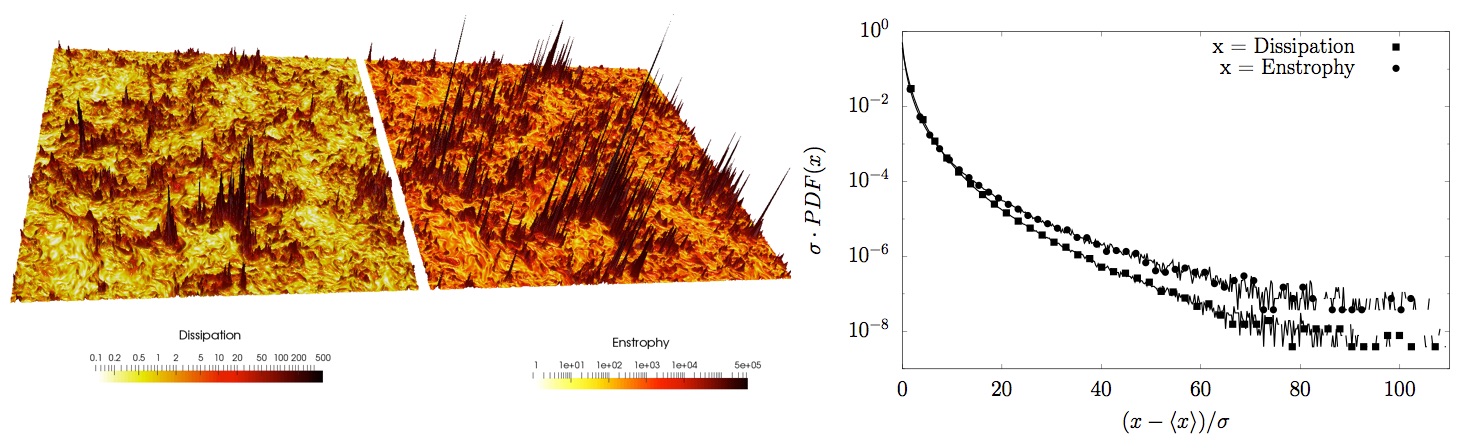}             %
\caption{2D snapshot of the local energy dissipation field  (left)  and the    %
enstrophy distribution (middle) from a 3D HIT flow. Notice the strong        %
burst-like structures typical of spatial intermittent configurations.          %
Right: standardised PDF of the two fields. Data are taken from a $1024^3$       %
    simulations (courtesy of M. Buzzicotti)  }\label{fig:intermittency}        %
\end{figure*}

Up to now, we have investigated cascades and transfer properties in turbulence by mainly using mean fluxes, i.e. by performing a global average over the configuration or Fourier representation, as summarized by  (\ref{eq:fluxbalance}) and  (\ref{eq:KHM}) in Sec. (\ref{sec:theoretical}). On the other hand, it is crucial to understand whether the cascade proceeds smoothly from the injection scale to the scale where it is dissipated or if strong spatial and/or temporal fluctuations arise. The possibility to develop highly non-homogeneous spatial fluctuations during the direct cascade process is already implicitly contained in Richardson's poem cited at the beginning of this review, evoking turbulence as the result of an infinite set of  nested structures made of smaller and smaller whorls. Indeed, it is well known that despite  the fact that the large-scale energy distribution is close to Gaussian, the energy dissipation in 3D turbulence is a highly fluctuating field \cite{yeung2015extreme}, with an instantaneous spatial distribution characterized by intense localized peaks in a sea of almost laminar regions. This is qualitatively and  quantitatively summarized in Fig. (\ref{fig:intermittency}) where we show the results for the local energy dissipation, $\nu  |\nabla \bu|^2$ and enstrophy  $|\bw|^2 $  on a 2D  cut of a 3D turbulent flow, together with their PDF. \\

The changing from a close-to-Normal statistics at the injection scale to a highly skewed and non-Gaussian distribution at the viscous scale is the phenomenon called turbulence {\it intermittency},   the tendency of some flow realization to develop stronger and stronger fluctuations by changing (usually decreasing) the scale, or by increasing Reynolds numbers. In order to have a proper definition, independent of the  flow properties at the injection scale,  we need first to define dimensionless  scale-dependent fields normalized with their rms fluctuations,  $\delta_r u \to \delta_r X =  \delta _r u/\langle (\delta_r u)^2\rangle^{1/2} $, where, e.g. we have used the longitudinal velocity increments as defined in (\ref{eq:sfdef}) and that we denote here  as $\delta_r u$ for simplicity.  Then, we define the realizations of the $X$ field to be {\it intermittent} if the PDF of its increments, $PDF(\delta_r X)$  depends on the scale $r$ and tends to develop larger  and larger tails by changing  $r$ in the inertial range. In particular, as we will see below, 3D turbulence in the forward cascade regime develops intermittency due to the existence of anomalous power laws for velocity structure functions, but the requirement to have power laws behaviour is not strict.

  The importance of {\it intermittency} is multi-fold.
First of all, theoretically speaking, it is believed to be the signature of the multi-fractal nature of the energy cascade \cite{Frisch,benzi1984multifractal,meneveau1991multifractal}.
It reflects the important phenomenological fact that the Richardson cascade does not proceed in a space-filling way and it tends to concentrate very intense dissipative events in a small sub-set of the 3D space.
Explaining the presence of intermittency, not to speak about the possibility to derive it from the original Navier-Stokes equations, is considered {\it the} theoretical problem for all basic turbulent flows, being a phenomenon ubiquitous to all flows in 3D. Up to now, all analytical attempts  to produce a systematic under-control protocol to calculate from first principle intermittent properties have failed (see \cite{Frisch,belinicher1998computing,yakhot2017multi,eyink1993lagrangian} for old and recent contributions in this direction). 
It is worth to mention a recent attempt based on a field-theoretical description \cite{oz2017spontaneous,eling2015anomalous}. In the latter, anomalous scaling is derived by  separating a background K41 field theory from  the fluctuations induced by the action describing the Nambu-Goldstone boson connected to the breaking of the dilatation symmetry. The final formula for the scaling exponent generalises the log-normal description: $ \zeta_n-n/3 = G(d) \zeta_n(1-\zeta_n)$, where $G(d)$ is the only free parameter and depends on the embedding dimension. The above formula is interesting because it has not the same drawbacks of the log-normal prediction (see \cite{Frisch} for a discussion about the inconsistency of log-normal distribution) and  leads to a square root dependency of $\zeta_n$ on $n$ for large moments. For low order moments is almost undistinguishable from the She-Leveque prediction \cite{she1994universal}   which is known to be a very good fit of empirical data for the anomalous exponents. 
Finally, the existence of strong non-Gaussian fluctuations is also a key problem in many applications where small-scale turbulence must be modelled as in the  Large Eddy Simulations approach (see Sec. \ref{sec:LES}).  Empirical observations support the idea that  the direct energy cascade intermittency is robust and universal,  independent of the way the flow is forced, if the injection is limited to a restricted set of scales \cite{arneodo1996structure,arneodo2008universal}. Moreover, intermittency is present both in the Eulerian and Lagrangian domain, i.e. also following velocity fluctuations along the evolution of tracer particles \cite{toschi2009lagrangian}. Intermittency  is observed in the direct enstrophy  cascade in 2D \cite{nam2000lagrangian,bernard2000influence,boffetta2002intermittency} but it is  absent for the inverse energy cascade \cite{boffetta2002intermittency,boffetta2000inverse}, the latter being partially understood invoking the fact that in the inverse transfer energy goes from faster to slower variables.   Furthermore, intermittency is not restricted to  the turbulent velocity field, being also observed in passive and active scalars \cite{kraichnan1994anomalous, falkovich2001particles, celani2002active, celani2004active, gotoh2015power} in magnetic fields \cite{greco2009statistical, veltri1999mhd, sorriso1999intermittency, muller2000scaling}, in bounded flows along homogeneous directions \cite{toschi1999intermittency,lohse2010small}, in compressible flows \cite{wang2012scaling,pan2009dissipative,benzi2008intermittency}  and in many other systems as summarized in \cite{Frisch,boffetta2008twenty,benzi1984multifractal}. \\
\paragraph{Parisi-Frisch Multifractal approach.} A popular way to quantify intermittency is based on the analysis of the scaling properties of velocity increments in the inertial range, the so-called structure functions already
introduced in Sec. (\ref{sec:configurational}). As expressed by  (\ref{eq:sf}), it is an empirical fact that the inertial range scaling properties in HIT are anomalous, i.e. longitudinal and transverse increments are characterized by a set of scaling exponents which do not follow the self-similar K41 prediction, i.e.  $ \zeta_n \neq n/3$.   A powerful phenomenological way to explain intermittency is based on the multifractal   theory as first introduced by Parisi and Frisch in \cite{parisifrisch} (see also \cite{frisch2016collective} for a recent historical review). The idea is based on the observation that the inertial terms of the NSE are invariant for a simultaneous rescaling operation where, $\bu \to \lambda^h \bu$, $\bx \to \lambda \bx $ and $ t \to \lambda^{1-h} t$ with an arbitrary exponent $h$. Unlike in Kolmogorov theory which assumes that $h=1/3$ everywhere in the 3D volume, Parisi and Frisch proposed that a fully developed  turbulent velocity field is characterized  by a continuum spectrum of local scaling exponents $h \in [h_{min},h_{max}]$ which appears in fractal sets of dimensions $D(h)$ in the embedding 3D volume:
\be
\delta_r u \sim (\frac{r}{\lin})^h,\; \text{with probability}\,  P_h(r) \sim (\frac{r}{\lin})^{3-D(h)},
\label{eq:multifractal}
\ee
where, here,  we do  not distinguish   among longitudinal or transverse increments for the sake of simplicity. Following the ansatz (\ref{eq:multifractal}) one might write for the generic longitudinal or transverse structure function:
\be
  S_n(r) = \langle  {(\delta_{r} u)^n} \rangle \propto  \int_{h_{min}}^{h_{max}} dh \, P_h(r) r^{hn} \sim r^{\zeta_n}   \qquad  \text{with} \qquad  \lnu \ll r \ll \lin
\label{eq:multifractalSF}
\ee
where  we have assumed $\lin=1$ for simplicity and
\be
\zeta_n = \min_h{(3-D(h) + nh)}
\label{eq:exponents}
\ee
are the scaling exponents obtained in the saddle-point approximation in the inertial range of scales,
supposing $r/\lin \ll 1$ in (\ref{eq:multifractalSF}). 
It is important to stress that (i) the formula (\ref{eq:exponents}) implies a non-linear concave shape for the scaling exponents $\zeta_n$ as a function of the order $n$ and that (ii) the skewness, and all hyper-skewness, based on normalized  odd moments  $m=2n+1$ with $n \ge 1$,  follow a power-law behaviour: 
\be
K_m(r) = \frac{S_m(r)}{(S_2(r))^{m/2}}  \sim (\frac{r}{\lin})^{\zeta_m-\frac{m}{2} \zeta_2} ; \qquad  \ell_\nu \ll r \ll \lin
\label{eq:fla}
\ee
which tend to diverge when $ r \to 0$ in a way that becomes faster and faster by increasing the order $m$ if the exponents are concave.
The growth of the  moments (\ref{eq:fla})  corresponds to  the existence of larger and larger  tails when $ r \to 0$ 
in the PDF of the normalized velocity increments, $\delta_r X$ defined earlier in this section, i.e. to an {\it intermittent} velocity distribution. \\ The normalized moments (\ref{eq:fla}) can also be considered as a measure of the out-of-equilibrium properties, because for $m$ odd we would have $K_m(r)=0$ at equilibrium.   
As a result, we can conclude that in the presence of {\it intermittency}, fluctuations of different intensities  enjoy different out-of-equilibrium properties.\\ 
\noindent For  completeness it is important to remark that empirical observations show that at small and moderate $Re$,
  longitudinal and transverse velocity increments  scale slightly differently \cite{benzi2008intermittency, ishihara2009study}, with a  slow tendency toward a recovery of similar scaling properties by increasing the Reynolds number \cite{iyer2017reynolds}. Indeed, theoretical arguments based on invariance under rotation of the NSE suggest that both  increments should have the same scaling behaviour for  large enough $\Reu$ and that the observed deviation must be due to some sub-leading effect (see \cite{biferale2005anisotropy} and Sec. \ref{sec:scalinganisotropic}). On phenomenological grounds, to describe different scaling properties it is enough to introduce two different spectra of fractal dimensions, $D_L(h)$ and $D_T(h)$ for the two sets of structure functions (\ref{eq:SFk41SF}).\\ From the structure of the Legendre transform (\ref{eq:exponents}), it is clear that the scaling exponents $\zeta_n$ will deviate from $n/3$ as soon as the distribution of local Holder exponents $h$  does not follow the K41 prediction and $D(h) \neq  3 \delta(h-1/3)$. 
The most recent state-of-the-art numerical data \cite{iyer2017reynolds} provide an estimate for the scaling exponents of longitudinal structure functions $\zeta_2^L = 0.720 (2), \zeta_4^L=1.300 (5), \zeta_6^L = 1.78 (1), \zeta_8^L = 2.18 (2)$ and slightly different for the transverse  ones. Most recent experimental data \cite{sinhuber2017dissipative}, exploits  Extended Self Similarity \cite{benzi1993extended} to extract with high precision the ratio $\zeta_4^L/\zeta_2^L = 1.86 (1) $  which is in agreement with the DNS data given above.  It is important to stress that the multi-fractal phenomenology can be easily extended to the energy dissipation distribution by introducing the Refined Kolmogorov Hypothesis \cite{Frisch}, which links the velocity increments to the average of the energy dissipation in a ball of size $r$ in the inertial range,
$\iepsilon_r(\bx)  = 1/r^3 \int_{|\bx-\by| \le r} d^3y \iepsilon_\nu(\by-\bx)$:
\be
(\delta_r u)^3/r  \sim \iepsilon_r. 
\ee
Doing that, one recovers a unified multi-fractal description for the energy dissipation measure and the velocity increments in the inertial range, which is in very good agreement with experimental and numerical data
\cite{iyer2015refined,benzi2009fully,eyink1995local}. Furthermore, descriptions based on the multi-fractal model have been successfully used to reproduce empirical data also for viscous-scale fluctuations, including velocity gradients PDFs \cite{benzi1991multifractality}, multi-scale fusion rules \cite{l1996fusion,benzi1998multiscale},  Lagrangian acceleration and Lagrangian structure functions \cite{biferale2004multifractal,chevillard2012phenomenological,arneodo2008universal} and many others observables. \\
\noindent In the presence of intermittency, gradient statistics develop a non-trivial dependency on the Reynolds number, $\langle (\partial_i u_j) ^n \rangle \sim Re^{\xi(n)}$, where the exponent $\xi(n)$ are connected to the scaling in the inertial range following the multi-fractal ansatz \cite{Frisch} (see also \cite{benzi2009fully} for a comparison between multifractal and other closures connecting viscous to inertial-range physics \cite{yakhot2004towards}). To date, the multi-fractal model remains the most efficient way to reproduce many of the intermittent properties observed in 3D turbulent flows, including non-Gaussian fat tails, long-range correlations and bridge relation among Eulerian and Lagrangian statistics \cite{borgas1993multifractal, kamps2009exact, biferale2004multifractal}.  There exist many stochastic approaches based on random multiplicative models to construct either the multi-fractal energy dissipation field or the inertial-range velocity increments. All rely on some basic assumption that cannot be derived from first principles from the NSE. Some of the most popular models can be found in the following references \cite{benzi1984multifractal, benzi1993random, she1994universal, chevillard2012phenomenological, meneveau1991multifractal}. Interestingly enough, recent rigorous progresses meant to prove the Onsager conjecture for the Euler equations have provided a first constructive way to build up weak solutions with a multi-fractal structure \cite{buckmaster2015anomalous, drivas2017onsager}. It remains to be clarified whether such a solutions are also relevant for the NSE.\\
It is important to stress that while anomalous power-law scaling implies intermittency, as shown by (\ref{eq:fla}) the reverse  is not true: one can have an intermittent flow even in the absence of any statistical power-law properties. As discussed earlier in this subsection, intermittency refers to the existence of a scale-dependent PDF for some  dimensionless observables. As such, it is a property that can be studied also in systems that break scaling invariance, as, e.g., in bounded flows.\\
Before concluding this section, let us stress again that intermittency seems to be an ubiquitous property of all direct cascades even for passive/active scalars and vectors, while it is not observed in inverse transfers. 
Intermittency is also present in other hydrodynamical systems as 
in Burgers' equations \cite{bec2007burgers}, 
{
in MHD turbulence \cite{politano1989h, grauer1994scaling, politano1995model, politano1995current, horbury1997structure,politano1998determination, muller2000scaling, mininni2006small, mininni2009finite}, 
in rotating and stratified turbulence \cite{rorai2014turbulence,feraco2018vertical},
quantum turbulence \citep{paoletti2008velocity,baggaley2011quantum},
}
in  shell models \cite{biferale2003shell,bohr2005dynamical} (see  Sec. \ref{sec:shell}) and 
it can be defined for anisotropic turbulence too as discussed in  subsection (\ref{sec:scalinganisotropic}).
Moreover, and more importantly, intermittency is fully understood for the case of Kraichnan's stochastic passive advection , where the linear structure of the equations allows to identify the origins of the anomalous scaling with  the existence of non-trivial zero modes of the inertial differential operator. In the latter case, anomalous intermittent exponents can be calculated perturbatively both in the isotropic and anisotropic sectors and non-perturbatively in shell models, proving their universality with respect to the large-scale injection mechanisms and their robustness at changing Reynolds numbers  \cite{kraichnan1994anomalous,falkovich2001particles}. 
%

\subsubsection{Turbulence under unconventional Galerkin truncation \label{sec:fractal}}
\begin{figure*}[htbp]                                                                      %
\centering                                                                                 %
\includegraphics*[width=0.92\textwidth,angle=0]{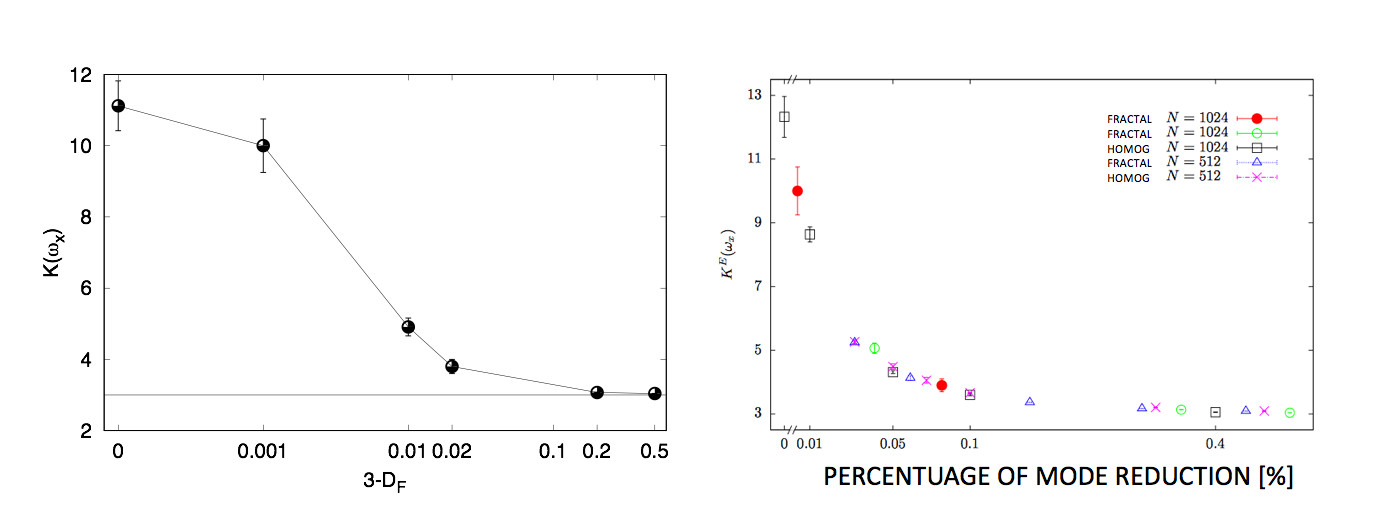}                   %
\caption{Left panel.  Flatness of one component of the vorticity in a DNS of HIT with      %
Galerkin decimation (\ref{eq:theta}) and  $\pi_k  \propto k^{D_F-d}$ as a function of the  %
fractal dimension of the Fourier support, $D_F$ (adapted from \cite{lanotte2015turbulence}).%
Right panel: same quantity plotted as a fraction of the total number of modes decimated    %
and comparing both fractal and homogeneous decimations                                     %
(adapted from \cite{buzzicotti2016lagrangian}). }\label{fig:flatnessfractal}               %
\end{figure*}

Galerkin truncation is conventionally introduced in pseudospectral direct numerical simulation to limit the number of active degrees of freedom that fall below some ultraviolet cutoff $k_c$, projecting the evolution on a finite dimensional phase-space made of all modes with $|\bk| < k_c$. Such a spherical truncation is considered harmless if viscosity is large enough to resolve  the Kolmogorov scale, $ k_\nu < k_c$. We have already discussed in Secs. (\ref{sec:inversehomochiral}) and (\ref{sec:discontinuous}) the  evolution of the 3D NSE under different {\it unconventional}  truncation protocols, e.g. ad-hoc removal of Fourier modes with a given helical component in the flow evolution. In this section, we briefly summarize some recent results obtained by performing a Galerkin truncation in  sparse {\it homogeneous}  or {\it fractal} sets in the whole Fourier space \cite{Frisch2012fractal}. The aim is to understand the effects of the removal of degrees of freedom on both the mean transfer properties and intermittency \cite{Frisch2012fractal, lanotte2015turbulence, lanotte2016vortex, ray2015thermalized, buzzicotti2016intermittency, fathali2017fractally}. The question goes back to the dichotomy among Fourier and configuration-space descriptions, often encountered in many physical phenomena. From one side,  intermittency is the result of  non-trivial alternating events of high and low dissipation in  configuration space, eventually induced by the presence of complex fluid structures \cite{chorin1994vorticity,aubry1988dynamics,hussain1986coherent}. On the other hand, analytical and phenomenological theories of turbulence are often -and mainly- based on Fourier space, where a proper decomposition in divergence-less degrees of freedom is available. Closures as the Eddy Dumped Quasi Normal Markovian Approximation \cite{Lesieur08}, Direct Interaction Approximation \cite{kraichnan1964decay} or the Renormalization Group \cite{yakhot1986renormalization,adzhemyan1999field,eyink1994renormalization} being three paradigmatic examples.  The signature of intermittency in Fourier space must be searched in the phase correlation, i.e. any field with random Fourier phase is Gaussian and with strictly zero energy transfer. Fourier variables are fully delocalised in the configuration space, and it is difficult to imagine what should be  the proper superposition to reconstruct the correct anomalous scaling properties. It is fair to say that we do not control the meaning of intermittency in Fourier variables. \\
A generic random Galerkin decimation is defined by introducing a projector on a vector field as follows:
\be
\hat \bg(\bx,t) \equiv {\mathcal P}  \bg(\bx,t) \equiv \sum_{\bk}  \gamma_{\bk} \, \bgtk e^{i \bk \cdot \bx}.
\label{eq:fractal}
\ee
The factors $\gamma_{\bk}$ are chosen to be  either 1 or 0 with the following probabilistic rule:
\begin{equation}
\label{eq:theta}
\gamma_{\bk} = \begin{cases}
1, & \text{with probability} \qquad  \pi_k  \\
0, & \text{with probability} \qquad  1-\pi_k.
\end{cases}
\end{equation}
Once defined, the set of factors $\gamma_{\bk}$  are kept unchanged, quenched in time.
Moreover, the factors $\gamma_{\bk}$ preserve Hermitian symmetry $\gamma_{\bk}= \gamma_{-\bk}$ so that ${\cal P}$ is a
self-adjoint operator.\\ The evolution of the NSE restricted to the random set of Fourier  modes is then given by:
\begin{equation}
\label{eq:decimNS}
\partial_t \hat {\bf u} = {\cal P}[- {\bf \bnabla}P -
(\hat {\bf u} \cdot {\bf \bnabla}  \hat{ \bf u })]  + 
  \nu \,\Delta \hat {\bf u} +  \hat {\bf f}\,. 
\end{equation}
where, similarly to (\ref{eq:GNS+}), the
nonlinear term must be projected on the quenched decimated set to
constrain the dynamical evolution to evolve on the same set of Fourier
modes at all times. Similarly, the initial condition and the external
forcing must have a support on the same decimated set of Fourier
modes. In the $L^2$-norm the self-adjoint operator ${\cal P}$ commutes with the gradient.
\begin{figure*}[htbp]                                                                     
\centering                                                                                
\includegraphics*[width=0.92\textwidth,angle=0]{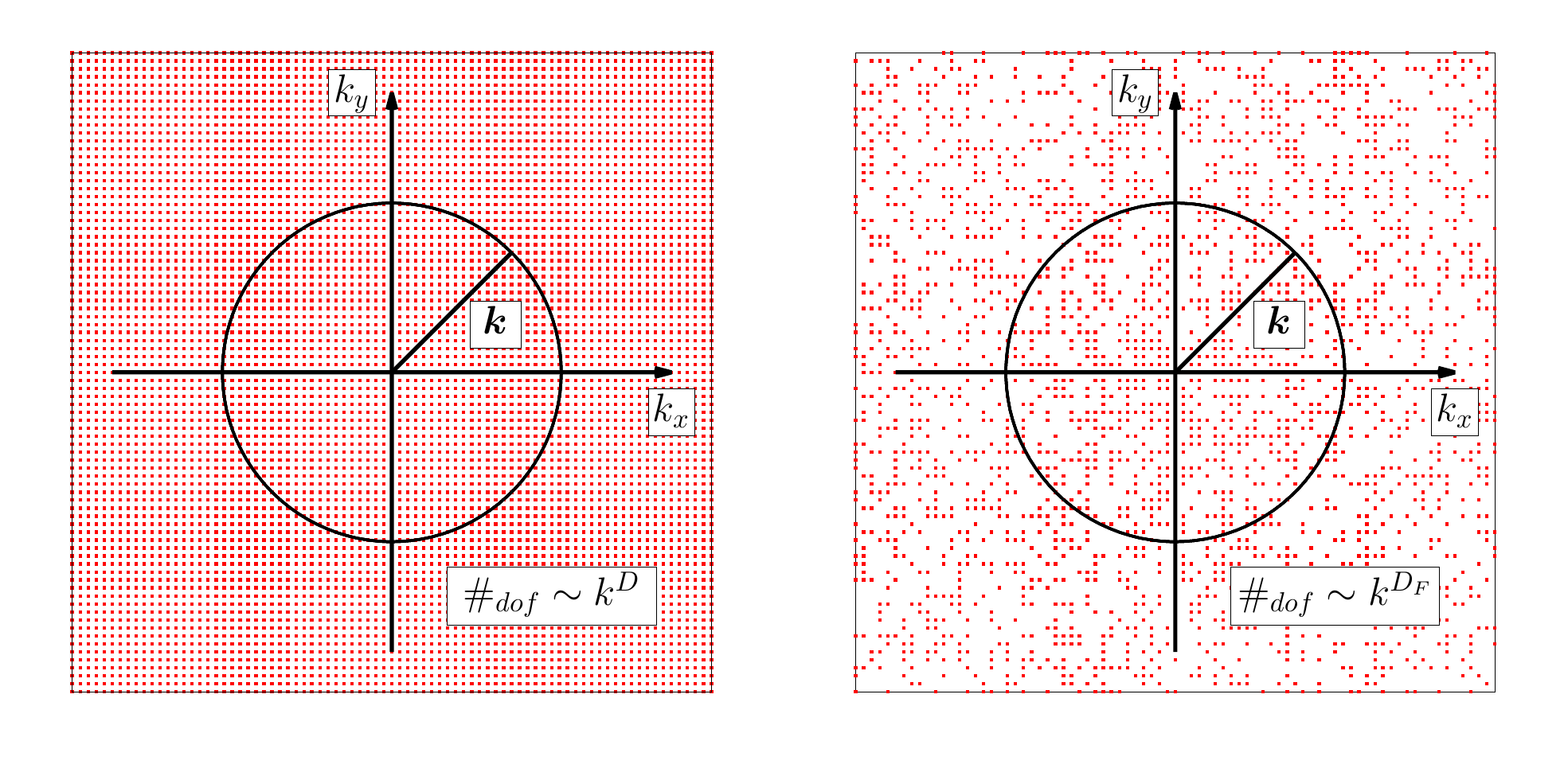}                          
\caption{ Left: distribution of Fourier modes for original undecimated Fourier space (red points). Rigth: 
an example of fractal Galerkin decimation for a 2D Fourier slice. }\label{fig:randomGalerkin}        
\end{figure*}                                                                             
It then follows that the inviscid invariants of the dynamics
are the same as of the original problem, namely energy and
helicity in $d=3$ and energy and enstrophy in $d=2$. \\ If the probability is chosen to be a power law, $\pi_k \propto (k/k_{in})^{D_F-d}$
where $d$ is the configuration-space dimension of the original problem we end up with NSE evolving on a fractal support of dimension $D_F$ in Fourier space (see Figure (\ref{fig:randomGalerkin}) for a graphical representation). \\ Effects of fractal decimation for the inverse energy cascade in 2D turbulence has been investigated in \cite{Frisch2012fractal}, where a tendency toward a flux-less equipartition state  for $D_F= 4/3$ has been observed, in agreement with the theoretical arguments that predict equipartition of enstrophy to coincide with the $-5/3$ inverse energy cascade spectrum for  such a fractal dimension. In a series of more recent papers \cite{lanotte2015turbulence,lanotte2016vortex,ray2015thermalized,buzzicotti2016intermittency,buzzicotti2016lagrangian} the effects of fractal decimation on the intermittency of the 3D direct cascade has been studied. Surprisingly, it was found that  a tiny reduction of the embedding space, $D_F <d$, is enough to strongly deplete the small-scale non-Gaussian turbulent fluctuations as summarized in Fig. (\ref{fig:flatnessfractal}) where in panel (a) we show that the flatness of one component of the vorticity field,
$K_\omega = \langle \omega_x^4\rangle/\langle \omega_x^2\rangle^2 $ is already close to the Gaussian value, $\sim 3$ for $D_F \sim 2.98$. In  panel (b) of the same figure we collect
all results for $K_\omega$  by presenting also data where the Galerkin projector produces a uniform random decimation in Fourier-space, $\pi_k = \alpha$ with $  0 < \alpha <1$.  The latter plot shows
that the reduction of intermittency is apparently independent of the way the decimation is performed,  the only relevant factor is given by the amount of degrees of freedom removed, at least for the decimation protocols investigated up to now. 
In conclusion, we can summarize that many of the aspects connected to {\it intermittency} are still not understood.  In particular, intermittent fluctuations in the forward energy cascade are highly sensitive to random decimation of Fourier degrees of freedom, suggesting that the signatures in Fourier space of strong non-Gaussian velocity and energy dissipation statistics is due to non-trivial correlations among phase variables involving many degrees of freedom. The latter observation supports the idea that small-scale burst-like structures in turbulence need a high degree of entanglement among Fourier modes explaining why  approximations based on quasi-Gaussian fluctuations as it is the case of
WWT theory described in Sec. (\ref{sec:WWT}) and the Eddy Dumped Quasi Normal Markovian (EDQNM) Approximation  models of Sec. (\ref{sec:EDQNMA}) do not develop any intermittency. 

 \subsubsection{Scaling in anisotropic flows}
\label{sec:scalinganisotropic}
 As we have discussed at length in this review, all flows are different in Nature. Even in the simplified assumptions of homogeneity, neglecting the effects induced by solid boundaries, the presence of confinement, rotation, stratification, coupling with other fields always introduces some anisotropy in the system. A very important technical issue is connected to the existence and quantification of scaling laws in such conditions.
 On one side, in the presence of some anisotropy, turbulent statistics cannot be fully universal, depending on the way the breaking of isotropy is introduced, e.g. by large scale shear, by buoyancy, rotation etc.  On the other hand, the notion that all turbulent flows tends to recover some universal {\it isotropic}  properties at small scales, regardless of the macroscopic details, has been a key concept for fundamental and applied turbulent research \cite{Kolmogorov,Frisch,pope2001turbulent,biferale2005anisotropy}. As such, anisotropic fluctuations are always connected to some degree of non-universality, i.e., dependence on the empirical set-up. \\
 It is clear that anisotropy prevents us from calculating scaling properties independently of the selected direction. Similarly, one cannot imagine that there might exists a continuum of
 different scaling exponents as a function of the angle selected for the analyzed direction. It is now well clarified that the only way to analyse scaling properties in the presence of anisotropic statistical fluctuations is to decompose the set of correlation functions in terms of the eigenvector of the group of rotation, SO(3)
 \cite{arad1998extraction,arad2001statistical,kurien2000anisotropic}  and \cite{biferale2005anisotropy} for a review. Just to fix the ideas, let us consider the  longitudinal velocity structure functions (\ref{eq:SFk41SF}) in the presence of some anisotropic statistics. In this case, we must consider the whole dependency on the vector increment $\br$ and not only on its magnitude  $r$. As a result,  each $S^L_n(\br)$, for any fixed $r$, can be decomposed in spherical harmonics $Y_{jm}(\theta,\phi)$, that forms an eigen-basis for  functions on the sphere, being invariant under the transformations of the group of rotation SO(3) in $D=3$:
 \be
 S^L_n(\br) = \langle  {(\delta_{\br} u)^n} \rangle =  \sum_{j=0}^\infty\sum_{m=-j}^j
 S_n^{L,jm}(r)Y_{jm}(\theta,\phi)
\label{eq:k41SFso3}
\ee
where $ S_n^{L,jm}(r)$ represent the projection on the $(j,m)$ SO(3) sector for any fixed scale separation. The decomposition (\ref{eq:k41SFso3}) is exact and labels more and more anisotropic fluctuations by increasing $j$, starting from the case $j=0,m=0$ that is the purely isotropic one. Using symmetry argument for the advection and diffusion operators in the NSE, one can argue that, for $Re \to \infty$, each projection in each different anisotropic sector will have its own scaling behaviour \cite{arad1999correlation}, opening the Pandora box of determining the scaling properties of
each separate anisotropic sectors too:
\be
S_n^{L,jm}(r) \propto C_n^{L,jm}  r^{\zeta_n^{L,j}},
\label{eq:so3}
\ee
where we have assumed that the exponents cannot depend on the $m$-eigenvalue because of the arbitrariness in the definition of the orientation axis and the prefactors $ C_n^{L,jm}$ are non-universal
dimensional constants.
It is easy to see that in terms of the sector dependent scaling exponents, $\zeta_n^{L,j}$,  the recovery of small-scale universal and isotropic fluctuations requires that a hierarchical organization exists, with exponents for high $j$-sectors being larger than the isotropic one, $\zeta_n^{L,0}$ for any $n$. This is what has been measured in some experimental  \cite{kurien2000anisotropic,shen2000anisotropy,warhaft2002higher,staicu2003turbulence,blum2011signatures} and  numerical data  \cite{biferale2001anisotropic,iyer2017multiscale}. Nevertheless, a proper and systematic analysis of anisotropic small-scale fluctuations in different flow set-ups is still lacking and we do not know whether the scaling exponents pertaining to high-order anisotropy are universal or not. Recently, a novel efficient algorithm to perform the decomposition has been proposed and tested on high-resolution numerical data in \cite{iyer2017multiscale}, opening the road to the application of this systematic technique to other data-sets too. Both recovery of small-scale isotropy and universality of scaling exponents can be proved rigorously in the case of passive scalar and magnetic advection by a Kraichnan flow \cite{falkovich2001particles,biferale2005anisotropy}. In the presence of a mean profile, the analysis is more complicated and less well-posed theoretically, due to the additional breaking of homogeneity too, see e.g. \cite{casciola2005scaling,arad1999disentangling,afonso2005inhomogeneous,musacchio2014turbulent}. Similarly, in the presence of very strong anisotropy, as in the case of almost two-dimensional flows in strongly rotating turbulence,
MHD, shallow layers or close to a solid boundary one might need to resort to the SO(2) decomposition in the homogeneous 2D directions to study anisotropic contribution in the plane \cite{biferale2002probing}.

\subsection{Modeling}
In this section we present a short survey of some of the main issues connected to cascades in turbulence models. In particular, we first briefly summarise the concept of {\it backscatter} in Large Eddy Simulations. Second,  we describe two different attempts to attack multi-scale high Reynolds numbers turbulence based on deterministic simplified dynamical shell models and on statistical closure of two-point correlation functions (EDQNM approximation). 
\subsubsection{Large Eddy Simulations \label{sec:LES}}
\begin{figure*}[htbp]                                                 
\centering                                                            
\includegraphics*[width=0.95\textwidth,angle=0]{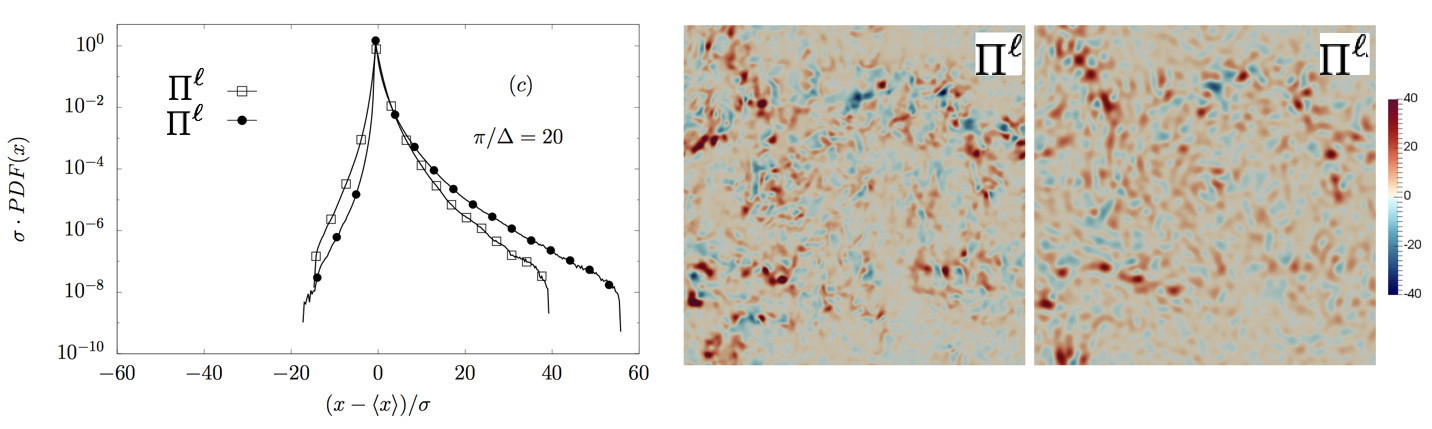}        
\caption{ Left: log-lin plot of the SGS energy transfer PDF using two 
different filter mechanisms: a sharp Fourier filter (empty squares)   
and a Gaussian filter (filled circles). Right: 2D    
snapshots of the SGS energy transfer as measured from a direct         
numerical simulation of HIT at $1024^3$ resolution for sharp-Fourier (left) 
and Gaussian (right) filters. Notice the presence of regions with     
positive and negative energy transfer. Data are adapted from          
\cite{buzzicotti2017effect}.   }\label{fig:LES}                       
\end{figure*}

A huge sector of research with many applications is represented by numerical methods introducing small-scale turbulent closures. The idea is to resolve the spatio-temporal evolution of the largest scales only, which goes under the name of  Large Eddy Simulation (LES) \cite{Smagorinsky63,Deardorff70,germano1991dynamic,MeneveauARFM,pope2001turbulent,Sagaut06,Sagaut08,Lesieur08,brasseur2010designing,zhou2001resolvable}. The motivations to resort to LES can be both applied and theoretical, in order to reduce the number of degrees of freedom that need to be simulated for the former or to understand the sensitivity of the turbulent motion on the ultraviolet physics for the latter. Theoretical motivations include the possibility to have a sort of {\it ideal}  infinite Reynolds number simulation, where the effects of an infinitely extended inertial range of unresolved scales  is perfectly reabsorbed in a suitable model. \\
The literature on the subject
is vast, starting from the early works in  \cite{lesieur1996new,germano1991dynamic,piomelli2003inner} and including recent 
attempts based on fully reversible closures (see \cite{fang2012time} and reference therein).\\
Here we briefly summarize the main points connected to the presence of the forward/backward energy transfer by referring to the representation already introduced in Sec. (\ref{sec:scalefiltered}).\\ In particular the key role is played by the modeling of the subgrid-scale (SGS) stress tensor in terms of the resolved fields, $\ou$:
\be 
\label{eq:lesmodel} 
\tau^\ell_{ij}(\bu,\bu) \to \tau^\ell_{ij}(\obu,\obu)\ee 
such as to be able to close the evolution  (\ref{eq:les}) and solve for  $\obu$ only on a less refined grid. The degree of success of such a strategy is of course dependent on the model used for the SGS stress. There is an extensive literature discussing the state-of-the-art. Models are typically  ad-hoc adapted  depending on the different turbulent realization
(see \cite{pope2001turbulent,Sagaut06,lesieur1996new,Lesieur08,MeneveauARFM,Stevens2014large,Akselvoll2014large} for a series of text books and reviews on fundamental and applied aspects). The most popular (and robust) approach
is based on the Smagorinsky closure \cite{Smagorinsky63} $\tau^\ell_{ij}(\obu,\obu) = -\nu_E(S)\oS_{ij} $ where $\oS_{ij}$ is the resolved stress tensor at the filter scale $\ell$ as defined after (\ref{eq:SGSE}), $\nu_E(S) = (c_S\ell)^2 \sqrt{\oS_{ij}\oS_{ij}}$ is a positive definite eddy viscosity and $c_S$ is a dimensionless number. The Smagorinsky model leads to a positive definite SGS energy transfer (\ref{eq:SGSE}), which cannot be exactly correct, as
  shown by  a direct comparison with data coming from DNS of the full NSE equations.
In Fig. (\ref{fig:LES}) we show the PDF of the exact SGS energy transfer (\ref{eq:SGSE}) as measured
using the full field from high resolution DNS \cite{buzzicotti2017effect} and using two different filters $\tilde G^\ell(\bk)$, a sharp cut-off or a Gaussian filter as introduced earlier in this review in the discussion of (\ref{eq:les}). It is clear that the local (in configuration space) SGS energy transfer has strong fluctuations
 with both positive and negative tails, reaching values as intense as 30-40 times the standard deviation. This high fluctuating field is also visualised in the two right  panels of the same figure. The presence of such  strong fluctuations, order of magnitudes larger then the spatial
mean,  is a serious problem for the physical interpretation  of the local  quantity $\Pi^\ell(\bx,t)$. It is difficult to interpret the sign changes and the variations in the intensity as  genuine  fingerprints of the energy cascade in  configuration space. This is because there is no clear meaning of local energy conservation and the statistics of $\Pi^\ell(\bx,t)$ is certainly strongly influenced by the advection term in (\ref{eq:sg-ene}) that will move kinetic energy across the volume. As a result, while the  averaged
properties over the whole volume have a clear physical meaning (\ref{eq:piaverage}), the  PDF shown in Fig. (\ref{fig:LES}) must be interpreted with caution. On one side, the presence of a left negative tail must be connected to the existence of some sort of local {\it backscatter} events. On the other hand, one cannot strictly speak about a local inverse energy cascade, being the latter uniquely defined only as an average over the whole volume.    This difficulty to build up a one-to-one correspondence between local properties of the SGS energy transfer and the cascade direction is particularly clear when one analyses the results from systems where both forward and backward cascades exist, as for rotating turbulence at small Rossby numbers. In the latter case, the PDFs of the SGS energy transfer show little differences in the two regimes, with the mean fluxes being always much smaller then the typical local fluctuations \cite{buzzicotti2018energy}. \\
Nevertheless, it is considered key to have SGS models for LES which go beyond the Smagorinsky approximation and
have a non-positive definite SGS energy transfer, in order  to be able to reproduce some of the complex features shown by the exact data in Fig. (\ref{fig:LES}). 

\subsubsection{Shell Models}\label{sec:shell}
Shell models have been introduced in the 70s as a suitable modeling of the Fourier-space dynamics of the Navier-Stokes equations \cite{gledzer1973system,desnianskii1974evolution}.
They can be thought as drastic simplifications of the
original Navier-Stokes equations by preserving only a few representative dynamical variables for spherical shells in Fourier space and keeping only local or quasi-local
interactions. As a result, the geometrical structure in configuration and Fourier representations  is lost, and only partial information among spectral properties is retained.  Despite  the drastic simplification, shell models have proved to share many non-trivial dynamical and statistical properties with the original Navier-Stokes equations, such as the presence of a forward  energy cascade for models of 3D turbulence, the existence of a dissipative anomaly, the presence of intermittency with anomalous scaling for the shell velocity statistics \cite{yamada1987lyapunov,jensen1991intermittency,l1998improved, biferale2017optimal}, the presence of an inverse energy cascade for specific 2D models \cite{gilbert2002inverse}, and for suitable helical submodels \cite{de2015inverse,biferale1995role}. There exist many extensions of hydrodynamical shell models to discuss also helical turbulence \cite{benzi1996helical,stepanov2015hindered,rathmann2017pseudo, plunian2013shell},  MHD \cite{gloaguen1985scalar,frick1998cascade,boffetta1999power,plunian2013shell}, thermal convection \cite{ching2008anomalous,benzi2010effect}, rotating turbulence \cite{andre1977influence} passive scalars \cite{jensen1992shell,wirth1996anomalous}, superfluids \cite{wacks2011shell, shukla2016multiscaling} and many other flow configurations \cite{Frisch, bohr2005dynamical, biferale2003shell, ditlevsen2010turbulence}. Shell models are also almost as difficult as the original NSE to be considered  from rigorous mathematical aspects. Indeed, only a few exact results are known about blow-up of the inviscid limit and regularity for all times, see e.g. \cite{cheskidov2008blow,barbato2006some,constantin2006analytic}.  In this section we simply summarize the general set-up for the simplest hydrodynamical back-bone. \\
As said, shell-models are a drastic simplification of the whole set of Fourier interactions of the original equations, reducing the embedding space to a logarithmically equispaced set of wavenumber shells,
representing all degrees of freedom within  $ k_n < k < k_{n+1}$, where $k_n =2^n k_0$, and  retaining only one (or a few) complex variables per shell, $u_n$. 
The equation of motion for one of the most popular models used nowadays  \cite{l1998improved} is given by:
\begin{equation}
\label{eq:shell}
\dot{u}_n = i k_n \left(2 a  u_{n+2} u_{n+1}^{*} \!+\! b u_{n+1} u_{n-1}^{*} \!+\! \frac{c}{2} u_{n-1} u_{n-2}\right)   - \nu k_n^2 u_n + f_n.
\end{equation}
where the set of variables is limited in the IR and in the UV by requiring that $u_0=u_{-1}=u_{N+1}=u_{N+2}=0$  if the velocity shells exist only for $n=1,\dots,N$. On the RHS of (\ref{eq:shell}) we recognize the viscous  operator,
$-\nu k^2_n$, the external forcing $f_n$ and the three non-linear quadratic terms with three free coefficients, $a,b,c$ which can be tuned such as   
when $\nu = f_n = 0$, the model  has two quadratic invariants.  The choice $a+b+c=0$ guarantees that the non-linear evolution  conserves the total energy $E = \sum_{n} |u_n|^2 \,.$
By changing $b,c$ and keeping $a=-b-c$, one can also obtain the conservation of the  total helicity (as in 3D) or of the total enstrophy (as in 2D). Among the advantages of using shell models we must cite the possibility to extend the inertial range by using a small number of degrees of freedom. It is indeed easy to realize that thanks to the introduction of only a few degrees of freedom for each shell, the number of shell variables grow only logarithmically as a function of Reynolds number: $ N \propto \log(Re)$. Because of that, shell models are often the only test-bed where proving quantitatively the robustness  of scaling properties as a function of $\Reu$, and for measuring 
 anomalous exponents with high accuracy. On the other hand, the restriction of the dynamics to  a one-dimensional Fourier-chain, is a strong constraint on the energy cascade.  For example, it is difficult to have models that are able to develop a robust  inverse energy cascade, due to the different role played by the equipartition solutions for models with one single complex variable per shell  (\ref{eq:shell}) \cite{gilbert2002inverse,de2015inverse}. The possibility to have a critical discontinuous transition from forward to inverse cascade at changing the model's parameters
  has been discussed by \cite{bell1977nonlinear,Giuliani2002critical}.  Shell models are also characterized by the presence of quasi-coherent instantonic solutions that bring energy from large to small scales \cite{dombre1998intermittency, l2002quasisolitons, mailybaev2013blowup, de2017chaotic}. As a result, the energy transfer  must be seen as the superposition and interaction of a statistical background plus quasi-coherent burst-like solutions travelling to small scales. This is similar to the  dichotomy among  random cascade  multifratcal models and coherent  structures in NSE. We cite also attempts to build up shell models with a more complex network of shell interactions, trying to  approach a closer description of turbulence dynamics also in configuration space. In particular, ultra-metric shell models have been proposed to reproduce the Richardson cascade in 3D \cite{benzi1997ultrametric,barbato2013dyadic} and  models considering anisotropy and/or real three dimensional embeddings  using nested polyhedra in Fourier space \cite{gurcan2017nested,gurcan2011anisotropic}.

\subsubsection{Eddy Dumped Quasi Normal Markovian Approximation}   
\label{sec:EDQNMA}
The  Eddy Damped Quasi Normal Markovian approximation is a statistical closure   that was developed to model isotropic turbulence \cite{orszag1970analytical,leith1971atmospheric}. A recent review can be found in \cite{Lesieur08}. It solves for the two-point correlation  (energy spectra) of a statistically homogeneous and isotropic turbulent flow by assuming that the relevant statistical fields have a quasi-normal behavior that allows to expand higher moments in terms of second-order moments and thus close the hierarchy of moments. Departures from normality (Gaussianity) is imposed by an eddy-damping term and a Markovianization step is used to ensure the realizability of the two-point correlation moments. 
For 3D homogeneous isotropic and mirror-symmetric (non-helical) turbulence the equation for the evolution of the energy spectrum $E(k,t)$ is given by \cite{andre1977influence}:
\be
\left(\partial_t +2\nu k^2\right) E(k,t) = \int \int \frac{k}{pq} \theta_{kpq}(t) b_3(k,p,q) \left[ k^2E(p,t)E(q,t) - p^2E(p,t)E(k,t)  \right] dpdq
\ee
where $b_3(k, p, q) = (p/k) (xy + z^3)$ is the geometrical coefficient introduced by Kraichnan ($x, y, z$ being the cosines of the angles opposite
to sides $k, p, q$ in the interacting triad) and 
\be
\theta_{kpq}(t)  = t/[1 + (\mu_k + \mu_p +\mu_q) t ],\quad \mathrm{with} \quad 
\mu_k = \nu k^2 + \lambda \left(\int_0^k p^2E(p) dp\right)^{1/2}
\ee
with $\lambda$ an adjustable parameter. The EDQNM equations have been extended for helical turbulence \cite{andre1977influence}, two dimensional turbulence for both inverse energy and direct enstrophy cascades \cite{Pouquet1975evolution},
MHD including the inverse cascade of magnetic helicity \cite{pouquet1976strong} to mention some of the applications. An anisotropic version of EDQNM approximation has also been applied to rotating flows \cite{cambon1989spectral}. 

Like the shell models examined in the previous section the EDQNM equations provide an oversimplification of the NSE. Nonetheless, they provide a tractable and very powerful tool to study high Reynolds number flows that go well beyond what can be attained with DNS nowadays. Unlike shell models, due to the assumed quasi-normality EDQNM can not develop any intermittency. On the other hand,  the EDQNM equations have the advantage to take account nonlocal interactions, thus they provide a more realistic model for flows where widely separated scales are coupled. Recently, the  EDQNM equations have been used to to study  the energy transfer properties of homochiral turbulence \cite{briard2017dynamics,briard2017closure}.   EDQNM can thus provide an indispensable tool to study transitions of cascades at high $Re$.

\subsection{{Cascades in bounded flows and with multi-scale injections} \label{sec:bounded}}  
In the presence of solid boundaries, the assumption of homogeneity and isotropy is lost and one needs to cope with
a series of new phenomena connected to the possibility to have a net transfer of momentum in the direction of the wall.
The literature concerning the energy-momentum balance in wall bounded flows is huge. A series of recent reviews of numerical, experimental and theoretical ideas can be found in \cite{jimenez2012cascades,smits2011high,adrian2007hairpin}.
Here we want to summarize only the major new problems arising in such a situation and to connect
them with the phenomenological set-ups  discussed in the main part of this review. \\
\noindent {\bf Channel flows}.  Let us fix the notation for this important set-up  by supposing to have a channel flow with periodic span-wise boundary conditions and two boundaries at $y= \pm L_0$. The  stream-wise stationary mean velocity profile averaged over a plane parallel to the wall at a given height $y$  will be denoted as $U(y)$ while the fluctuating  stream-wise, wall-normal and span-wise components  as $u=u_1$, $v=u_2$ and $w=u_3$  respectively.  One can then write a height-dependent kinetic energy balance \cite{pope2001turbulent}, generalising the global balance (\ref{eq:globalbalance2}):
\be
\label{eq:kineticwall}
-\langle uv \rangle \frac{dU}{dy} -\frac{1}{2} \frac{d}{dy} \langle u_iu_i v\rangle +\frac{\nu}{2} \frac{d^2 \langle u_iu_i \rangle }{dy^2} -\frac{d \langle pv \rangle}{dy} - \langle \iepsilon(y)\rangle 
\ee
where all averages $\langle \bullet \rangle$ in this section are meant only along the homogeneous directions $x,z$ and $\iepsilon(y) = \nu \langle
\partial_i u_j \partial_i u_j \rangle $ is the dissipation at the given height. By introducing the definition of
the turbulent kinetic energy flux: $\phi(y) =   \frac{1}{2} \langle u_iu_i v\rangle -\frac{\nu}{2} \frac{d \langle u_iu_i\rangle }{dy} +
\langle pv \rangle$ one can rewrite the above equations in a conservative form:
\be
\label{eq:balancewall}
\frac{ d \phi(y)}{dy} =  -\langle uv \rangle \frac{dU}{dy} - \langle \iepsilon(y) \rangle 
\ee
from where it is clear that energy is also transported to/from the wall due to the possible local imbalance between  the production and the dissipative term on the RHS of (\ref{eq:balancewall}). The main source of difficulties is in the coupling among the mean profile and turbulent stress, $\langle uv \rangle \frac{dU}{dy} $,  which is a multi-scale source of energy,
dependent on the distance from the wall. On top of this spatial transfer, we must consider the transfer among scales, which couples mean profiles to smaller and smaller scales at any given height. As a result, one speaks about a scale-by-scale energy budget \cite{marati2004energy,cimarelli2016cascades} which is now position dependent, making the global phenomenology
much more complicated than for the homogeneous cases.   Following \cite{hill2002,marati2004energy} one can derive the generalization of  the von K{\'a}rm{\'a}n-Howarth-Monin relation (\ref{eq:KHM}) to this non-homogeneous and anisotropic case. The resulting scale-by-scale balance between  the  transfer, production and dissipative terms has been extensively studied as a function of the distance from the wall $y$ using DNS in the recent years (see
\cite{cimarelli2016cascades}
and references therein).
Two different kinds of scale energy fluxes have been  identified: that  connected  to the transport of  energy in physical space, and
that related to the transfer  across the spectrum of turbulent scales. The former is the scale-by-scale  extension of the classical turbulent transport across the
channel. The latter describes  different forms of energy transfer which occur in a wall-bounded turbulent flow, thus generalizing the concept of the energy cascade in HIT. A certain scale at a given distance from the wall  receives  energy by three mechanisms: (i)  from the spatial flux (ii)  from the inter-scale non-linear transfer and (iii) by the interaction with the mean flow or by possible external stirring mechanisms. In steady conditions, the above dynamics must be balanced by the local energy dissipation. Furthermore, each kind of scale-energy flux is influenced by two  different contributions: inertial and diffusive which have different relative importance in the  viscous layer, buffer region, log-layer or  bulk region. In particular,  the buffer layer is the region where energy is produced and  transferred  to adjacent zones (towards and outwards the wall).  In the log-layer, production and dissipation are in scale-balance, in the sense that at each given scale, there is no important exchange of energy in the wall-normal direction. On the other hand, the spatial flux is key for turbulence in the bulk region. In fact, the excess production in the buffer layer crosses the log-layer to reach the bulk of the flow. The latter can be seen as a sort of inverse energy cascade, from small eddies in the buffer layer to large eddies in the bulk. In the space of scales,  the production range is always followed by a nearly classical Kolmogorov transfer range, ended by  dissipation at the local dissipative scales.
The  presence of the energy transfer to/from the wall and across different scales is at the basis of the celebrated Townsend attached eddy hypothesis  \cite{townsend1980structure},
where the simultaneous energy and momentum transfer across space and scale is phenomenologically described in terms of a superposition of a  forest of eddies of size comparable to their distance
from the wall.   The attached eddies would then be responsible for the  inverse energy cascade from small eddies (the ones close to the wall) to larger ones, with the double  effect of moving energy to larger scales and away from the wall. This model has been revisited theoretically and studied both numerically and  experimentally in a series of recent works \cite{woodcock2015statistical,marusic2010predictive,marusic2013logarithmic,meneveau2013generalized} where clear evidences concerning the existence of log-laws also for high-order moments in the turbulent boundary layer have been presented, including the possibility to extend the scaling beyond the log-layer, i.e. in  bulk and buffer regions, using Extended Self Similarity \cite{yang2016extended}.   \\
A channel flow is only one of the many instances where turbulence is fed by multi-scale injections, other paradigmatic examples are given by natural convection \cite{lohse2010small}, turbulence with power-law  \cite{sain1998turbulence,biferale2004anomalous} or fractal forcing \cite{hurst2007scalings} and atmospheric flows with simultaneous energy injections due to three dimensional small-scale  or two-dimensional large-scale  instabilities \cite{vallis2017atmospheric,gage1979evidence,lindborg2006energy,cencini2011nonlinear}.\\
\\
\noindent {\bf Convection}.
Natural convection is a paradigmatic case where the buoyancy term injects kinetic energy at all scales. For standard convective set-ups, driven by a hot boundary on the bottom and a cold one on the top,  the energy budget is similar to that described for bounded flows, with an  additional term due to the coupling with temperature fluctuations. The small-scale phenomenology  is dependent on the dimension of the  embedding space. In 2D, energy dissipation is vanishingly small for large Reynolds numbers, and the energy flux is balanced by the buoyancy term, leading to a  Bolgiano scaling (\ref{eq:spestrat}) as discussed in the section devoted to unstably stratified flows  \cite{monin2013statistical,lohse2010small}. In 3D, the forward energy cascade is  dominant in the bulk, leading to a classical local Kolmogorov $-5/3$ spectrum (\ref{eq:kolmostra}). Close to the walls, the Bolgiano length becomes of the order of the distance from the wall and
the horizontal velocity and temperature fluctuations go back to a Bolgiano 2D-like regime \cite{calzavarini2002evidences}. Kolmogorov forward cascade (in 3D) or Bolgiano (in 2D) scaling are measured also for purely homogeneous convection, in the absence of boundaries and with an imposed mean temperature profile  in the whole cell or for the similar non-stationary set-up of Rayleigh-Taylor turbulence \cite{celani2002scaling,celani2006rayleigh,boffetta2012bolgiano,chertkov2003phenomenology}.\\
\noindent {\bf Power-law stirring}. Randomly stirred fluids with a power law energy injection rate have been mainly studied for theoretical reasons. Renormalization Group techniques are based on perturbative expansions around a solution where forcing is dominating at
all scales \cite{forster1977large,antonov2006renormalization,mejia2012nonperturbative}. Depending on the  scale-by-scale energy injection intensity, the non-linear 3D cascade can be leading or sub-leading. In the latter case, scaling exponents  change and the turbulent spectrum follows the scaling imposed by the forcing power law correlation \cite{sain1998turbulence,biferale2004effects}. If the non-linear K41-like transfer is leading, spurious sub-leading effects induced by the multi-scale forcing might nevertheless lead to complex
Reynolds number dependent  scaling properties \cite{biferale2004anomalous,mitra2005multiscaling,pandit2009statistical}.  \\
Atmospheric flows are another paradigmatic case where energy can (is) injected at multiple  scales by different mechanisms. Some of them, connected to stratification have already been discussed at length in this review. In 3D layers, as for the atmosphere, 3D and 2D phenomenologies are both relevant. Aircraft measurements \cite{nastrom1984kinetic} reveal that energy spectra in the troposphere develop two power laws, $-5/3$  at mesoscale wave numbers and $-3$  at larger --synoptic--
scales, even though both regimes should already be quasi-2D. In a pure 2D framework, the presence of a $-3$ spectrum at large scales, followed by a $-5/3$ at small scales, can be interpreted as the superposition of two fluxes due to two injection mechanisms \cite{cencini2011nonlinear}: one at small scales producing the inverse 2D energy cascade and one at large scales producing a direct enstrophy cascade. The final combination results on  the existence of an intermediate range of scales where the two opposite fluxes coexist.  

\subsection{{Guided tour across different turbulent systems}}       
\label{sec:guided}
In this section we give a short ``guided tour'' of different systems that display a turbulent cascade that go beyond turbulence described by the Navier-Stokes equations. The aim of this section is only
to point to different relevant research directions and give some key references without attempting in any way to present a complete review of these fields.

\subsubsection{Gravito-Capillary surface waves }           

Gravito-capillary waves are observed at the surface of any fluid due to the combined effect of gravity and surface tension and represent  another system where wave turbulence theory can be applied.
The wave dispersion relation is:
\[ \omega_\bk^2 = g k + \gamma k^3/\rho  \]
where $g$ is the gravity $\rho$ is the density of the fluid and $\gamma$ is the surface tension.  
In principle gravity waves and capillary waves can be treated separately if different scales are considered ($k\ll \sqrt{g\rho/\gamma }$ and $\sqrt{g\rho/\gamma } \ll k$) 
so that either gravity or surface tension is dominant. Indeed different energy spectra have been derived for gravity-wave \cite{zakharov1982kinetic,zakharov1967energy,zakharov2012kolmogorov} and for capillary turbulence  \cite{zakharov1967weak,pushkarev1996turbulence}. An important difference between gravity-wave  and capillary turbulence is that for the former
three wave interactions are absent and the flow is dominated by four-wave interactions. As a result there is an additional invariant  result related to the wave action that is 
predicted to cascade inversely \cite{zakharov1967energy,korotkevich2008simultaneous}. In the recent years, gravito-capillary turbulence has been tested in various limits in carefully designed experiments \cite{falcon2007observation, denissenko2007gravity, falcon2007observation2, falcon2008fluctuations, hassaini2017transition,berhanu2013space, cobelli2011different, herbert2010observation}.
To isolate gravity from capillary effects large basins \cite{aubourg2015nonlocal, campagne2018impact,  aubourg2017three} or zero gravity environments
\cite{falcon2009capillary} were used, respectively. These experiments provided means to compare with the theoretical predictions and against numerical simulations \cite{korotkevich2008simultaneous, deike2014direct, pan2014direct}. Finally, we note that evidence for a split energy cascade has been observed in gravito-capilary turbulence \cite{abdurakhimov2015bidirectional,abdurahimov2015formation} on the surface of liquid Helium. A recent review of these experimental results can be found in \cite{falcon2010laboratory}.

\subsubsection{Quantum turbulence}                        
Quantum turbulence describes  the turbulent motion of quantum systems such as superfluid helium and atomic Bose-Einstein condensates, which show the presence of quantized vorticity, superfluidity, and  two-fluid behavior for non-zero temperature.  There are three main characteristics that distinguish quantum from normal fluids: (i)
at any finite temperature the system is composed of  a  normal and superfluid component,  (ii) the superfluid  can flow without the effect of viscous forces, and (iii) the local vorticity  is concentrated in thin 
vortex lines with quantized circulation such that $ \oint_C \bu_s \cdot {\bf dr} = h/m$  where with $\bu_s$ we denoted the quantum component,  $h$ is the Planck's constant, $m$ the mass of the boson and $C$ a closed loop around the vortex line.
There are three main physical set-ups where quantum turbulence is observed: $^4He$, the B-phase of $^3He$ and ultracold atoms. The physical problem is old, being observed for the first time in 1957 by  \cite{vinen1957mutual}. The are two main mechanisms to dissipate energy in quantum flows, either by mutual friction with the normal component, $\bu_n$ or by Kelvin waves along vortex lines which might lead to acoustic emissions at large  $k$. In the weak amplitude limit, Kelvin waves can be  studied by weak wave-turbulence discussed in section (\ref{sec:WWT}). \\
Different equations are used to describe quantum fluids and quantum turbulence depending on the physical set-up.
The Gross-Pitaevski (GP)  model is considered a good description for the zero temperature limit of an interacting  Bose-Einstein  condensate, introducing an additional quantum-pressure effect.
GP equations have been used to describe atomic condensates \cite{kobayashi2007quantum,white2010nonclassical} including the effects of a dual cascade of energy and helicity triggered by Kelvin wave interactions \cite{di2015spatiotemporal,di2017dual}. At mesoscale, an approach based on a vortex filament model is also often used \cite{schwarz1988three} where at zero temperature a vortex line moves due to the self-induced Biot-Savart law plus some algorithmic rule to allow reconnection with close-by vortex lines \cite{kondaurova2014structure}. Finally a macroscopic two-fluid approach developed by   Hall-Vinen-Bekarevich-Khalatnikov is employed to describe the large-scale evolution of both normal and superfluid components at any non-zero temperature, coupled by a mutual friction term \cite{boue2015}. Both viscosities and densities strongly depend on temperature for quantum flows. As a result, different regimes are possible. For very small temperature, both $^4He$ and $^3He$ are mainly  made by a single superfluid component, energy is dissipated  by vortex reconnection \cite{zuccher2012quantum} which triggers  acoustic emissions due to a forward cascade of Kelvin waves \cite{kozik2004kelvin,l2007bottleneck,l2010spectrum,walmsley2007dissipation,yepez2009superfluid,krstulovic2011energy,fonda2014direct}. At higher temperature, the normal component of $^3He$ has a relatively high viscosity. As a result, Kelvin waves can be adsorbed by the mutual friction with the normal fluid and at temperature high enough turbulence is completely suppressed \cite{boue2012temperature,eltsov2014quantum}. 
High temperature $^4He$ is described by two fluids that are both turbulent, energy is exchanged among the two components in a highly non-trivial way by  mutual friction, leading to an  enhancement of intermittency in  some temperature ranges  \cite{biferale2018turbulent}. Because of the complicated superposition of different physical mechanisms, the spectral properties of quantum turbulent flows are still  debated. In many experimental and numerical cases,   a $-5/3$ slope is observed as for the classical turbulence \cite{nore1997kolmogorov,maurer1998local,skrbek2006energy,salort2011mesoscale,skrbek2012developed}. For  $^3He$, the presence of a strongly damped normal component might induce non-local effects in the energy transfer with a modification of the spectral slope as recently shown by numerical simulations in \cite{bifquantum}.  A recent experimental theoretical and numerical review on this relatively young field can be found in \cite{barenghi2014introduction}.

\subsubsection{Elastic wave turbulence}            

Elastic wave turbulence refers to the evolution of deformations in elastic materials \cite{landau1986theory} in 2D and 3D systems. Such materials sustain waves that interact and transfer their energy at different scales much like hydrodynamic turbulence. The energy is composed of the kinetic and the elastic energy that is not quadratic in the amplitude of the deformation, but higher-order terms are also present. In  recent years a lot of research has focused on the evolution of elastic wave turbulence in elastic plates. The reason for this interest is that recent experimental advancements using profilometry techniques \cite{cobelli2009global, cobelli2009space} allowed to measure deformations in the entire plate. This technique gives access to  full space information contrary to
the previous state-of-the-art restricted to  few-point-measurements. Thus experimental data can be Fourier transformed to obtain spatial energy spectra in exactly the same way as in numerical simulations. Elastic wave turbulence thus provides a unique opportunity where theoretical work based on weak wave turbulence 
can be tested \cite{during2006weak}. Relations similar to von K{\'a}rm{\'a}n-Howarth-Monin relations have been derived
in \cite{during2018exact} and compared  with experimental data in \cite{boudaoud2008observation,auliel2015wave, deike2013nonlinear, miquel2011nonlinear, miquel2013transition, miquel2011nonstationary, mordant2010fourier} and with numerical simulations  in \cite{during2006weak,miquel2013transition,ducceschi2014dynamics,yokoyama2014identification,yokoyama2014single, bilbao2015conservative,yokoyama2017integrated}. This crossing of theory, experiments and numerical simulations has brought up number of issues such as the effect of finite dissipation coefficients and the role of 
coherent structures in the formation of the spectrum that are often neglected in theoretical work.

\subsubsection{Compressible and Relativistic turbulence}      
A vast area of research in fluid mechanics is devoted to compressible flows, that we have only partially addressed in this review concerning the study of passive scalars in Sec. (\ref{compadvection}). The argument is so vast that cannot be usefully  summarized here. We briefly mention the potential applications to fusion \cite{welser2013two}, supersonic aircraft design \cite{smits2006turbulent}, reactive flows at high temperature \cite{pope1985pdf} and in many astrophysical problems, from stars formation to the interstellar medium \cite{falceta2014turbulence,higdon1984density,price2010density}. Moreover, all relativistic fluids are also obviously compressible as those encountered in gamma-rays burst, pulsars and quark-gluon plasmas \cite{narayan2009turbulent,bucciantini2005relativistic,teaney2010viscous}. The main theoretical issue in compressible turbulence is connected to the existence of anomalies for  viscous dissipation  and for  entropy production.
Recent results have revealed exact relations for compressible turbulence \cite{konstandin2012statistical, galtier2011exact, banerjee2016alternative, banerjee2013exact, banerjee2014kolmogorov} and a flux-loop condensate state (see $\ref{def15}$) has been observed in \cite{falkovich2017vortices}. Furthermore a novel and comprehensive theoretical analysis  of both compressible and relativistic turbulence has been presented in \cite{eyink2018cascades,eyink2018cascadesb}. For compressible turbulence, the analysis is based on  a coarse-grained version of the Euler equations  as defined in Sec. (\ref{sec:scalefiltered}) to show that  standard conservation laws are broken by turbulent anomalies and that the kinetic energy is dissipated by a cascade of  a different mechanism called pressure defect.  Anomaly in the entropy conservation is also argued to exist due to an input of negative entropy by pressure work and a successive cascade to small scales. For relativistic flows, It is possible to  show \cite{eyink2018cascades,eyink2018cascadesb} that the Lorentz covariance is broken by the regularization induced by the scale-filtered fields and that it is restored by sending the size of the filter $\ell \to 0$ in (\ref{eq:filtering}). Also, one can show that the anomalous heat input into the internal energy for the  relativistic equations  coincides with the anomalous dissipation in the non relativistic  where the speed of light $\to \infty$. The reader is referred to the two papers \cite{eyink2018cascades,eyink2018cascadesb} and references therein for a deeper analysis of this emerging fields. For a series of recent simulations and analysis on the multiscaling properties of compressible flows see  \cite{kritsuk2007statistics,benzi2008intermittency,pan2009dissipative,wang2013cascade,wang2012scaling,wang2017spectra,zhao2018inviscid}. Finally we mention the recent results on  general relativity  \cite{galtier2017turbulence} where wave turbulence has been applied to gravitational waves and an alternative mechanism to inflation has been proposed.

\subsubsection{Active matter}                                 
  The study of motile (self-propelled) living micro-organisms or artificial colloids   has many important connections with fluid dynamics  since the pioneering works of \cite{taylor1951analysis}. In many biological and engineering applications, the motion develops at very small Reynolds numbers and the active agents are diluted. Nevertheless, many important issues are considered open, in particular connected to the influence of the flow on the behavior of the micro-organisms both at level of single swimmers or for an entire population, e.g. concerning the combined effects of flow and gradient-sensing for chemotaxis  or for the optimization of colloids motion in micro-suspensions of active particles \cite{howse2007self}. 
In dense active suspensions, non-linear feedback on the flow can be triggered and controlled by the activity of the swimmers, leading to the so-called {\it bacterial-turbulence} regime, resulting in complex rheological properties depending on the propulsion mechanism. It is important to stress that the word {\it turbulence} it is sometimes abused, being -in most cases- the complexity the result of the active swimmers action  more than the outcomes of a genuine non-linear flow transfer.
Pusher, like bacteria, tend to decrease the fluid viscosity  while pullers, like bi-flagellate algae, enhance it \cite{rafai2010effective}. When the concentration of swimming micro-organisms becomes sufficiently large, the individual motion self-organize in macroscopic patterns typical of fluid dynamics, including vortices and turbulent motion \cite{stenhammar2017role, linkmann2018phase} with typical velocities larger than the swimming speed of the single swimmer. The study of these regimes is at its infancy and we do not have quantitative validated models able to describe on the continuum the evolution of the complex flows. Evidence of inverse energy transfer  with helical properties have been reported both using experiments and numerical approaches \cite{slomka2017spontaneous}.  Recently a continuum model has been proposed for bacterial turbulence at high concentration \cite{bratanov2015new} and derived from the microscopic dynamics  \cite{reinken2018derivation}, which qualitatively reproduces several features observed in the experiments, while models for semi-dilute suspensions are still not fully developed. Recent reviews for this rapidly evolving field can be found in \cite{marchetti2013hydrodynamics,bechinger2016active}\\

\section{Conclusions and open problems \label{sec:conclusions}} 

  In this review we have classified many different turbulent systems in terms of their common and distinguished  cascade properties, making clear what the shared phenomena behind the presence of direct/inverse cascades, split and multiple cascades, bi-directional transfer, flux-loop, condensates and quasi-equilibria states are. This grouping has  been made possible thanks to the precise  set of definitions and classifications given in Secs. (\ref{sec:definitions},\ref{sec:Classification}) that we hope will provide a common glossary for the different communities working in each particular application. Using this unified approach we have reviewed recent numerical, experimental and theoretical results in seven different key turbulent configurations, listed in  Sec. (\ref{sec:applications}), and in Sec. (\ref{sec:further}) we have also briefly mentioned applications to other emerging fields where non-linear cascades develop . To conclude our work,  we list a series of open  problems that have risen  based on the new knowledge that we have reviewed in this work.

\begin{itemize}
\item {\bf Critical transitions.} 
  For most of the applications discussed in Sec. (\ref{sec:applications}) that displayed a change in the cascade direction, the transition was shown to occur at a critical point, either as a 1st order, e.g. in Sec. (\ref{sec:discontinuous}), or in a continuous way as for the cases discussed in Secs. (\ref{sec:Thin}, \ref{sec:Rotation}, \ref{sec:stratification}, \ref{sec:MHD2D}). We note however that this conclusion is only  based on very few numerical evidences. Convergence to a critical transition has been demonstrated  only in  \cite{Seshasayanan2014edge, Seshasayanan2016critical, benavides2017critical, sahoo2017discontinuous} where the existence of   a critical point was supported by the comparison of  larger and larger box-sizes and in the limit of  larger and larger $\Red,\Rhd$. Even in those cases, the values of the Reynolds number examined were rather moderate due to computational limitations in 3D simulations  or they were restricted to examine 2D-configurations to be able to reach larger scale separation.   It would then  be very desirable to have a more firm demonstration of criticality from numerical simulations and/or experiments for all cases discussed in Sec. \ref{sec:applications}.  

Most importantly, there is a lack of theories that support the existence of criticality, explain quantitatively how these out-of-equilibrium (phase) transitions take place and predict their exponents and their universality class (if any). In all the examined cases that displayed  critical behaviour there was a competition between different  {\it channels},  some that transfer  energy to large scales and some to  small scales. One paradigmatic example being rotating turbulence where the competing  different channels are the 2D slow and 3D fast manifolds and the homo and heterochiral triads as discussed in Secs. (\ref{sec:Rotation},\ref{sec:depforcing},\ref{sec:Helicity}). 
{Furthermore, as was suggested in Sec. (\ref{sec:CritThin}) by analysing a model system, the dominance of a forward or backward transfer is restricted to some special regions in  configuration space  
and the interactions between these competing transfer processes display predator-prey dynamics,} as has recently been  claimed for the transition to turbulence in Couette and Poiseuille flow  \cite{barkley2015rise, goldenfeld2017turbulence, lemoult2016directed, sano2016universal}, and thus some of these cases could fall in the universality class of directed percolation \citep{obukhov1980problem} as suggested by \cite{Pomeau1986} for sub-critical instabilities in turbulence. Nonetheless all these possibilities remain at a speculative level and they  need to be further  explored.     

Finally, 
although the two different limiting behaviors (forward and inverse cascading) have been confirmed experimentally for some important systems, including thick layers and rotating turbulence  \cite{Shats2010turbulence, Byrne2011robust, potherat2014why, Moisy2014direct}  we still lack a systematic study of the energy flux behaviour close to the critical point. An experimental realization of such a study for a model system would provide invaluable help for confirming the existence and the understanding of critical phenomena  in out-of-equilibrium  turbulent flows.  


\item{\bf Phase space diagram of rotating and stratified turbulence.} Rotating and stratified turbulence is perhaps the most important applied turbulent system that calls for a deeper understanding. The reason being that most stellar and  planetary (including our earth's atmosphere and ocean) flows have a non-negligible rotation and stratification.  To this day, we  have a partial control  of some limiting situations only, like the zero rotation or zero stratification limit, and for some other intermediate (relatively isolated) cases. Nonetheless, the phase-space is very large with $Ro, Fr, H/\lin, Re$ being only a few of the parameters involved.   A systematic coverage of this large parameter space would be invaluable for any large-scale atmospheric and oceanic modelling.

\item{\bf Anisotropic cascades.} An issue that has come up often in this review is concerned with the difficulty to have a precise  description of a cascade in the presence of anisotropy both for quantifying the energy spectra and the flux in different directions. As of now,  this has been attained only for limiting cases where the cascade is constrained to mainly develop either on $k_\|$ or $k_\perp$ as in weak turbulence, in MHD (see Sec. \ref{sec:AnisMHD}) and rotating turbulence (Sec. \ref{sec:decomp}). 
The general anisotropic case remains ambiguous. Attempts to describe anisotropic cascades by means of a critical balance fall short in many respects in giving precise predictions. It is fair to say that in most anisotropic cases 
as discussed in Secs. (\ref{sec:Thin}-\ref{sec:MHD})
we do not have complete quantitative predictions of the 2D spectra $E(k_\perp,k_\|)$ or of the fluxes in the perpendicular and parallel directions except, perhaps, for some asymptotic limits when $k_\perp,k_\| \to \infty$, which are  difficult to realize in realistic empirical set-ups.

\item {\bf Non-quadratic invariants.} Most of our discussions in this review has been focused on quadratic invariants like energy, enstrophy, helicity etc. Nonetheless in many cases non-quadratic invariants also exist like all the high moments of vorticity in 2D turbulence (sec. \ref{sec:2DinvCsd}) and the second and higher moments of potential vorticity in stratified flows (sec.\ref{sec:stratification},\ref{sec:RotationStratification}). Furthermore for non-Boussinesq or compressible flows even the energy ceases to be a quadratic invariant as it depends on the product of the density and the square of the velocity. For cubic or quartic  invariants a Fourier description of the cascade is not adequate, as the invariant is not any more diagonal in Fourier space and it is expressed in terms of products of modes at different wavenumbers.  Most descriptions of such cases are thus limited to configuration space \cite{aluie2011joint, eyink2018cascades, eyink2018cascadesb, galtier2011exact}. How these higher-order invariants influence the transfer  processes is still at large unknown and extensive research in this direction is required. 

\item{\bf Condensates and thermal equilibrium states.} In many of the examined flow configurations we came across systems that are close to an equilibrium state, often including  the formation of condensates (see sec. \ref{sec:2DinvCsd}).  To what extent these systems can be described by equilibrium statistical physics such as a Gibbs ensemble is still only conjectural. Furthermore,  many of the systems developing a condensate saturate to  a flux-loop mechanism (see Secs. \ref{sec:finitethin},\ref{sec:finitehorizontalsize}). These cases, although they have almost zero flux, they are far from an equilibrium state. Further research in these directions would certainly be beneficial for our understanding of such configurations that arise in many flows in Nature. 

\item {\bf Quasi-2D Flows} Flows close to be exactly 2D are met in thin layers (see Sec. \ref{sec:Thin}), rotating turbulence (Sec. \ref{sec:Rotation}) or in the presence of  strong magnetic fields (Sec. \ref{sec:MHD}). {In these cases, on can prove analytically that below a critical $Re_c$ the flows will be  exactly 2D with all 3D perturbations decaying exponentially. In the presence of rotation, strong magnetic fields, or simply for very thin layers this critical $Re_c$ however can be very large so that the 2D field has fully developed turbulence. 
For $Re$ close to $Re_c$ the 3D field exhibits  a strong spatio-temporal intermittent behaviour such that 3D turbulent {\it puffs} appear locally in space and time. The 3D energy $\cE_{3D}$ in this case scales scales like a high power-law exponent $\beta$ with the deviation from the onset $\cE_{3D} \propto (Re-Re_c)^\beta$.  This exponent $\beta$ still lacks any quantitative theoretical explanation.}

\item {\bf Intermittency.} We could not close this section without mentioning intermittency whose origin still remains poorly understood.  In this review we have mainly focused our attention to low-order moments, such as the spectra and the fluxes, because these are believed to be  enough to control the basic properties of the flow behaviour.  Still, one cannot exclude that close to a critical point rare but intense intermittent fluctuations might influence  the way the transition happens. The latter is a fully open  problem that would need to be studied in highly simplified models first. We  only briefly discussed  in Sec. (\ref{sec:intermittency}) the problem of controlling the statistical properties of high-order moments, a fact that is  also important concerning the quantification of  the out-of-equilibrium contents of intense rare fluctuations. Although many of the turbulent intermittent properties can be described by the multi-fractal phenomenology,  a derivation from first principles of the origins and properties of anomalous scaling and intermittency is far from being achieved. Moreover, very few numerical and experimental studies exist, which  quantify intermittency in anisotropic flow configurations, as the ones discussed in the application section of this review (\ref{sec:applications}) and in the methodology section concerning anisotropic scaling properties in (\ref{sec:scalinganisotropic}).   
\end{itemize}

We hope this review will trigger new studies in this fascinating world at the border between  non-linear fluid mechanics and out-of-equilibrium statistical mechanics, where turbulent flow configurations  key for geophysical, astrophysical and engineering applications display properties  common to many other systems in Nature, coexisting in different statistical or dynamical phases and developing critical transitions among different asymptotic behaviour at changing the control parameters. Understanding such a criticality is the first step toward controlling and optimising.

\section{Acknowledgements}

We acknowledge long-lasting  interactions and collaborations with many friends and colleagues, which have contributed to shape our little understanding and control of this difficult subject. In particular, we want to thank  R. Benzi, M. E. Brachet, G. Boffetta, M. Cencini, A. Celani, G. Eyink, S. Fauve, U. Frisch, B. Gallet,  C. Gissinger, R. Marino, P. Mininni, I. Procaccia, A. S. Lanotte, S. Musacchio, F. Petrelis, A. Pouquet,  M. Sbragaglia and   F. Toschi for the continuous inspiration during recent and past collaborations. We also want to thank our younger collaborators and/or  former students, which have also deeply contributed to our work:  S. Benavides, F. Bonaccorso, M. Buzzicotti,  A. Cameron, S. Colabrese,  P. Clark di Leoni, V. Dallas, M. De Pietro, K. Gustafsson,   K. Iyer,  A. van Kan,  M. Linkmann, S. Malapaka, I. M. Mazzitelli,  P. Perlekar,  G. Sahoo,  R. Scatamacchia and  K. Seshasayanan. We thank our friends and colleagues P. Augier, M. Buzzicotti, A. Campagne, B. Gallet, A. Pouquet, J. Herault, T. Gotoh, W. Irvine, N. Ouellette, D. Rosenberg and  A. Sozza for providing some of the beautiful figures we included in this work. 
Finally, we are deeply indebted to M. Buzzicotti, M. Linkmann and A. Pouquet for a critical reading of the manuscript and for the many corrections and suggestions which have strongly improved our presentation.  L.B. acknowledges funding from the European Research Council under the European Unions Seventh Framework Programme, ERC Grant Agreement No 339032. A.A  acknowledges funding from the French National Research Agency (ANR DYSTURB project No. ANR-17-CE30-0004). 


\newpage
\bibliographystyle{elsarticle-num} 







\end{document}